\documentclass[preprint2]{aastex}
\usepackage{pdflscape}
\usepackage{epstopdf}
\usepackage{graphicx}

\shorttitle{Planet Migration Halting}
\shortauthors{PLAVCHAN \& BILINSKI}

\begin{document}

\title{Stars Don't Eat Their Young Migrating Planets -- Empirical Constraints On Planet Migration Halting Mechanisms}

\author{Peter Plavchan\altaffilmark{1} and Christopher Bilinski\altaffilmark{2}}

\altaffiltext{1}{NASA Exoplanet Science Institute, California Institute of Technology, M/C 100-22, 770 South Wilson Avenue, Pasadena, CA 91125}
\altaffiltext{2}{University of Arizona}

\begin{abstract}
The discovery of ``hot Jupiters'' very close to their parent stars confirmed that Jovian planets migrate inward via several potential mechanisms.  We present empirical constraints on planet migration halting mechanisms.  We compute model density functions of close-in exoplanets in the orbital semi-major axis -- stellar mass plane to represent planet migration that is halted via several mechanisms, including the interior 1:2 resonance with the magnetospheric disk truncation radius, the interior 1:2 resonance with the dust sublimation radius, and several scenarios for tidal halting.  The models differ in the predicted power law dependence of the exoplanet orbital semi-major axis as a function stellar mass, and thus we also include a power law model with the exponent as a free parameter.  We use a Bayesian analysis to assess the model success in reproducing empirical distributions of confirmed exoplanets and Kepler candidates that orbit interior to 0.1 AU.   Our results confirm a correlation of the halting distance with stellar mass.  Tidal halting provides the best fit to the empirical distribution of confirmed Jovian exoplanets at a statistically robust level, consistent with the Kozai mechanism and the spin-orbit misalignment of a substantial fraction of hot Jupiters.  We can rule out migration halting at the interior 1:2 resonances with the magnetospheric disk truncation radius and the interior 1:2 resonance with the dust disk sublimation radius, a uniform random distribution, and a distribution with no dependence on stellar mass.   Note, our results do not rule out Type II migration, but rather eliminate the role of a circumstellar disk in stopping exoplanet migration.  For Kepler candidates, which have a more restricted range in stellar mass compared to confirmed planets, we are unable to discern between the tidal dissipation and magnetospheric disk truncation braking mechanisms at a statistically significant level.  The power law model favors exponents in the range of 0.38--0.9.  This is larger than that predicted for tidal halting (0.23--0.33), which suggests that additional physics may be missing in the tidal halting theory.
\end{abstract}


\section{Introduction}

Many ``hot Jupiter'' planets have been discovered to orbit very close to their central stars \citep[e.g.,][]{planetref1,planetref2,keplerref}.   It is well-established that these planets must form further out from their host stars, likely beyond the snow line, and either migrate embedded in a primordial disk or via dynamical interactions \citep[e.g., ][]{lin96,lubow10,kozai,spinorbitref}. Observations of the Rossiter-McLaughlin effect to identify stellar spin -- planet orbit misalignment show that a significant fraction of ``hot Jupiters'' are aligned, and a significant fraction are also misaligned \citep[]{morton,spinorbitref,spinorbitref2}.  The mis-aligned planets are likely directed inwards via planet-planet scattering, the Kozai mechanism, secular chaos, or analogous mechanisms \citep[e.g.,][]{kozai,wukozai,scatterref2,scatterref3,secularchaos}.  Recent work by \citet[]{dawson1,dawson2} suggests that less than 15\% of hot Jupiters undergo migration via the Kozai mechanism, instead favoring planet-planet scattering.

For aligned close-in planets, planet migration embedded in a primordial disk is suspected to explain the observed planet location.  Most disk migration models involve similar physical processes, but contain differences in the underlying assumptions about the structure and properties of the primordial disk and the planet \citep[e.g., viscosity, density, scale height, temperature, dissipation time-scale, toroidal magnetic fields embedded in the disk, etc., ][]{lin96,lubow10,migrateref3,migrateref4,migrateref5,toroidalref}.  Current hypotheses often combine a few models together in an attempt to explain observed exoplanet mass and semi-major axis distributions.  The justification cited is that conditions within the disk and the planet mass and density change over time, resulting in different models being applicable at different times in the planet migration process.  

Type I migration assumes that the density structure of the disk is not affected by planets.  Instead, turbulence determines the density structure of the disk.  For this reason, this form of migration is most applicable to small mass planets.  In the case of type II migration, a gap is formed between the disk and a high mass planet.  This gap is the result of the tidal torques from the planet becoming stronger than the viscous torques of the disk.  Initial theoretical models of Type I and Type II migration suggested a rapid planet migration rate that could result in planet destruction by dispersing the accreted material \citep[]{ward,ward2}.  Various mechanisms are thought to decrease the planet migration rate, including eccentric planet orbits and disk turbulence \citep[]{lubow10,migrateref4}.  In all cases, planet-disk migration must take place while the primordial disk of gas and dust is still present during the classical T Tauri phase, or first $\sim$5 Myr, of the host star's life \citep[]{disklifetimeref,disklifetimeref2}.  

Type II migration offers a mechanism to transport gas giants that must form beyond the snow line inward to their host star.   However, it is not well-constrained how planet migration is halted once started, lest the planet be tidally disrupted by the host star.  Possible braking mechanisms include -- tidal circularization \citep[]{tidalref4,tidalref5,tidalref6, tidalref,tidalref2,tidalref3}, trapping the exoplanet in the 1:2 interior orbital resonance with the magnetospheric truncation radius \citep[]{eisner05}, or trapping the planet in the 1:2 interior orbital resonance with the dust sublimation radius \citep[e.g., ][]{kuchner02}.  For an approximately solar mass star, the gas disk truncation radius is comparable to the dust sublimation radius for a typical T-Tauri star magnetic field strength of $\sim$2 kGauss \citep[]{eisner05}, but that is not the case for lower and higher mass stars.  For lower-(higher-)mass T Tauri stars, the dust sublimation radius can be interior (exterior) to the estimated magnetospheric truncation radius (${\S}$3).  

In this work we investigate exoplanet distributions as a function of semi-major axis and host stellar mass as a test for migration halting mechanisms. The increasing number of exoplanet discoveries provides sufficiently large ensembles of close-in (e.g. $<$0.1 AU) exoplanets over a range of host stellar masses to discern which mechanism may be responsible for halting exoplanet migration.  In ${\S}$2, we outline our empirical samples.  In ${\S}$3, we present each migration halting mechanism model and its corresponding prediction for the exoplanet distribution density function.  In ${\S}$4, we present our methodology to evaluate the success of each migration halting model at reproducing empirical distributions, and in ${\S}$5 we present the results of these statistical tests.  In ${\S}$6, we present our conclusions.

\section{Empirical Samples}

We make use of two empirical samples -- confirmed exoplanets as of February 2012 with $M_{pl}<30 M_J$ \citep{exoplanetarchive,wright}, and the third tabulation of Kepler planetary candidates, also known as Kepler Objects of Interest \citep[KOI, ][]{batalha, keplerref,howard}.   We further sub-divide the confirmed planets by mass M$_{pl}$ (or m$sin$i) into mass bins of M$_{pl}$$<$10 M$_\oplus$, 10 M$_\oplus$$<$M$_{pl}$$<$ 0.2 M$_J$, and M$_{pl}$ $>$ 0.2 M$_J$.  We also further sub-divide the Kepler candidates by estimated planet radius into three radius bins with $R_{pl}< 2 R_\oplus$, $2 R_\oplus < R_{pl}< 6 R_\oplus$, and $R_{pl}>6 R_\oplus$.   These sub-divisions are chosen to approximate the terrestrial, super-Earth / Neptune, and Jovian planet mass/radius boundaries.  Estimates of stellar mass are culled from the literature from the NASA Exoplanet Archive and exoplanets.org for the confirmed exoplanets.  The revised stellar masses from \citet[]{batalha} are utilized for the Kepler candidates, rather than the Kepler Input Catalog \citep[]{kic}.  These six empirical samples are shown in Figure 1.

We exclude all exoplanets and candidates with orbital semi-major axes $>$0.1 AU to focus on close-in planets most likely to have undergone some form of migration in their orbital evolution, rather than forming in-situ at their present locations.  We also constrain our samples to stellar masses between 0.1 and 1.5 M$_\odot$.  The upper limit of 1.5 M$_\odot$ is chosen to exclude planets around higher mass stars that can be evolved sub-giants \citep[e.g.,][]{johnson}.  The final exoplanet counts in each of our samples are: 203 confirmed exoplanets and 1199 Kepler candidates, including 115 KOIs with $R_{pl}>6 R_\oplus$, 434 with $2 R_\oplus < R_{pl}< 6 R_\oplus$, and 650 with $R_{pl}< 2 R_\oplus$.

Inherent in these samples are many survey biases and incompleteness.  We discuss these briefly.  The frequency of planets as a function of stellar mass is highly dependent on the survey sample selection criteria of ongoing searches.  For example, the Kepler Input Catalog was selected to focus on FGK-type stars, with a paucity of M dwarfs \citep[]{batalha2,batalha3,keplerref}, and visible radial velocity searches initially focused on similar solar-mass stars but now include smaller samples of lower and higher mass stars \citep[][]{planetref1}.  Additionally, there are differences in the planet frequency as a function of planet mass and stellar mass that are not yet well constrained \citep[e.g.,][]{howard}.  Thus, in our analysis that follows we fix our models to match the empirical frequency of exoplanets as a function of stellar mass (within 0.1 AU for a given data set).  Further, we do not draw any conclusions about the planet frequency as a function of stellar mass.

The transit and radial velocity detection methods, responsible for the discovery of most close-in confirmed exoplanets, are highly biased towards the detection of short-period orbits.   Additionally, at a fixed semi-major axis, planets around a lower-mass star will have a longer orbital period.  This introduces bias towards a higher detection efficiency towards higher stellar masses at a fixed semi-major axis.  However, most ground-based surveys are reasonably complete to within our semi-major axis cut of 0.1 AU, corresponding to an orbital period of $\sim$16 days for a solar-mass star.  After more than two years of operation, Kepler is also complete out to 0.1 AU, down to some nominal terrestrial planet size \citep[]{howard,batalha}.   Thus, we do not expect detection completeness as a function of semi-major axis to significantly impact our analysis.  

For the confirmed exoplanet sample, we do not apply any minimum constraint on planet mass (or planet mass limit in the case of radial velocity detected exoplanets).  While terrestrial planets are more likely to have formed in-situ rather than migrated inward to their present locations, most radial velocity confirmed exoplanets are Neptune-massed or larger, with a few Super-Earths.  The presence of a few low-mass planets in our sample does not significantly impact our analysis.  

Finally, we also do not correct for the false-positive rate in the Kepler exoplanet candidate list, which is thought to be between 10--35\%, but may be particularly high for orbital periods of $<$3 days, where background eclipsing binaries are more likely to mimic the signature of a hot Jupiter \citep[]{howard,keplerref, morton,plavchan2,santerne}.   Future releases of the Kepler candidates including follow-up identification of false positives and improved completeness at short orbital periods will improve our analysis presented herein, but accounting for the false positives in the current KOI list is beyond the scope of this work.

\section{Migration Halting Models}

For each migration halting mechanism model we present in this section, our goal is to generate a reasonably simple prediction for the density of exoplanets as a function of stellar mass and semi-major axis within 0.1 AU.  We will present the model for each mechanism in turn, but first we outline commonalities in our methodology across models.   

With the exception of the uniform random and exponential decay distribution models, our prescription involves first identifying a single 1:1 curve in the semi-major axis -- stellar mass plane.  The specification of these curves are outlined in subsections ${\S}$3.1-3.4 for each particular model.   We next convolve that curve with a Gaussian kernel in the log of the orbital semi-major axis to arrive at a predicted probability density function ($PDF$) to compare with an empirical distribution.  The Gaussian width is a free parameter optimized for each model and empirical distribution combination we test.  The use of a Gaussian kernel is an ad hoc step in our model generation, but its use is motivated by its simplicity and the observed scatter of exoplanet orbital semi-major axes.

Each model is next divided by the empirical frequency of exoplanets (within 0.1 AU) as a function of stellar-mass to correct for sample survey biases which are not relevant to our investigation herein (Figure 1).  Finally, for computational simplicity, the $PDF$ for each model is evaluated numerically on a 500 by 500 grid evenly spaced in stellar mass from 0.1 to 1.5 M$_\odot$ and in the log of the orbital semi-major axis in AU from log(a)=-2 to -1.  The model density functions are summed over all pixels and the sum is normalized to 1.

\subsection{Tidal Dissipation}

Planet-planet scattering, secular chaos, and in particular the Kozai cycle are theoretical mechanisms proposed to migrate a Jovian exoplanet inwards towards its host star \citep[]{kozai,scatterref2,scatterref3,secularchaos,wukozai}.  These mechanisms invoke tidal forces acting on the exoplanet to lower the semi-major axis and eccentricity until the orbit is circularized. Some of these mechanisms are likely required to explain the known fraction of spin-orbit misaligned close-in exoplanets \citep[]{morton,spinorbitref,albrecht,albrecht2,winn}.  The association of the observed spin-orbit misalignment with the Kozai and similar migration mechanisms relies on the assumption that the stellar spin axis is aligned with the primordial disk rotation axis. This star-disk alignment is intuitively expected from the process of star formation \citep[][and references therein]{prato,hale,watson}, although mechanisms such as an external perturber are proposed to alter this alignment \citep[]{kaib,thies}.  In order for tidal circularization to occur, the exoplanet must get within several stellar radii of the parent star, i.e. within a few tenths of an AU  \citep[]{tidalref}.   Tidal circularization predicts  closer-in orbits around lower mass stars with deeper convective atmospheres \citep[]{kozai,wukozai,tidalref}.    

We include two theoretical predictions for the stopping radius as a function of stellar mass from tidal circularization.  First, \citet[]{tidalref4} predicts a minimum allowable envelop for the exoplanet semi-major axis scaling from the Roche radius, with a scaling with stellar mass of $a \propto M_*^{1/3}$.  \citet[]{tidalref6} suggests that the particular proportionality constant should be increased over that in \citet[]{tidalref4}.    Second, \citet[][Eqn 6]{tidalref5} predicts the final semi-major axis itself, scaling with a slight different dependence on stellar mass of  $a \propto M_*^{3/13}$ and weakly dependent on the unknown planet resonant Q factor.   In both of our models used in our analysis, we allow for the proportionality constant $C$ as a free parameter, only fixing the power law exponent ($\alpha = 1/3$ or 3/13).  The proportionality constant and Gaussian kernel used to generate the exoplanet density function in these models can be interpreted to represent a degenerate range of exoplanet and/or stellar densities/Q's  and initial conditions about these assumed mean values.  

\subsection{The Interior 1:2 Orbital Resonance with the Accretion Disk Truncation at the Magnetospheric Radius}

The magnetosphere of a T Tauri star has long been thought to truncate the inner accreting primordial gas disk \citep[e.g.,][]{chiang,meyer}, and recent interferometric observations of young stars confirm these inner holes exist \citep[]{eisner05}.  A Jovian exoplanet undergoing Type II migration in a primordial disk could halt after it enters into this inner clearing, as has been proposed \citep[]{eisner05,kuchner02,lin96}.   In this scenario, a migrating Jovian exoplanet interacts with a protoplanetary disk at Lindblad resonances, transferring angular momentum via torques to the disk as the planet migrates inward.  When the 2:1 Lindblad resonance site enters the evacuated portion of the disk, the planet can no longer transfer angular momentum to the disk and the migration is hypothesized to halt.  The exoplanet continues to orbit the host star at a period equal to one-half of the Keplerian orbital period of the magnetospheric gas disk truncation radius.   A related but distinct halting mechanism is proposed for smaller planets undergoing Type I migration in \citet[]{tsang}.

The magnetospheric radius can be approximated by \citep[]{eisner05,konigl91}:
\begin{eqnarray}
R_{mag} = 2.27R_{1} \left[ \frac{  (B_{0}/1\mbox{kG})^4 (R_{*}/R_{\odot})^5 } { (M_{*}/M_{\odot})(\dot M / (10^{-7} M_{\odot} \mbox{yr}^{-1}))^2   } \right]^{1/7}
\end{eqnarray}

\noindent where $B_{0}$ is the stellar magnetic field strength.  A field strength of 2 kG is typical for T Tauri stars \citep[]{johns03}, and is a free parameter in our model density function.  $\dot M$ is the stellar accretion rate in units of $M_{\odot}$yr$^{-1}$, which can vary by several orders of magnitude for young stars.  \citet[]{muzerolle03} derives an approximate power law relationship between stellar mass and accretion rate of $\dot M\propto M^2$.  We estimate a proportionality constant of 10$^{-8.5}$$M_{\odot}$yr$^{-1}$ that yields appropriate accretion rates for solar type stars as inferred from \citet[ Figure 8]{muzerolle03}.  For a 2 kG field, the magnetospheric truncation radius is thus approximately given by:

\begin{equation}
R_{mag} = 9.05R_{\odot} \left(\frac{R_{*}}{R_{\odot}}\right) ^{12/7}\left(\frac{M_{\odot}}{M_{*}}\right) ^{5/7}\approx 9R_{*}
\end{equation}

The 1:2 interior orbital resonance with this inferred magnetospheric radius is thus located at $\sim$6 R$_{*}$ for a 2 kG stellar magnetic field, and our full migration halting semi-major axis location, $a$, is given by:

\begin{equation}
a=9.05 R_{\odot} \left(\frac{B_0}{2\:\mbox{kG}}\right)^{4/7}\left(\frac{R_{*}}{R_{\odot}}\right)^{\frac{12}{7}} * \left(\frac{M_{\odot}}{M_{*}}\right)^{\frac{5}{7}} * \frac{1}{2}^{2/3} \propto M^{\frac{1}{7}}
\end{equation}

Finally, to arrive at our model analytic curve, we assume $\mbox{log}\:g = 4$ as is typical for T Tauri stars that are still contracting onto the main sequence \citep[]{ttauriref}.  With these assumptions, we can express this migration halting radius as a function of only $B_0$ and $M_*$.  The Gaussian kernel that we use to generate the model exoplanet density function from the analytic curve in Equation 3 can be interpreted to represent the degenerate range in magnetic field strengths, stellar accretion rates, and/or stellar surface gravity about the assumed values.  For example, the range of observed T Tauri accretion rates \citep[$\sim$100, ][]{muzerolle03} would correspond to a range for $a$ in Equation 1 of a factor of $\sim$2.  Additionally, the pre-main sequence contraction times imply that the median magnetic field strength should vary as a function of stellar mass due to the range of different evolutionary states at a fixed proto-stellar age.  Finally,  exoplanet Type II migration may also preferentially take place at different stellar ages as a function of stellar mass \citep[]{lubow10,lin96}.   In our analysis that follows, however, we assume that we are only varying the magnetic field strength, and we keep the median magnetic field strength constant as a function of stellar mass.  At this time, we do not overcomplicate our model to account for these various degenerate factors (${\S}$5).

\subsection{The Interior 1:2 Orbital Resonance with the Dust Disk Sublimation Radius}

The truncation of the dust disk at the dust sublimation radius is also proposed as a mechanism to halt the inward migration of Jovian planets for solar type stars \citep[]{kuchner02,lin96}.  This scenario for migration halting is identical to that for the magnetospheric truncation model in ${\S}$3.1, except the planet is hypothesized to halt instead at the interior 1:2 orbital resonances with the dust disk sublimation radius.   While the primordial (gas and dust) disk for a typical T Tauri star is thought to dissipate by a stellar age of $\sim$5 Myr, a debris dust disk from the collision of planetesimals can persist for much longer.  For a solar-mass star, the dust sublimation radius is approximately equal to the expected magnetospheric gas disk truncation radius, and thus solar-type stars alone are a poor discriminator of migration halting mechanisms.  However, the dust sublimation radius has a significantly different dependence on stellar mass when compared to the magnetospheric gas disk truncation radius. 

The dust sublimation radius is given approximately by the expression \citep[]{jura98}:

\begin{equation}
R_{subl}=\frac{1}{2}R_{*}\left(\frac{T_{*}}{1500}\right)^2
\end{equation}

\noindent where $T_{*}$ is the effective temperature of the host star in Kelvin, $R_{*}$ is the radius of the host star, and we have assumed a dust sublimation temperature of 1500 K.    Equation 4 assumes that the dust can be approximated by a blackbody in local thermal equilibrium, and that the dust is optically thin to the incident stellar radiation.  Assuming to the contrary for both factors would decrease the dust sublimation radius.  Viscous heating in a primordial disk can  increase the dust sublimation radius, and more sophisticated treatments yield a stronger dependence of the sublimation radius on the stellar temperature \citep[]{robitaille06,robitaille07,dalessio}.  In our analysis we retain the approximation in Equation 4 for simplicity, but also because of the lack of success for this particular model (${\S}$5).

We adopt a temperature--radius relation using a \citet[]{siess} 10 Myr isochrone to express the migration halting radius, $a$, as a function of only the stellar radius, e.g.:

\begin{equation}
a=\frac{1}{2}^{5/3} \left(\frac{T_{*\mbox{Siess}}(R_*)}{1500}\right)^2 R_{*}
\end{equation}

\noindent where the extra factor of $1/2^{\frac{2}{3}}$ comes from Kepler's laws and the 1:2 interior resonance location with respect to the dust sublimation radius in Equation 5.  Our results are not significantly altered if we instead use a 10 Myr isochrone from \citet[]{baraffe}.

We adopt the age isochrone of 10 Myr with the assumption that the exoplanet migration under this scenario must take place early in a star's evolution.  An older stellar age will decrease the modeled migration halting radius in Equation 5, since the star will continue to contract onto the main sequence.  Finally, to arrive at our model analytic curve, we again assume log $g = 4$ in order to express this migration halting radius as a function of only $M_*$.  The Gaussian kernel that we use to generate the model exoplanet density function from Equation 5 can be interpreted to represent a degenerate range in stellar surface gravity and/or dust sublimation temperatures about the assumed values.

Since the stellar effective temperature varies little over the main sequence lifetime of the host star, we can also estimate the distance $a$ in Equation 5 for older main sequence stars.   For main sequence M dwarfs, we note that the semi-major axis in Equation 5 for this orbital resonance is less than $\sim$2.2 R$_{*}$.  This orbital separation falls within the estimated Roche radius of $\sim$2.4 R$_{*}$ (${\S}$3.1), and a planet at this distance would likely be tidally disrupted.    We conclude that no migrating Jovian planets would survive around M dwarfs if the orbital resonance with the dust sublimation radius is responsible for braking Jovian planet migration, and if the migration takes place after the M dwarfs have reached the main sequence.  The expected exoplanet detection frequency for this scenario is zero for M dwarfs.   While this is an interesting potential mechanism to explain the observed lack of M dwarf hot Jupiters relative to solar-mass stars \citep[]{endl,plavchan}, we do not find this scenario likely given our results in ${\S}$5.

\subsection{Power Law and Other Models}

The models described in the previous sections predict a distribution of exoplanets in the stellar mass -- semi-major axis plane that fits a particular choice for the exponent $\alpha$ in a power law model.  To ensure we are considering a more complete range of models, we also carry out a separate analysis with a power law model where the exponent $\alpha$ is a free parameter.  Again, we also include a Gaussian kernel width $\sigma$ and a proportionality constant $C$ as free parameters in this model.

Next, we include three additional models as a sanity check on our analysis.  The first assumes that the exoplanets within 0.1 AU are uniform randomly distributed as a function of stellar mass and semi-major axis.  The second assumes a uniform random dependence on stellar mass, and an exponential decaying dependence on semi-major axis -- e.g., favoring smaller semi-major axes in a fashion that is uniform random in log(a).  The third model assumes that exoplanets halt at a constant semi-major axis $A$ that is independent of the stellar mass of the star -- e.g. a power law model with an exponent of zero.   A Gaussian kernel width is included as a free parameter to generate the final $PDF$ as with previous models.

%Fourth and finally, we consider the hypothesis that the exoplanets we observe are piling up at the Roche radius of host stars given by:

%\begin{equation}
%a=R_{*} (2 \rho_{*}/\rho_{pl})^{\frac{1}{3}}
%\end{equation}

%\noindent where $\rho_{*}$ and $\rho_{pl}$ are the mean stellar and exoplanet densities respectively.  To arrive at our model analytic curve, we assume that log $g = 4.5$ to estimate the stellar density.  We also assume an exoplanet density equal to 0.8 g/cm$^3$, which is less than the density for Jupiter of 1.3 g/cm$^3$.  We adopt this lower density, which lies between Jupiter and the density of HD 209458b of 0.37 g/cm$^3$ \citep[]{exoplanetarchive}.  A higher planet density, e.g. for terrestrial planets, would decrease the migration halting radius in Equation 6.  This estimated halting radius is a few times smaller than the magnetospheric disk truncation halting radius, even though it has the same dependence on stellar mass, allowing us to distinguish between these two mechanisms.  In the case of the third and fourth models, a Gaussian kernel width is included as a free parameter to generate the final $PDF$ as with previous models.

\section{Methodology for Evaluating Model Success}

We use a variety of different free parameters for our models as described in ${\S}$3.  To evaluate which migration braking mechanism is best at reproducing the different empirical distributions, and to determine optimal parameters, we use two approaches.  We discuss each in turn.

\subsection{Bayesian Evaluation}

We use a Bayesian analysis to estimate the posterior probability $P\left(\left.H_i\right|D\right)$ of obtaining a given empirical data set $D=\{D_j\}$ for a particular migration braking model hypothesis $H_i$.  Our approach is similar to that in \citet[]{marshall}.  Explicitly, we re-state Bayes' Theorem:

\begin{eqnarray}
P\left(\left.H_i\right|D\right)=\frac{P\left(H_i\right)P\left(D\left|H_i\right.\right)}{P(D)}
\end{eqnarray}

\noindent To optimize the parameter selection for a given empirical data set $D$ and model $H_i$, we maximize the posterior probability $P\left(\left.H_i\right|D\right)$ for the same model over a range of its free parameters.

Following the Bayesian technique, we obtain the likelihoods $P\left(D\left|H_i\right.\right)$ by multiplying together the individual probability density functions $PDF_{H_i}(D_j)$ to calculate the probability of obtaining a single $D_j$:  

\begin{eqnarray}
 P\left(D\left|H_i\right.\right) = \prod_j  PDF_{H_i}(D_j)
\end{eqnarray}

\noindent where $D_j$ is the semi-major axis and stellar host mass value pair for an individual exoplanet.  We use the probability densities as described in ${\S}$3 to estimate each individual $PDF_{H_i}(D_j)$, linearly interpolating from the 500x500 model grid.

For the prior terms, $P\left(H_i\right)$, we make some assumptions.  First, we use an uninformed Jeffrey's prior of $\frac{1}{\sigma}$ (e.g. $P(H_i)\propto \frac{1}{\sigma_i}$) for the Gaussian width $\sigma$ parameter present in a number of the models.  To normalize a given model prior such that $\sum _i P\left(H_i\right)=1$ in the discrete limit, we have that $P(H_i) = \frac{C}{\sigma_i}$ where $C = 1/\sum_i\frac{1}{\sigma_i}$, and $\{\sigma_i\}$ are the discrete set of widths evaluated in our analysis from 0.02 to 1.02 in steps of 0.02.  Next, we also assume uninformed Jeffrey's priors for the log of $B_0$, $C$, and $A$ parameters in the magnetospheric disk truncation, power law, and stellar mass independent models respectively, discretely evaluated with 51 steps between a = 0.01 and 0.11.   Finally, we assume a uniform random prior for $\alpha$ in the power law model, discretely evaluated between 0, and 1 with steps of 0.02.  The bounds for our parameter space exploration are not preferred as optimal values for our models, with the exception of the value of $\sigma$ for the dust sublimation halting model which is particularly inadequate in describing the empirical data sets.

%  (e.g.  $\frac{\partial(P\left(H_i\right))}{\partial \mbox{log}(\{B_0,C,A\})} = \frac{1}{\{B_0,C,A\}}\frac{\partial(P\left(H_i\right))}{\partial \{B_0,C,A\} } $ )

The last term in Bayes' theorem needed to compute the posterior probabilities (evidences) is the marginal probability, $P(D)$.  However, we can rewrite this term as
\begin{eqnarray}
P(D)=\sum _iP\left(D\left|H_i\right.\right)P\left(H_i\right),
\end{eqnarray}
where we sum over all hypotheses, $i$.   We do not have an exhaustive (complete) list of models.  Additionally, we include an arbitrary normalization $N$ in our analysis to avoid the double data type machine precision limit. Thus, we do not obtain absolute posterior probabilities, and we can end up with relative posterior probabilities greater than 1.  Nevertheless, we know that the value of $N/P(D)$ is the same in all of our computations, so we can ignore it and compute accurate relative posterior probabilities when we are comparing parameters for a given model or between models.  In other words, the quantity of interest is:

\begin{eqnarray}
\frac{P\left(\left.H_1\right|D\right)}{P\left(\left.H_2\right|D\right)} = \frac{N\:P\left(H_1\right)P\left(D\left|H_1\right.\right)}{P(D)} \times \frac{P(D)}{N\:P\left(H_2\right)P\left(D\left|H_2\right.\right)} = \frac{P\left(H_1\right)P\left(D\left|H_1\right.\right)}{P\left(H_2\right)P\left(D\left|H_2\right.\right)}
\end{eqnarray}

\noindent This approach also enables a relative comparison of models that factors in the degrees of freedom and the ranges of explored parameter space for a given model.

\subsection{Chi-Squared Tests}

In order to test the predicted exoplanet density function against the empirical planet distributions using the chi-squared test, we first generate a density function for the empirical planet distributions.  This was not necessary for the Bayesian analysis, which directly tested the empirical distribution of exoplanets against the model density function.

The density function is calculated for each data set using a kernel density estimation method analogous to that in \citep[]{wasserman}.  Each exoplanet (or candidate) in the stellar mass -- log orbital semi-major axis plane is convolved with a Gaussian kernel.  The width of the Gaussian kernel is set to 0.165 in stellar mass in solar units, and 0.165 in the log of the orbital semi-major axis in AU.  These particular widths are chosen from the median separation between exoplanets in the confirmed exoplanet empirical distribution, which were identified to be 0.167 log AU  in the log of the semi-major axis and 0.164 M$_\odot$ in the stellar mass.   The same kernel width is used for all empirical data sets.  The kernels are summed to produce the empirical density functions shown in Figure 1.  As was done for the migration halting models, each empirical distribution is evaluated numerically on a 500 by 500 grid evenly spaced in stellar mass from 0.1 to 1.5 M$_\odot$ and in the log of the orbital semi-major axis in AU from log(a)=-2 to -1.  This enables a direct subtraction of the model probability density function from the empirical density function to calculate the reduced $\chi^2$ statistic.  

Since there are no uncertainties in our empirical and theoretical distributions, to compute the reduced $\chi^2$ statistic we divide the square of the difference between the observations and model by the model value instead of dividing by the squared uncertainty.  The model value can approximate the square of the uncertainty when the model is normalized such that the assumption of Gaussian statistics is appropriate.

\section{Results}

Our results are presented in Tables 1 through 3.  Table 1 presents the optimal parameters for a given model and empirical data set from our Bayesian analysis.   Table 2 summarizes the relative posterior probabilities for a given model.  Table 3 summarizes the corresponding reduced $\chi^2$ values.  Figures 2-10 show our best fit model density functions and residuals as a function of the model.  

Excluding the power law model for the moment, migration halting due to tidal circularization provides the best evidence and fit to the data for every data set using both the Bayesian and $\chi^2$ analysis, with one exception.    The evidence ratio is largest for the confirmed Jovian exoplanets, which span a larger dynamic range in stellar mass relative to the KOIs.  The lone exception -- Neptune radii KOIs with the $\chi^2$ analysis -- slightly favors the magnetospheric disk truncation halting mechanism.  However, the difference is not statistically significant.     

Comparing the 1/3$^{rd}$ exponent tidal halting model to the magnetospheric hole halting model, the evidence ratios are 3.9, 8.3, 61, 28, 4.9 and 2.5 for confirmed exoplanets with M$_{pl}$$<$10 M$_\oplus$, 10 M$_\oplus$$<$M$_{pl}$$<$ 0.2 M$_J$, and M$_{pl}$ $>$ 0.2 M$_J$, and KOIs with $R_{pl}< 2 R_\oplus$, $2 R_\oplus < R_{pl}< 6 R_\oplus$, and $R_{pl}>6 R_\oplus$, respectively.  Thus, our Bayesian analysis disfavors the models of halting interior to the magnetospheric dust truncation radius, halting interior to the dust sublimation radius, halting at a constant radius independent of stellar mass, and a uniform random distribution in both the semi-major axis and the log of the semi-major axis.

The power law model has the largest posterior probabilities for the confirmed exoplanet sub-samples, and is within a factor of two of the most favored models (tidal circularization) for the KOI sub-samples.  The power law model is clearly favored over the stellar-mass independent model, with posterior probability ratios ranging from 1.5--10$^4$.  This result is weakest for the Jovian KOIs, but strongest for the confirmed Jovian exoplanets, and we again attribute that to the lack of dynamic range in stellar mass for the KOIs but also may be partially attributable to a high false-positive rate for Jovian KOIs.

For both the confirmed exoplanets and KOIs, smaller exoplanets favor \textit{both} steeper power laws and larger dispersion (Gaussian Kernel width) than Jovian planets, and in all cases the exponent is larger than either the magnetospheric disk truncation and exoplanet tidal excitation model exponents.  This result is intriguing, and we speculate that there may be additional physics in the tidal theories to halt planet migration as a function of exoplanet mass/density.  

\citet{howard} report the decreasing planet frequency towards smaller orbital separations within $\sim$0.04 AU.  Our analysis confirms the observed dearth of Kepler exoplanet candidates at these small orbital radii by rejecting the uniform random and exponential decay halting models. % Thus our analysis also disfavors migration halting at the Roche radius of the host star. 

The subtle differences between the best models is not clear ``by eye'' in Figures 2--5.  However, the $\chi^2$ test demonstrates a slight preference for the tidal circularization halting models for all data sets save the Neptune-like KOIs.  Our Bayesian analysis does not rely on a Gaussian kernel estimate of the empirical density function used in the $\chi^2$ analysis (${\S}$4.2), and this may partially account for the weaker statistical significance between the two approaches.

The best fit magnetospheric disk truncation model magnetic field strengths range from 1--4.4 kGauss for each of our data sets, consistent with observed T Tauri magnetic field strengths given our assumptions about the mass accretion rates in ${\S}$3.2 \citep[]{johns03,eisner05}.  However, this is likely coincidental. For the model of halting at a constant semi-major axis independent of the stellar mass, the preferred mean value of $a$ ranges from 0.044 to 0.098 AU as expected for the close-in exoplanets.    

Finally, we find that the best fit values for the Gaussian kernel width $\sigma$ in the log of the semi-major axis provide a reasonable prescription for the data as evidenced by the reduced $\chi^2$ values. We do not attempt to explain the additional non-Gaussian sub-structure in the estimated empirical probability density function that can be seen in Figure 1 for all data sets and in the model residuals in the subsequence figures, but note that such sub-structure could point to multiple migration-halting mechanisms operating.

\section{Conclusions}

We use the empirical distribution of confirmed exoplanets and Kepler planet candidates in the host stellar mass -- exoplanet orbital semi-major axis plane as a diagnostic for migration halting mechanisms.  Migration halting from tidal circularization provides the best posterior probabilities for all empirical samples, favored by factors of 2.5--61 for the different sub-samples investigated herein when compared to halting at the 1:2 interior resonance with the magnetospheric disk truncation radius.  We can rule out migration halting at the 1:2 interior resonance with the dust sublimation radius,  a uniform random halting radius, and an exponential decay halting radius as viable models for the majority of observed close-in exoplanets.  Our generalized power law model favors a dependence of the halting distance with stellar mass that is stronger than predicted from tidal dissipation theories, and clearly rules out the independence of the halting distance on stellar mass, with posterior probability ratios ranging from 1.5--10$^4$.  The favoring of a stronger power law dependence than predicted for the tidal halting model suggests future theoretical work may be needed to better reproduce the observed sub-structure in the empirical distribution of exoplanets as a function of semi-major axis and stellar mass.

The authors would like to thank the referee Dan Fabrycky for their patience, corrections and comments to our analysis, which substantially improved this work.    This research has made use of the NASA Exoplanet Archive, which is operated by the California Institute of Technology, under contract with the National Aeronautics and Space Administration under the Exoplanet Exploration Program.  This research has made use of the Exoplanet Orbit Database and the Exoplanet Data Explorer at exoplanets.org.   The authors would like to thank Thayne Currie for his useful (and rapid turnaround) comments on the manuscript.  We would also like to acknowledge utility of the Bayesian statistics class taught by John Johnson at Caltech.

\clearpage
\begin{landscape}
\begin{deluxetable}{lp{2.2cm}p{2.2cm}p{2.2cm}p{2.2cm}p{2.2cm}p{2.2cm}}
 \tabletypesize{\scriptsize}
% \rotate
\tablewidth{0pc}
\tablecolumns{5}
\tablecaption{Model Best Fit Parameter Values\tablenotemark{a}}
\tablehead{ \colhead{Model} & \colhead{ M$_{pl}<10$M$_{\oplus}$} & \colhead{10M$_{\oplus}<$M$_{pl}<0.3$M$_{J}$} & \colhead{M$_{pl}>0.3$M$_{J}$} & \colhead{R$_{pl}$ $<2R_\oplus$} & \colhead{$2R_\oplus \leq $R$_{pl}$$\leq6R_\oplus$} & \colhead{R$_{pl}$$>6R_\oplus$}}
\startdata
$M^{1/3}$ Tidal Model & $C=0.052$,\newline $\sigma=0.34$ & $C=0.072$,\newline $\sigma=0.14$ & $C=0.042$,\newline $\sigma=0.16$ & $C=0.07$,\newline $\sigma=0.3$ & $C=0.102$,\newline $\sigma=0.26$ & $C=0.056$,\newline $\sigma=0.24$ \\
$M^{3/13}$ Tidal Model & $C=0.05$,\newline $\sigma=0.36$ & $C=0.07$,\newline $\sigma=0.14$ & $C=0.044$,\newline $\sigma=0.16$ & $C=0.068$,\newline $\sigma=0.3$ & $C=0.1$,\newline $\sigma=0.26$ & $C=0.056$,\newline $\sigma=0.24$ \\
Magnetospheric Truncation & $B_0=1.1547$ kG,\newline $\sigma=0.38$ & $B_0=2.4075$ kG,\newline $\sigma=0.16$ & $B_0=1.0683$ kG,\newline $\sigma=0.16$ & $B_0=2.2884$ kG,\newline $\sigma=0.3$ & $B_0=4.338$ kG,\newline $\sigma=0.26$ & $B_0=1.6292$ kG,\newline $\sigma=0.24$ \\
Dust Sublimation & $\sigma=1.02$ & $\sigma=1.02$ & $\sigma=0.86$ & $\sigma=1.02$ & $\sigma=1.02$ & $\sigma=1.02$ \\
Power Law & $\alpha=0.9$,\newline $C=0.062$,\newline $\sigma=0.34$ & $\alpha=0.58$,\newline $C=0.078$,\newline $\sigma=0.14$ & $\alpha=0.68$,\newline $C=0.042$,\newline $\sigma=0.16$ & $\alpha=0.6$,\newline $C=0.074$,\newline $\sigma=0.3$ & $\alpha=0.38$,\newline $C=0.104$,\newline $\sigma=0.26$ & $\alpha=0.4$,\newline $C=0.058$,\newline $\sigma=0.24$ \\
Uniform Random & \nodata & \nodata & \nodata & \nodata & \nodata & \nodata \\
Exponential Decay & \nodata & \nodata & \nodata & \nodata & \nodata & \nodata \\
Stellar Mass Independent & $A=0.048$,\newline $\sigma=0.5$ & $A=0.07$,\newline $\sigma=0.18$ & $A=0.044$,\newline $\sigma=0.16$ & $A=0.066$,\newline $\sigma=0.3$ & $A=0.098$,\newline $\sigma=0.26$ & $A=0.056$,\newline $\sigma=0.24$ \\
% Roche Radius & $\sigma=1.02$ & $\sigma=1.02$ & $\sigma=1.02$ & $\sigma=1.02$ & $\sigma=1.02$ & $\sigma=1.02$ \\

\enddata
\tablenotetext{a}{$\sigma$ values are the Gaussian kernel widths in units of log(AU) that are convolved with the analytic 1:1 curves to produce the model density functions as described in ${\S}$3.  The units of log$(A)$ are also log(AU).  The second through fourth columns are for the confirmed exoplanet sub-samples, and the last three columns correspond to the KOIs.}
\end{deluxetable}
 
\begin{deluxetable}{lllllll}
\tabletypesize{\scriptsize}
% \rotate
\tablewidth{0pc}
\tablecolumns{5}
\tablecaption{Relative Bayesian Posterior Probabilities (Unnormalized)\tablenotemark{a}}
\tablehead{ \colhead{Model} & \colhead{ M$_{pl}<10$M$_{\oplus}$} & \colhead{10M$_{\oplus}<$M$_{pl}<0.3$M$_{J}$} & \colhead{M$_{pl}>0.3$M$_{J}$} & \colhead{R$_{pl}$ $<2R_\oplus$} & \colhead{$2R_\oplus \leq $R$_{pl}$$\leq6R_\oplus$} & \colhead{R$_{pl}$$>6R_\oplus$}}
\startdata
$M^{1/3}$ Tidal Model	& $1.2190 \times 10^{-01}$ & $6.9976 \times 10^{07}$ & $3.8776 \times 10^{43}$ & $2.3146 \times 10^{123}$ & $4.3350 \times 10^{164}$ & $8.7087 \times 10^{28}$ \\
$M^{3/13}$ Tidal Model	& $7.9400 \times 10^{-02}$ & $3.6697 \times 10^{07}$ & $6.1489 \times 10^{42}$ & $6.7614 \times 10^{122}$ & $3.4503 \times 10^{164}$ & $7.7497 \times 10^{28}$ \\
Magnetospheric Truncation	& $3.1100 \times 10^{-02}$ & $8.4200 \times 10^{06}$ & $6.4026 \times 10^{41}$ & $8.3070 \times 10^{121}$ & $8.9229 \times 10^{163}$ & $3.5485 \times 10^{28}$ \\
Dust Sublimation	& $3.6000 \times 10^{-03}$ & $7.7860 \times 10^{-04}$ & $6.1275 \times 10^{11}$ & $1.2069 \times 10^{34}$ & $5.6110 \times 10^{35}$ & $4.7453 \times 10^{10}$ \\
Power Law	& $2.1550 \times 10^{-01}$ & $7.2185 \times 10^{07}$ & $2.4825 \times 10^{44}$ & $3.5253 \times 10^{123}$ & $2.2953 \times 10^{164}$ & $6.3967 \times 10^{28}$ \\
Uniform Random	& $7.5600 \times 10^{-02}$ & $3.1060 \times 10^{05}$ & $1.5013 \times 10^{19}$ & $2.8514 \times 10^{110}$ & $1.0866 \times 10^{139}$ & $2.1531 \times 10^{25}$ \\
Exponential Decay	& $7.6600 \times 10^{-02}$ & $3.0870 \times 10^{-01}$ & $5.4692 \times 10^{12}$ & $9.1955 \times 10^{49}$ & $6.3116 \times 10^{53}$ & $1.3048 \times 10^{14}$ \\
Stellar Mass Independent	& $3.2300 \times 10^{-02}$ & $5.9364 \times 10^{06}$ & $2.5710 \times 10^{40}$ & $1.0893 \times 10^{121}$ & $9.7962 \times 10^{163}$ & $4.2242 \times 10^{28}$ \\
%Roche Radius	& $7.3107 \times 10^{-04}$ & $4.8581 \times 10^{-05}$ & $3.3891 \times 10^{07}$ & $4.7831 \times 10^{19}$ & $8.4426 \times 10^{20}$ & $2.9812 \times 10^{07}$ \\
\enddata
\tablenotetext{a}{The second through fourth columns are for the confirmed exoplanet sub-samples, and the last three columns correspond to the KOIs.}
\end{deluxetable}

\begin{deluxetable}{lllllll}
 \tabletypesize{\scriptsize}
% \rotate
\tablewidth{0pc}
\tablecolumns{6}
\tablecaption{Reduced $\chi^2$ Values\tablenotemark{a}}
\tablehead{ \colhead{Model} & \colhead{ M$_{pl}<10$M$_{\oplus}$} & \colhead{10M$_{\oplus}<$M$_{pl}<0.3$M$_{J}$} & \colhead{M$_{pl}>0.3$M$_{J}$} & \colhead{R$_{pl}$ $<2R_\oplus$} & \colhead{$2R_\oplus \leq $R$_{pl}$$\leq6R_\oplus$} & \colhead{R$_{pl}$$>6R_\oplus$}}
\startdata
$M^{1/3}$ Tidal Model	&	0.51739	&	0.32685	&	0.89054	&	1.0987	&	1.2792	&	0.35351	\\
$M^{3/13}$ Tidal Model	&	0.54308	&	0.41063	&	1.6656	&	1.1376	&	1.2416	&	0.35557	\\
Magnetospheric Truncation	&	0.57783	&	0.39862	&	2.4932	&	1.2987	&	1.2235	&	0.37129	\\
Dust Sublimation	&	0.78871	&	1.8692	&	5.4704	&	13.1747	&	18.9103	&	2.7711	\\
Power Law	&	0.5709	&	0.39799	&	0.55665	&	1.1233	&	1.3821	&	0.37203	\\
Uniform Random	&	0.88506	&	0.7415	&	4.7696	&	3.7036	&	3.8442	&	1.1901	\\
Exponential Decay	&	0.72727	&	1.6272	&	5.5312	&	11.078	&	15.8296	&	2.4884	\\
Stellar Mass Independent	&	0.62176	&	0.51691	&	5.3648	&	1.518	&	1.4775	&	0.42532	\\
% Roche Radius	&	0.87309	&	2.0973	&	6.1121	&	15.5875	&	22.0302	&	3.2525	\\
\enddata
\tablenotetext{a}{The second through fourth columns are for the confirmed exoplanet sub-samples, and the last three columns correspond to the KOIs.}
\end{deluxetable} 
\end{landscape}

\clearpage
\begin{landscape}
\begin{figure}
\centering
\includegraphics[width=0.20\textwidth,clip=true,trim=0cm 0cm 0cm 0cm]{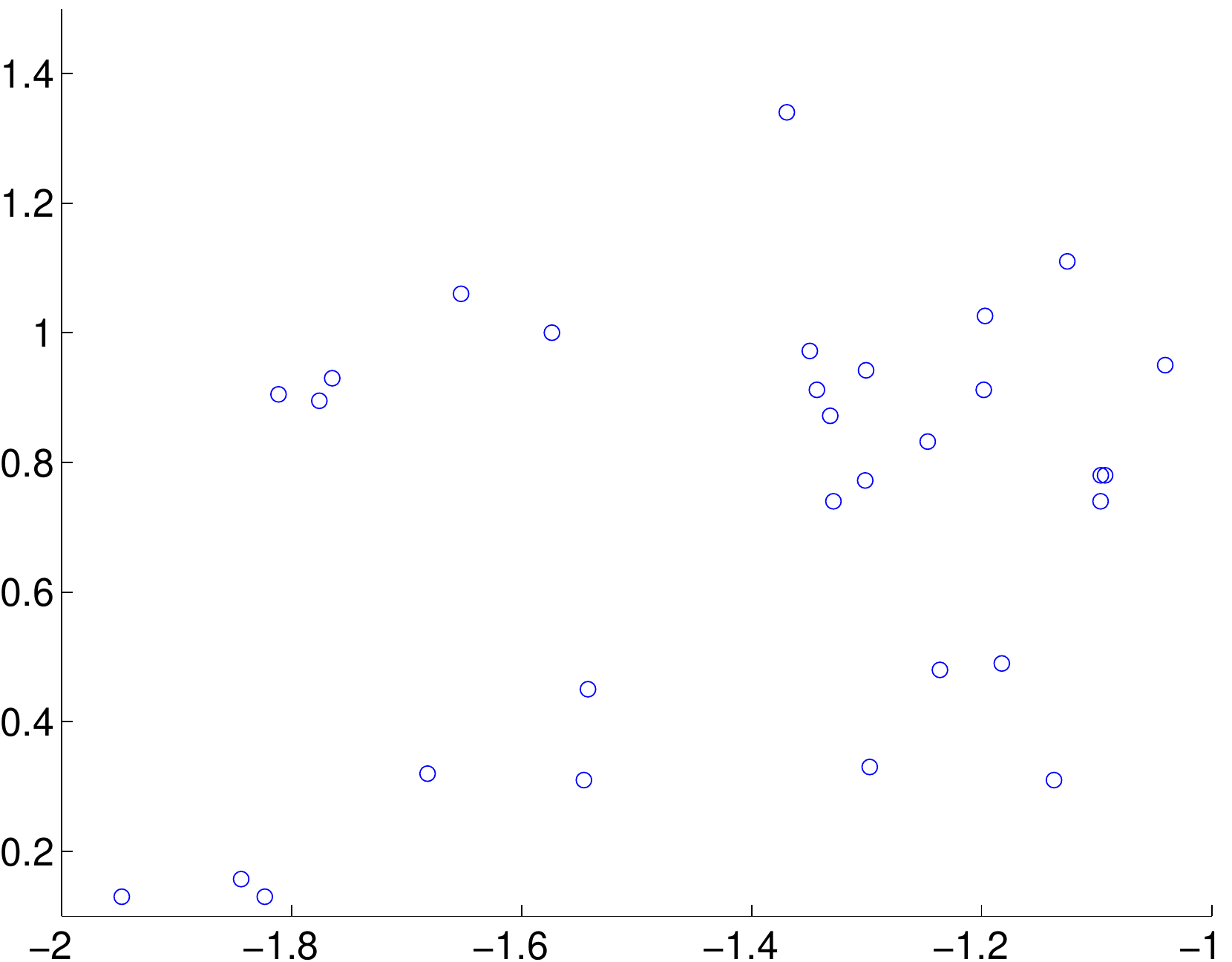} 
\includegraphics[width=0.20\textwidth,clip=true,trim=0cm 0cm 0cm 0cm]{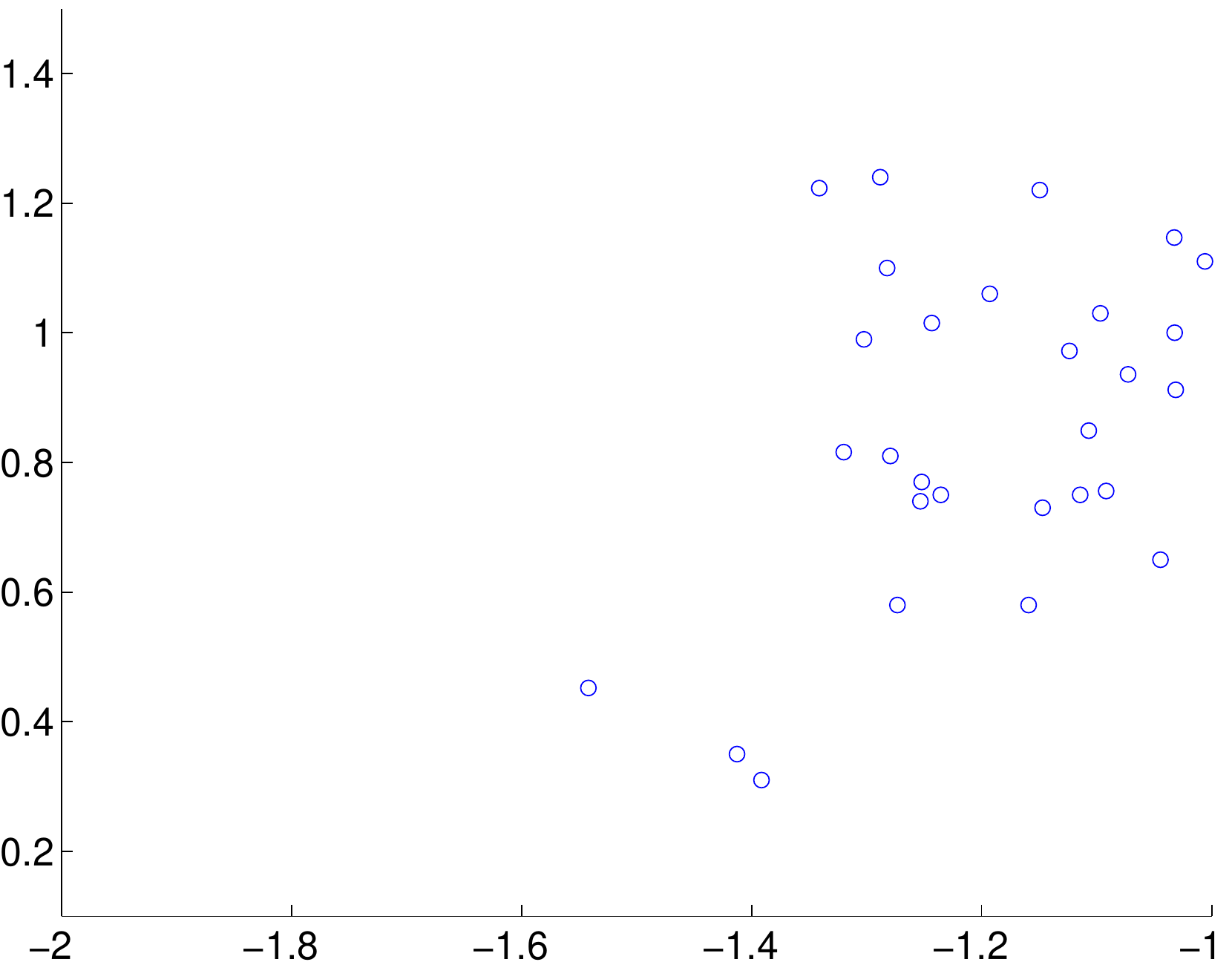} 
\includegraphics[width=0.20\textwidth,clip=true,trim=0cm 0cm 0cm 0cm]{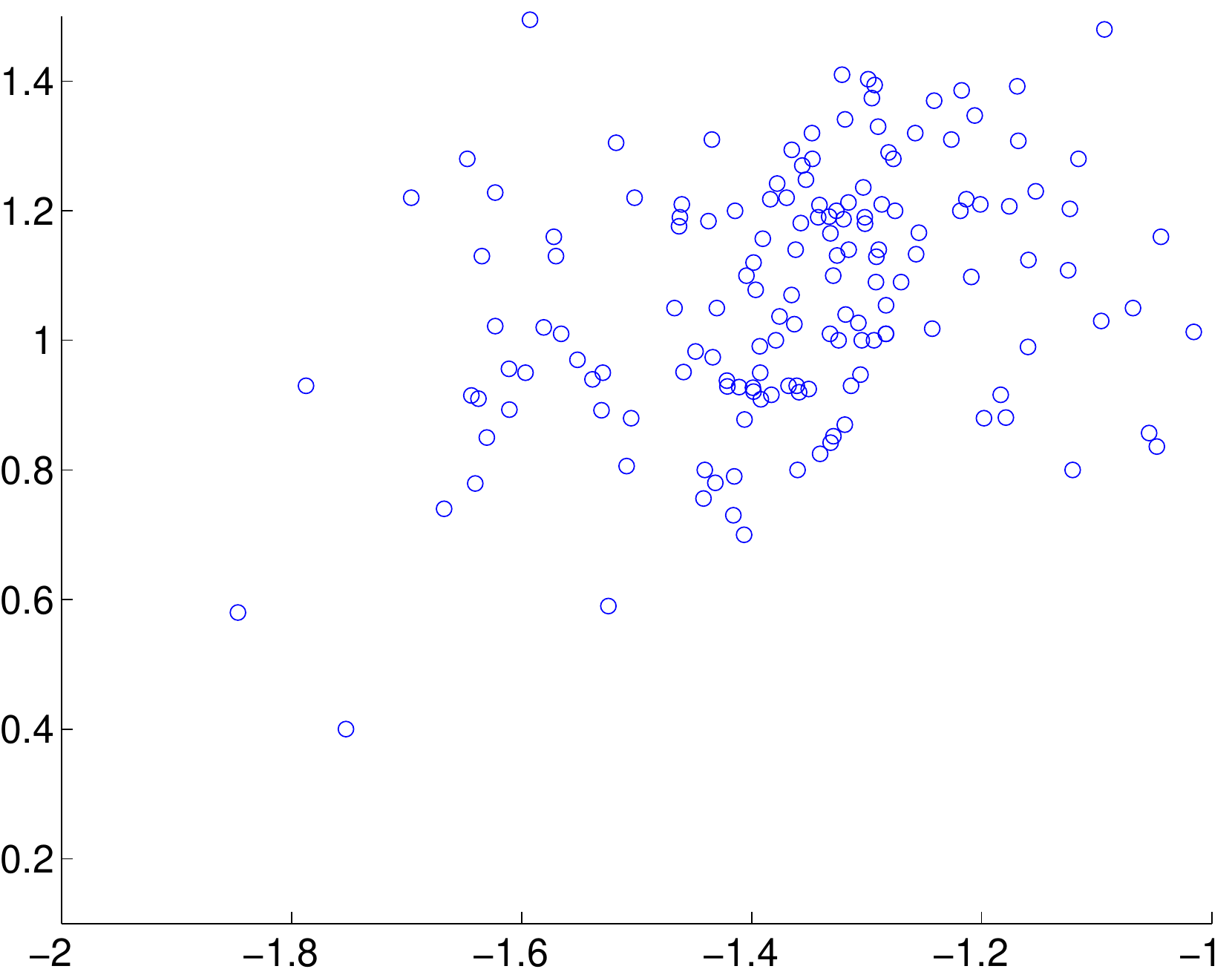} 
\includegraphics[width=0.20\textwidth,clip=true,trim=0cm 0cm 0cm 0cm]{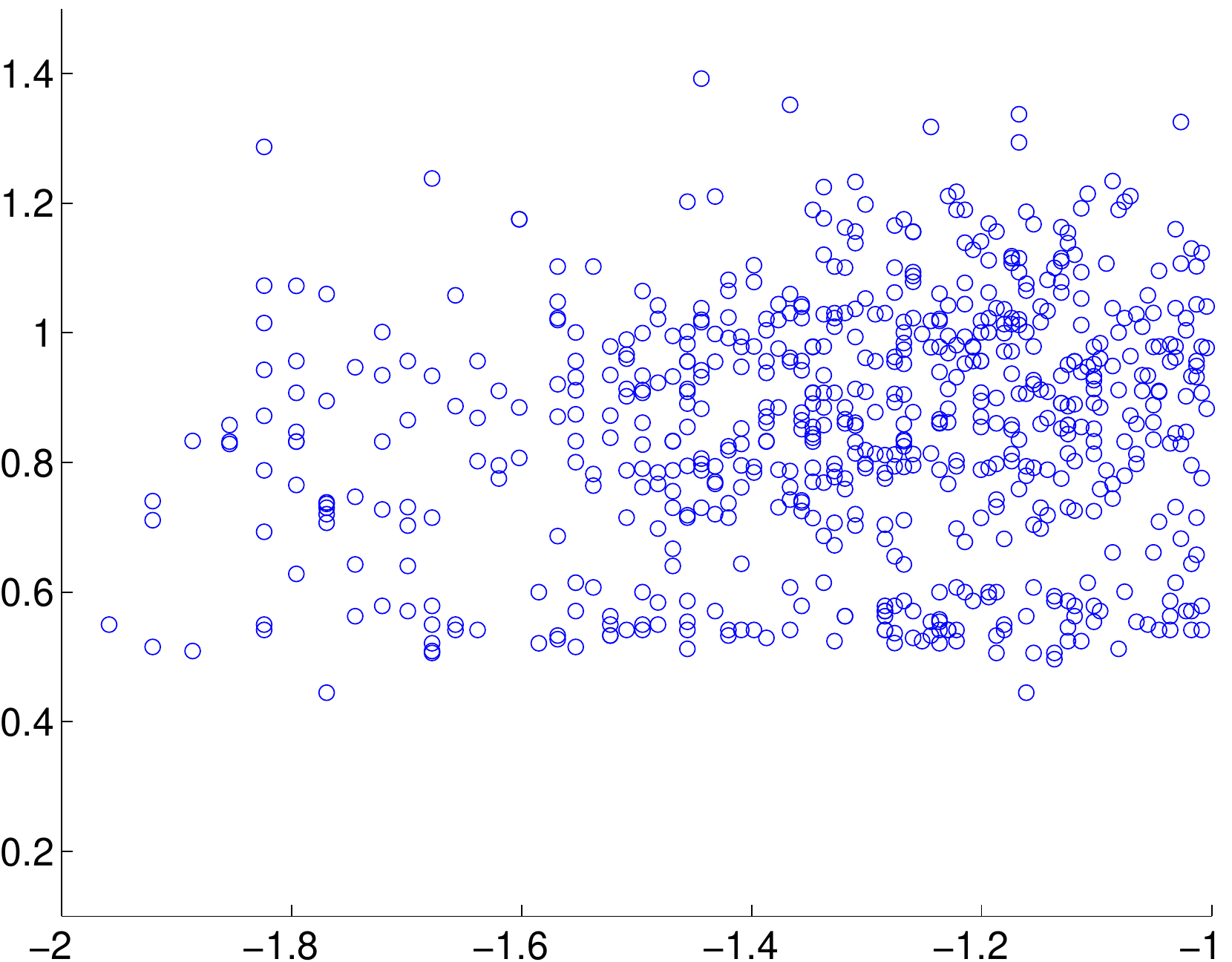} 
\includegraphics[width=0.20\textwidth,clip=true,trim=0cm 0cm 0cm 0cm]{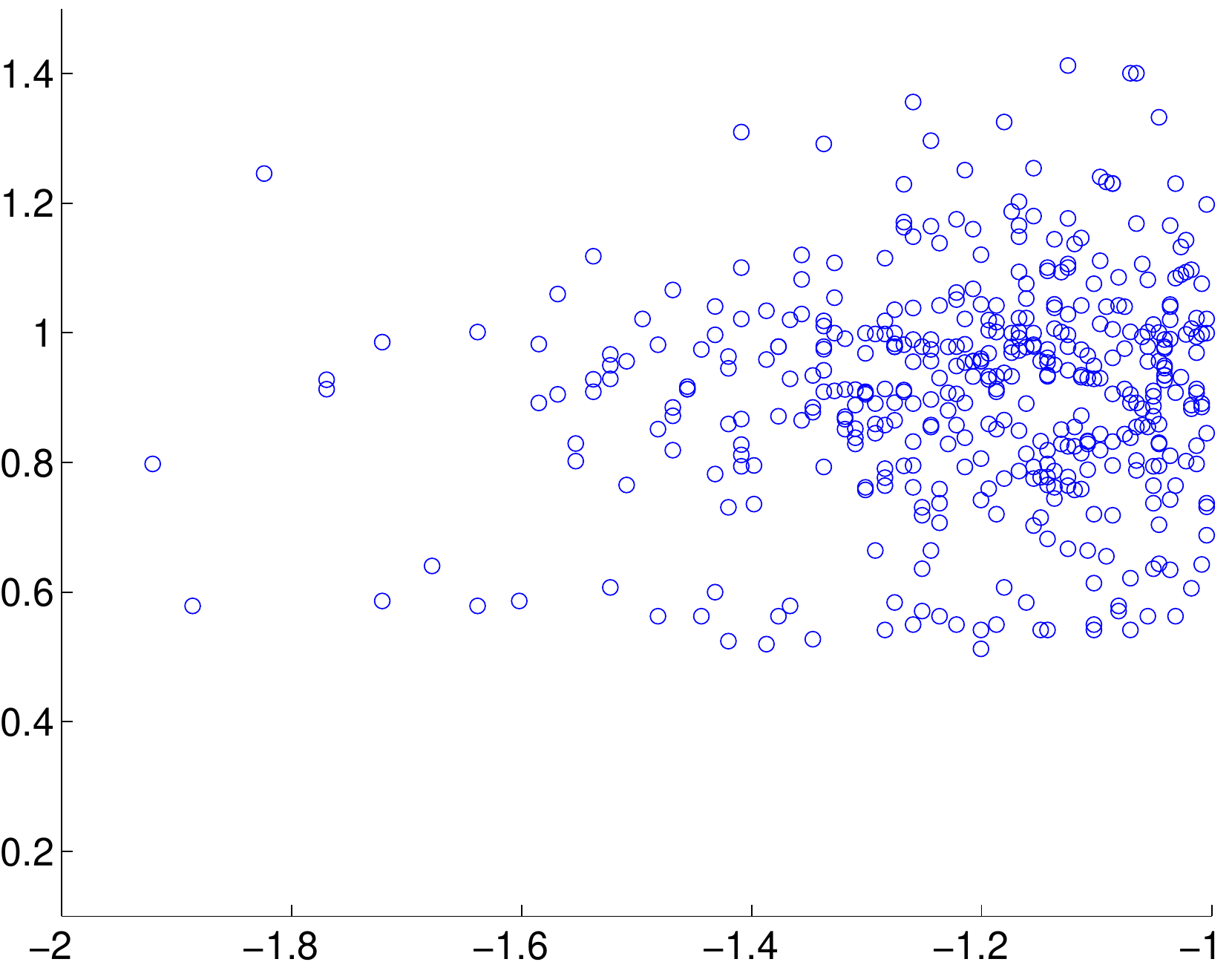} 
\includegraphics[width=0.20\textwidth,clip=true,trim=0cm 0cm 0cm 0cm]{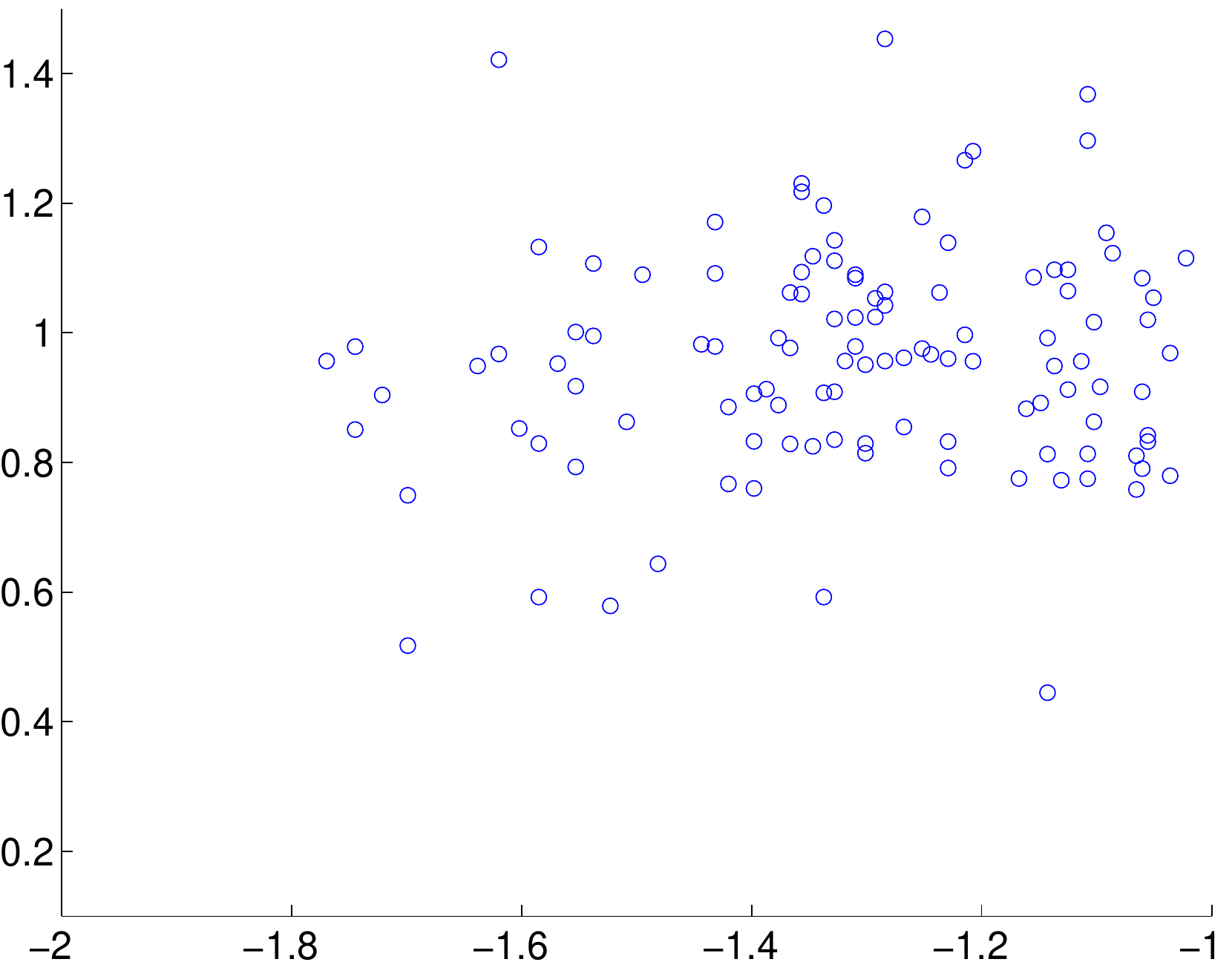} \\
\includegraphics[width=0.20\textwidth,clip=true,trim=0cm 0cm 0cm 0cm]{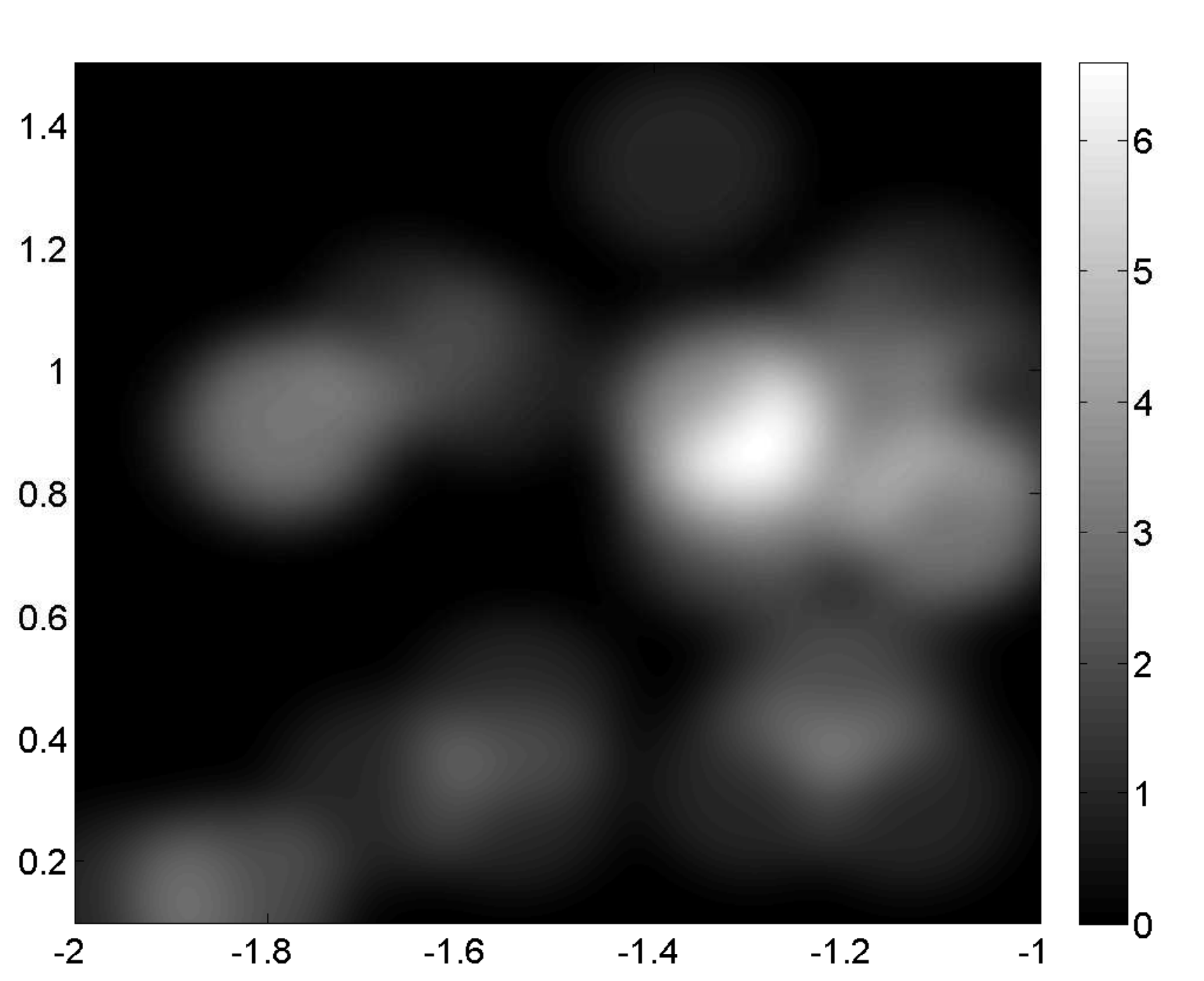} 
\includegraphics[width=0.20\textwidth,clip=true,trim=0cm 0cm 0cm 0cm]{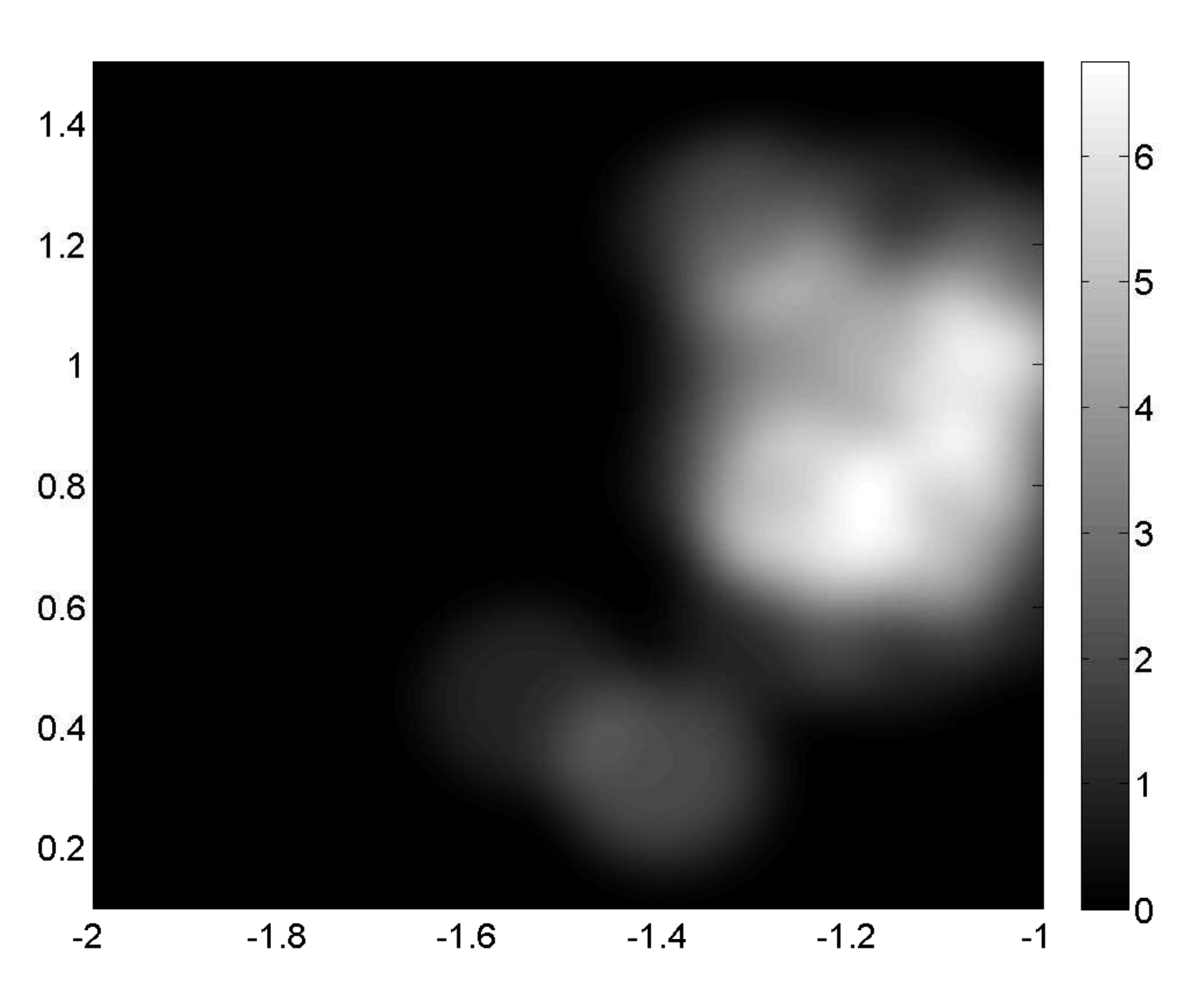} 
\includegraphics[width=0.20\textwidth,clip=true,trim=0cm 0cm 0cm 0cm]{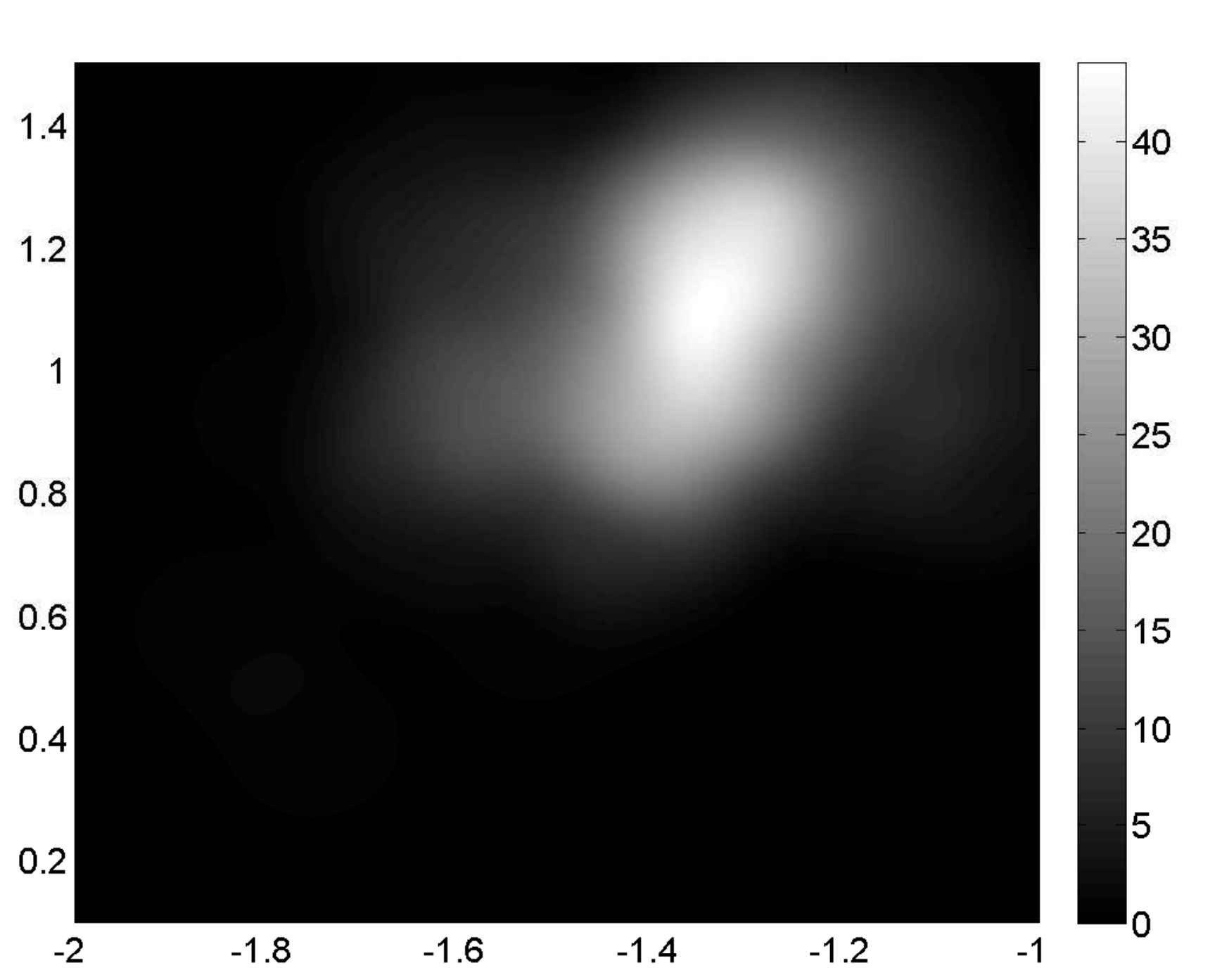}
\includegraphics[width=0.20\textwidth,clip=true,trim=0cm 0cm 0cm 0cm]{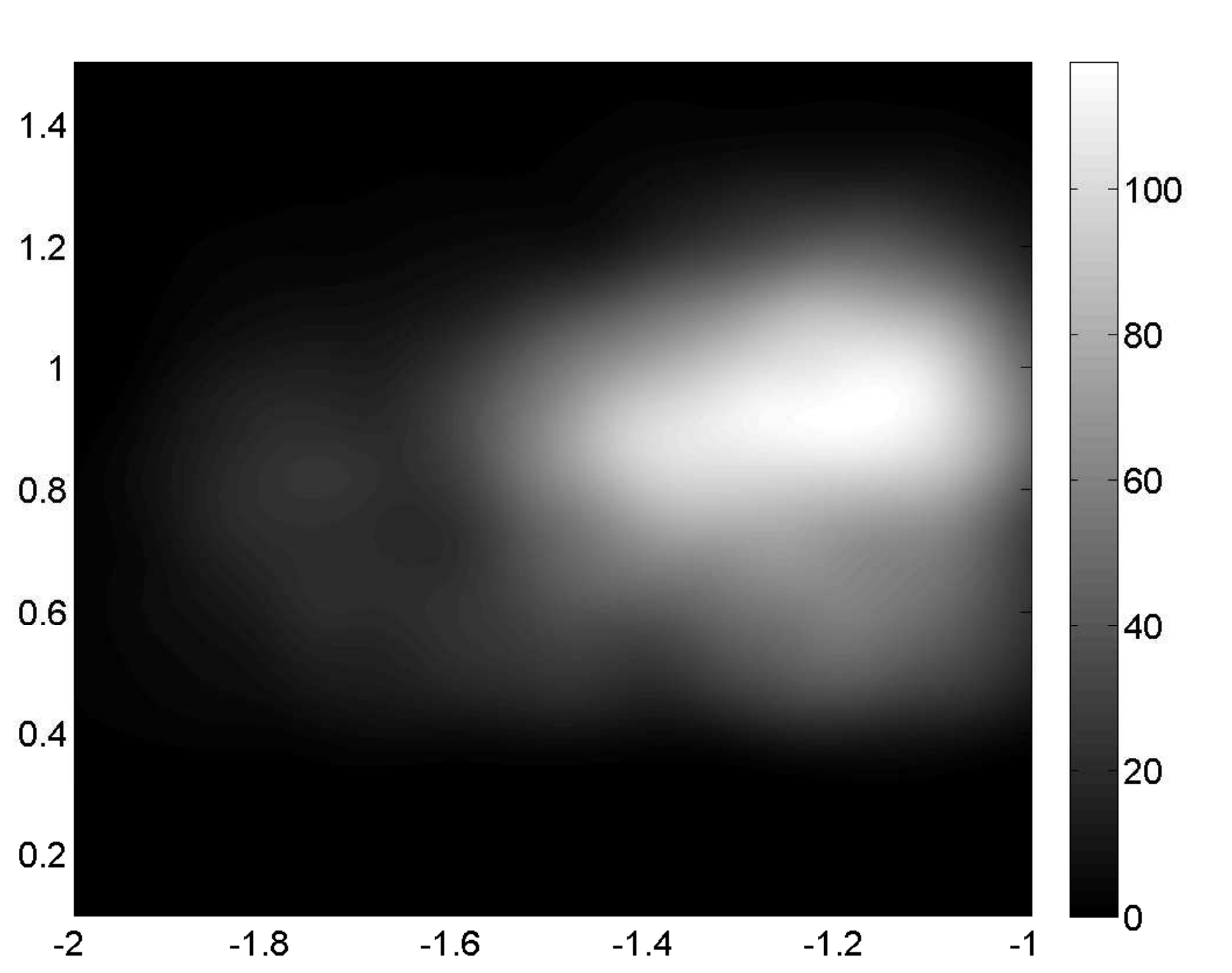} 
\includegraphics[width=0.20\textwidth,clip=true,trim=0cm 0cm 0cm 0cm]{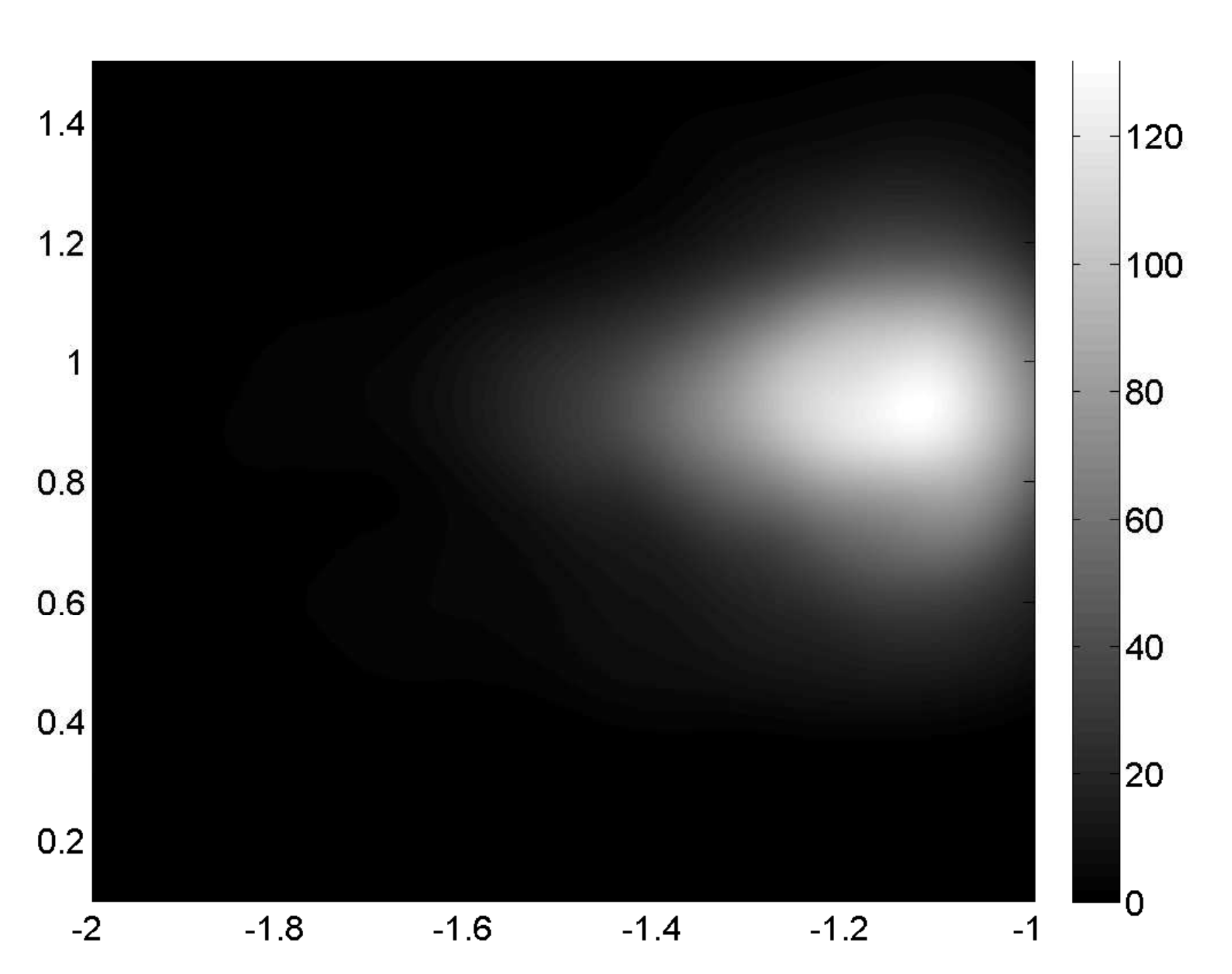} 
\includegraphics[width=0.20\textwidth,clip=true,trim=0cm 0cm 0cm 0cm]{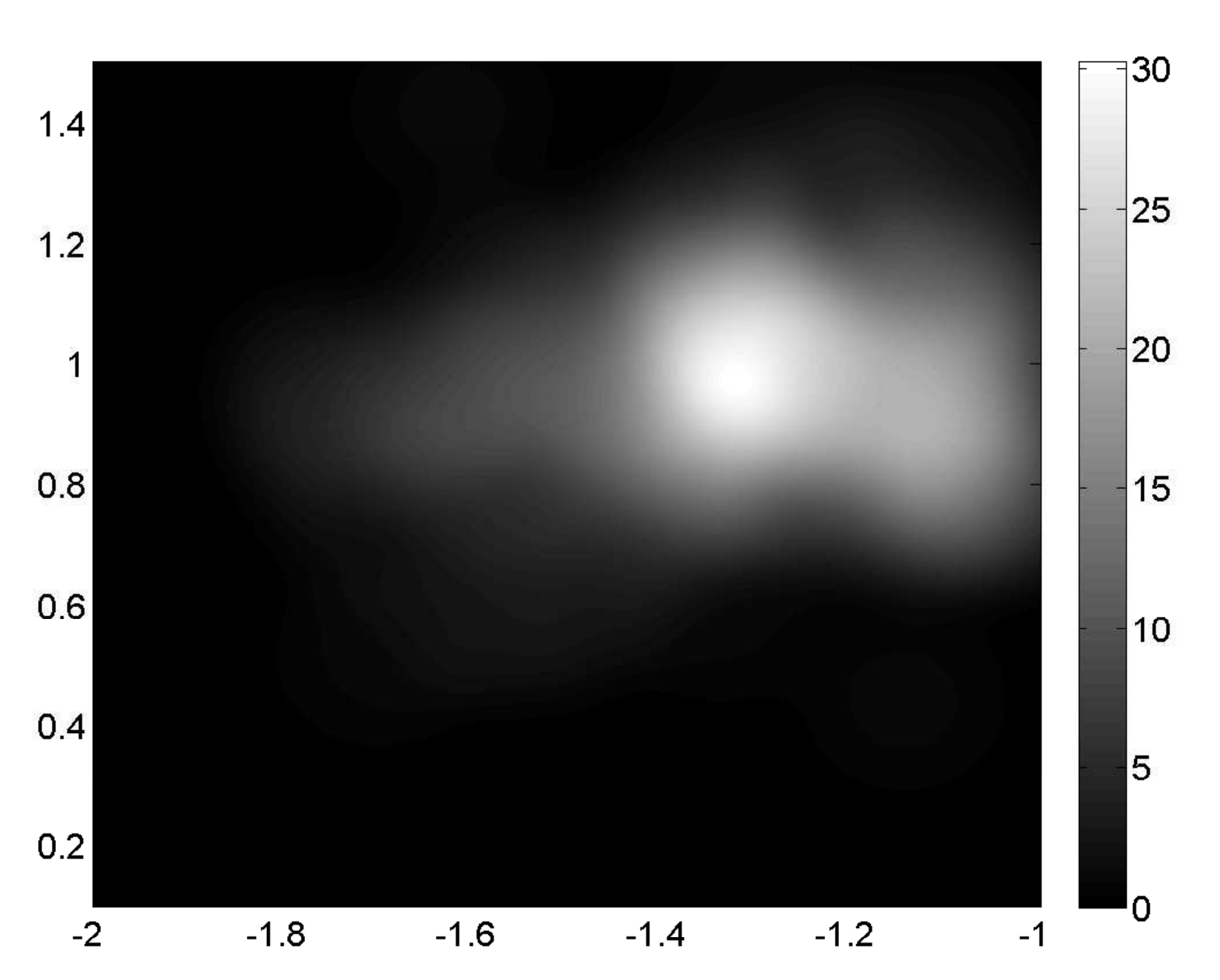} \\
\includegraphics[width=0.20\textwidth,clip=true,trim=0cm 0cm 0cm 0cm]{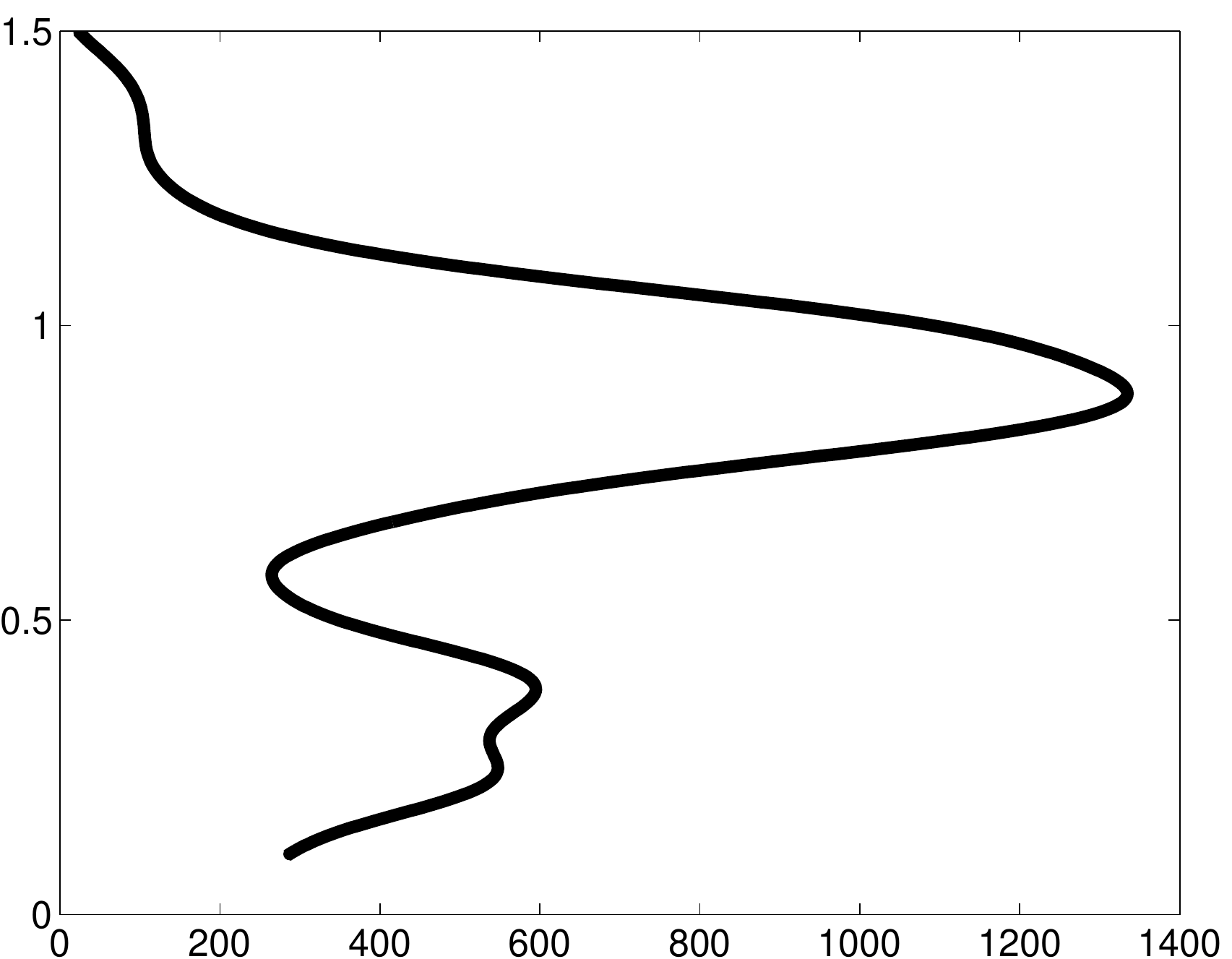} 
\includegraphics[width=0.20\textwidth,clip=true,trim=0cm 0cm 0cm 0cm]{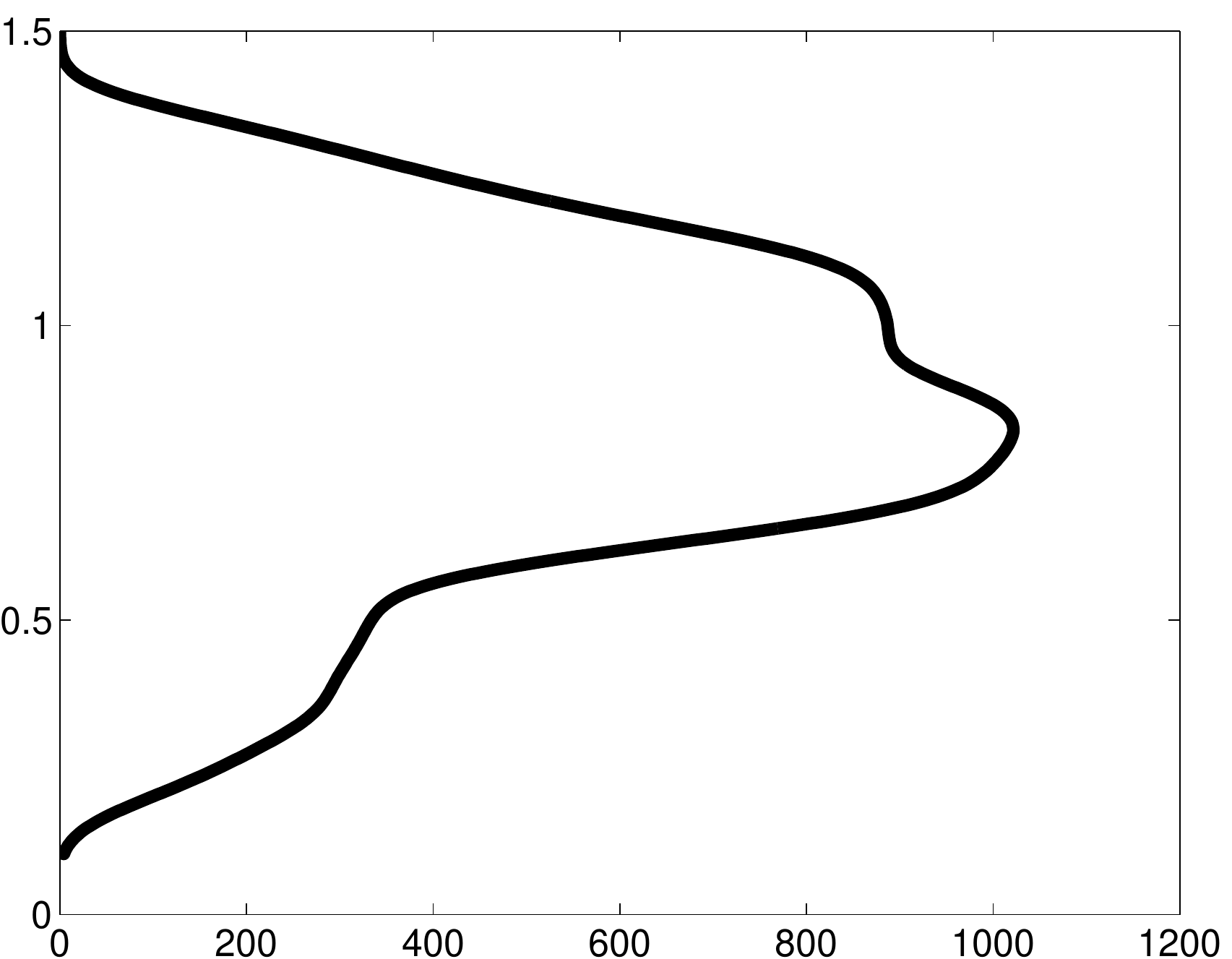} 
\includegraphics[width=0.20\textwidth,clip=true,trim=0cm 0cm 0cm 0cm]{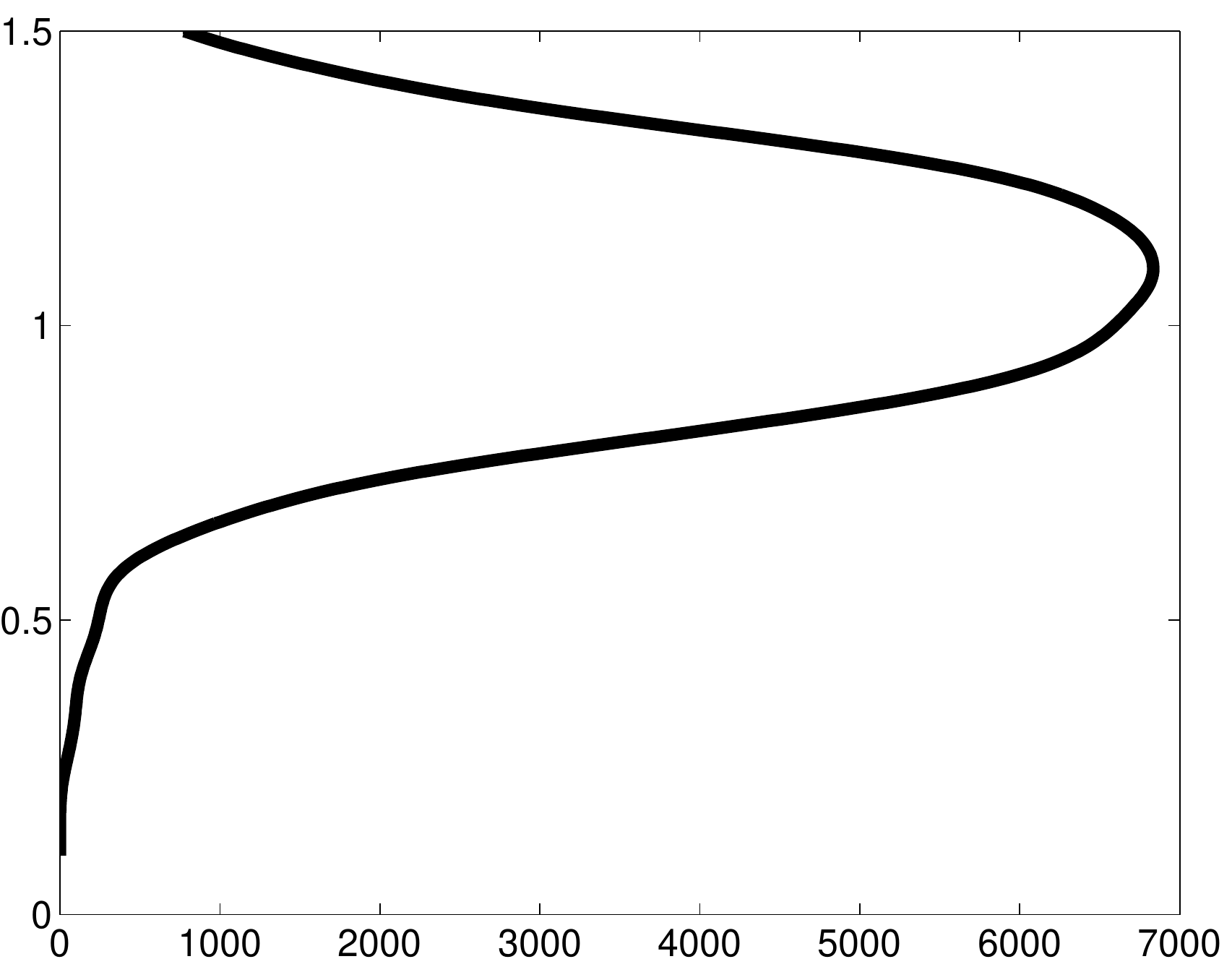} 
\includegraphics[width=0.20\textwidth,clip=true,trim=0cm 0cm 0cm 0cm]{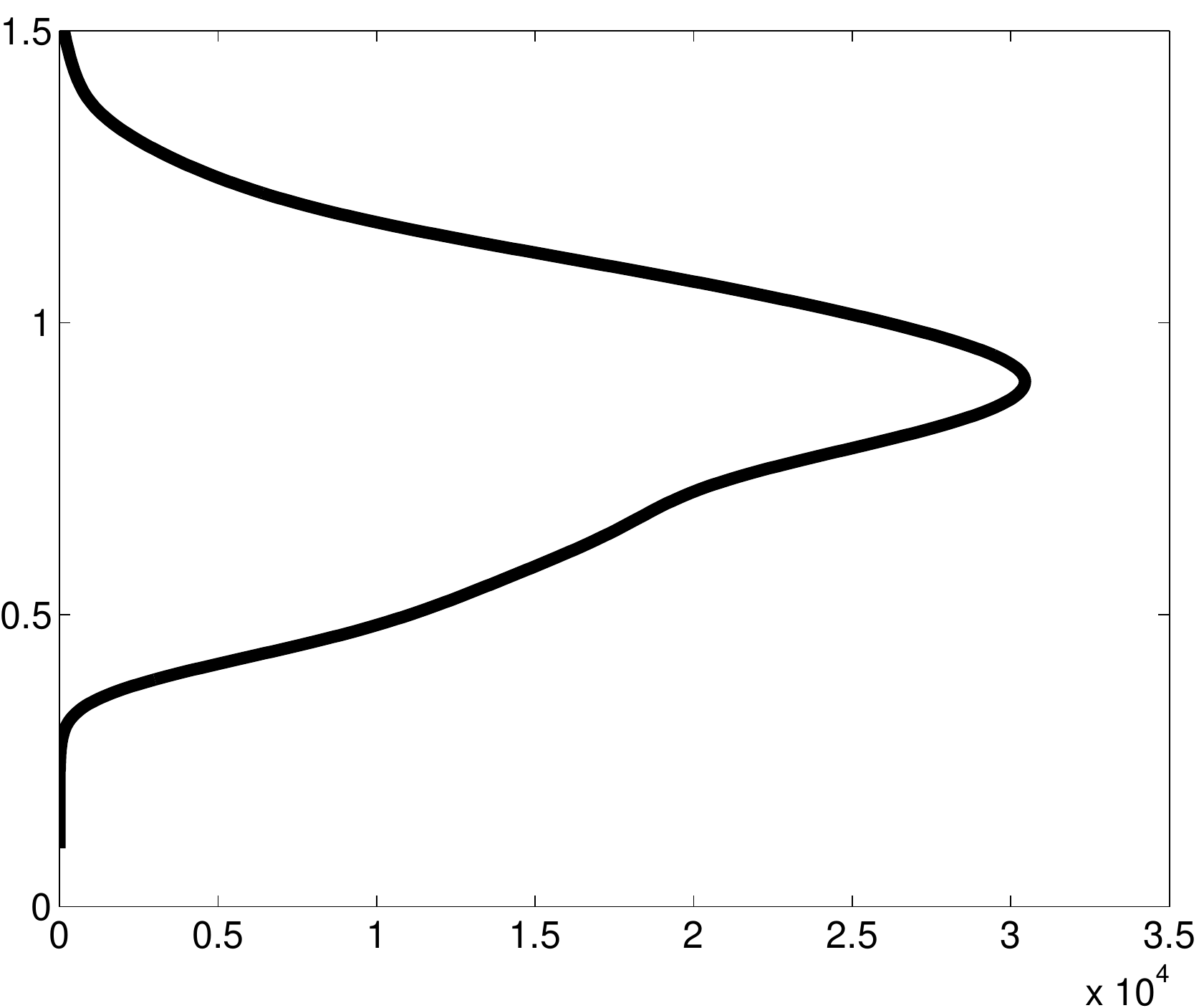} 
\includegraphics[width=0.20\textwidth,clip=true,trim=0cm 0cm 0cm 0cm]{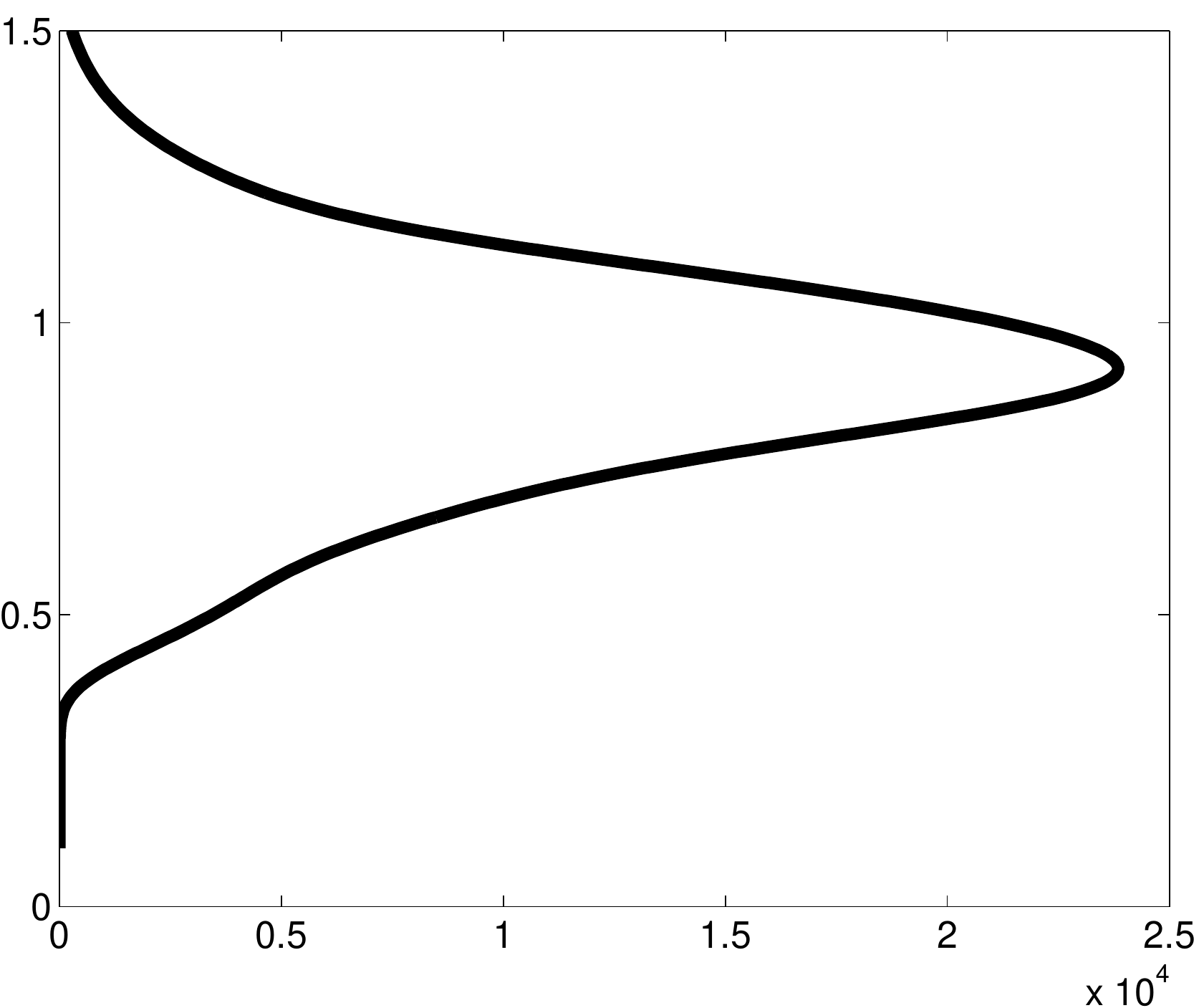} 
\includegraphics[width=0.20\textwidth,clip=true,trim=0cm 0cm 0cm 0cm]{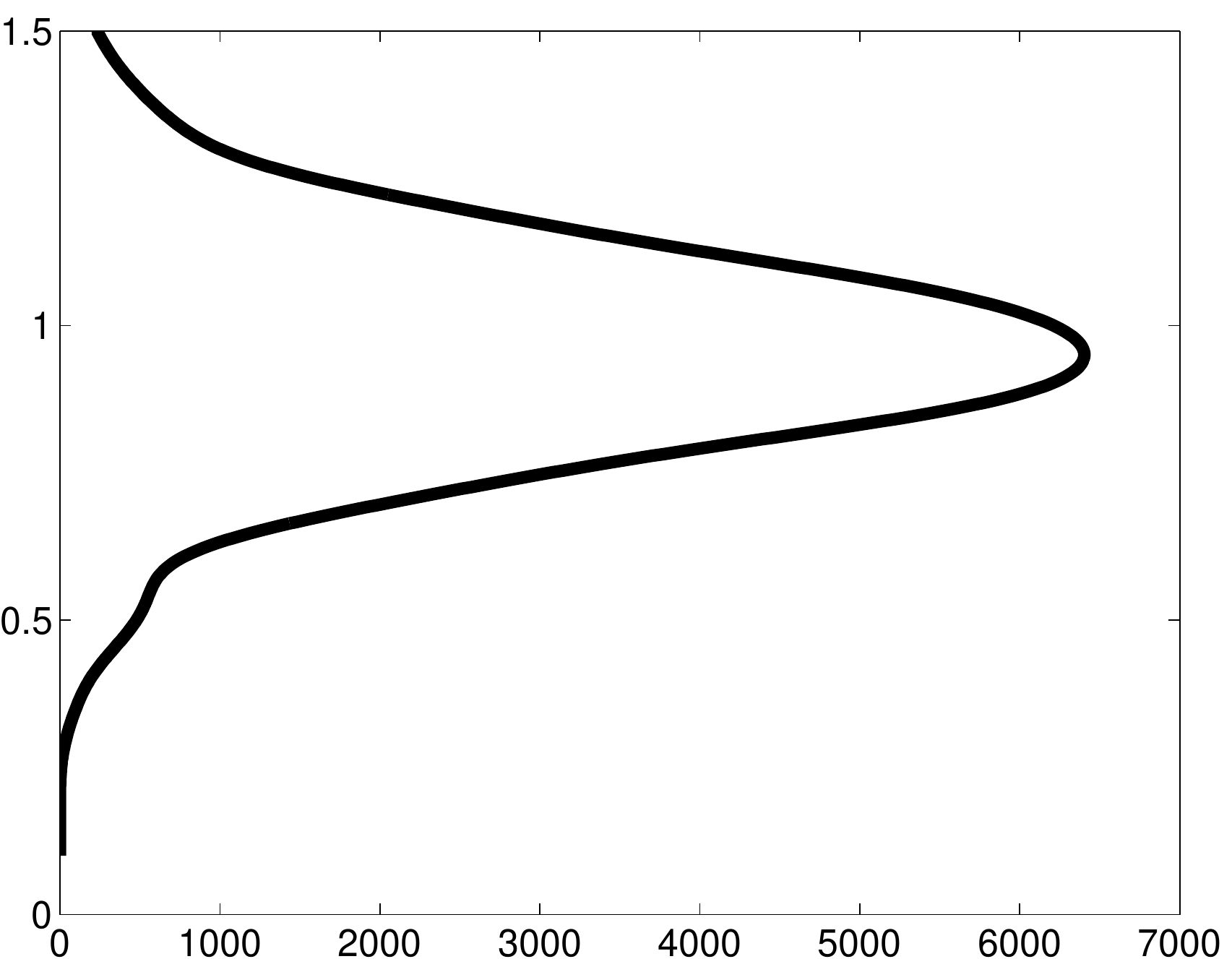} 
\caption{First row: Empirical distributions of exoplanet semi-major axes and host star masses for (from left to right): confirmed exoplanets as of February 2012 (NASA Exoplanet Archive) with M$_{pl}$$<$10 M$_{\oplus}$, confirmed exoplanets with 10M$_{\oplus}<$M$_{pl}<0.3$M$_{J}$, confirmed exoplanets with M$_{pl}>0.3$M$_{J}$, Kepler Objects of Interest from \citet[]{batalha} with R$_{pl}$ $<2R_\oplus$, Kepler Objects of Interest with $2R_\oplus \leq $R$_{pl}$$\leq6R_\oplus$, and Kepler Objects of Interest with R$_{pl}$$>6R_\oplus$.  Second row: Empirical density functions generated for each data set using a method analogous to that in \citep[]{wasserman}.  Third row: Planet frequency as a function of stellar mass for each data set in the first row, shown with arbitrary normalization.  This is used as a normalization for our migration braking models to remove survey sample selection effects as a function of stellar mass.}
\end{figure}
\end{landscape}

\clearpage
\begin{landscape}
\begin{figure}
\centering
\includegraphics[width=0.20\textwidth,clip=true,trim=0cm 0cm 0cm 0cm]{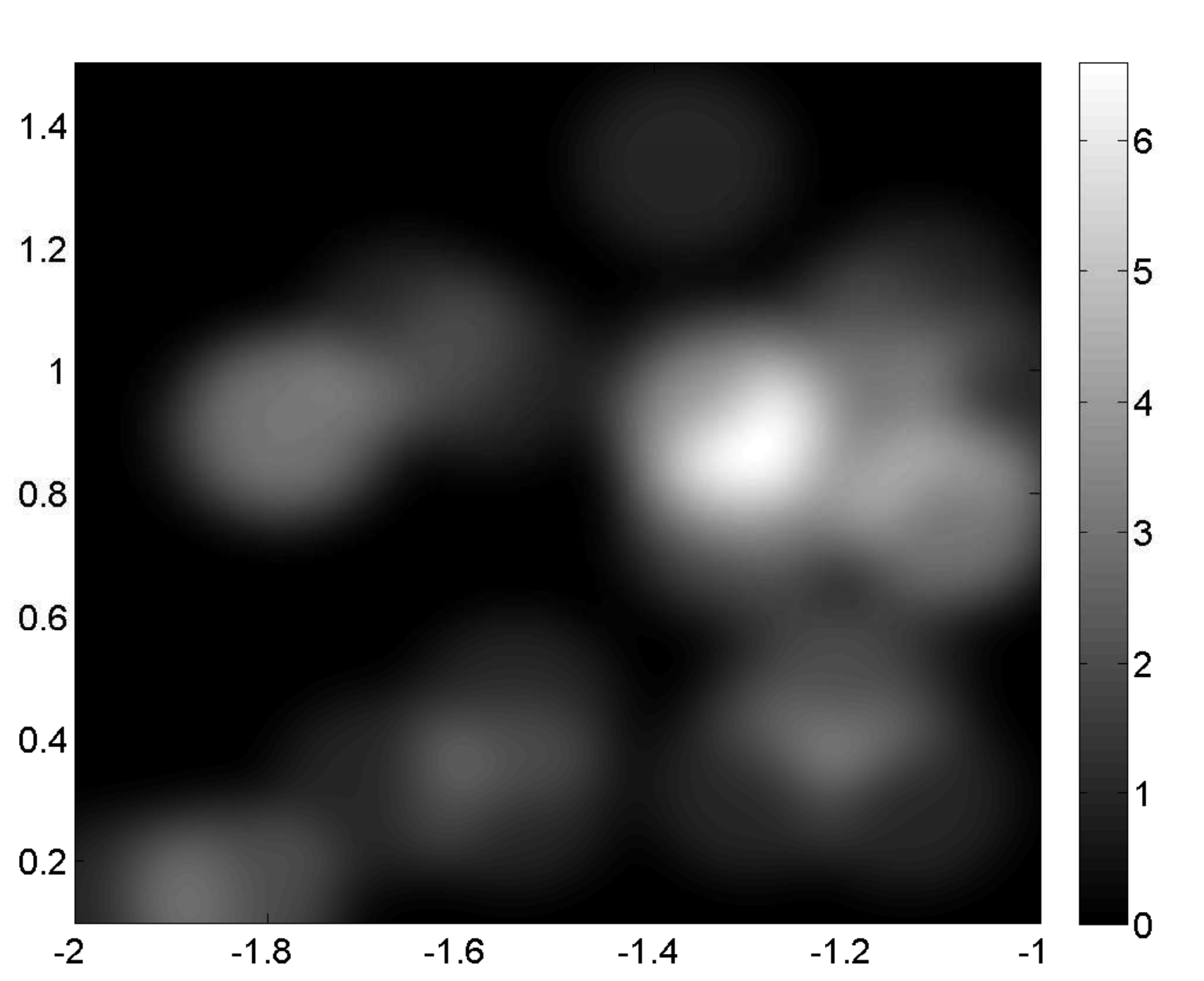} 
\includegraphics[width=0.20\textwidth,clip=true,trim=0cm 0cm 0cm 0cm]{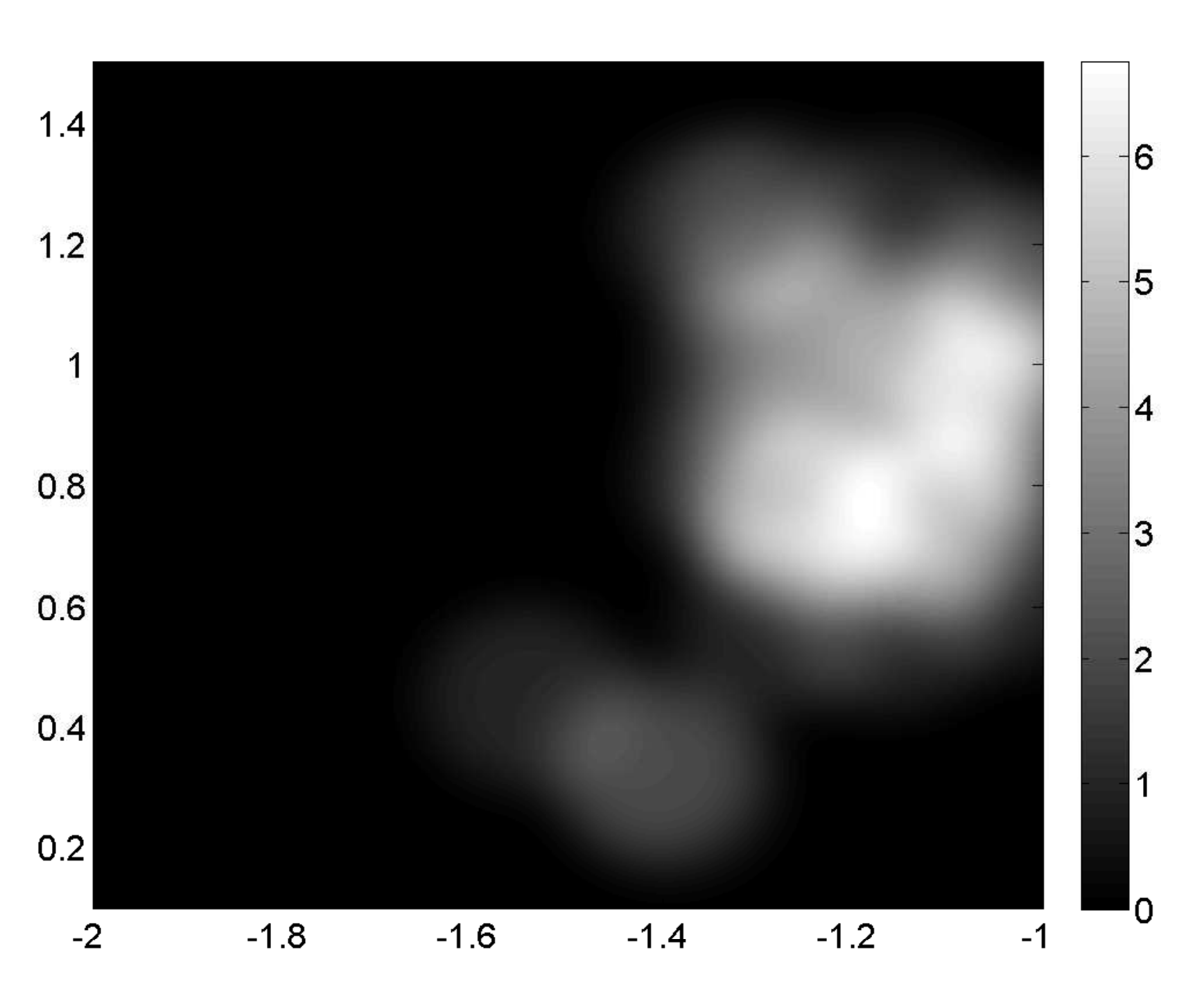} 
\includegraphics[width=0.20\textwidth,clip=true,trim=0cm 0cm 0cm 0cm]{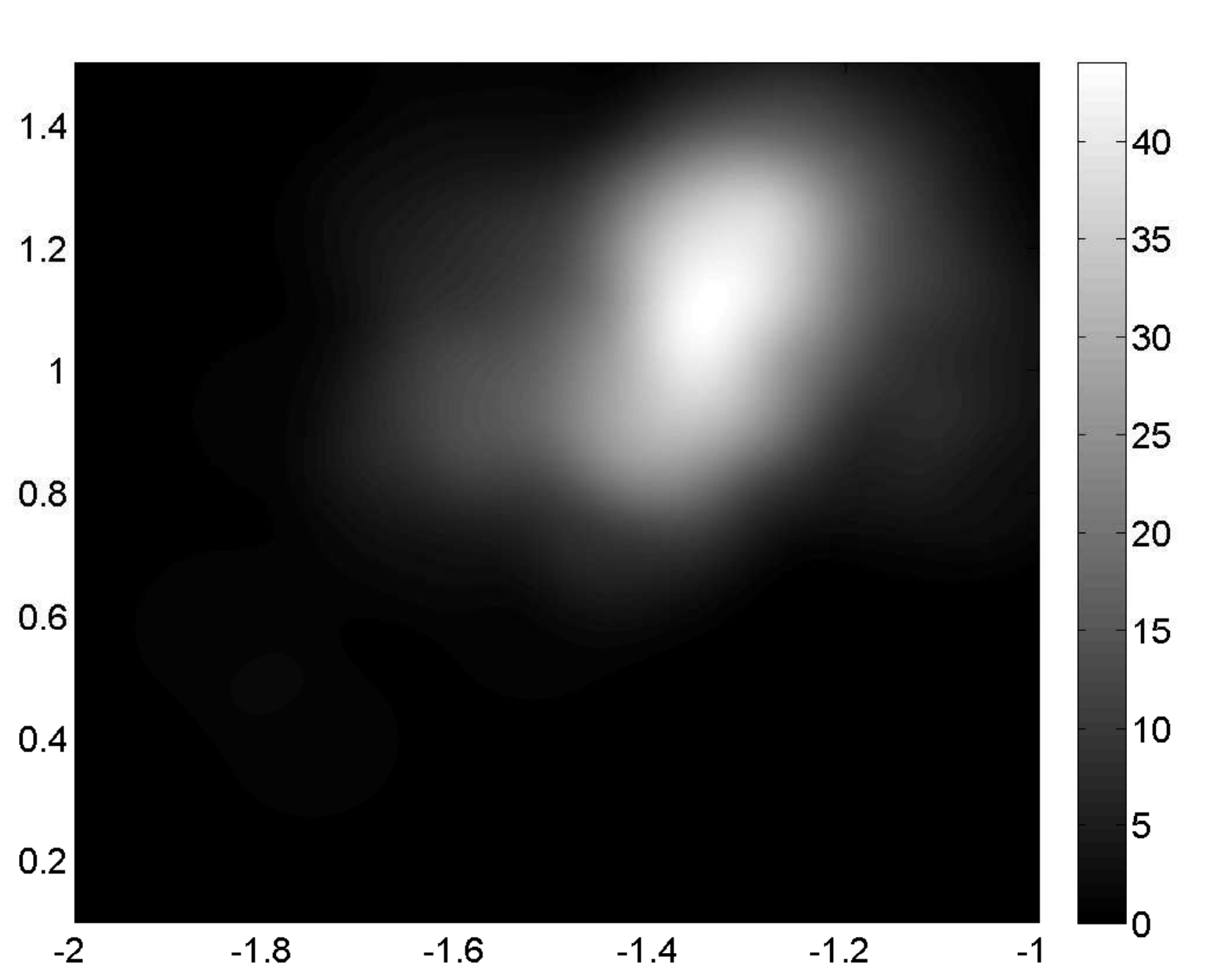} 
\includegraphics[width=0.20\textwidth,clip=true,trim=0cm 0cm 0cm 0cm]{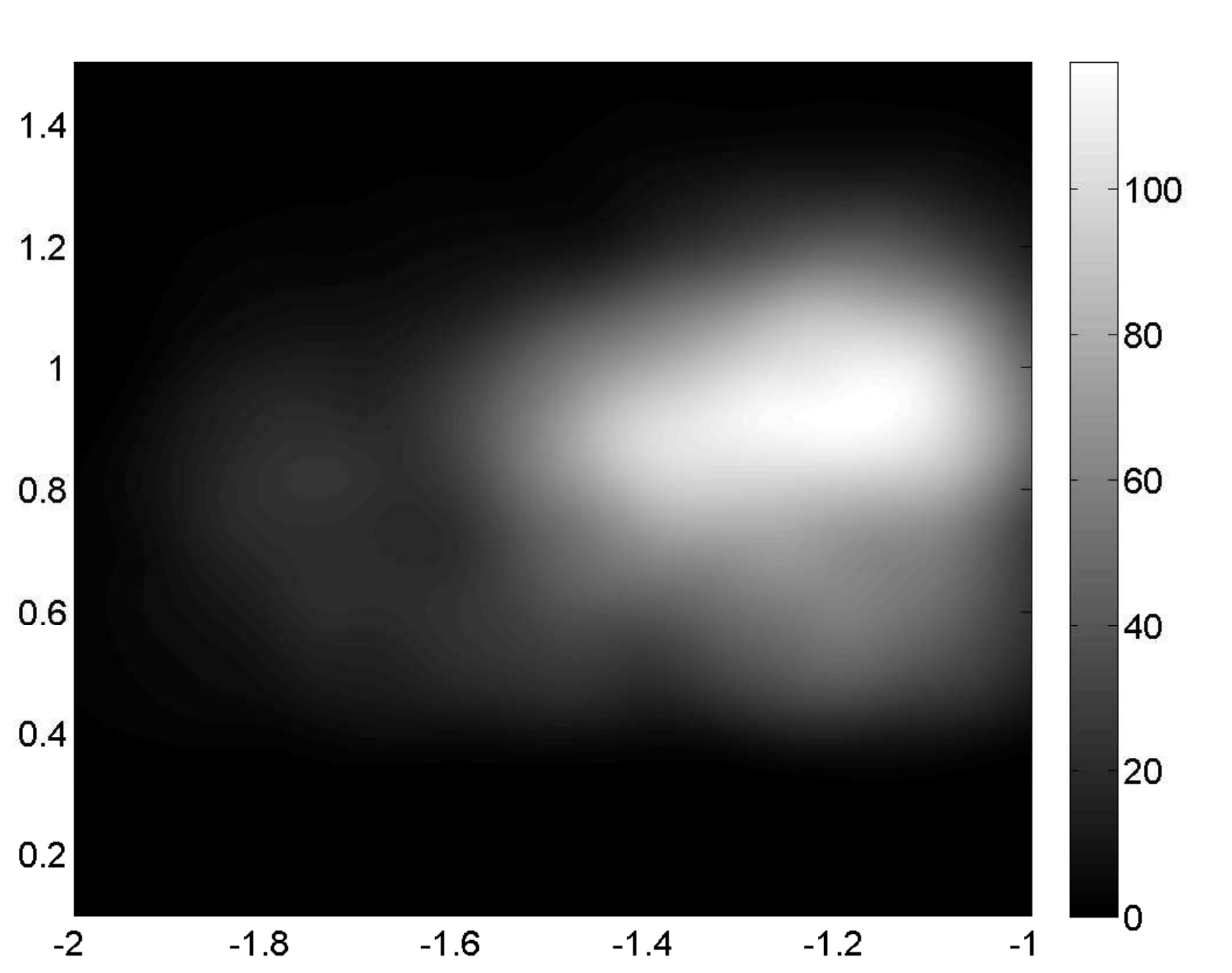} 
\includegraphics[width=0.20\textwidth,clip=true,trim=0cm 0cm 0cm 0cm]{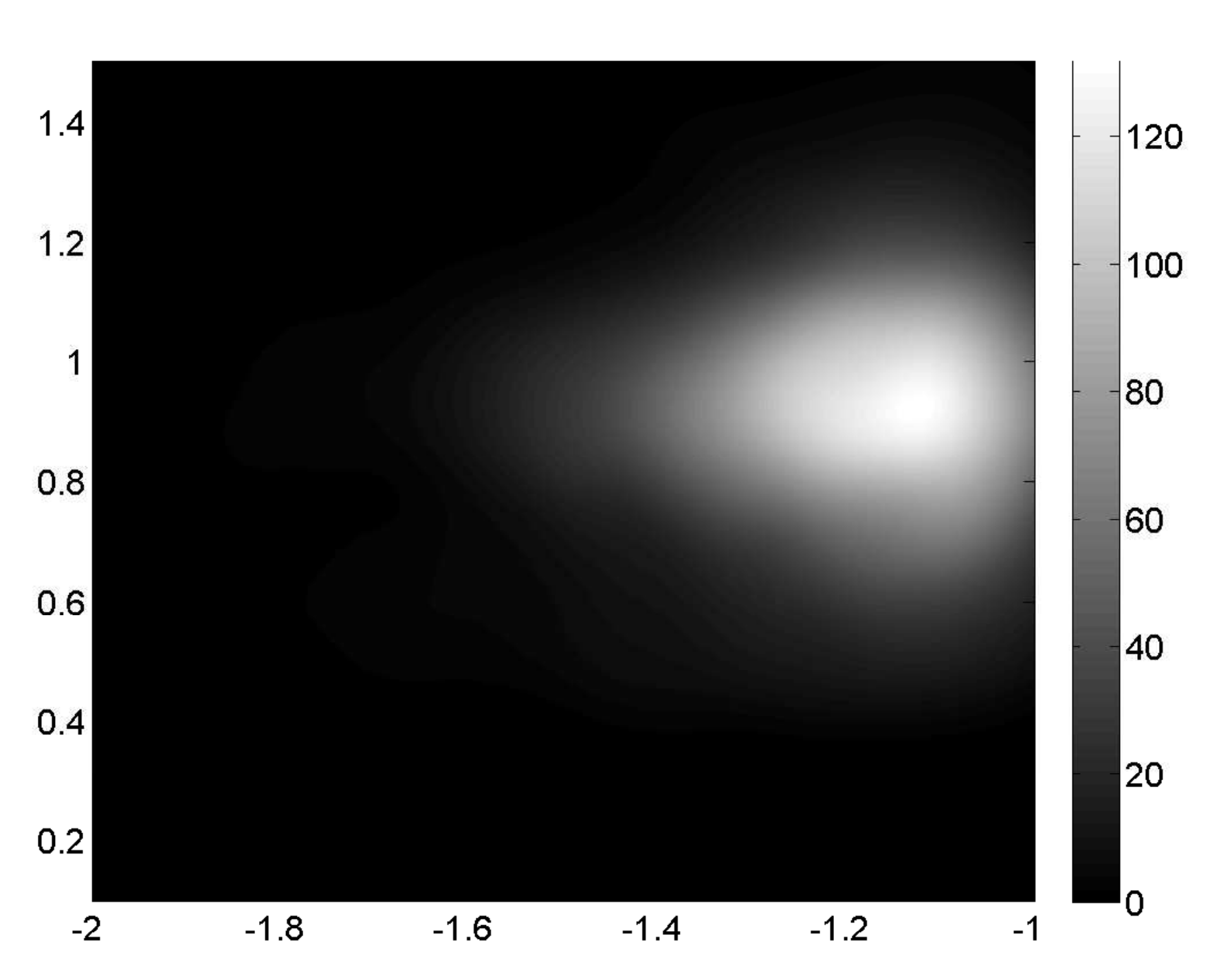} 
\includegraphics[width=0.20\textwidth,clip=true,trim=0cm 0cm 0cm 0cm]{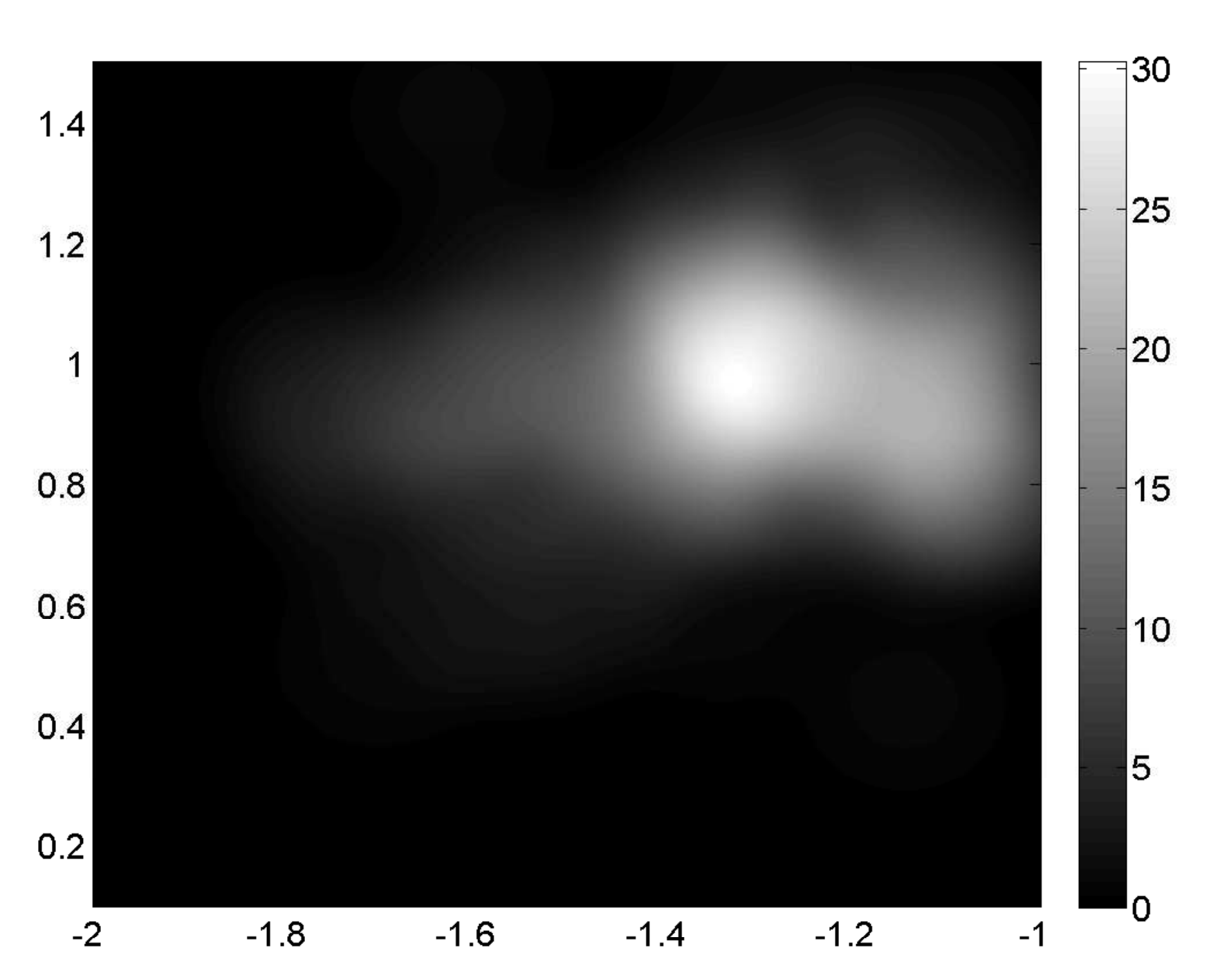} \\
\includegraphics[width=0.20\textwidth,clip=true,trim=0cm 0cm 0cm 0cm]{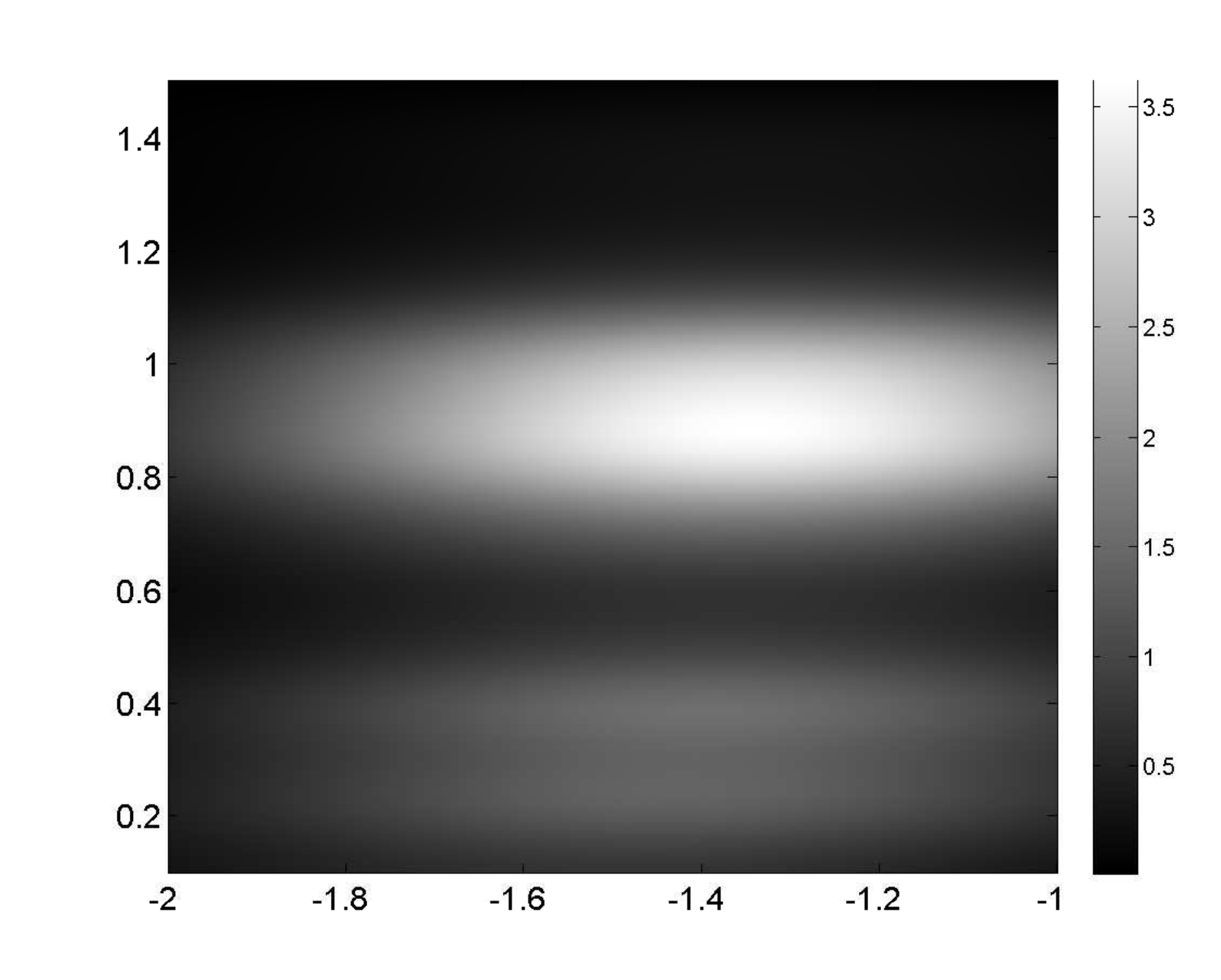} 
\includegraphics[width=0.20\textwidth,clip=true,trim=0cm 0cm 0cm 0cm]{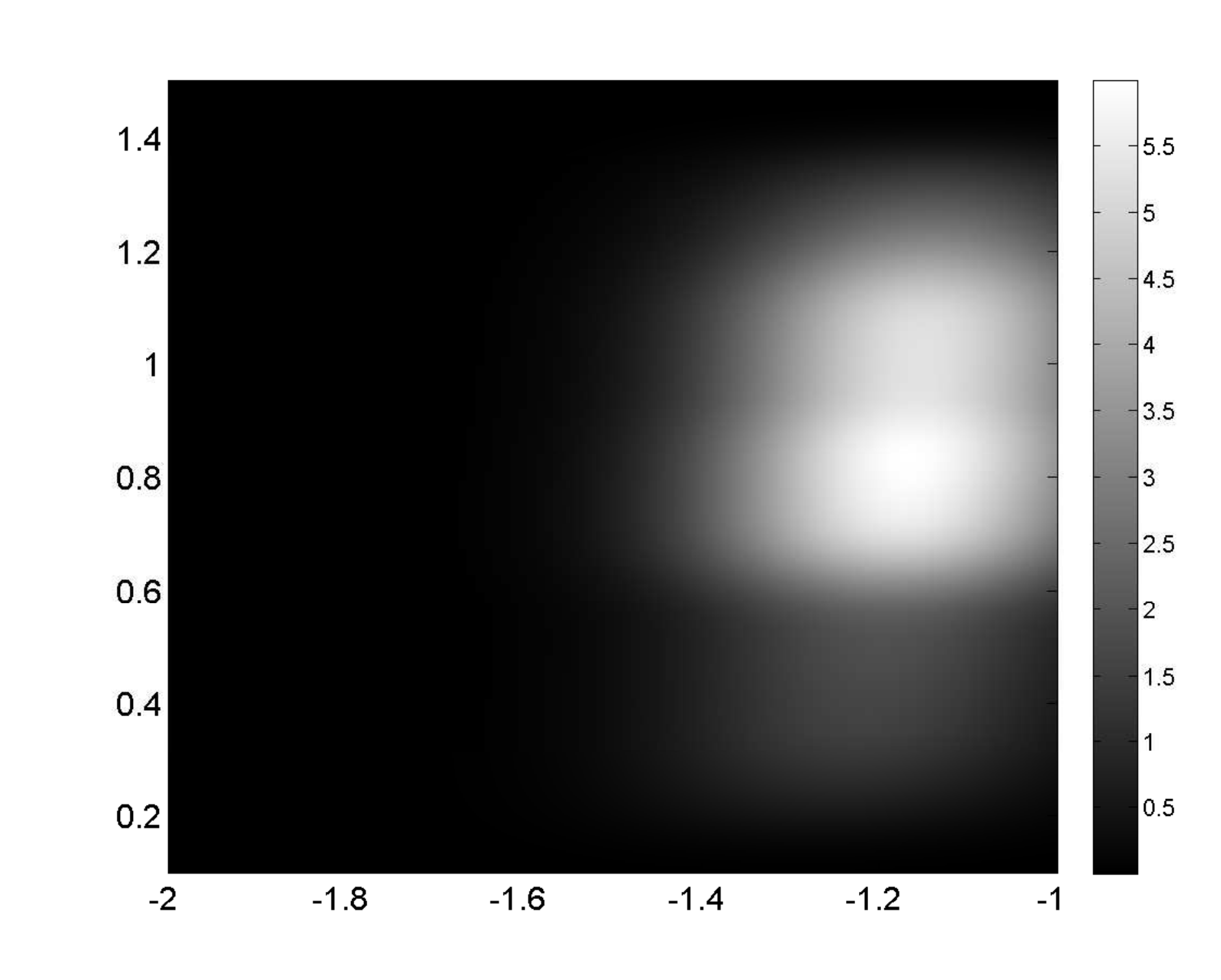} 
\includegraphics[width=0.20\textwidth,clip=true,trim=0cm 0cm 0cm 0cm]{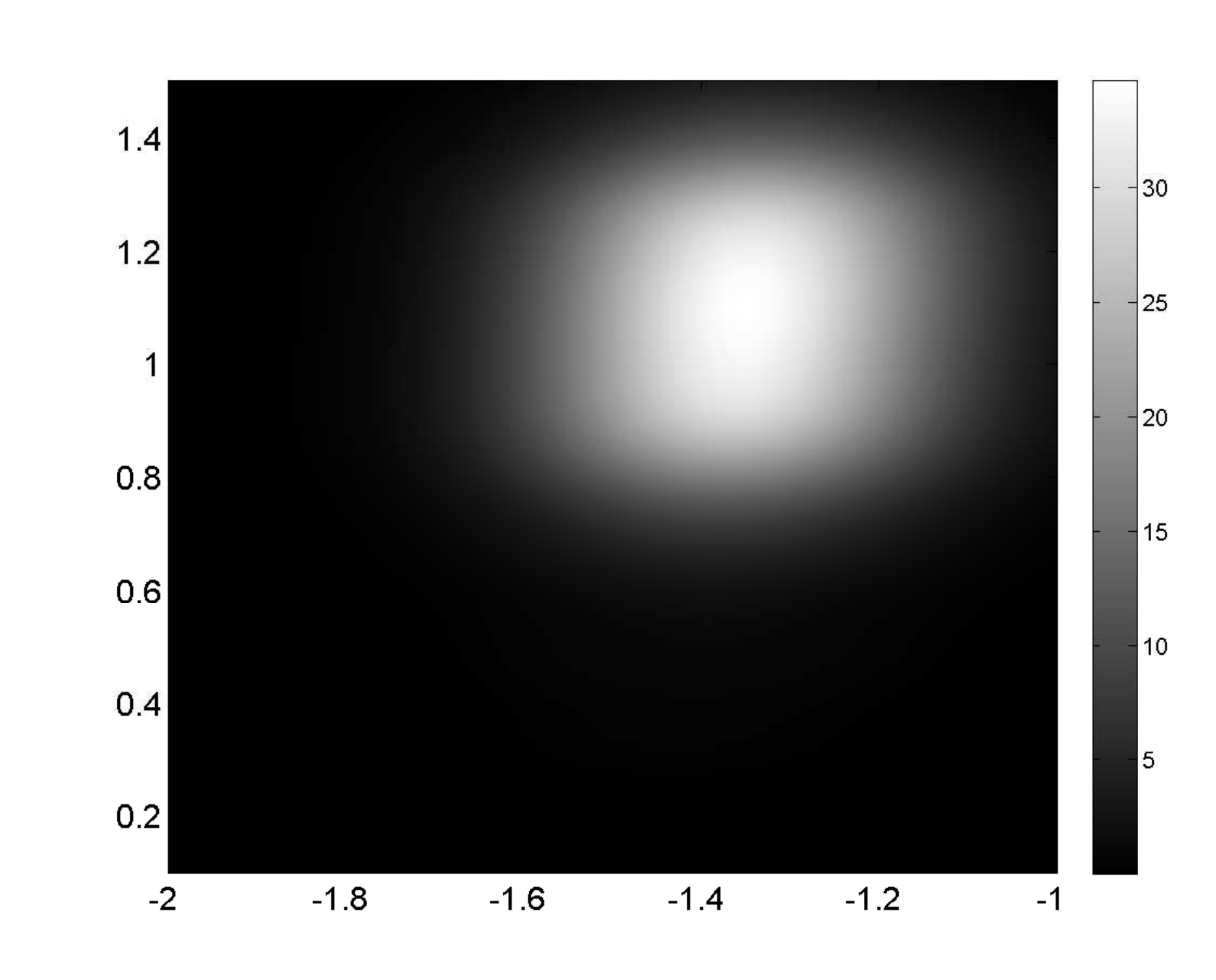}
\includegraphics[width=0.20\textwidth,clip=true,trim=0cm 0cm 0cm 0cm]{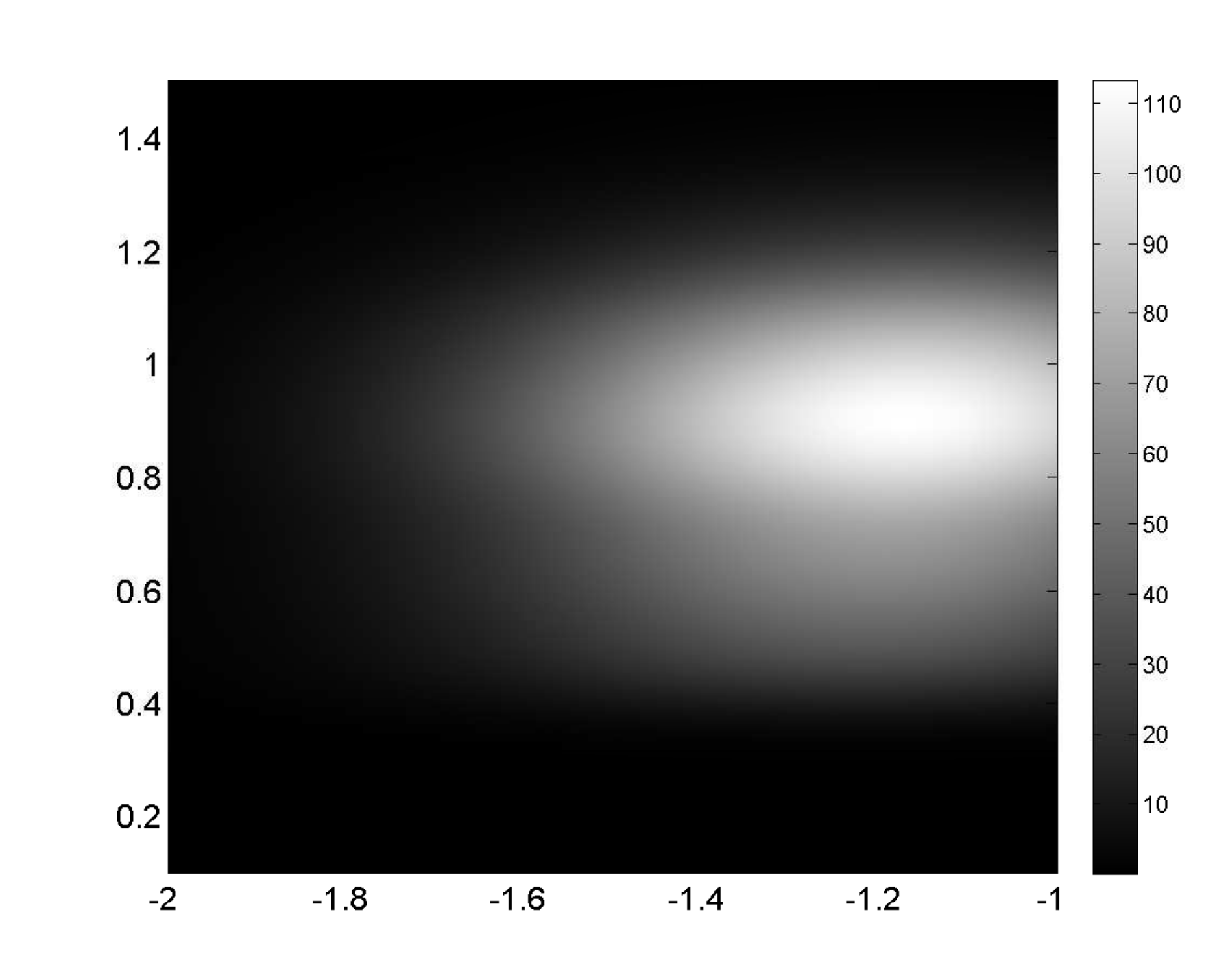} 
\includegraphics[width=0.20\textwidth,clip=true,trim=0cm 0cm 0cm 0cm]{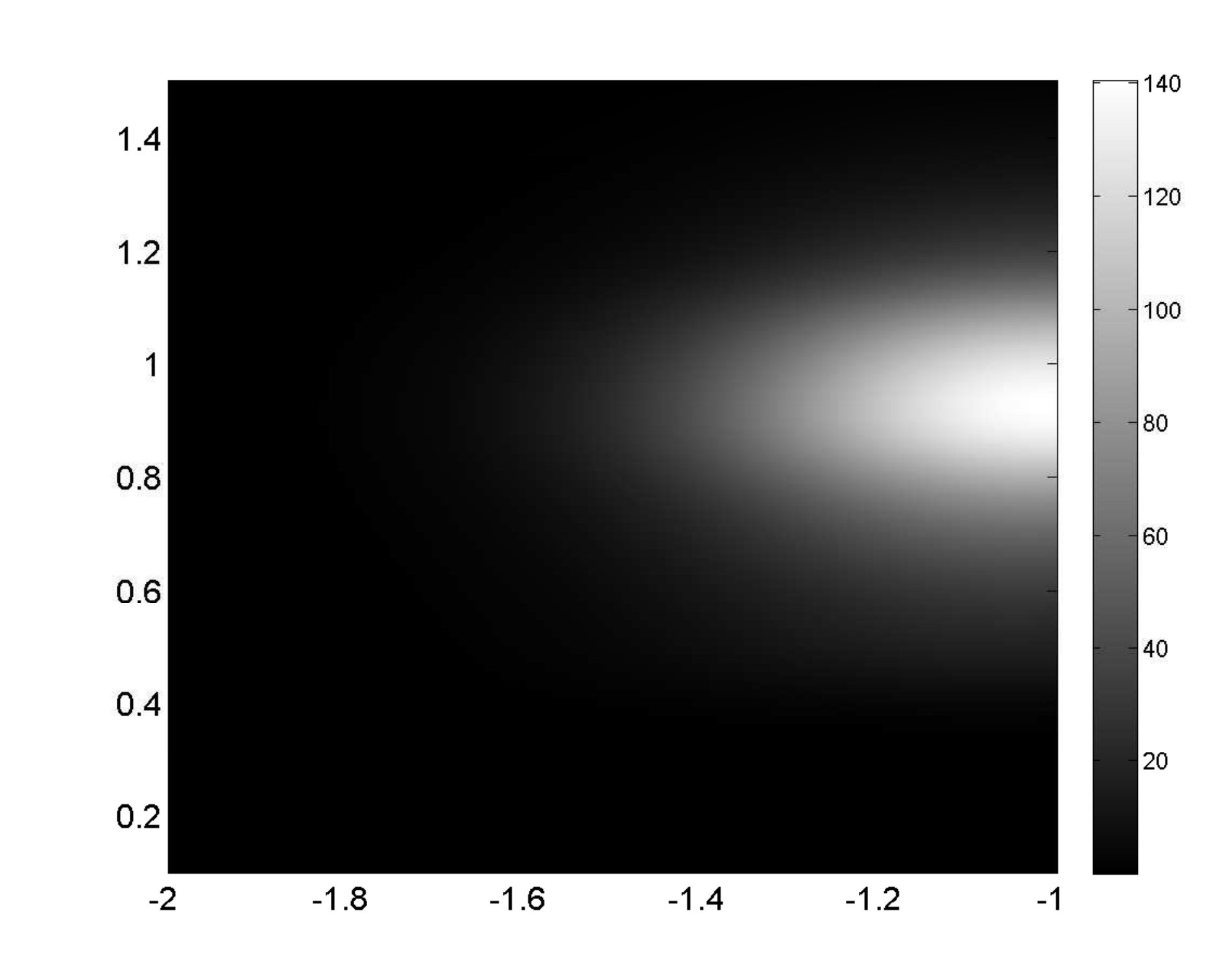} 
\includegraphics[width=0.20\textwidth,clip=true,trim=0cm 0cm 0cm 0cm]{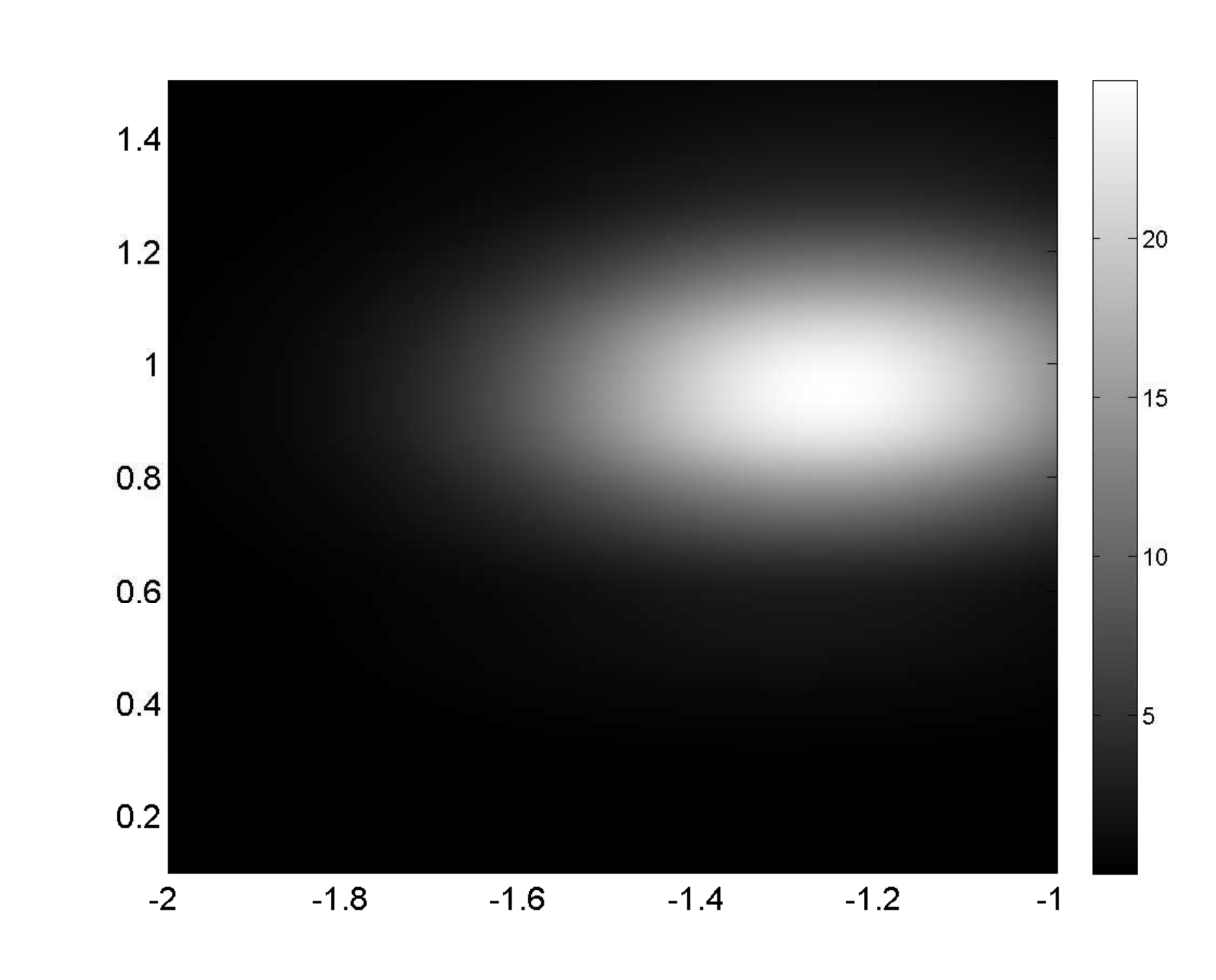} \\
\includegraphics[width=0.20\textwidth,clip=true,trim=0cm 0cm 0cm 0cm]{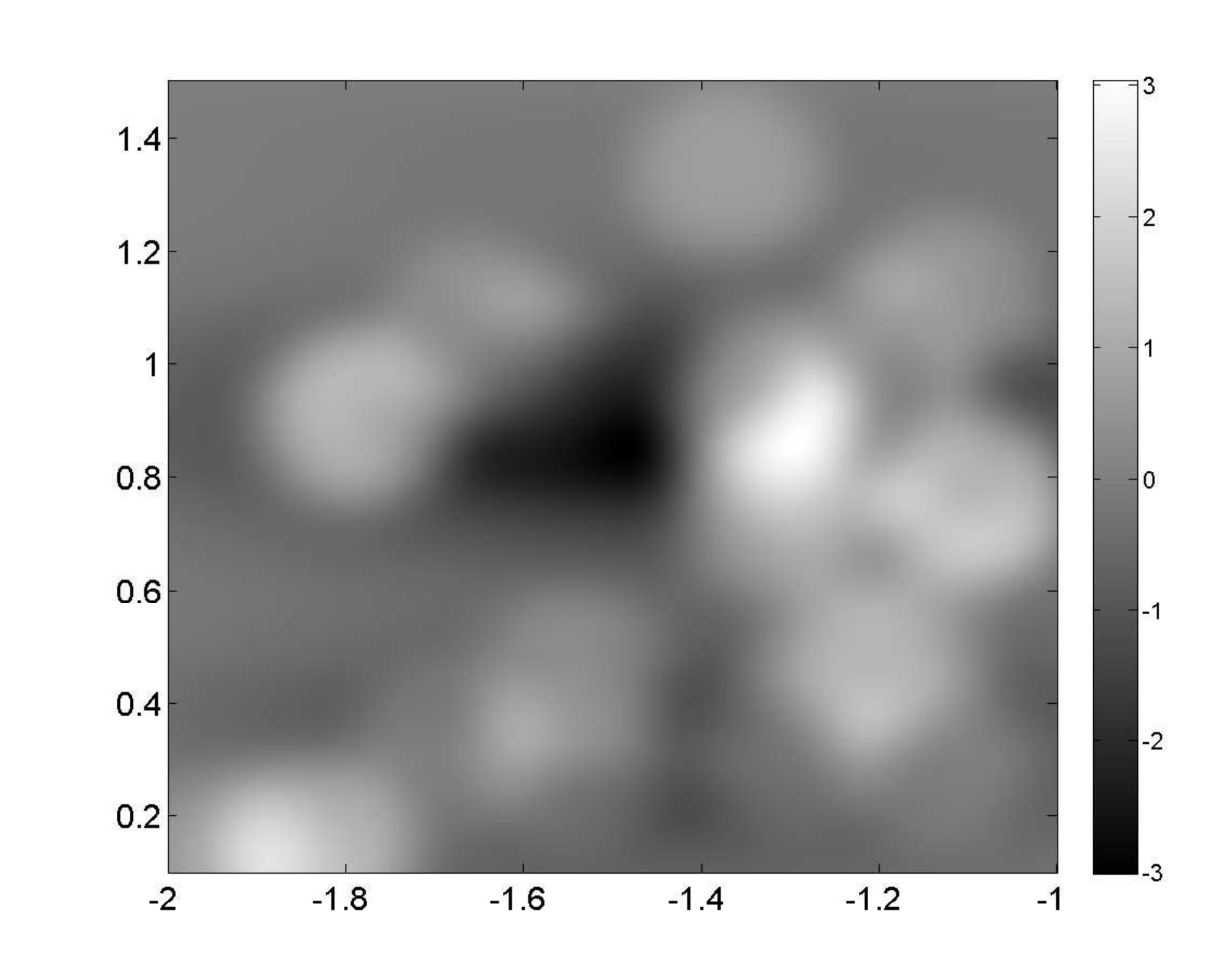} 
\includegraphics[width=0.20\textwidth,clip=true,trim=0cm 0cm 0cm 0cm]{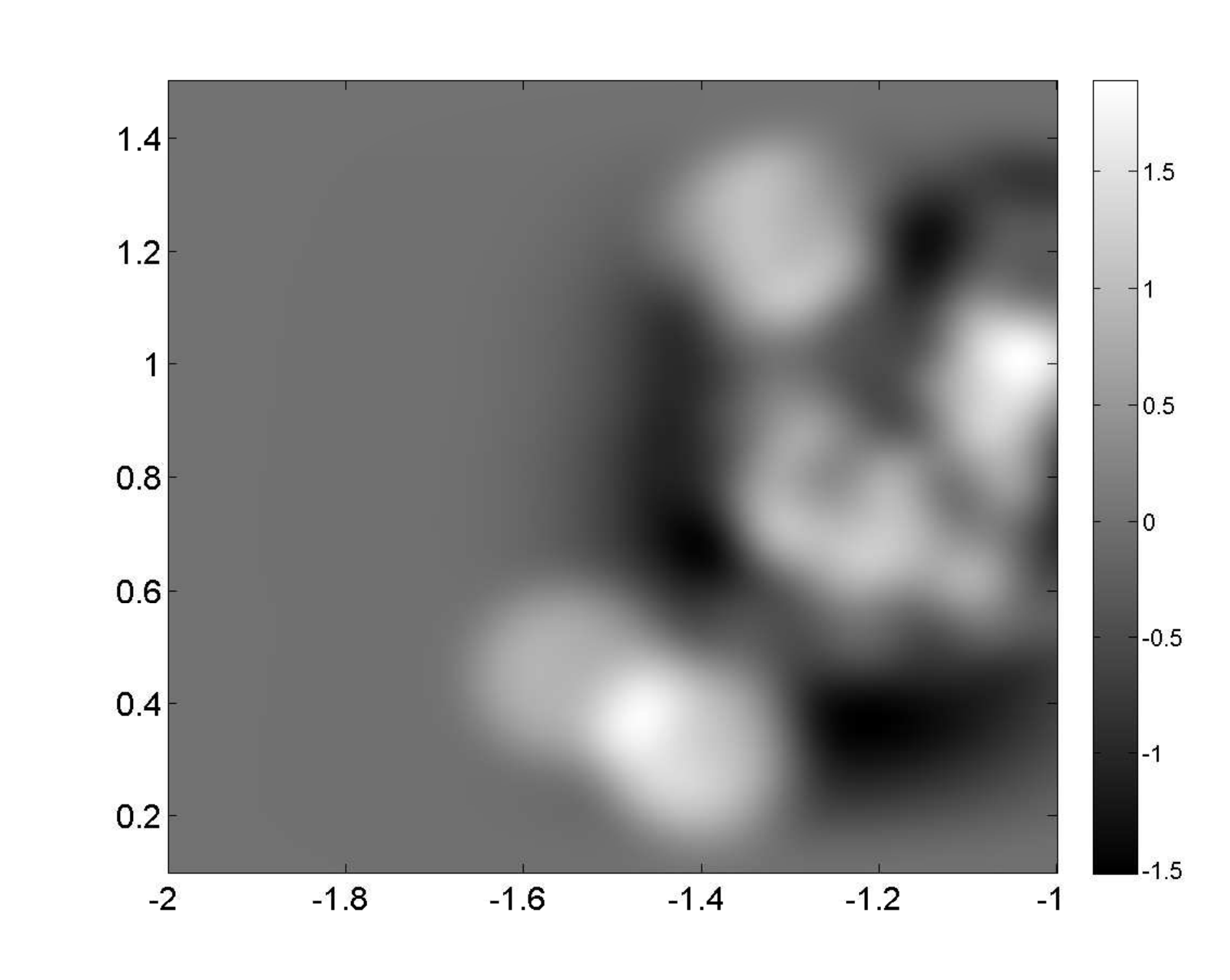} 
\includegraphics[width=0.20\textwidth,clip=true,trim=0cm 0cm 0cm 0cm]{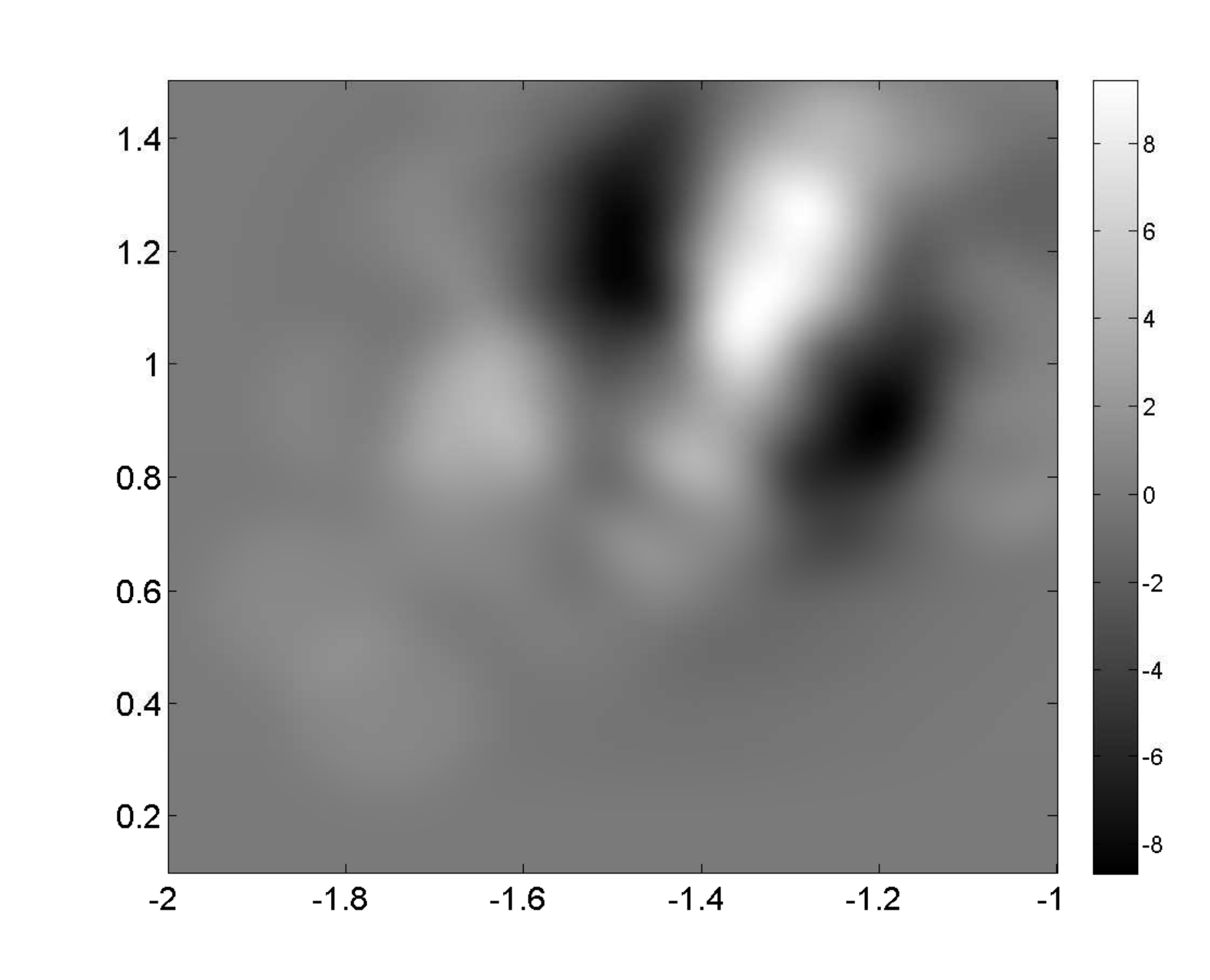} 
\includegraphics[width=0.20\textwidth,clip=true,trim=0cm 0cm 0cm 0cm]{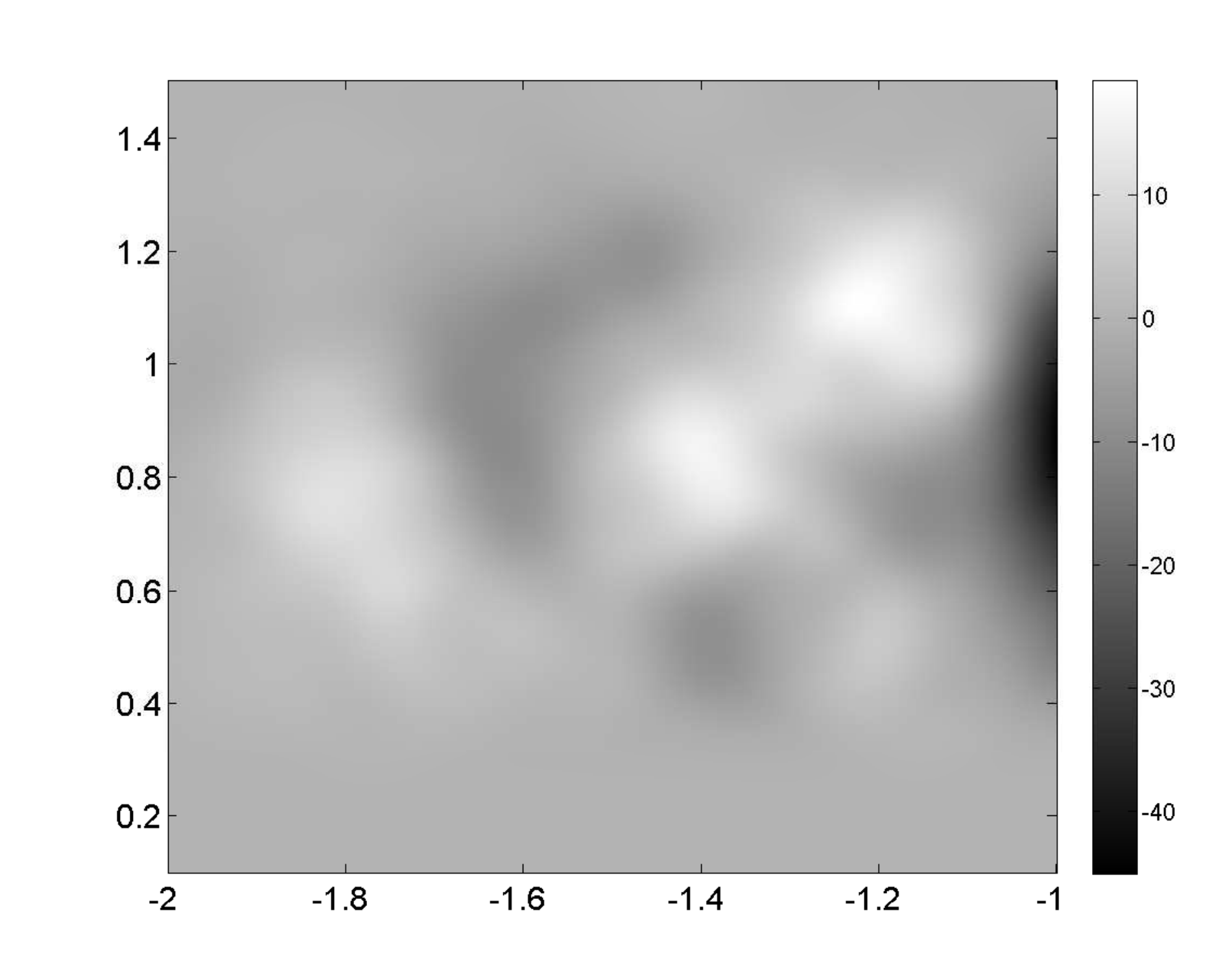} 
\includegraphics[width=0.20\textwidth,clip=true,trim=0cm 0cm 0cm 0cm]{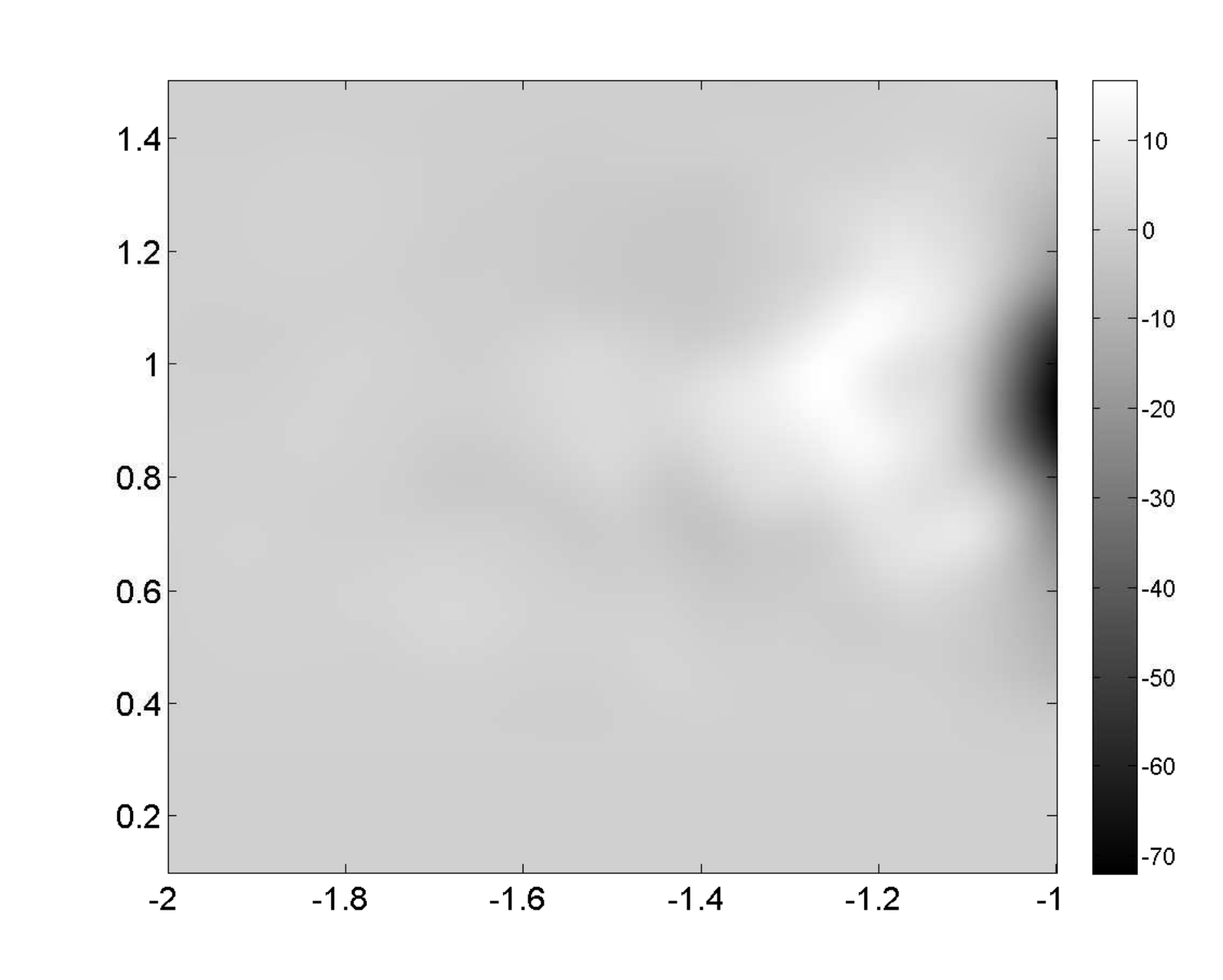} 
\includegraphics[width=0.20\textwidth,clip=true,trim=0cm 0cm 0cm 0cm]{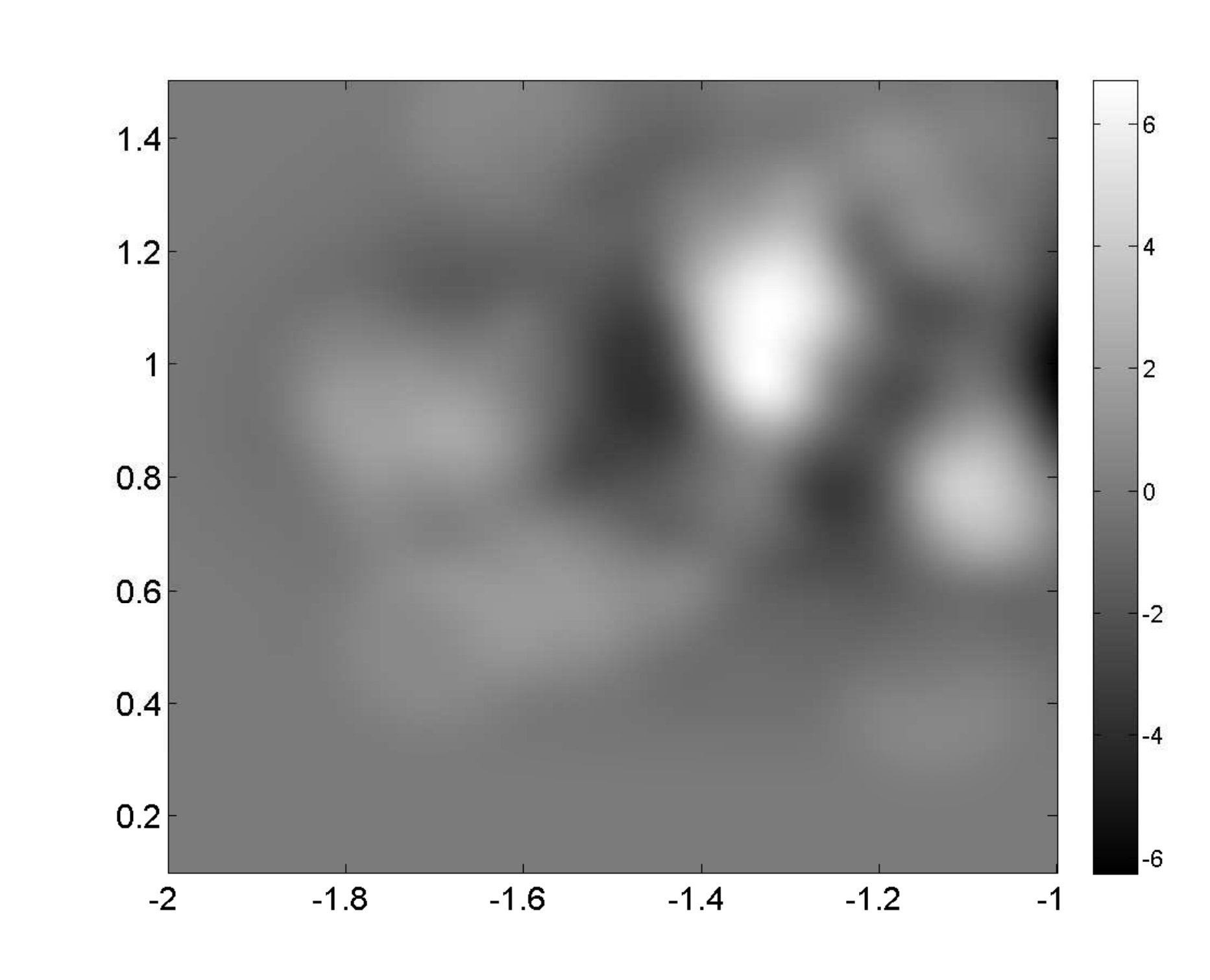} 
\caption{Top: Empirical density functions as in the second row of Figure 1.  Middle: Model probability density function with optimal parameters in Table 1 for the halting model at the interior 1:2 orbital resonance with the magnetospheric disk truncation radius.  Bottom: Subtracted residuals.}
\end{figure}
\end{landscape}

\clearpage
\begin{landscape}
\begin{figure}
\centering
\includegraphics[width=0.20\textwidth,clip=true,trim=0cm 0cm 0cm 0cm]{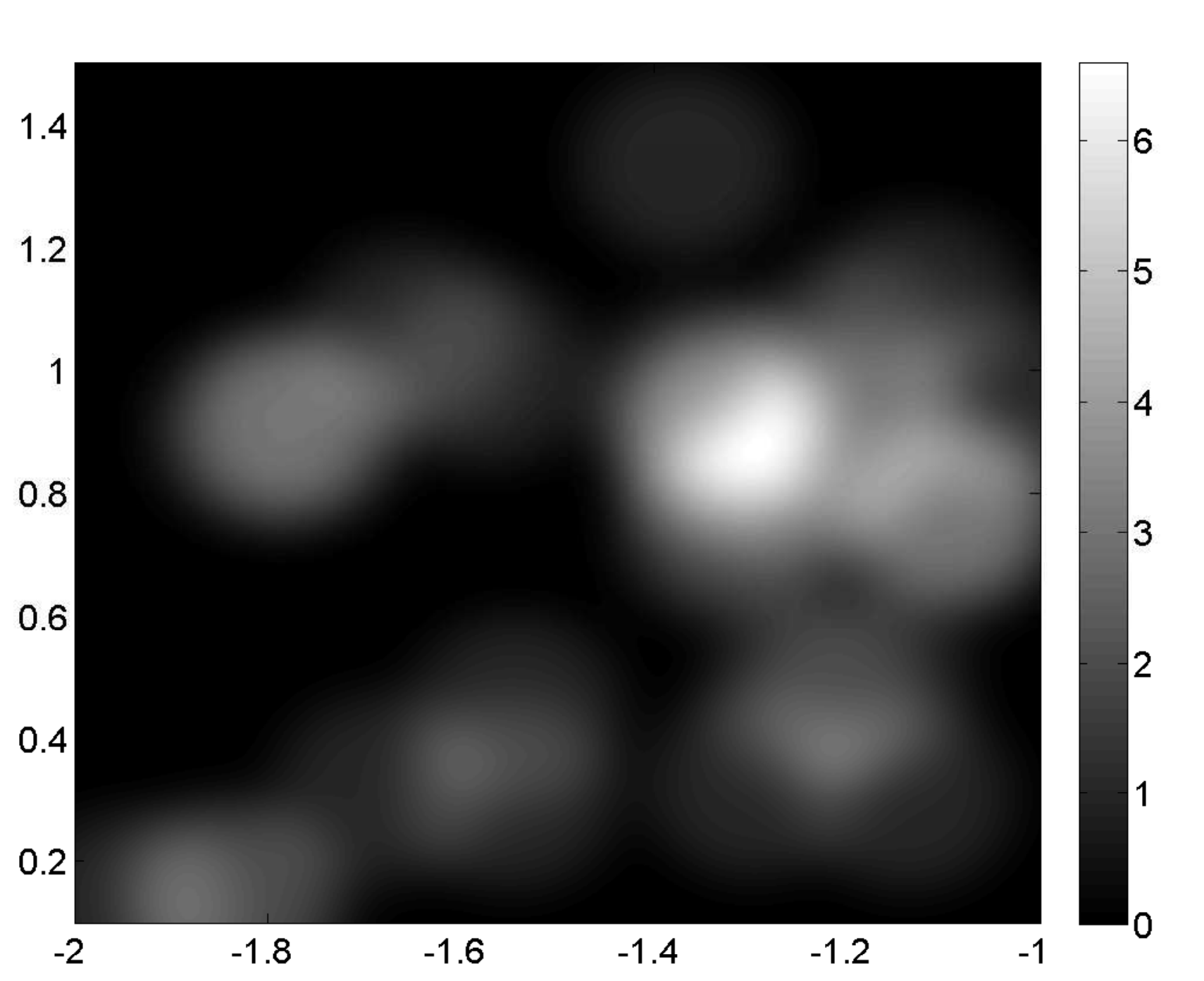} 
\includegraphics[width=0.20\textwidth,clip=true,trim=0cm 0cm 0cm 0cm]{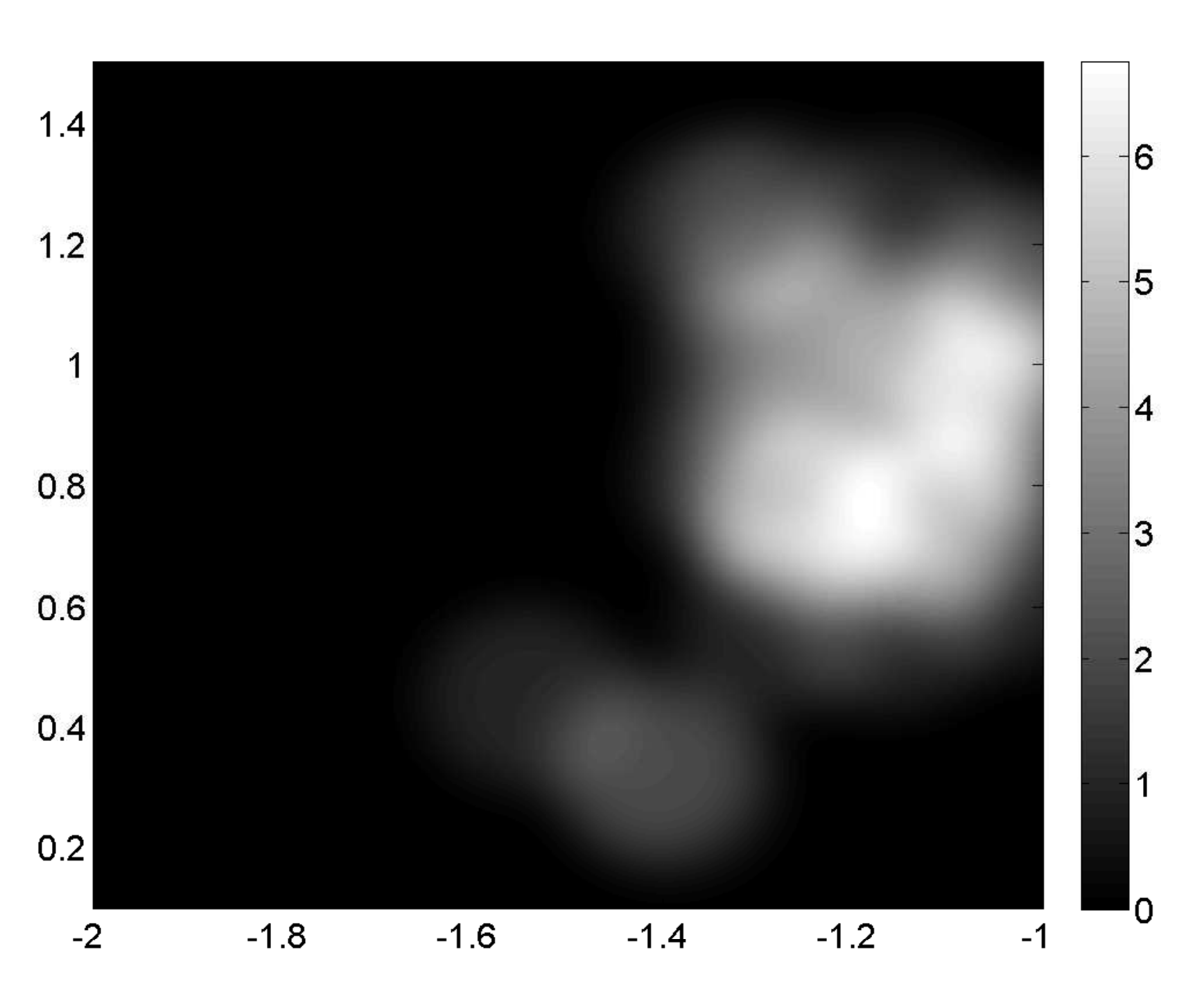} 
\includegraphics[width=0.20\textwidth,clip=true,trim=0cm 0cm 0cm 0cm]{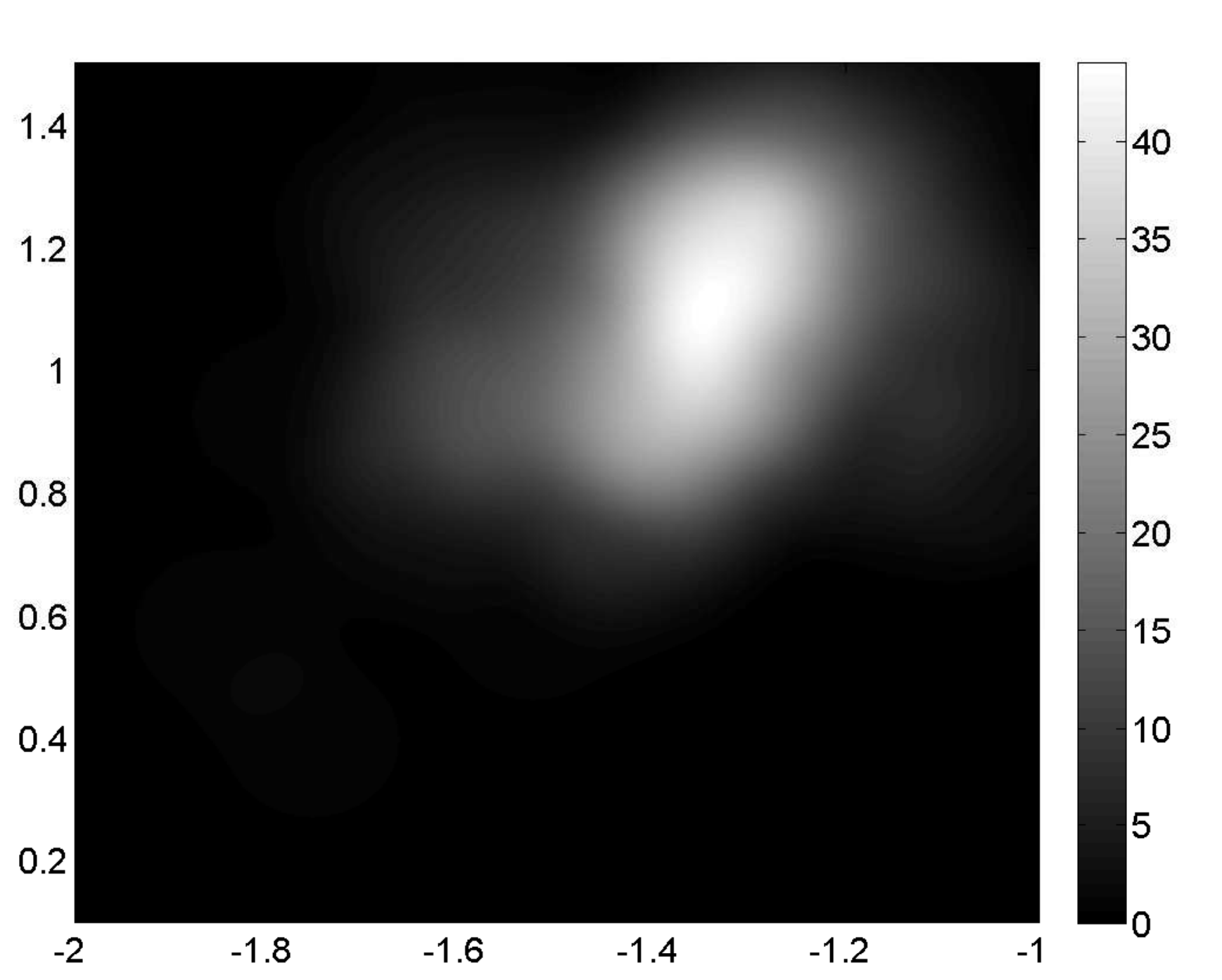} 
\includegraphics[width=0.20\textwidth,clip=true,trim=0cm 0cm 0cm 0cm]{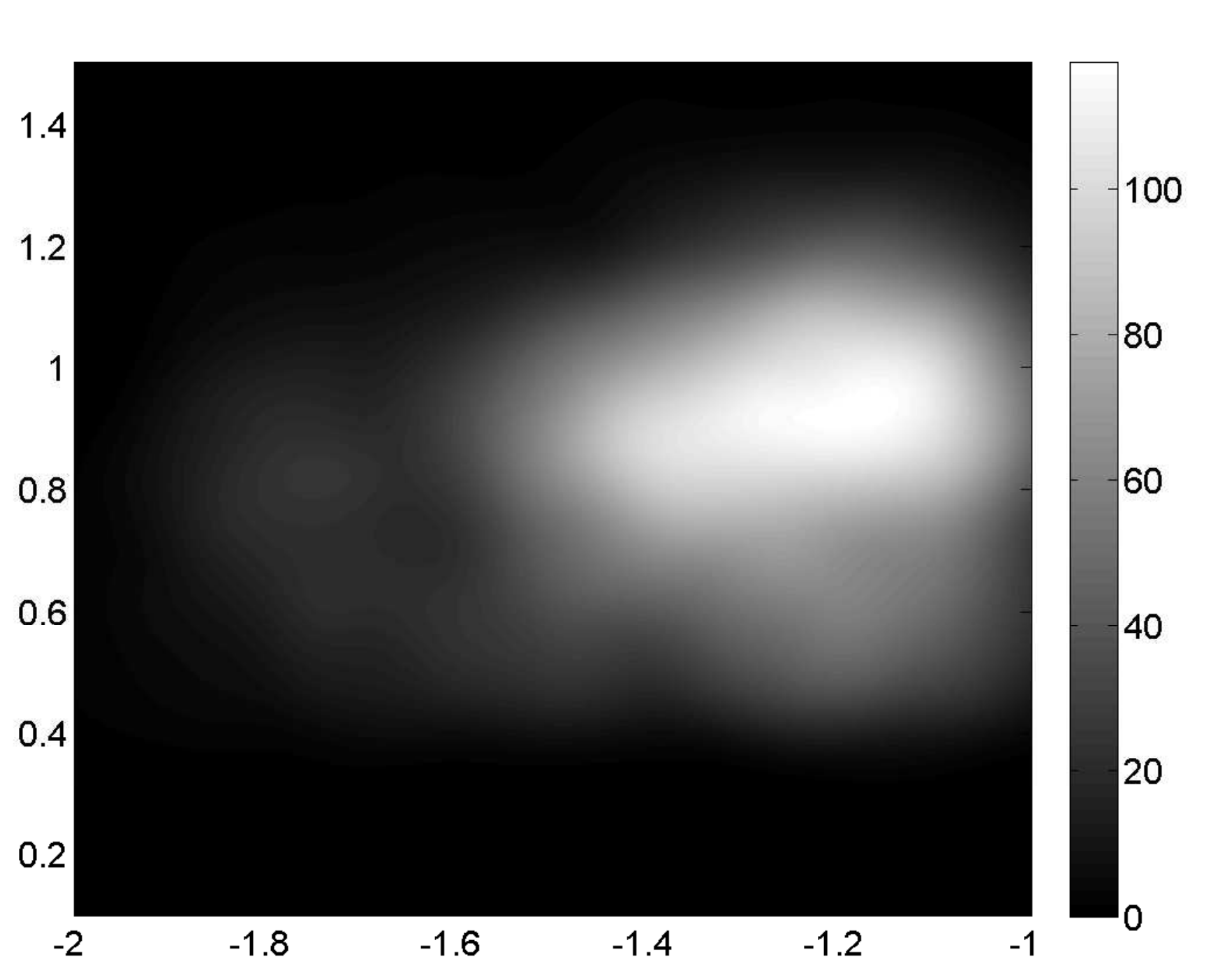} 
\includegraphics[width=0.20\textwidth,clip=true,trim=0cm 0cm 0cm 0cm]{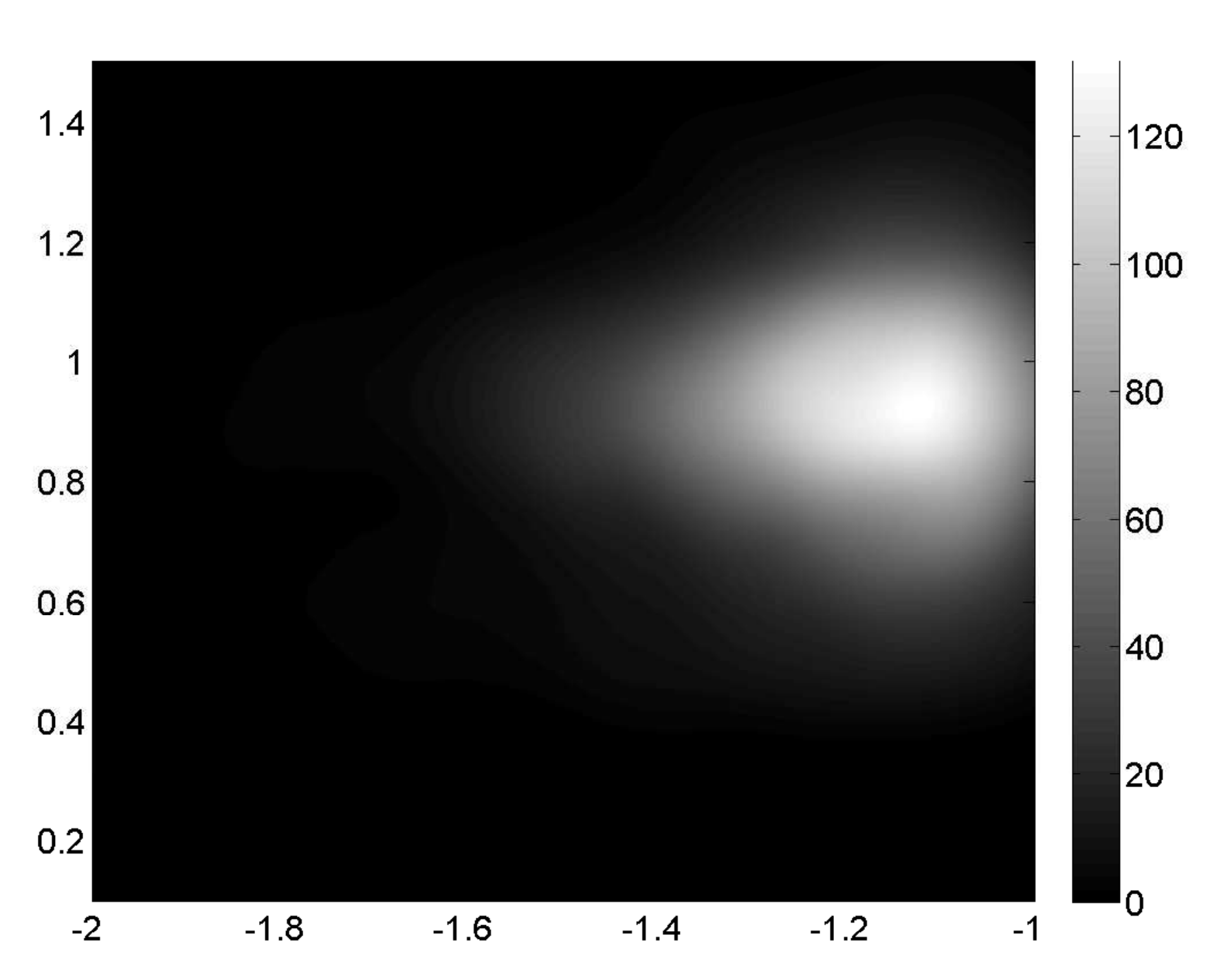} 
\includegraphics[width=0.20\textwidth,clip=true,trim=0cm 0cm 0cm 0cm]{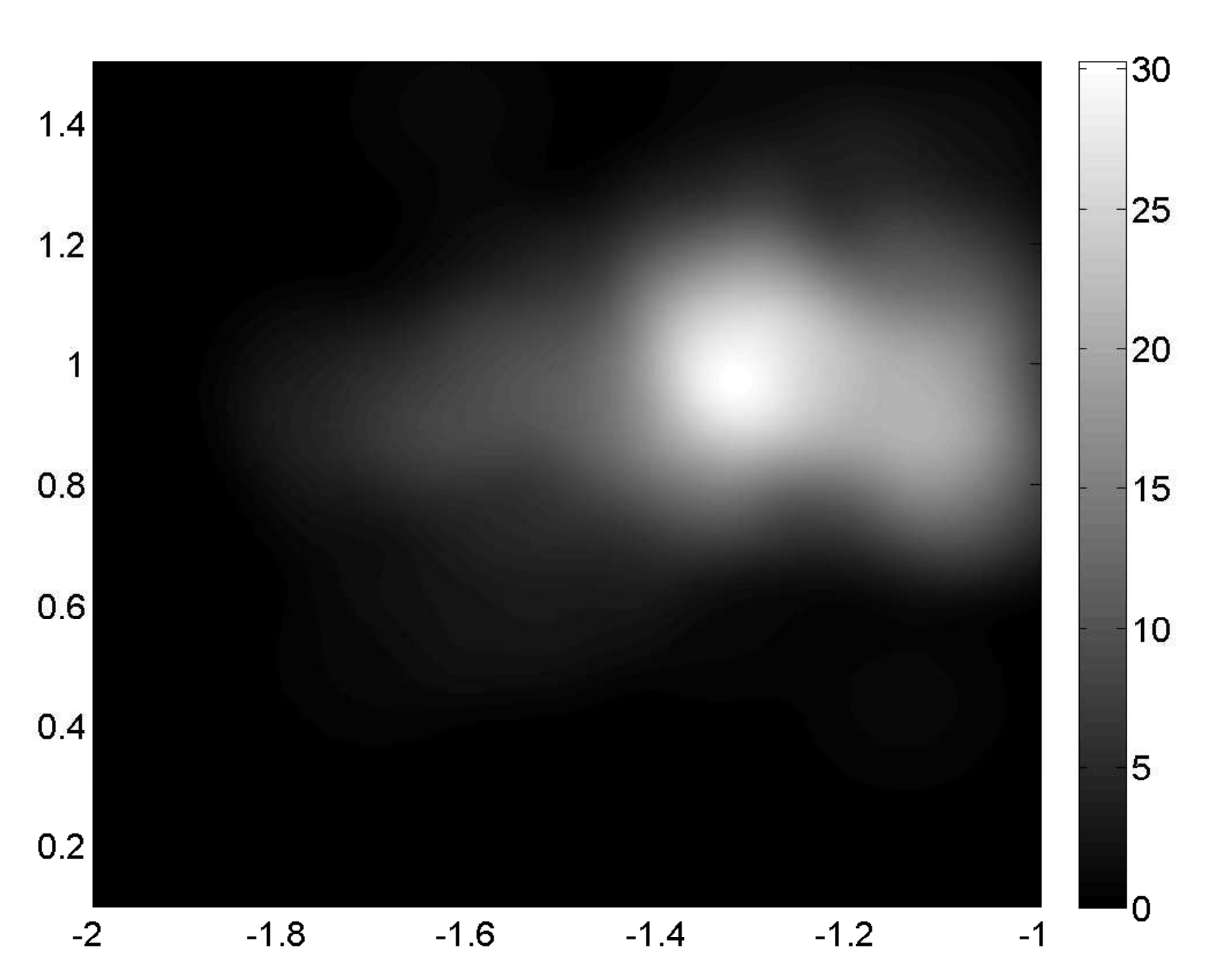} \\
\includegraphics[width=0.20\textwidth,clip=true,trim=0cm 0cm 0cm 0cm]{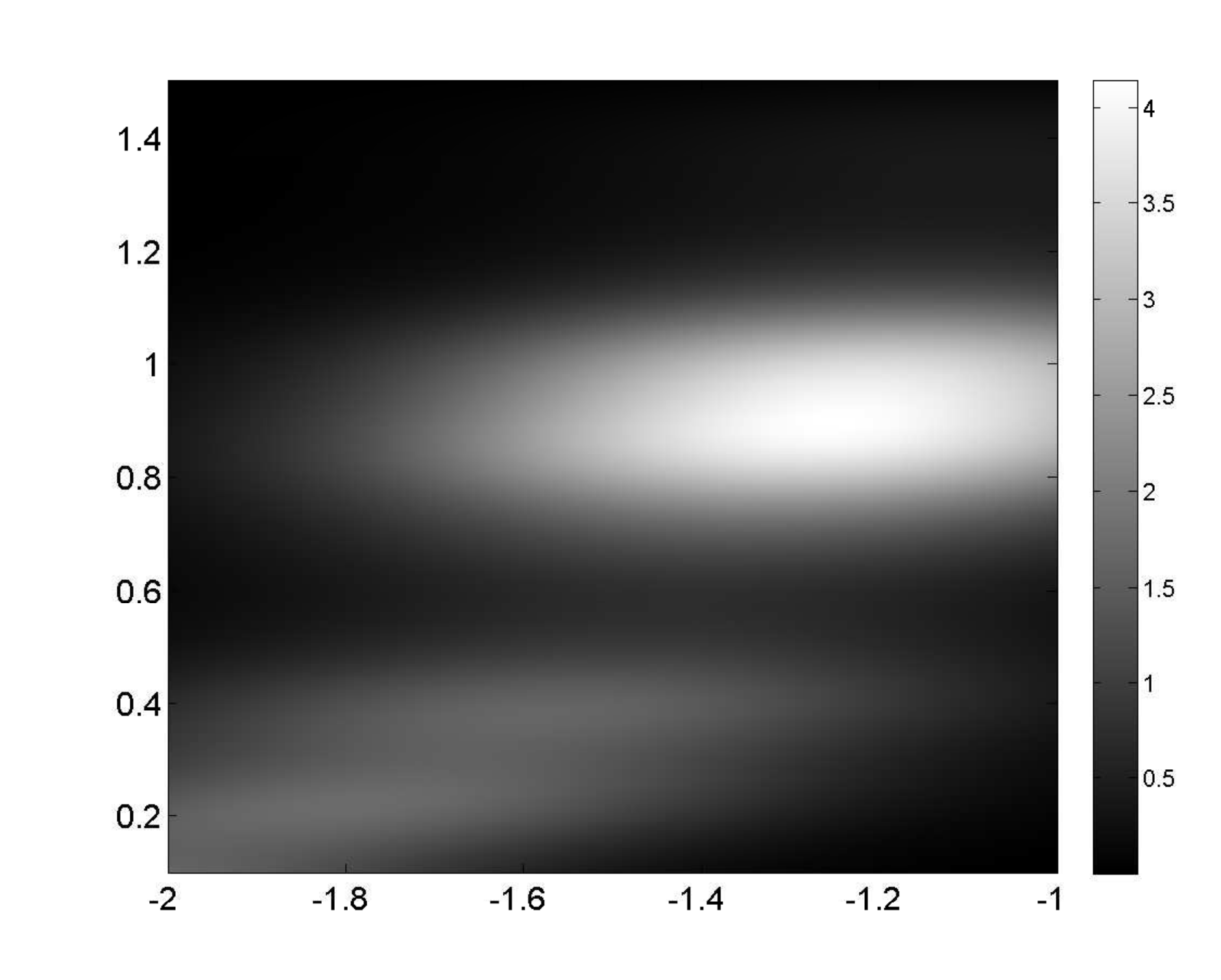} 
\includegraphics[width=0.20\textwidth,clip=true,trim=0cm 0cm 0cm 0cm]{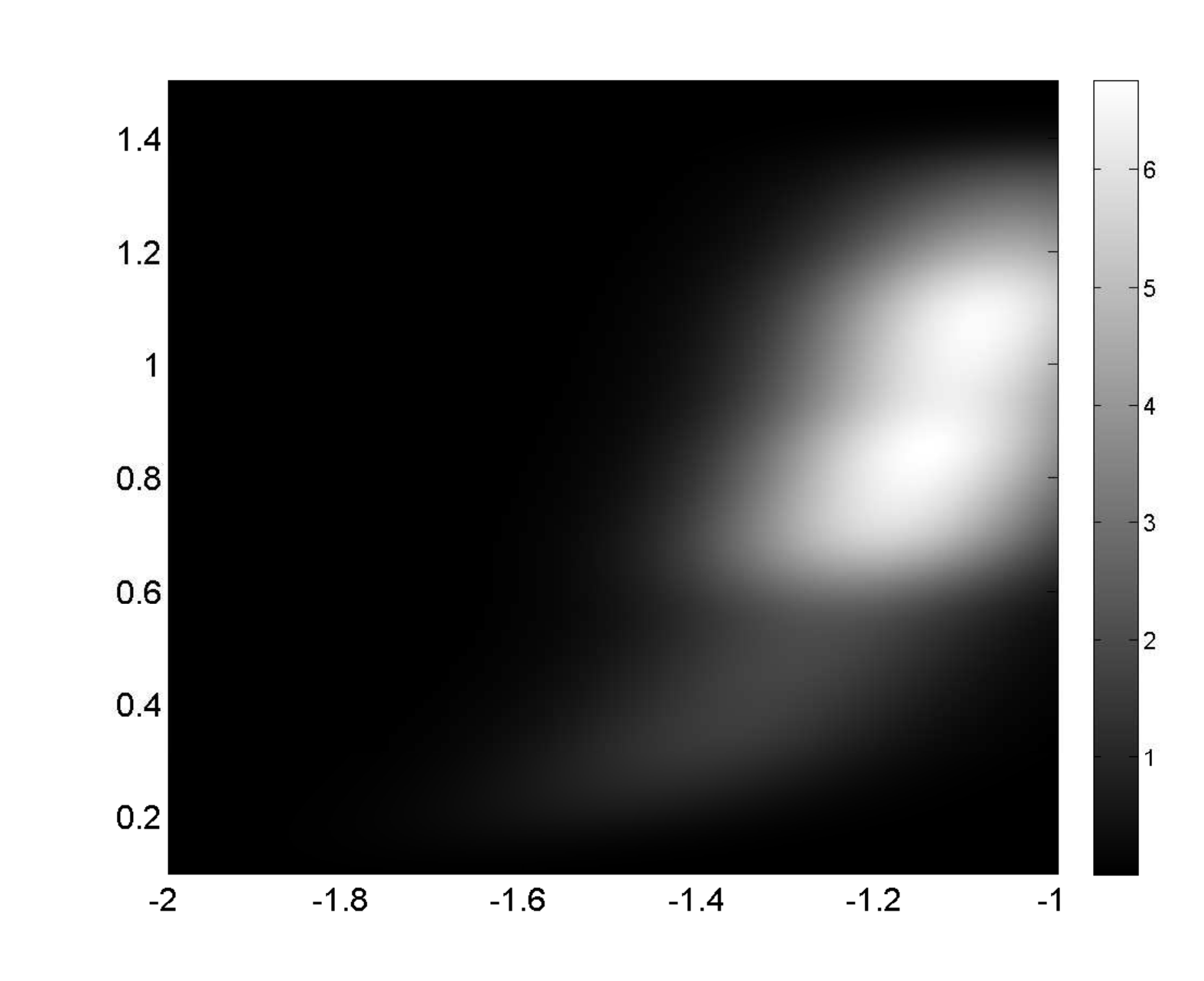} 
\includegraphics[width=0.20\textwidth,clip=true,trim=0cm 0cm 0cm 0cm]{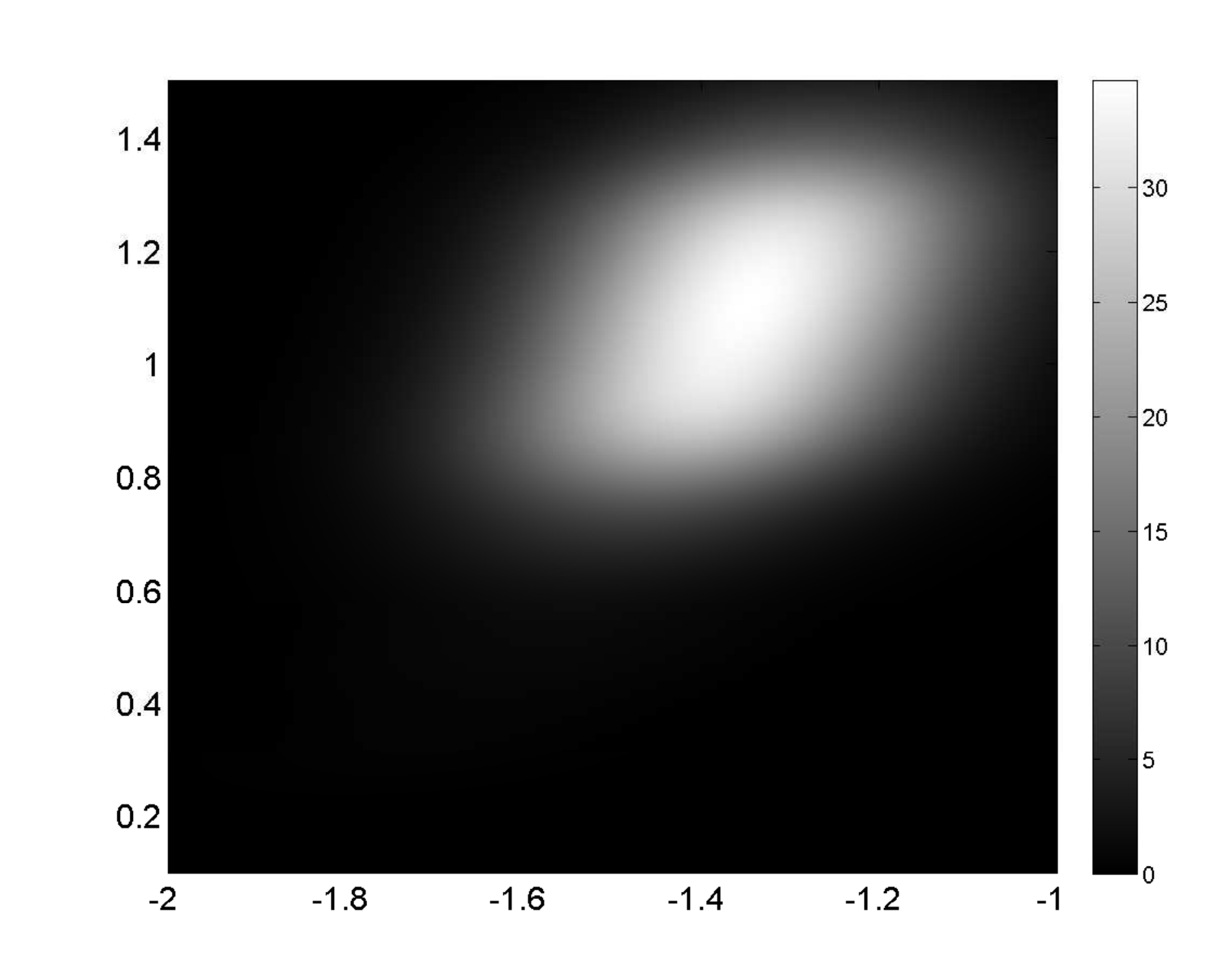}
\includegraphics[width=0.20\textwidth,clip=true,trim=0cm 0cm 0cm 0cm]{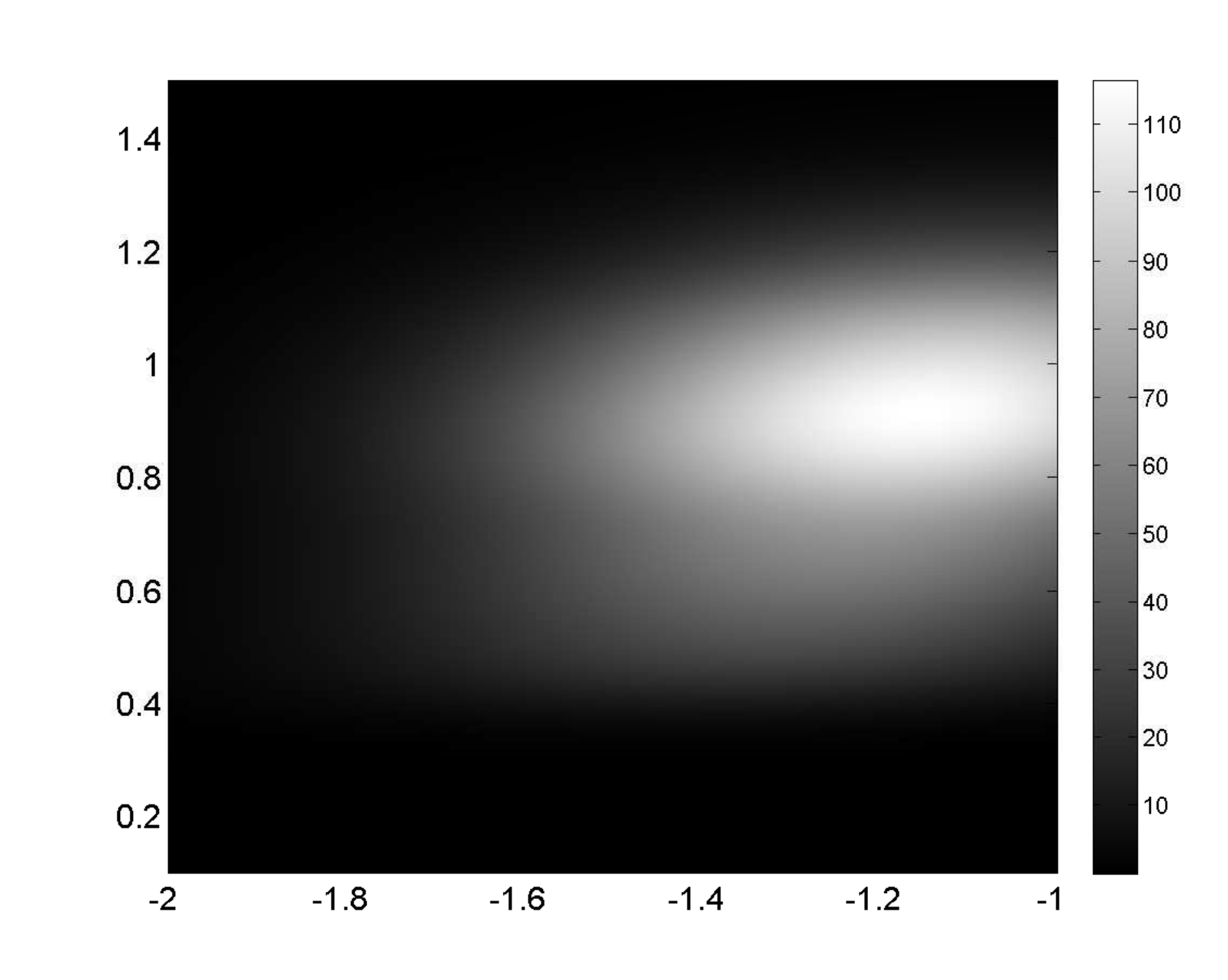} 
\includegraphics[width=0.20\textwidth,clip=true,trim=0cm 0cm 0cm 0cm]{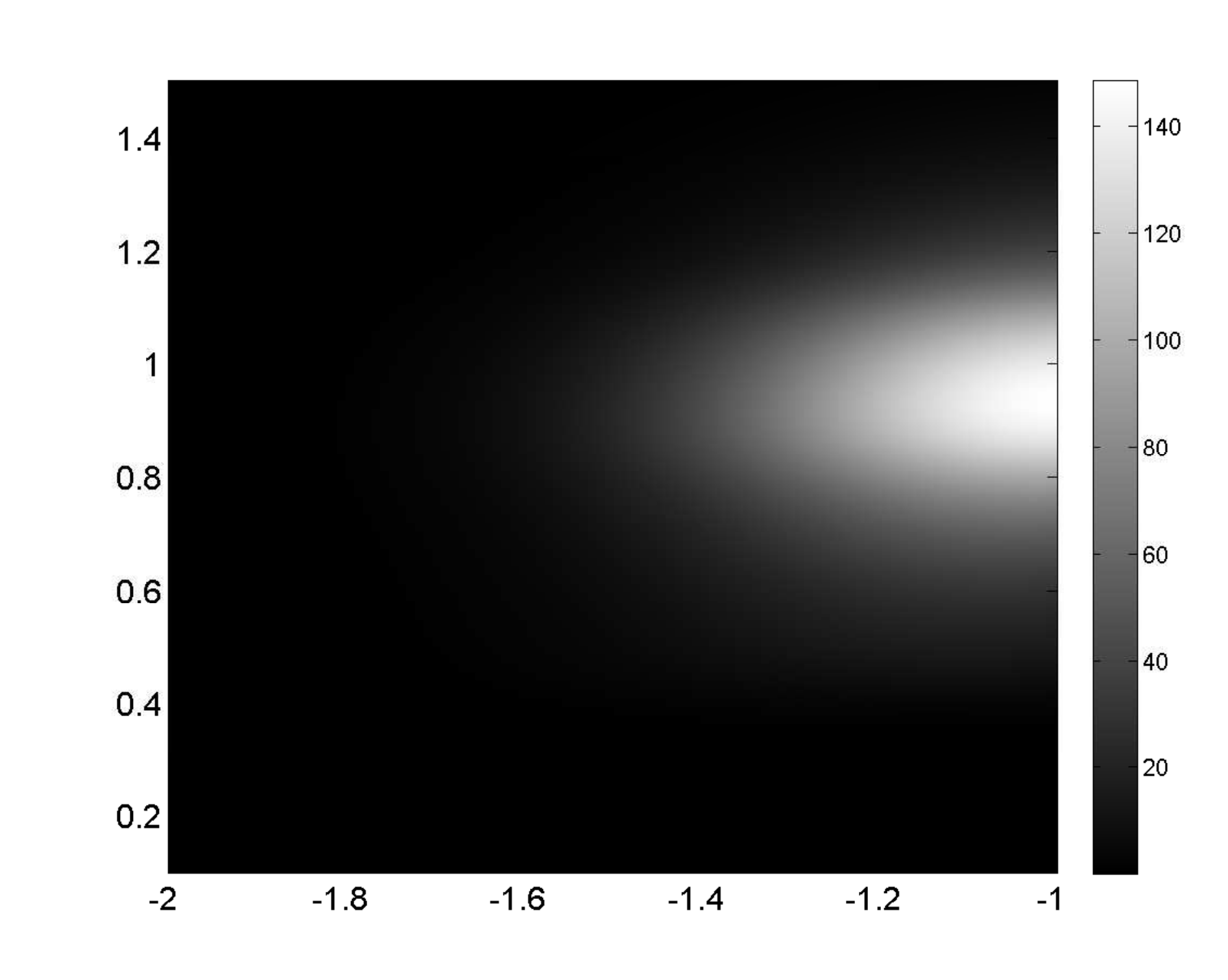} 
\includegraphics[width=0.20\textwidth,clip=true,trim=0cm 0cm 0cm 0cm]{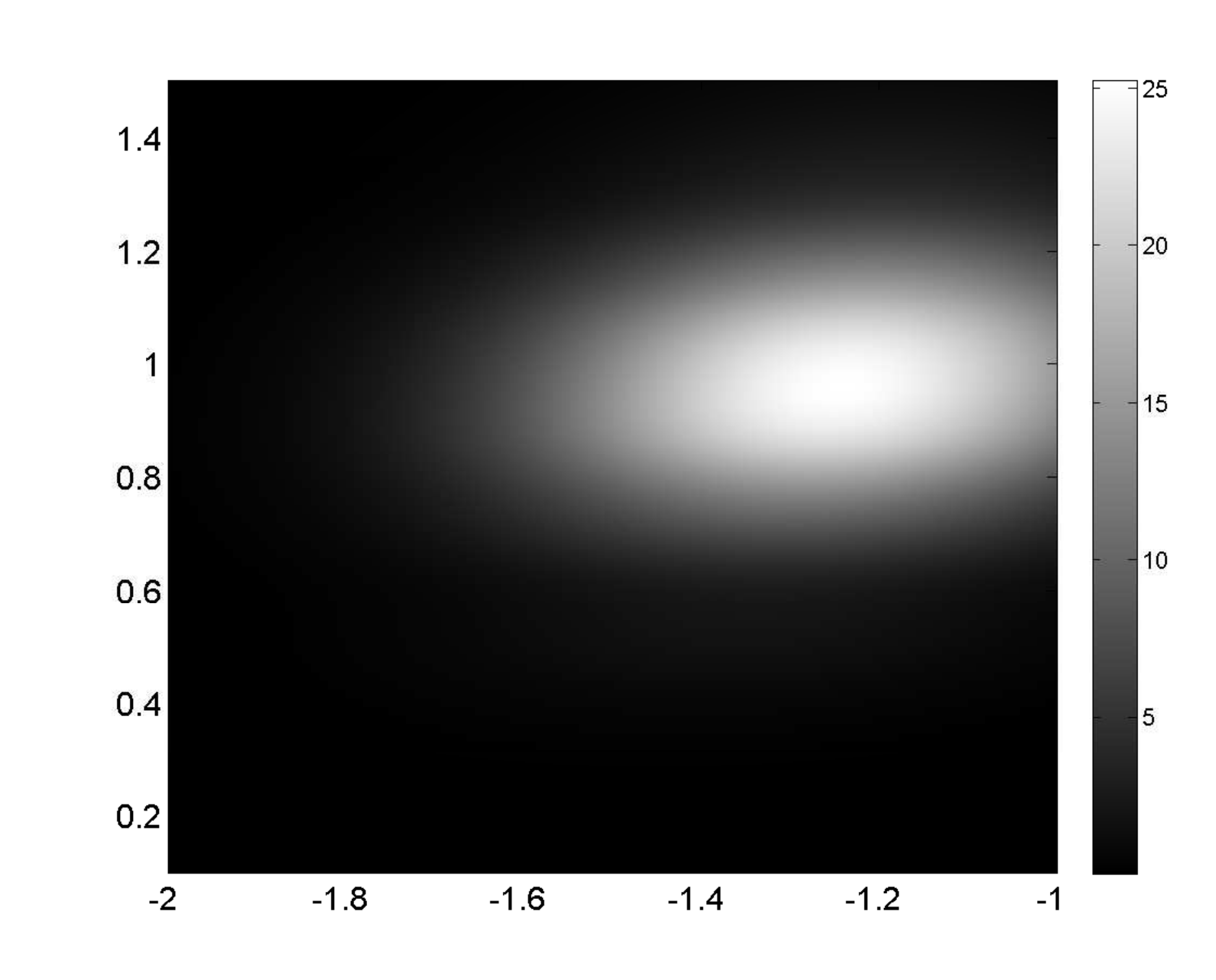} \\
\includegraphics[width=0.20\textwidth,clip=true,trim=0cm 0cm 0cm 0cm]{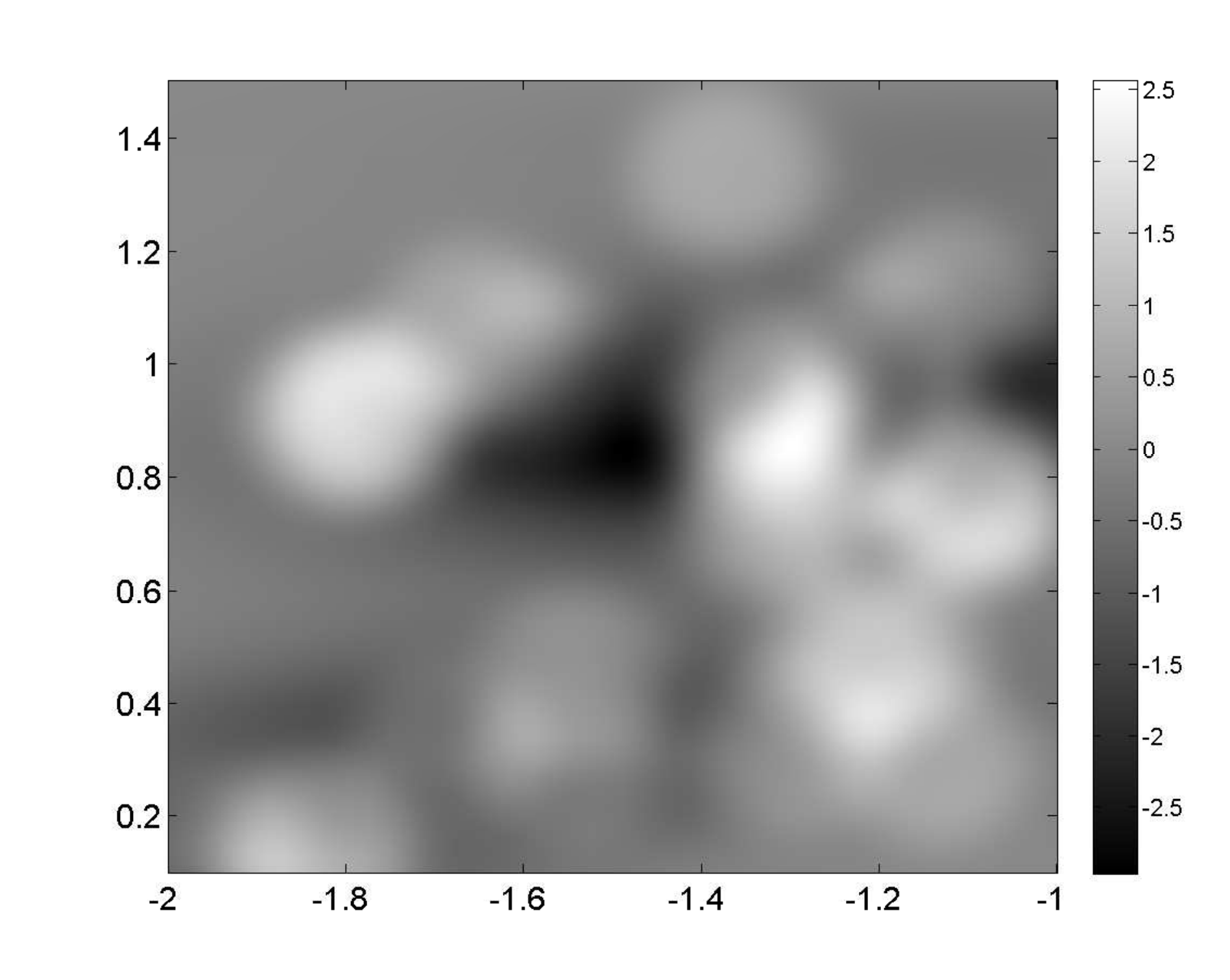} 
\includegraphics[width=0.20\textwidth,clip=true,trim=0cm 0cm 0cm 0cm]{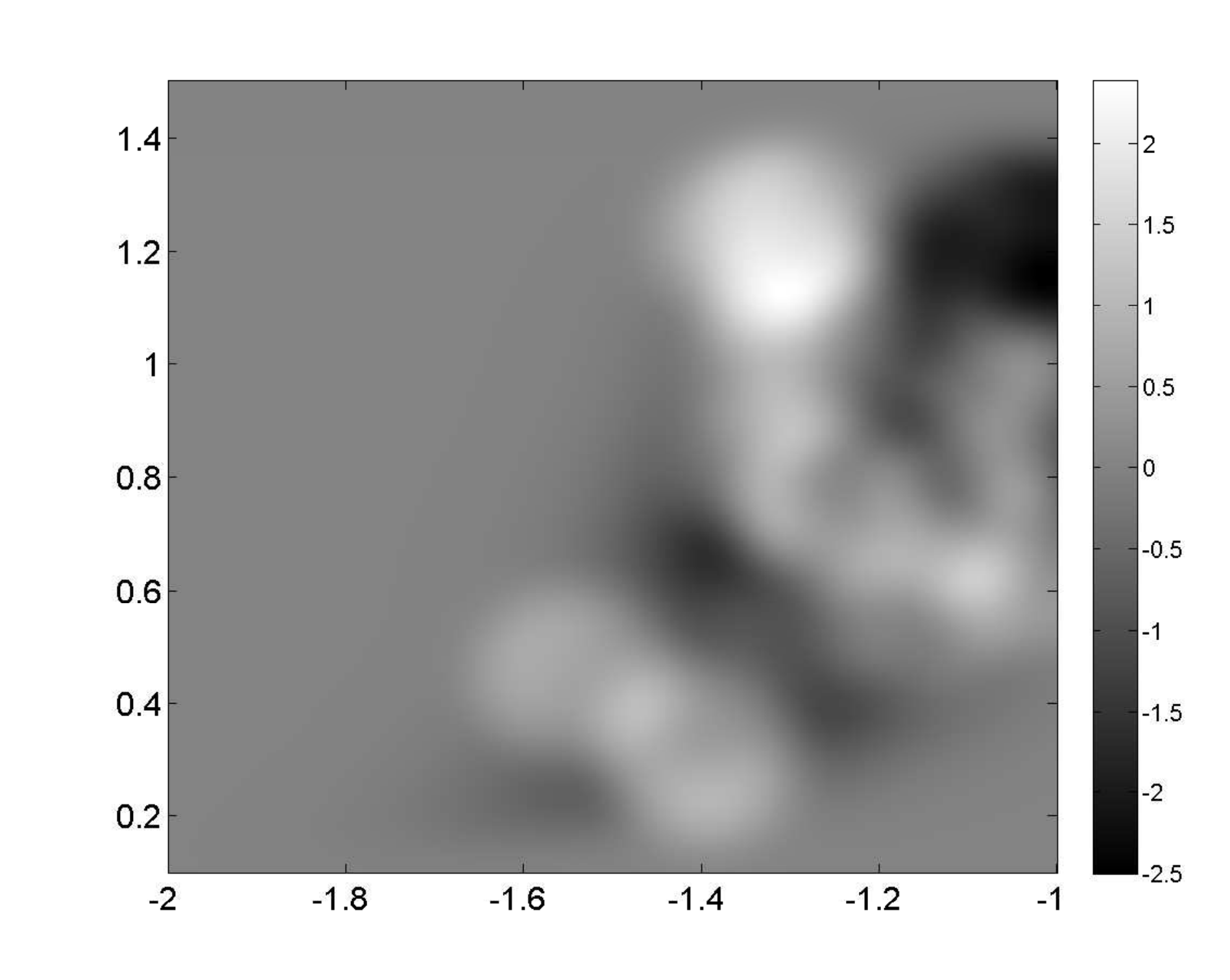} 
\includegraphics[width=0.20\textwidth,clip=true,trim=0cm 0cm 0cm 0cm]{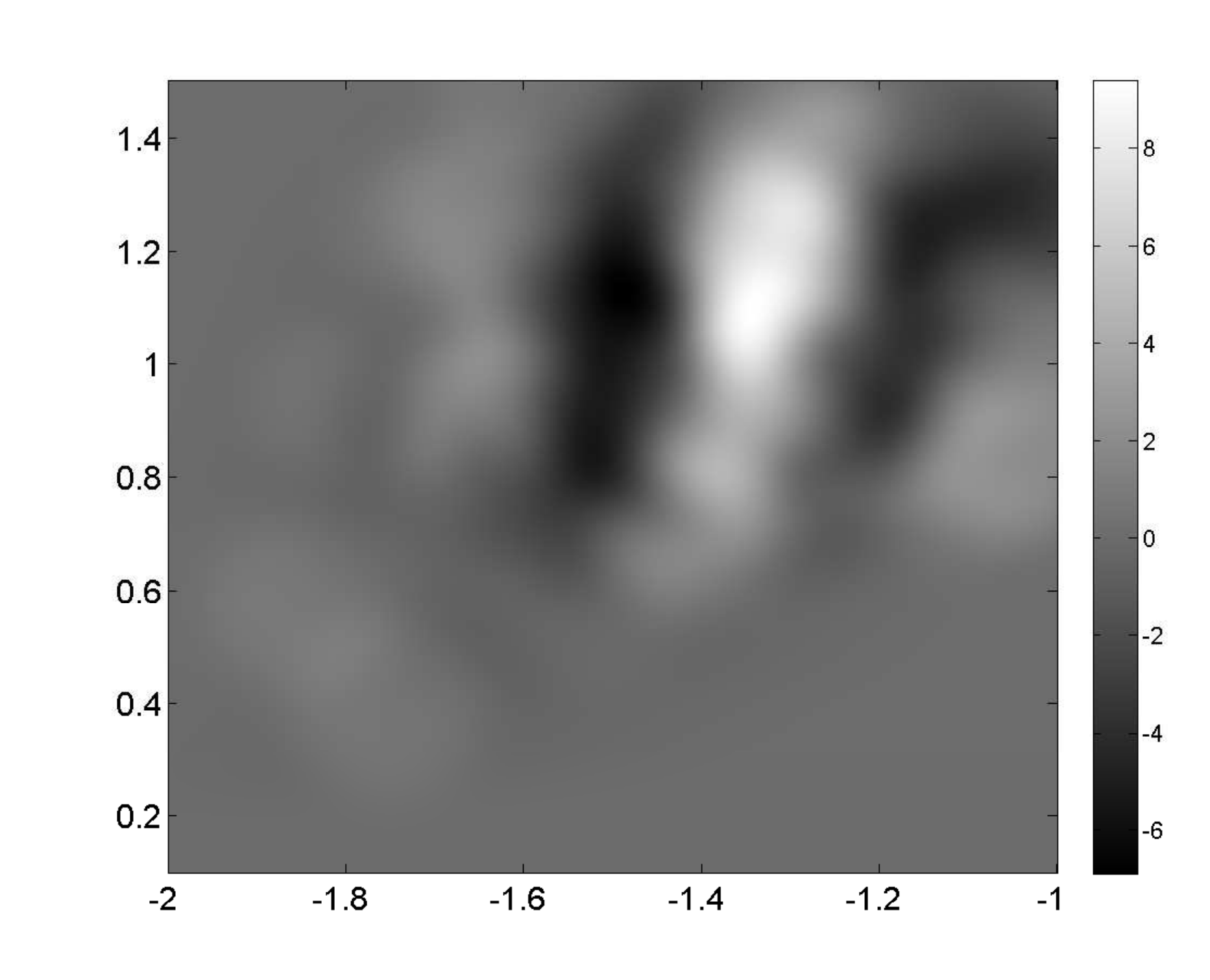} 
\includegraphics[width=0.20\textwidth,clip=true,trim=0cm 0cm 0cm 0cm]{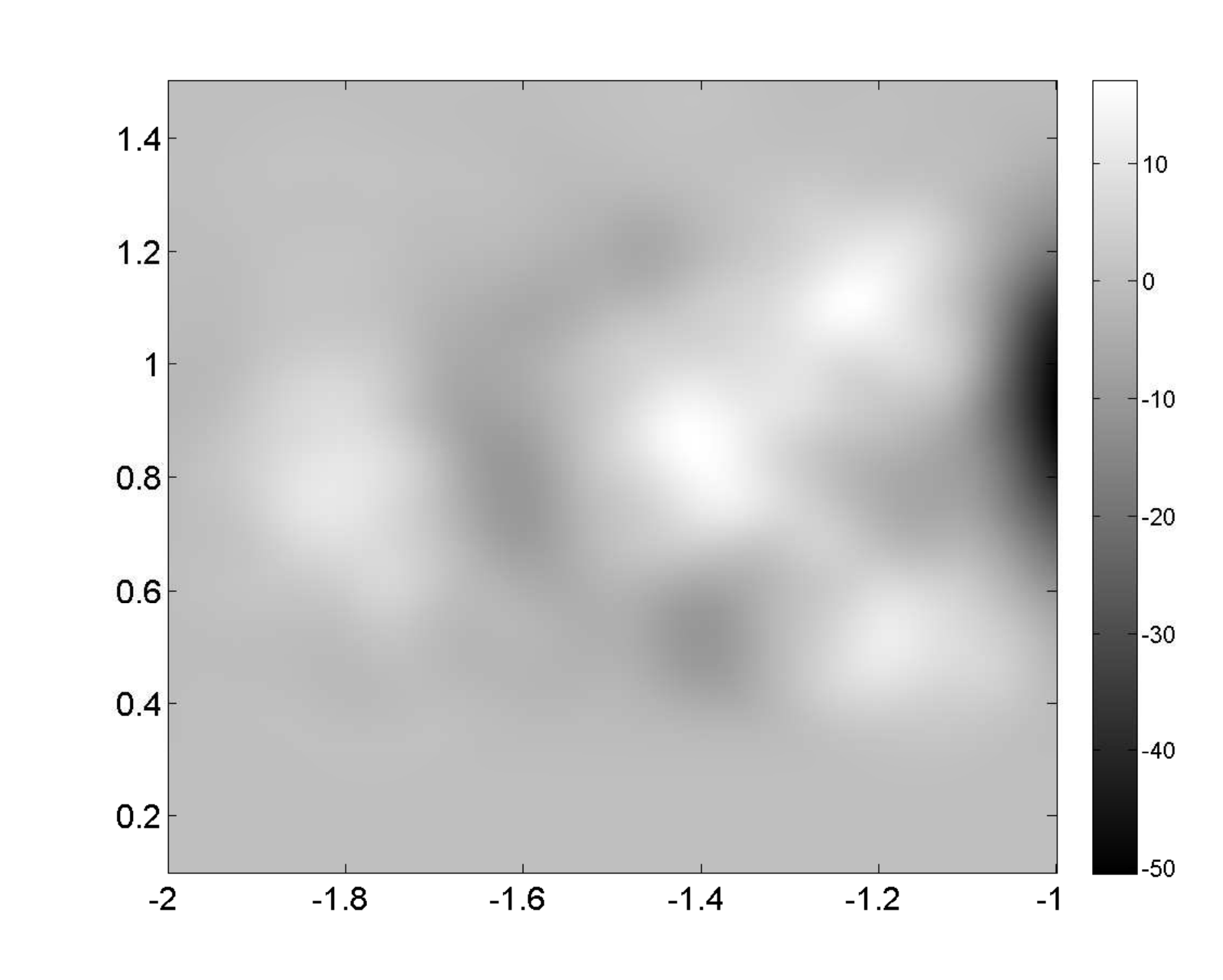} 
\includegraphics[width=0.20\textwidth,clip=true,trim=0cm 0cm 0cm 0cm]{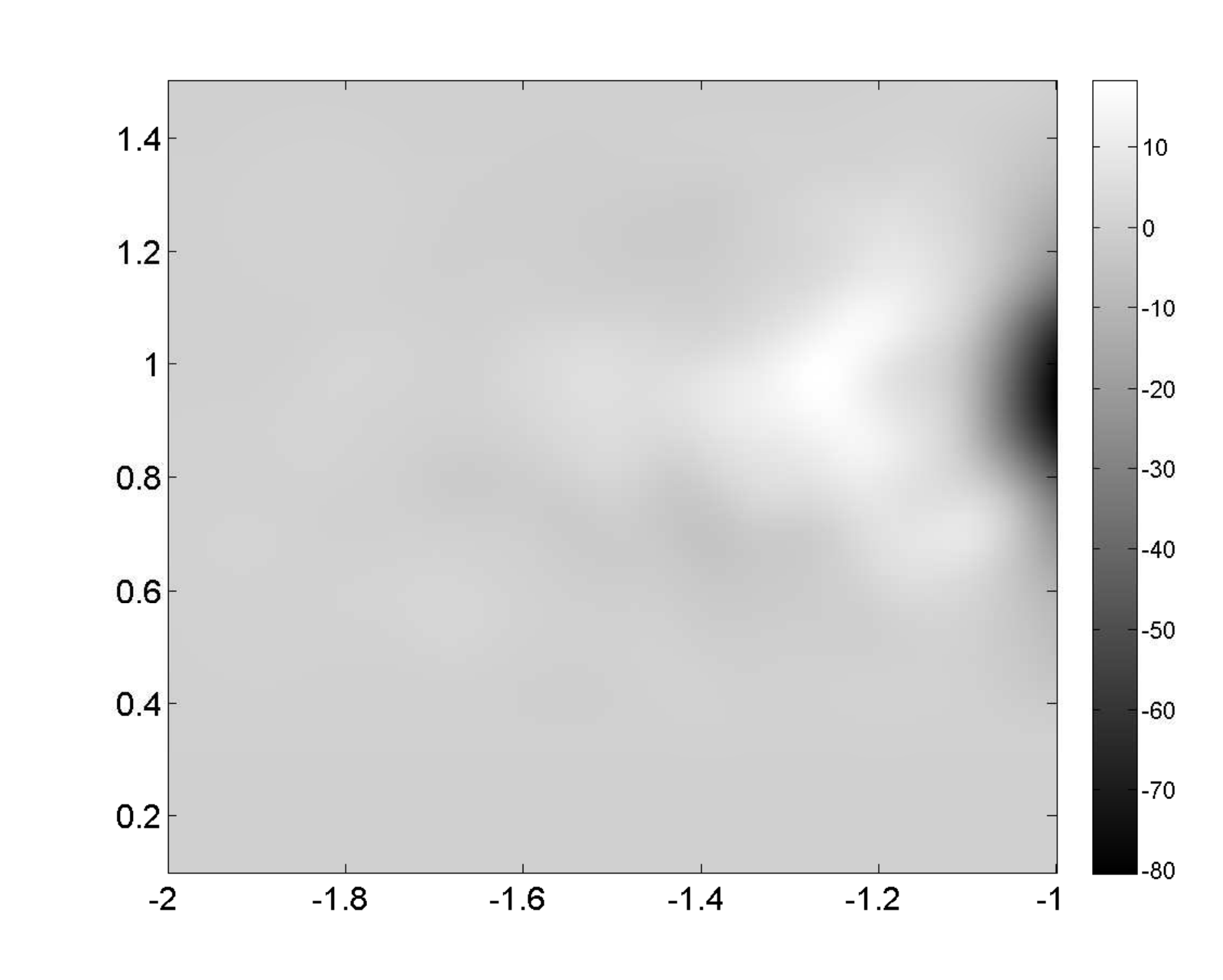} 
\includegraphics[width=0.20\textwidth,clip=true,trim=0cm 0cm 0cm 0cm]{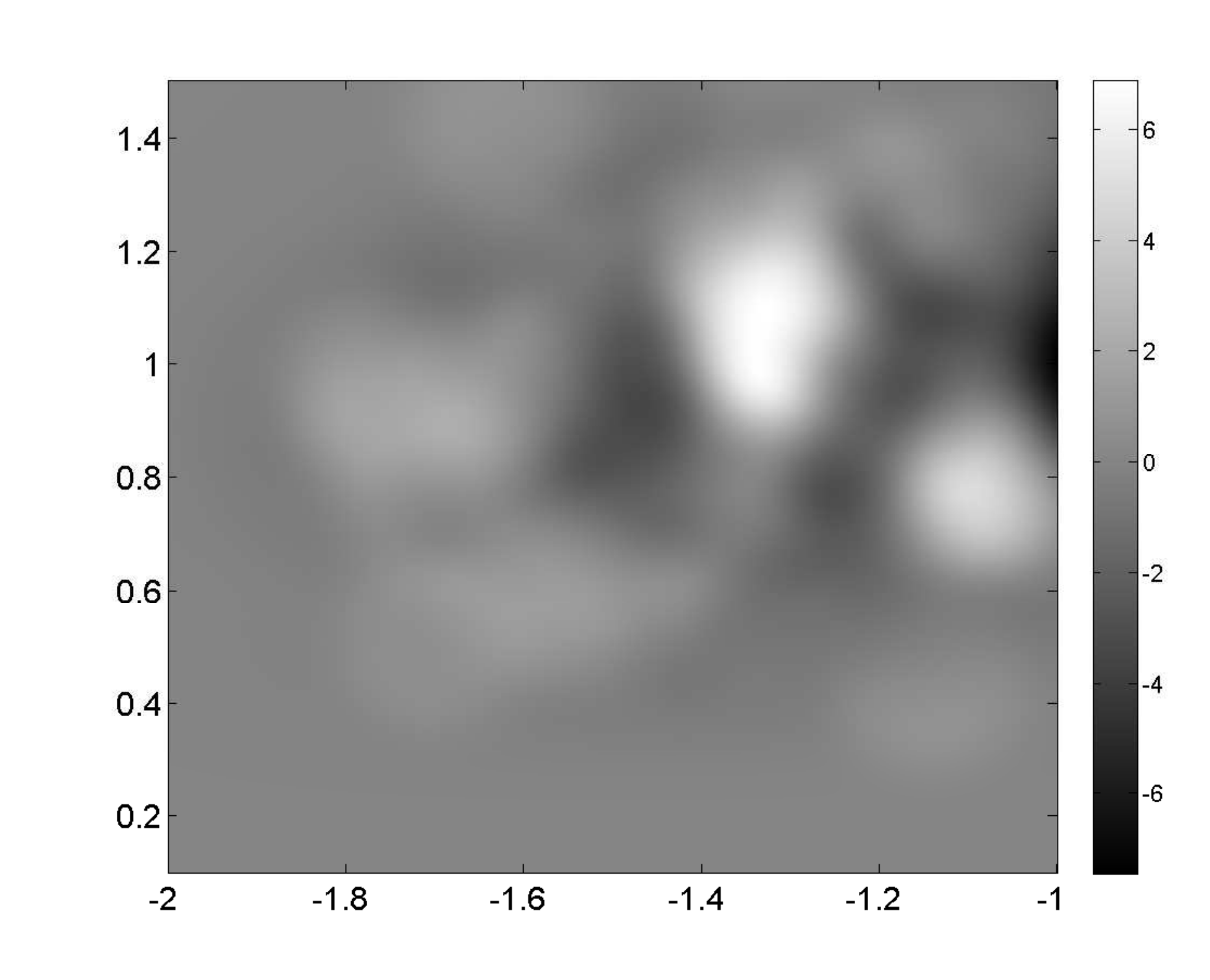} 
\caption{Same as Figure 2 for the Power Law halting model, with an optimized exponent free parameter.}
\end{figure}
\end{landscape}

\clearpage
\begin{landscape}
\begin{figure}
\centering
\includegraphics[width=0.20\textwidth,clip=true,trim=0cm 0cm 0cm 0cm]{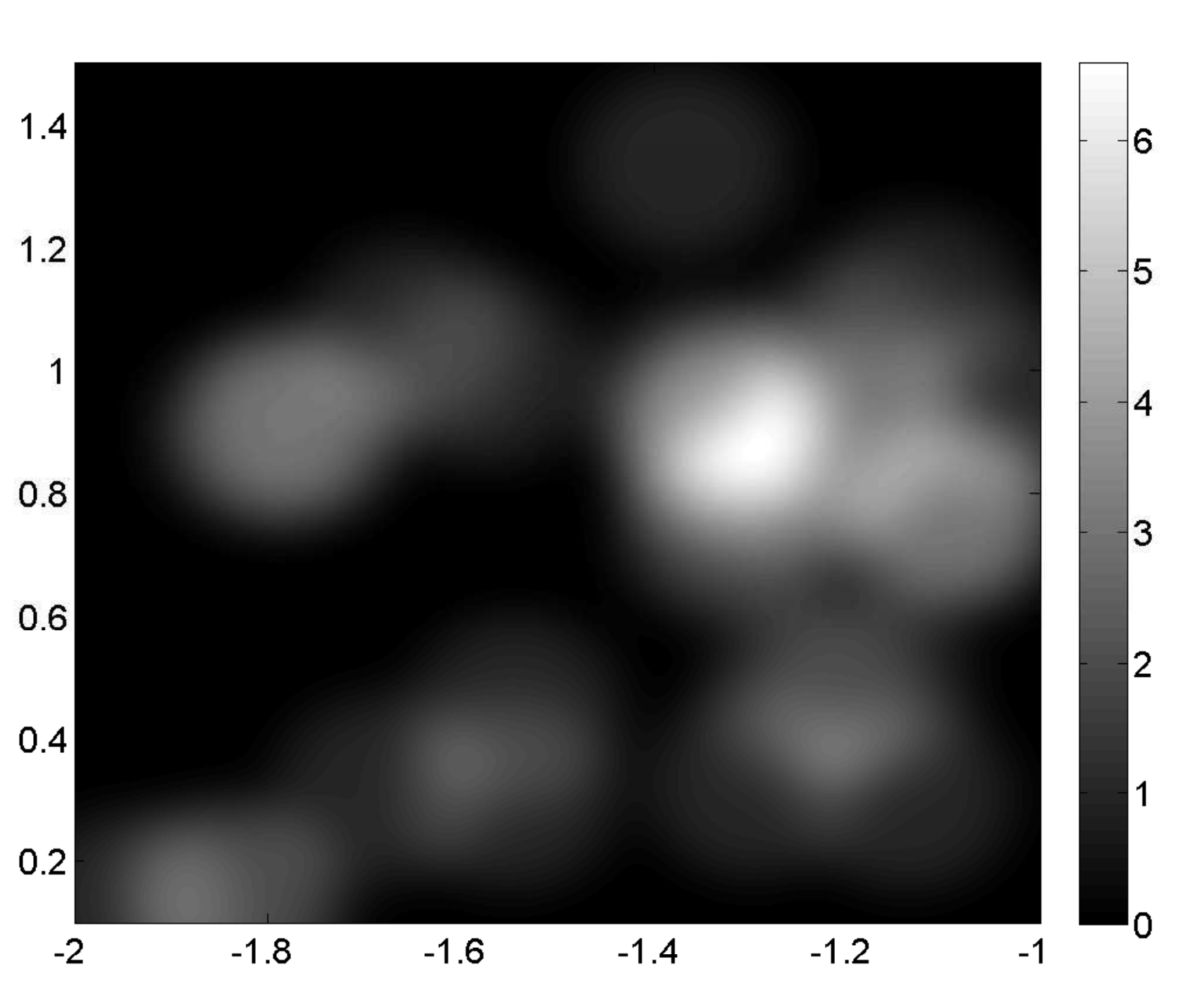} 
\includegraphics[width=0.20\textwidth,clip=true,trim=0cm 0cm 0cm 0cm]{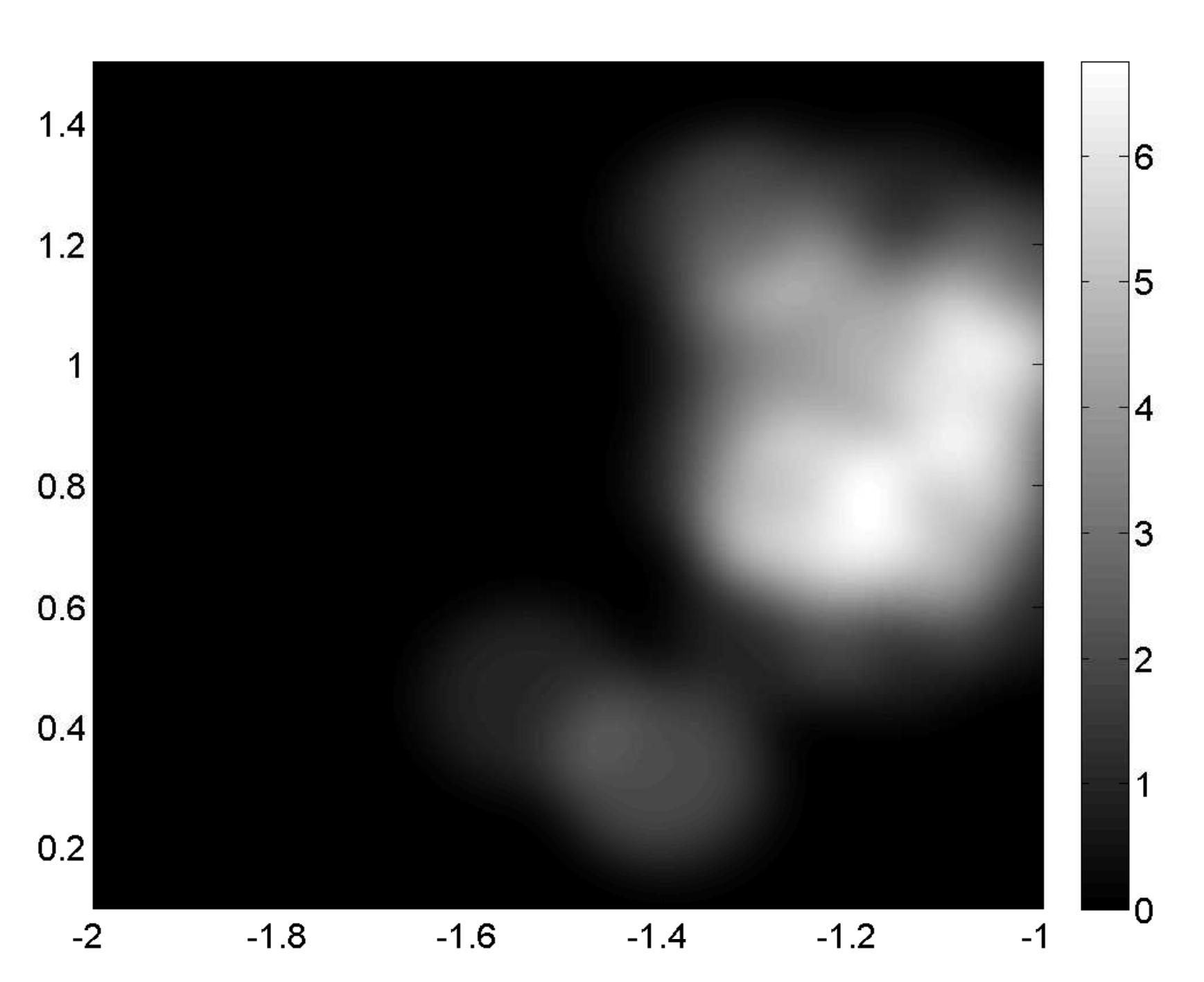} 
\includegraphics[width=0.20\textwidth,clip=true,trim=0cm 0cm 0cm 0cm]{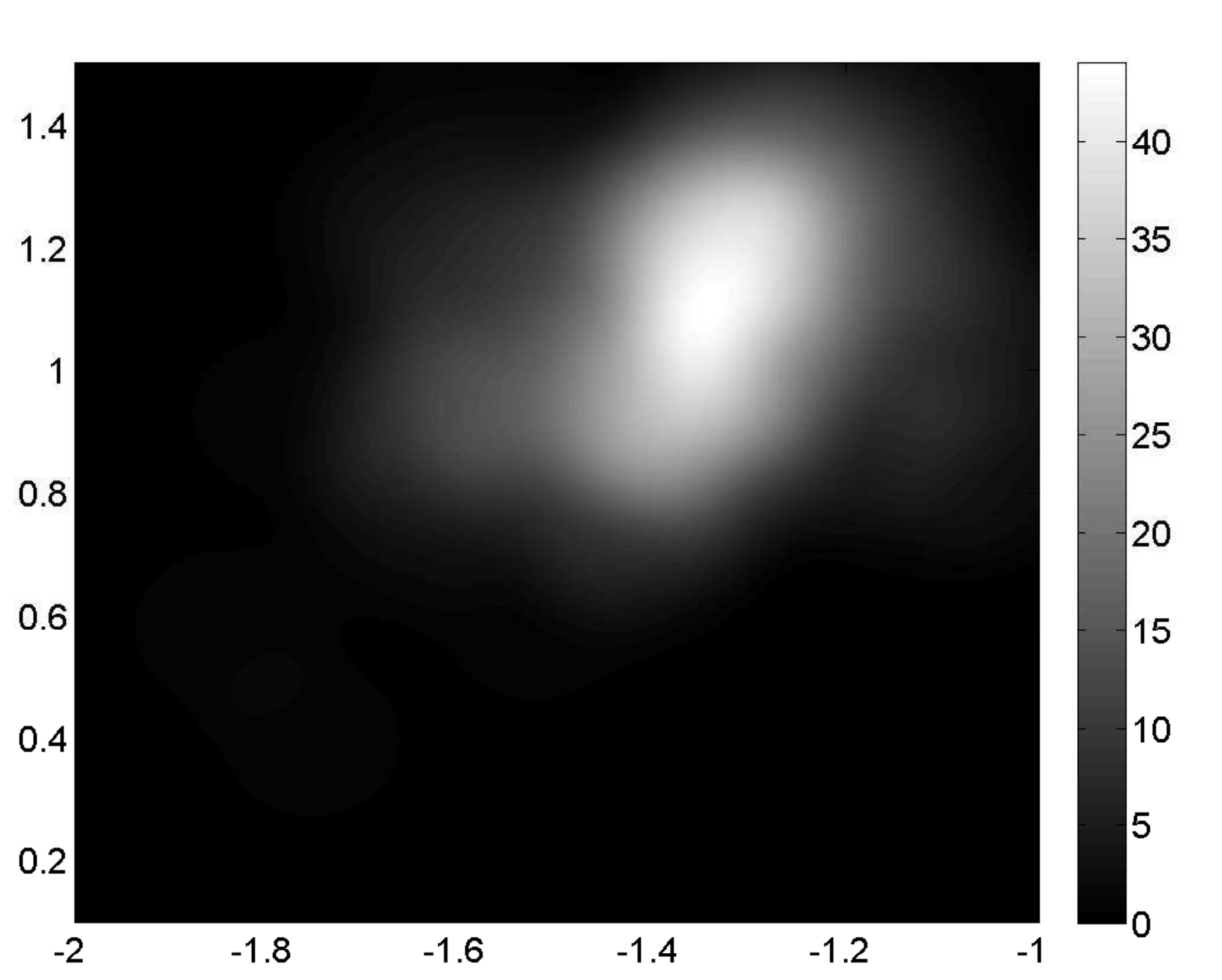} 
\includegraphics[width=0.20\textwidth,clip=true,trim=0cm 0cm 0cm 0cm]{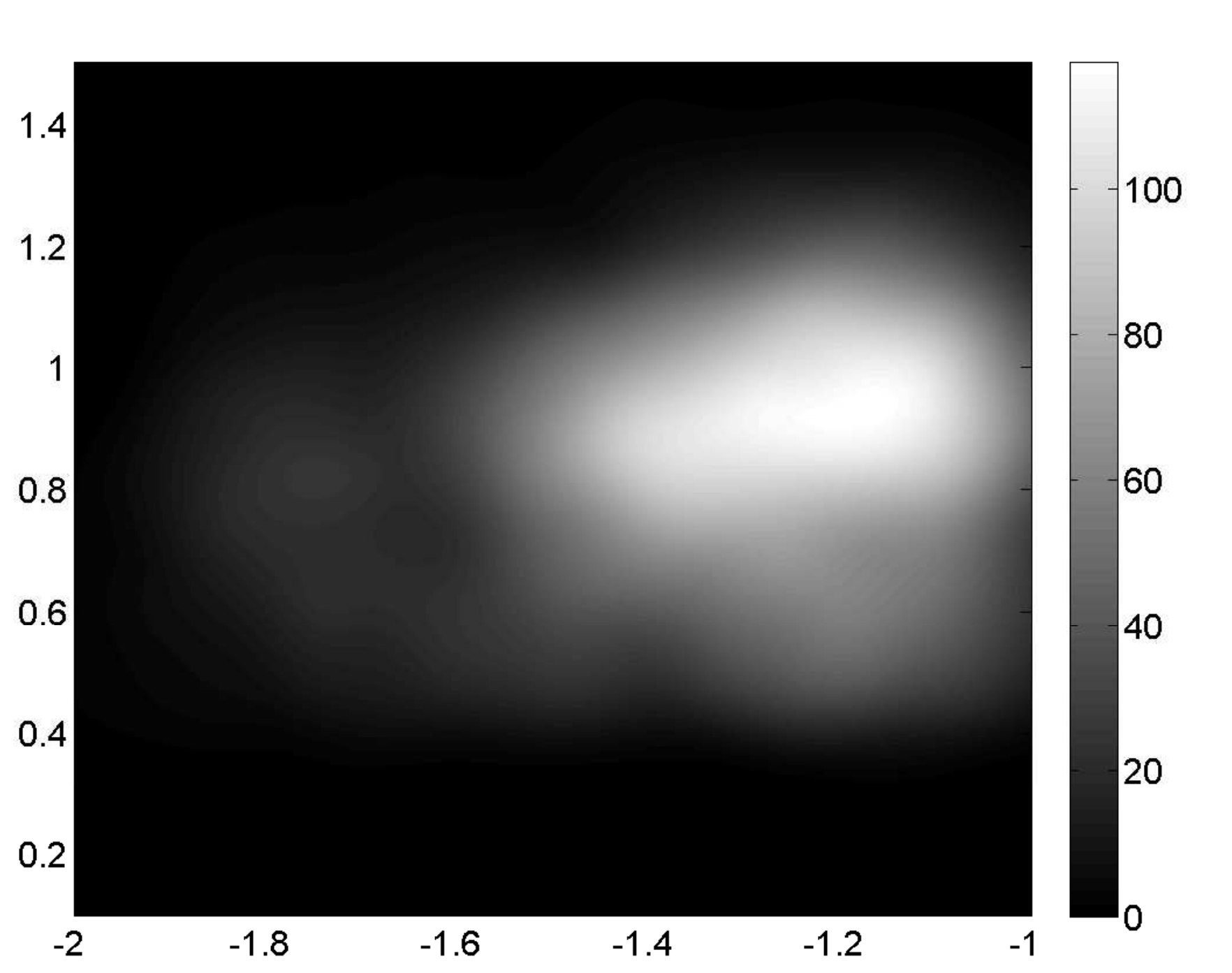} 
\includegraphics[width=0.20\textwidth,clip=true,trim=0cm 0cm 0cm 0cm]{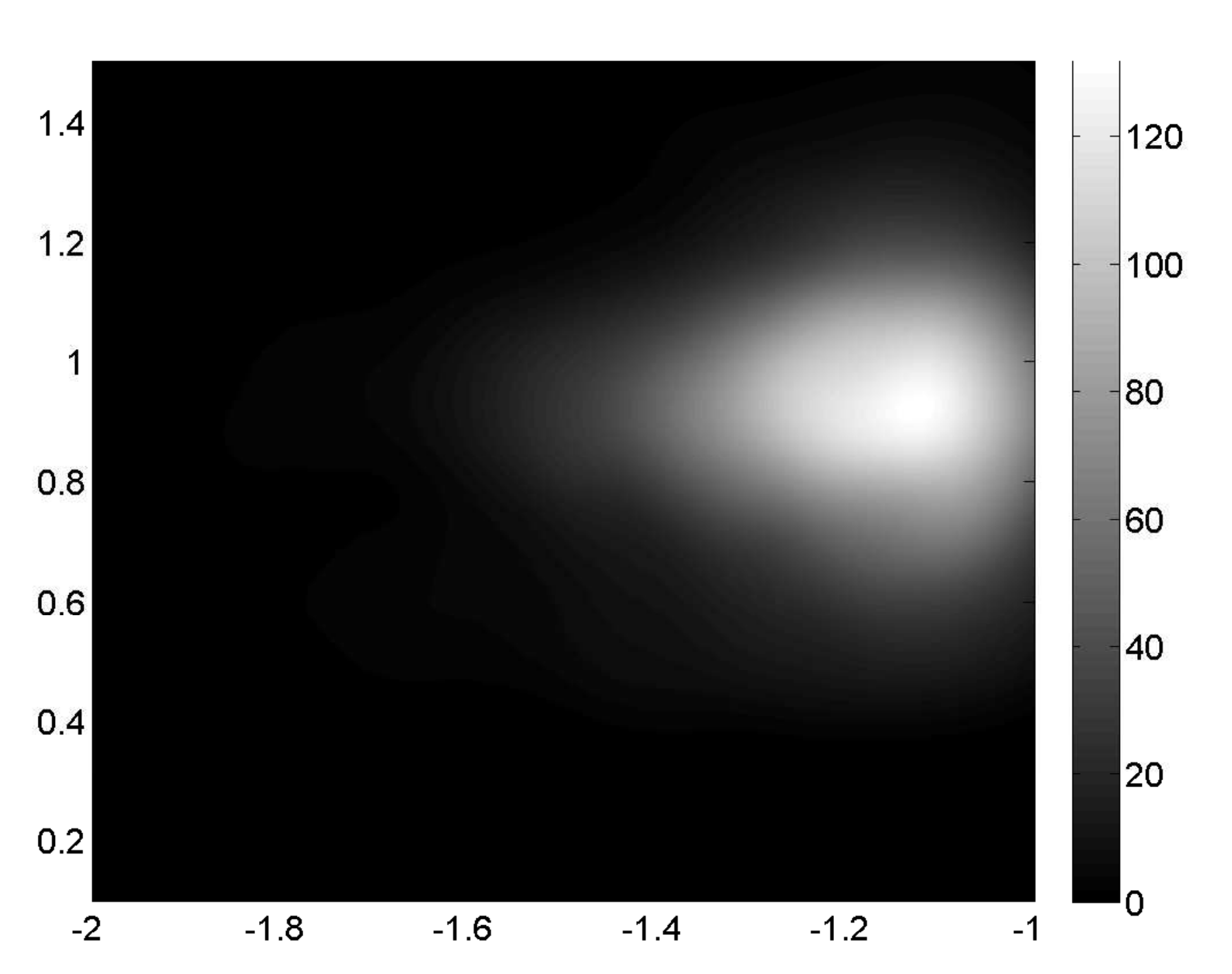} 
\includegraphics[width=0.20\textwidth,clip=true,trim=0cm 0cm 0cm 0cm]{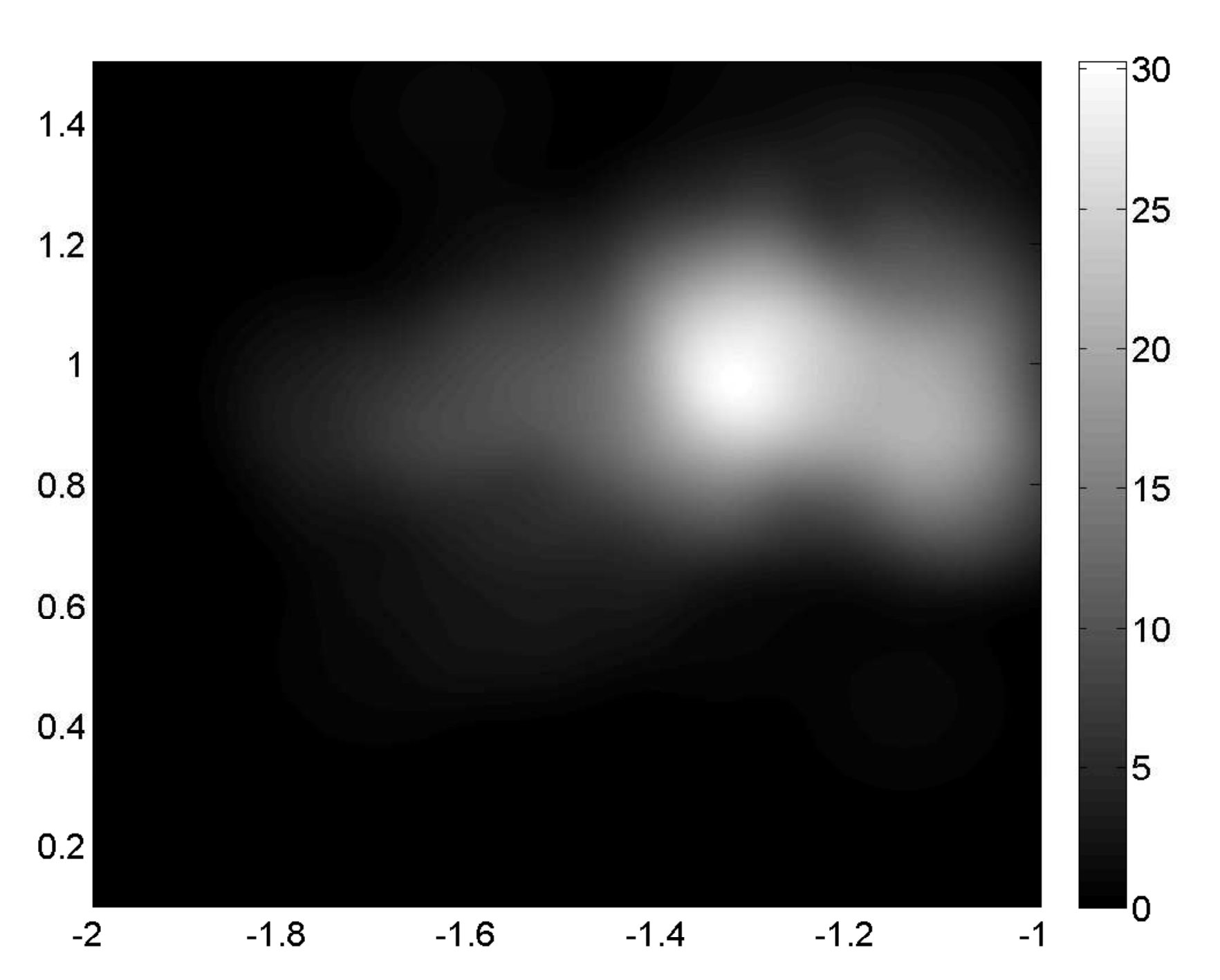} \\
\includegraphics[width=0.20\textwidth,clip=true,trim=0cm 0cm 0cm 0cm]{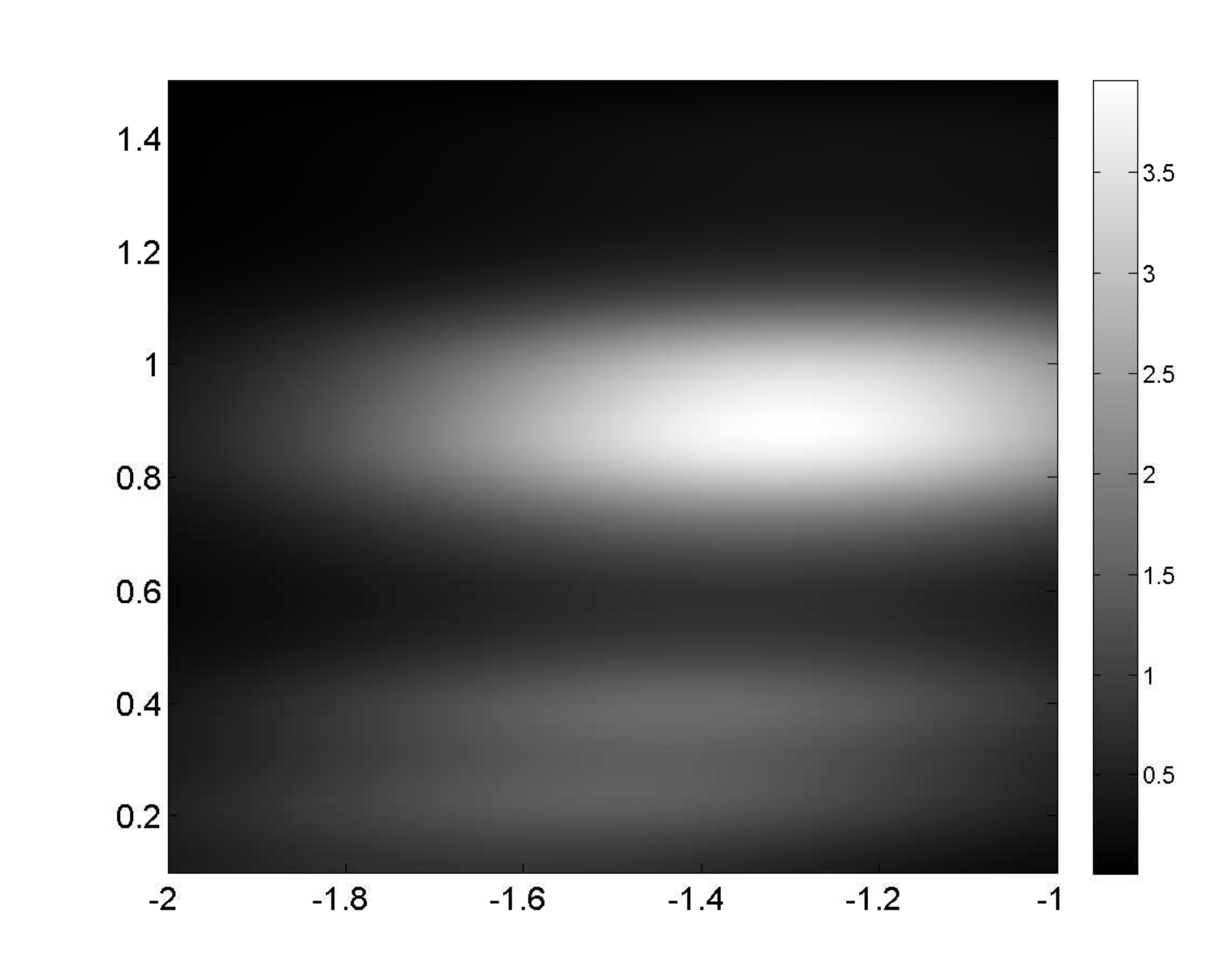} 
\includegraphics[width=0.20\textwidth,clip=true,trim=0cm 0cm 0cm 0cm]{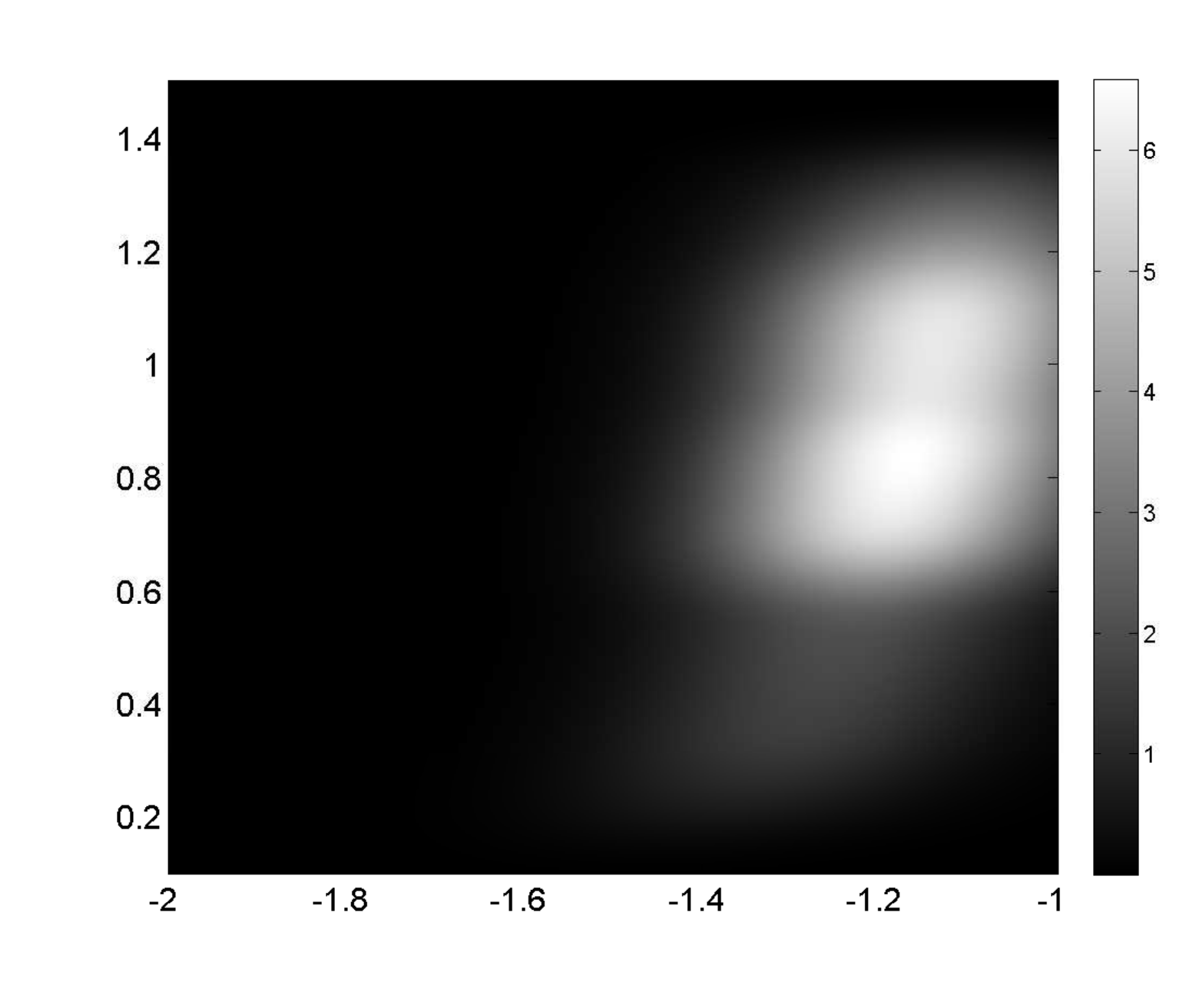} 
\includegraphics[width=0.20\textwidth,clip=true,trim=0cm 0cm 0cm 0cm]{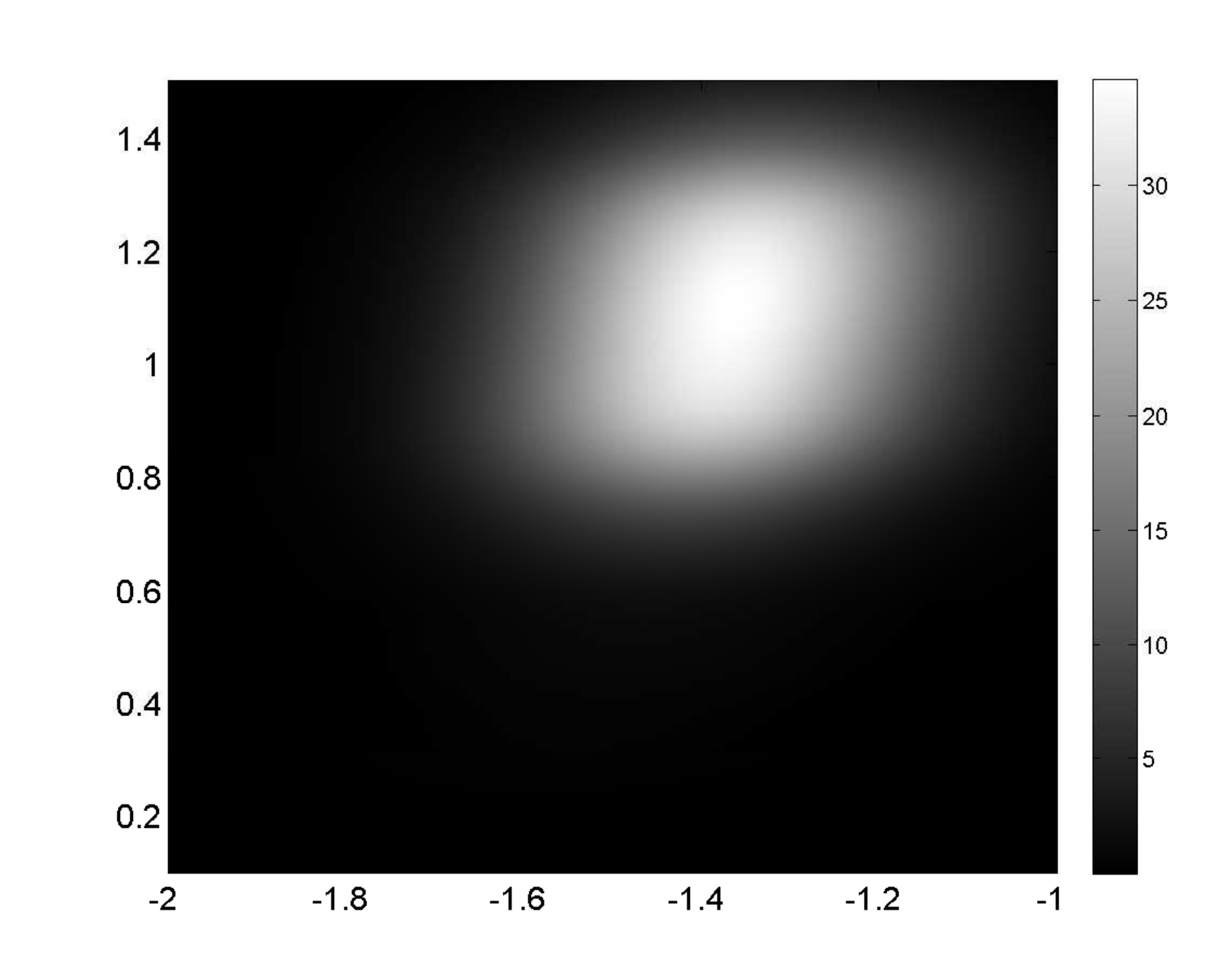}
\includegraphics[width=0.20\textwidth,clip=true,trim=0cm 0cm 0cm 0cm]{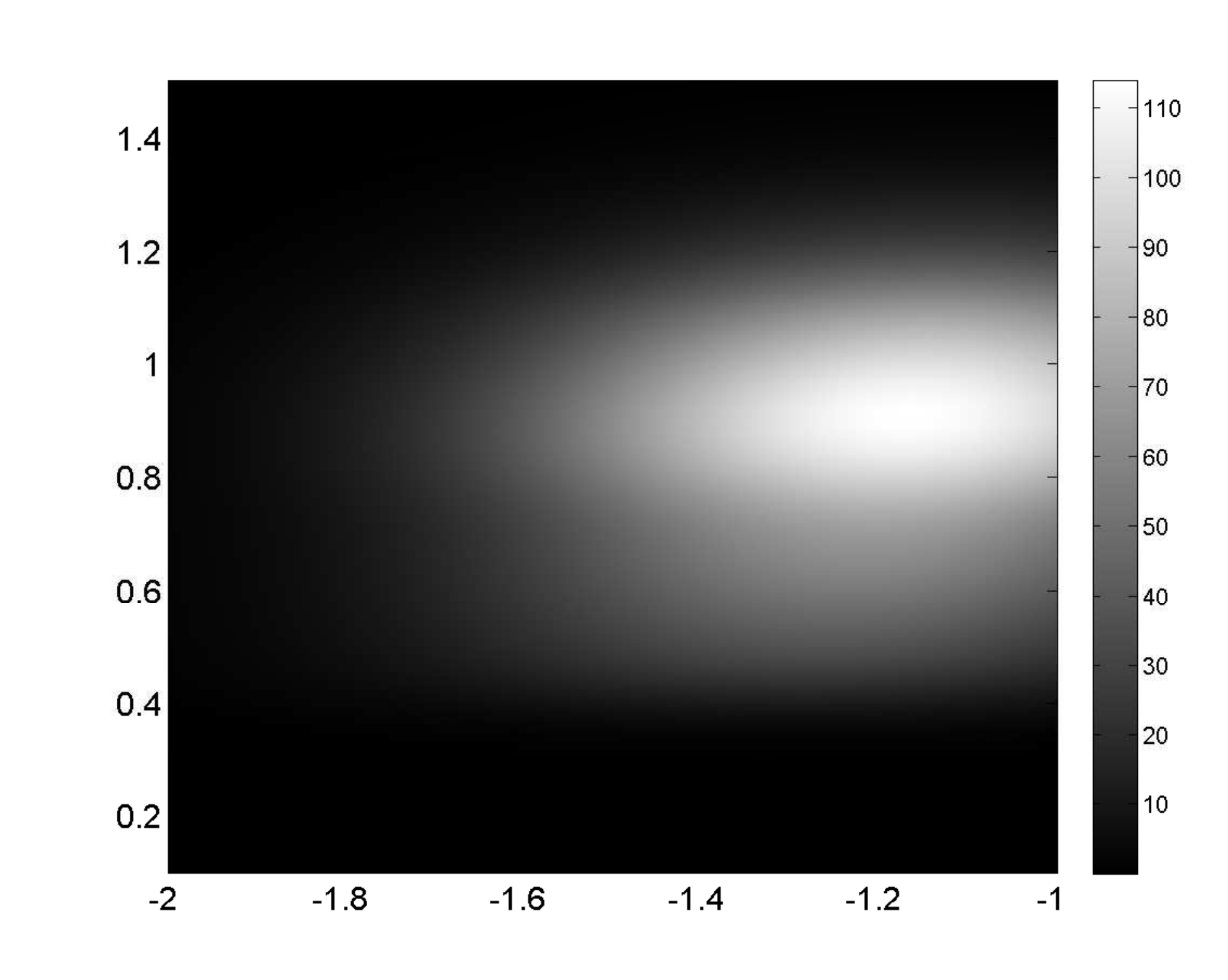} 
\includegraphics[width=0.20\textwidth,clip=true,trim=0cm 0cm 0cm 0cm]{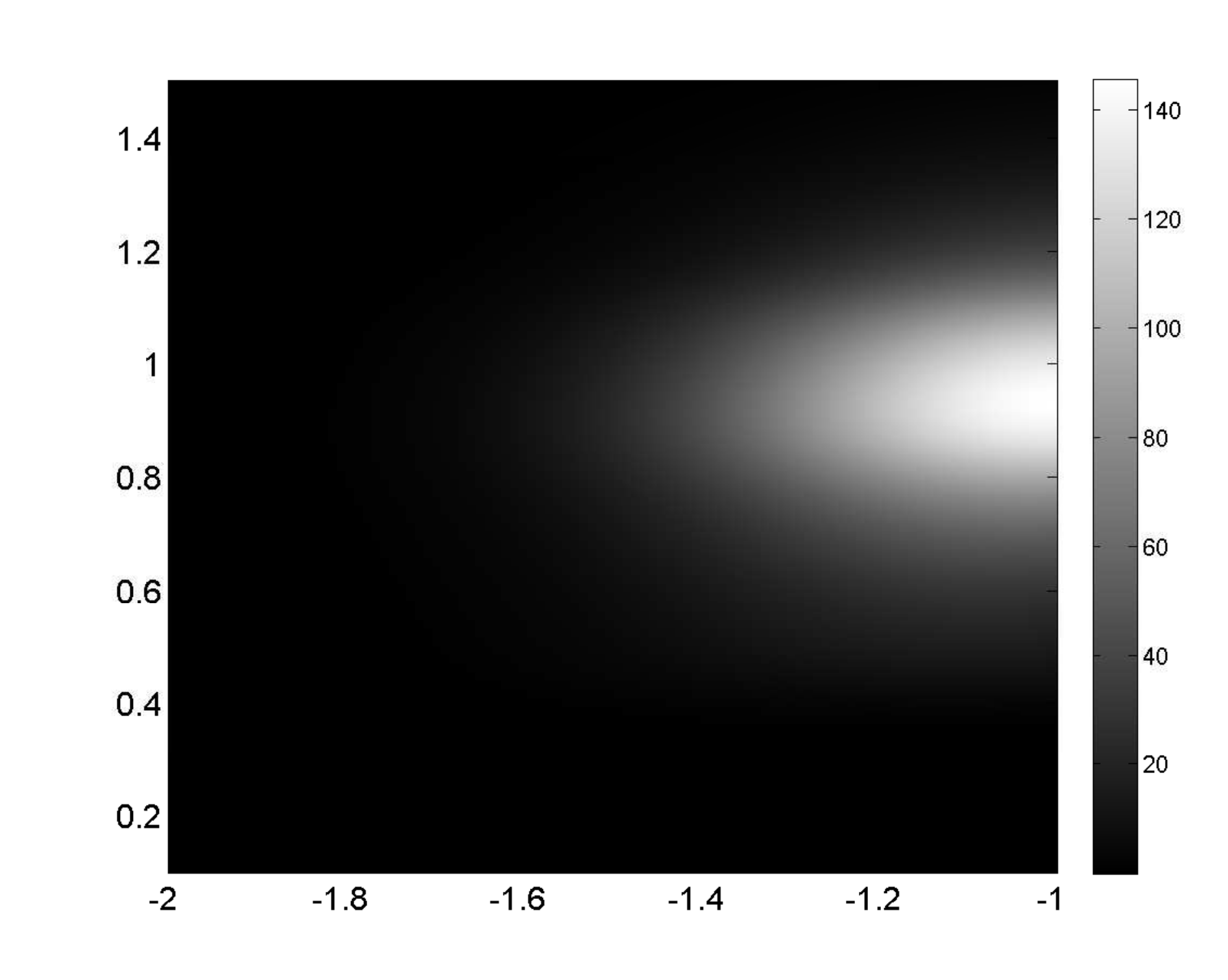} 
\includegraphics[width=0.20\textwidth,clip=true,trim=0cm 0cm 0cm 0cm]{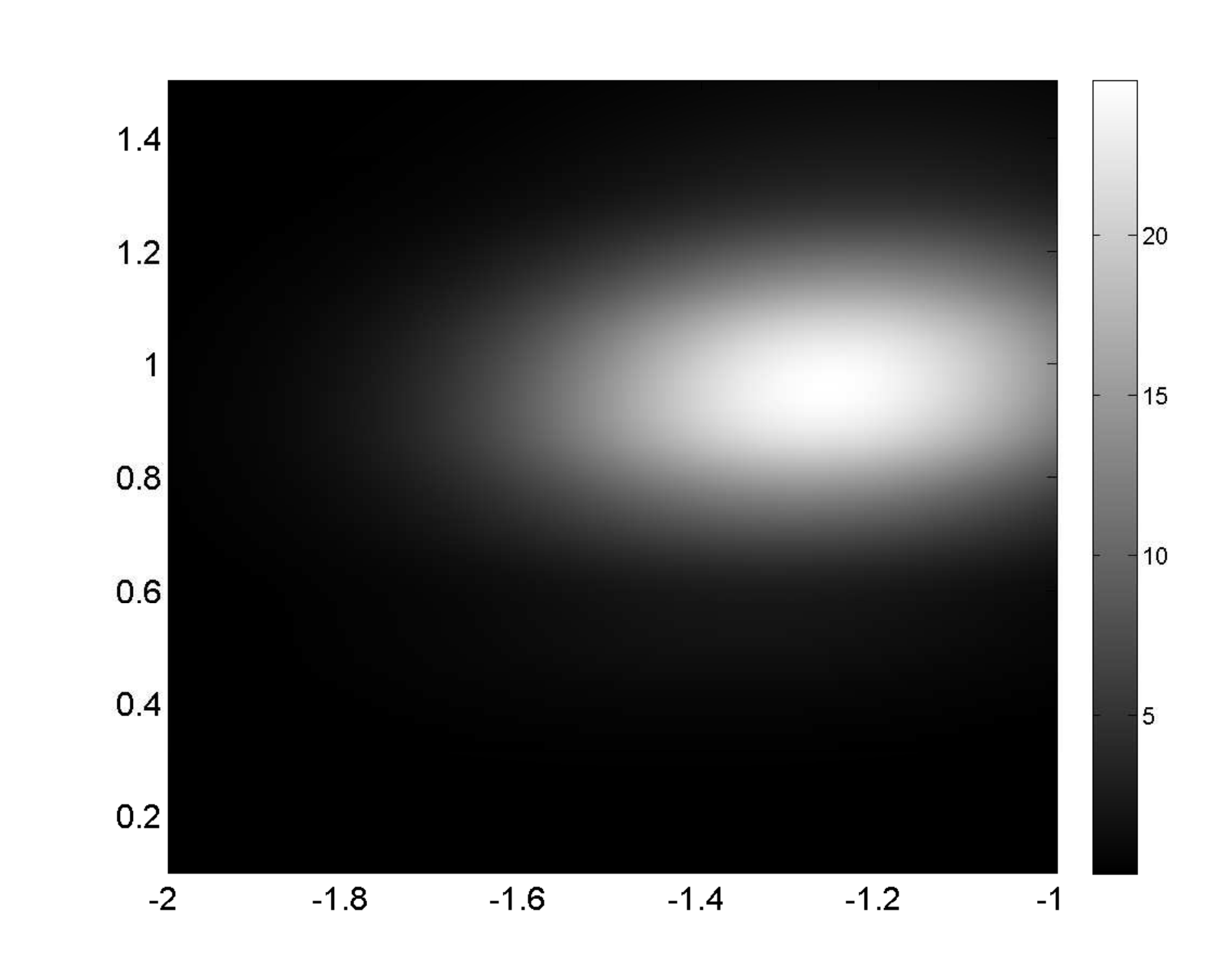} \\
\includegraphics[width=0.20\textwidth,clip=true,trim=0cm 0cm 0cm 0cm]{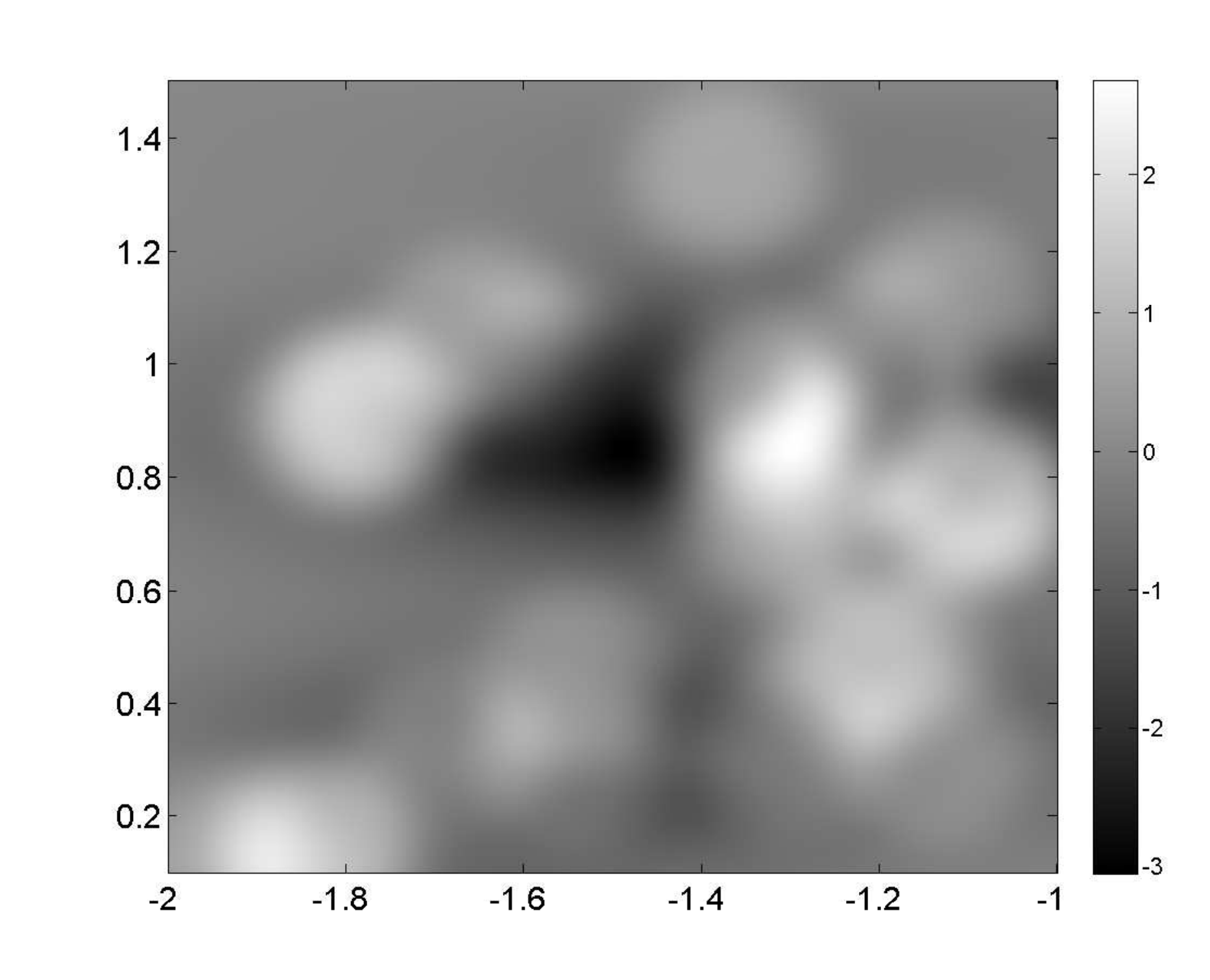} 
\includegraphics[width=0.20\textwidth,clip=true,trim=0cm 0cm 0cm 0cm]{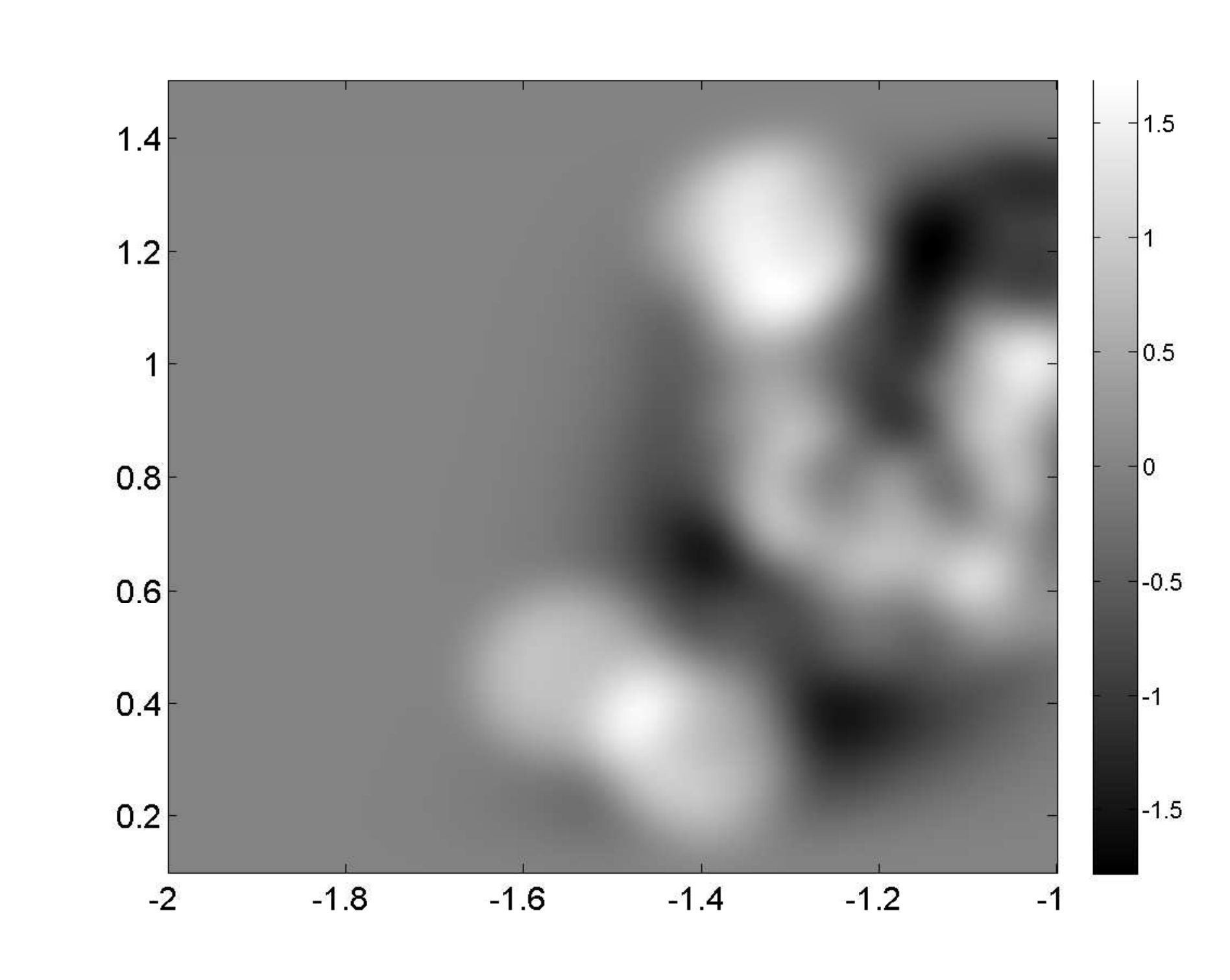} 
\includegraphics[width=0.20\textwidth,clip=true,trim=0cm 0cm 0cm 0cm]{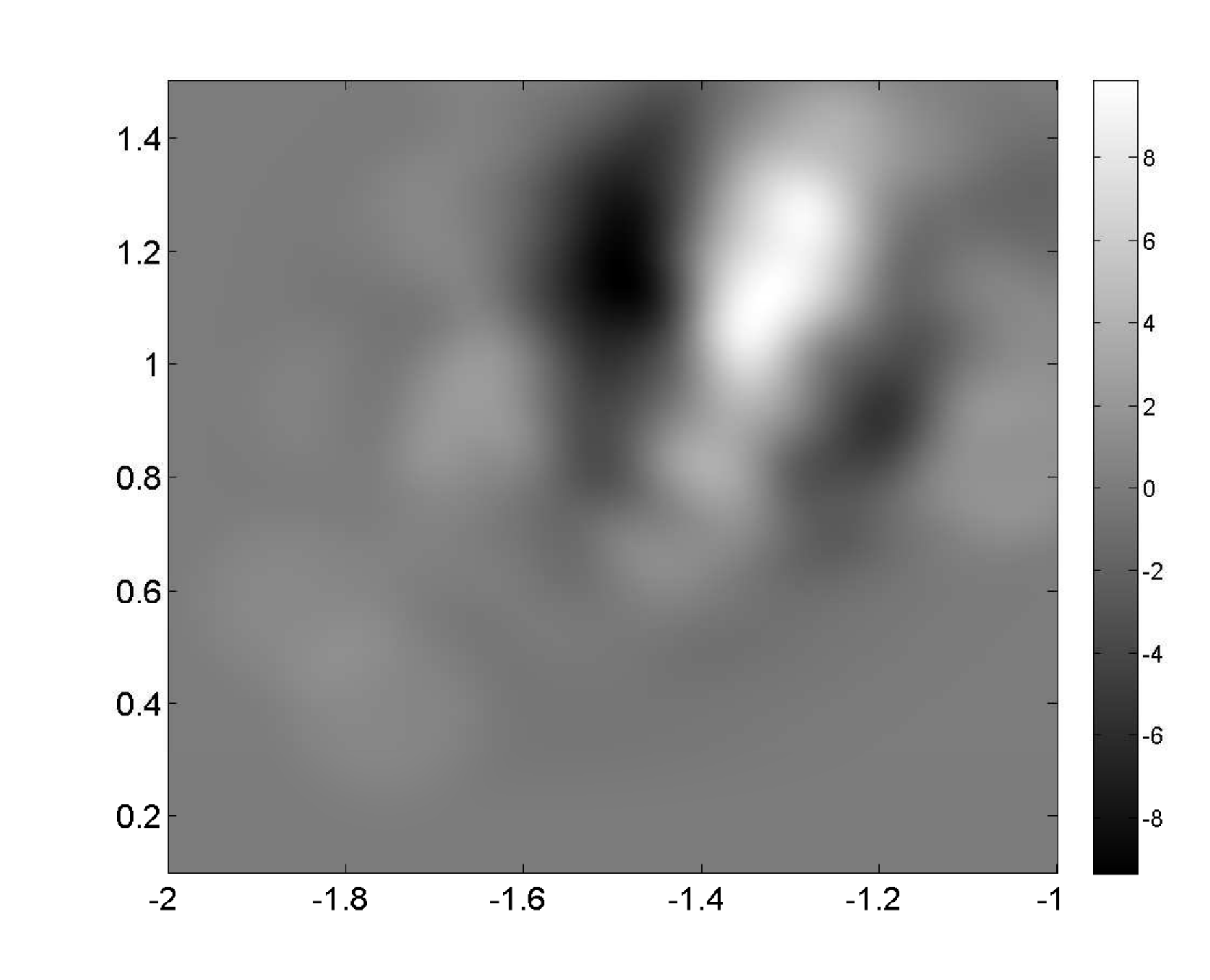} 
\includegraphics[width=0.20\textwidth,clip=true,trim=0cm 0cm 0cm 0cm]{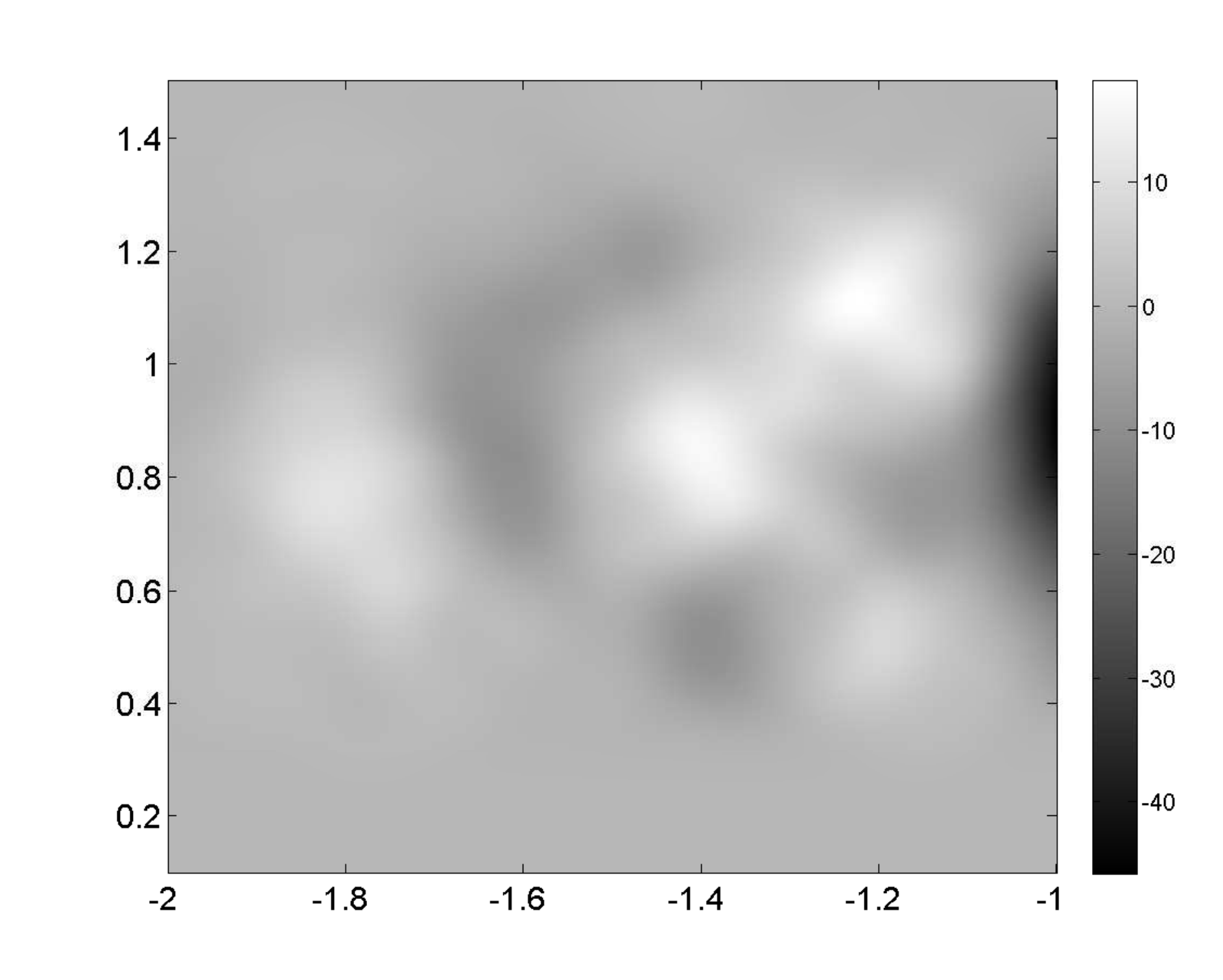} 
\includegraphics[width=0.20\textwidth,clip=true,trim=0cm 0cm 0cm 0cm]{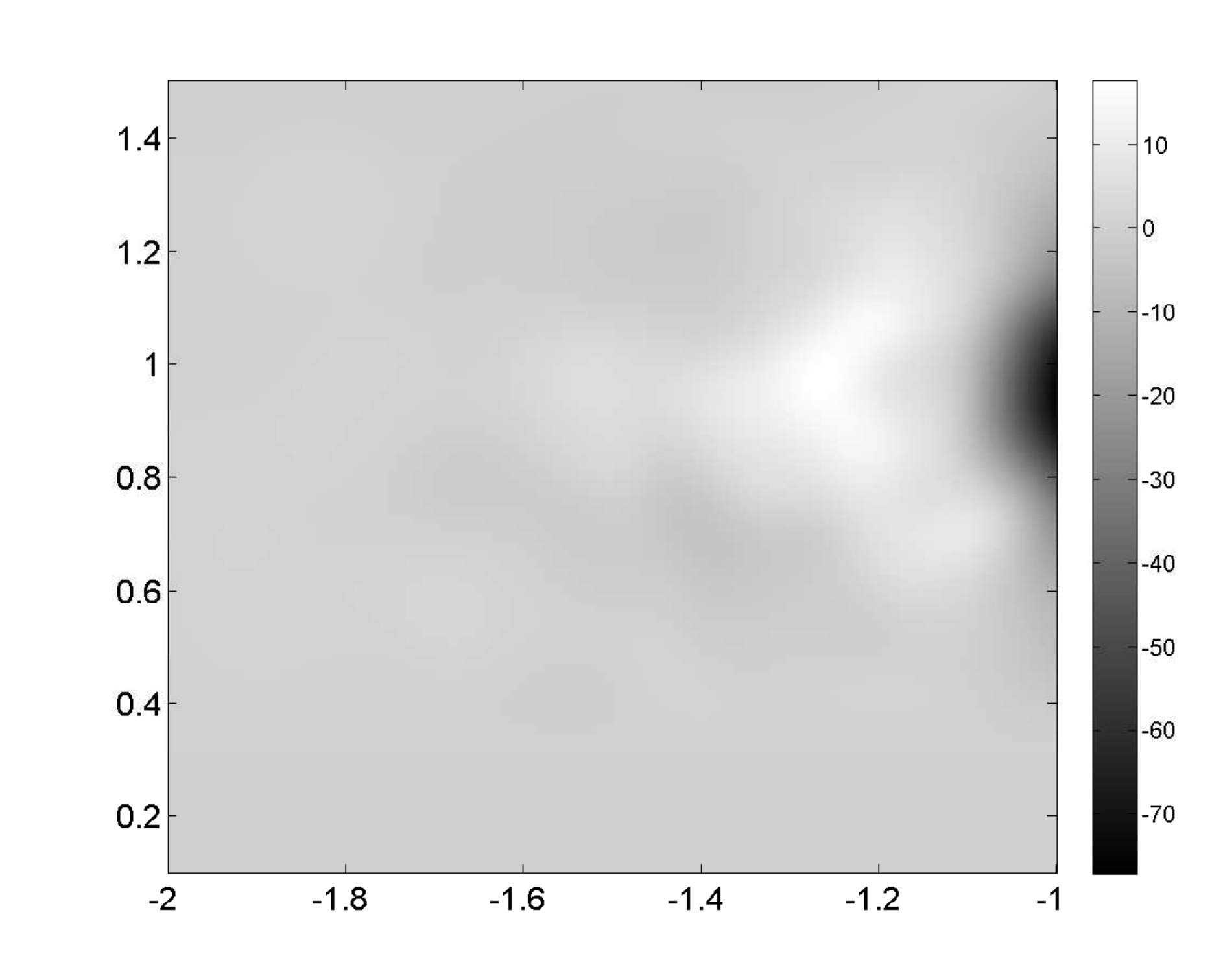} 
\includegraphics[width=0.20\textwidth,clip=true,trim=0cm 0cm 0cm 0cm]{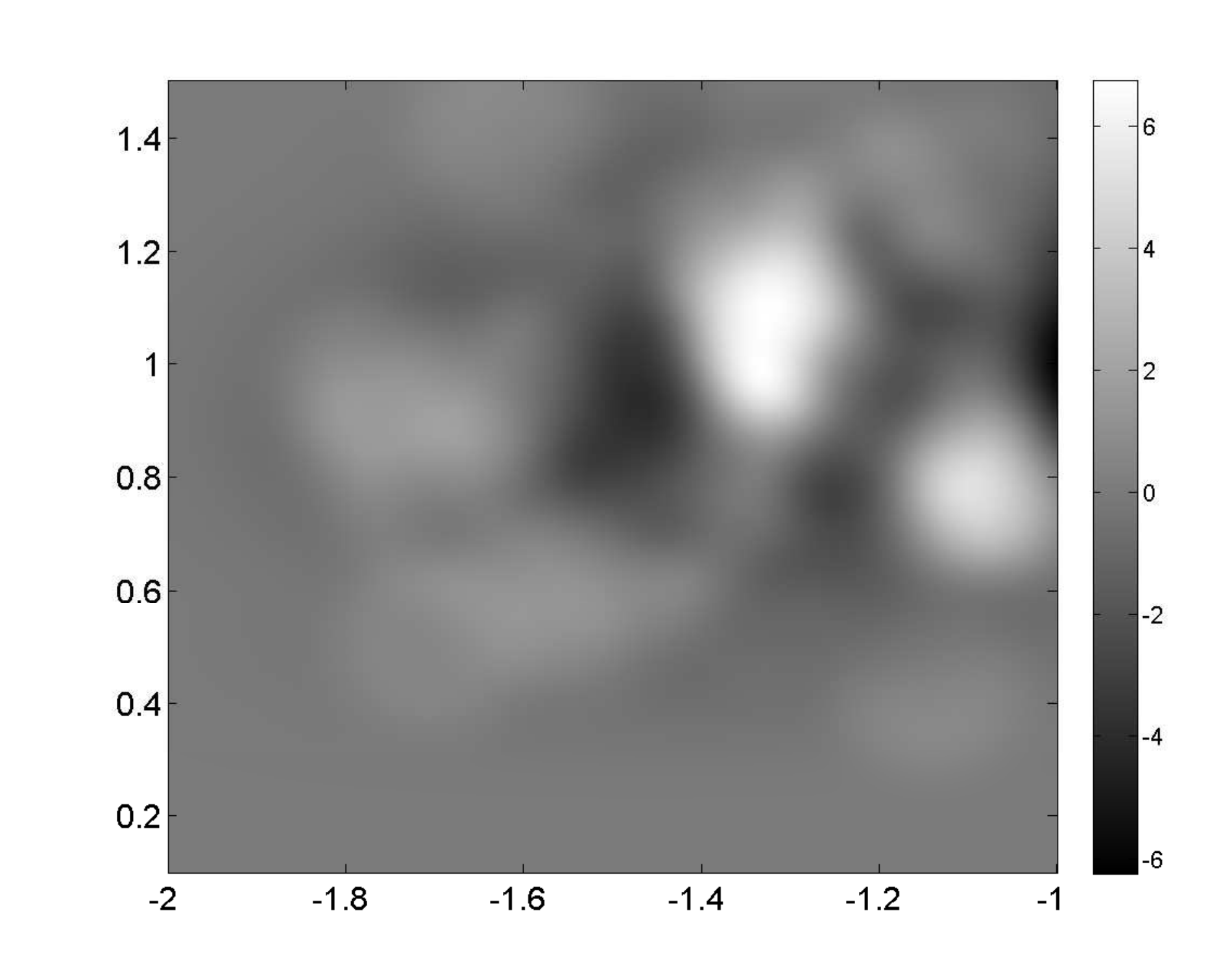} 
\caption{Same as Figure 2 for the Roche radius a $\propto$ M$^{1/3}$ tidal halting model.}
\end{figure}
\end{landscape}

\clearpage
\begin{landscape}
\begin{figure}
\centering
\includegraphics[width=0.20\textwidth,clip=true,trim=0cm 0cm 0cm 0cm]{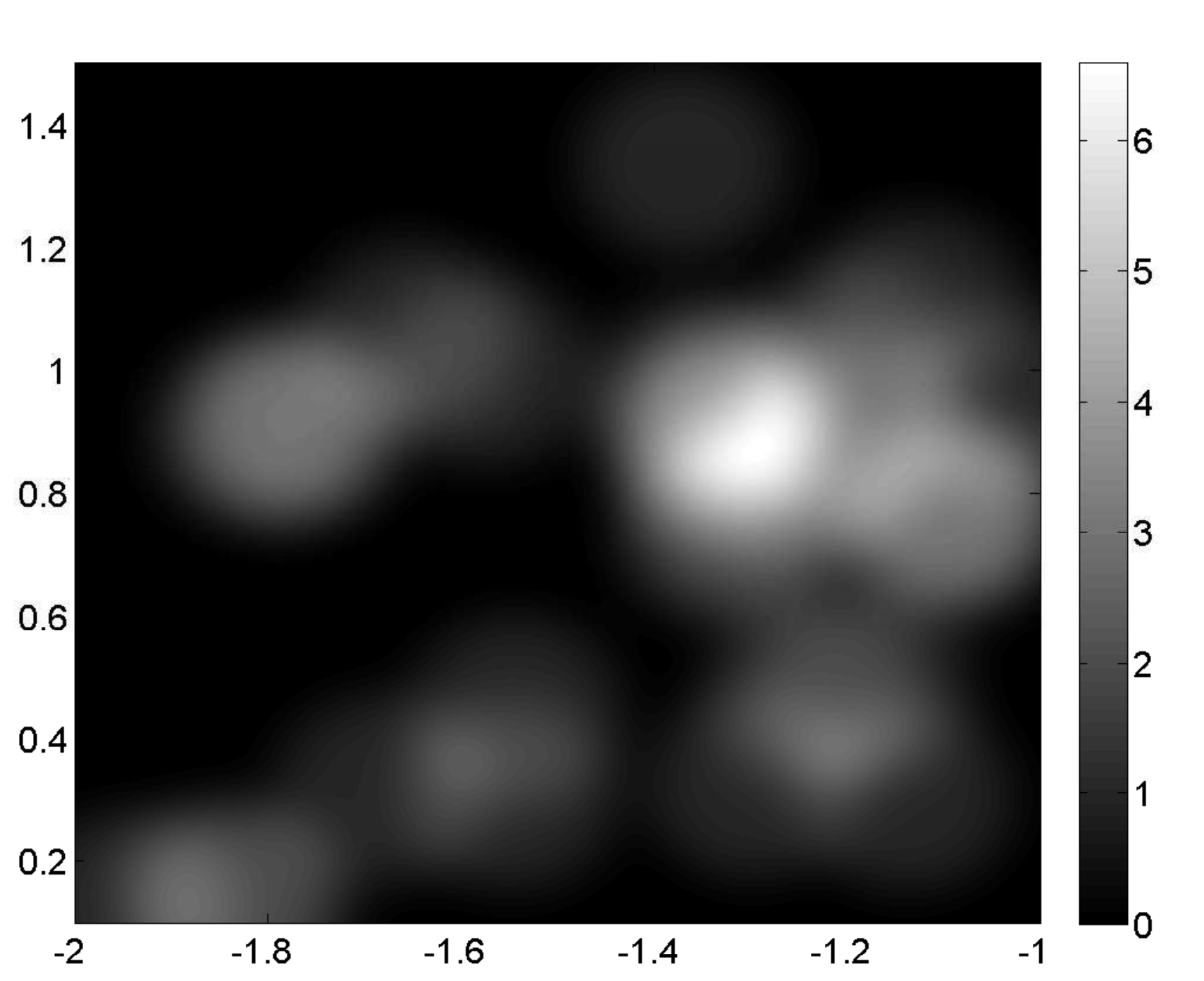} 
\includegraphics[width=0.20\textwidth,clip=true,trim=0cm 0cm 0cm 0cm]{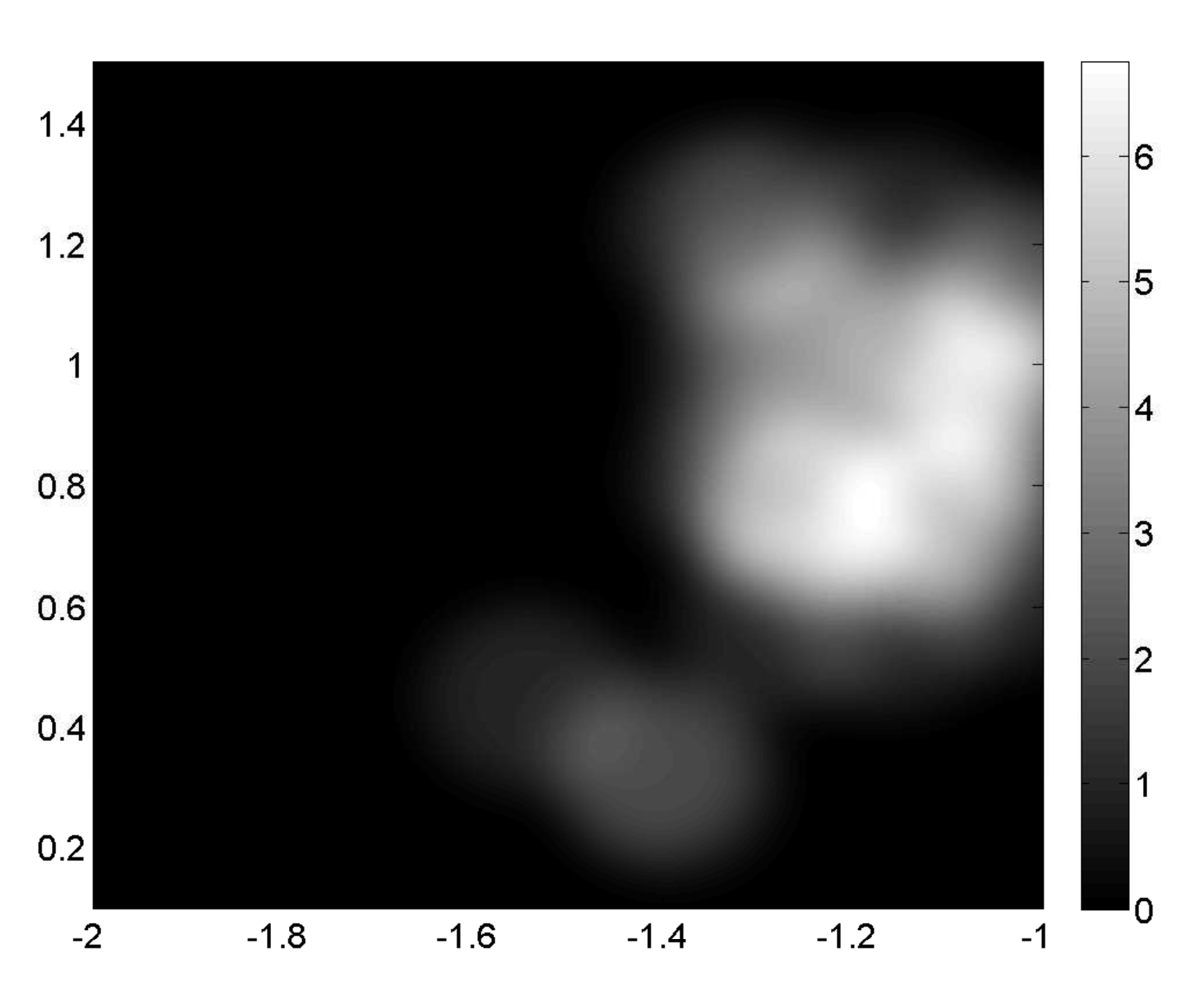} 
\includegraphics[width=0.20\textwidth,clip=true,trim=0cm 0cm 0cm 0cm]{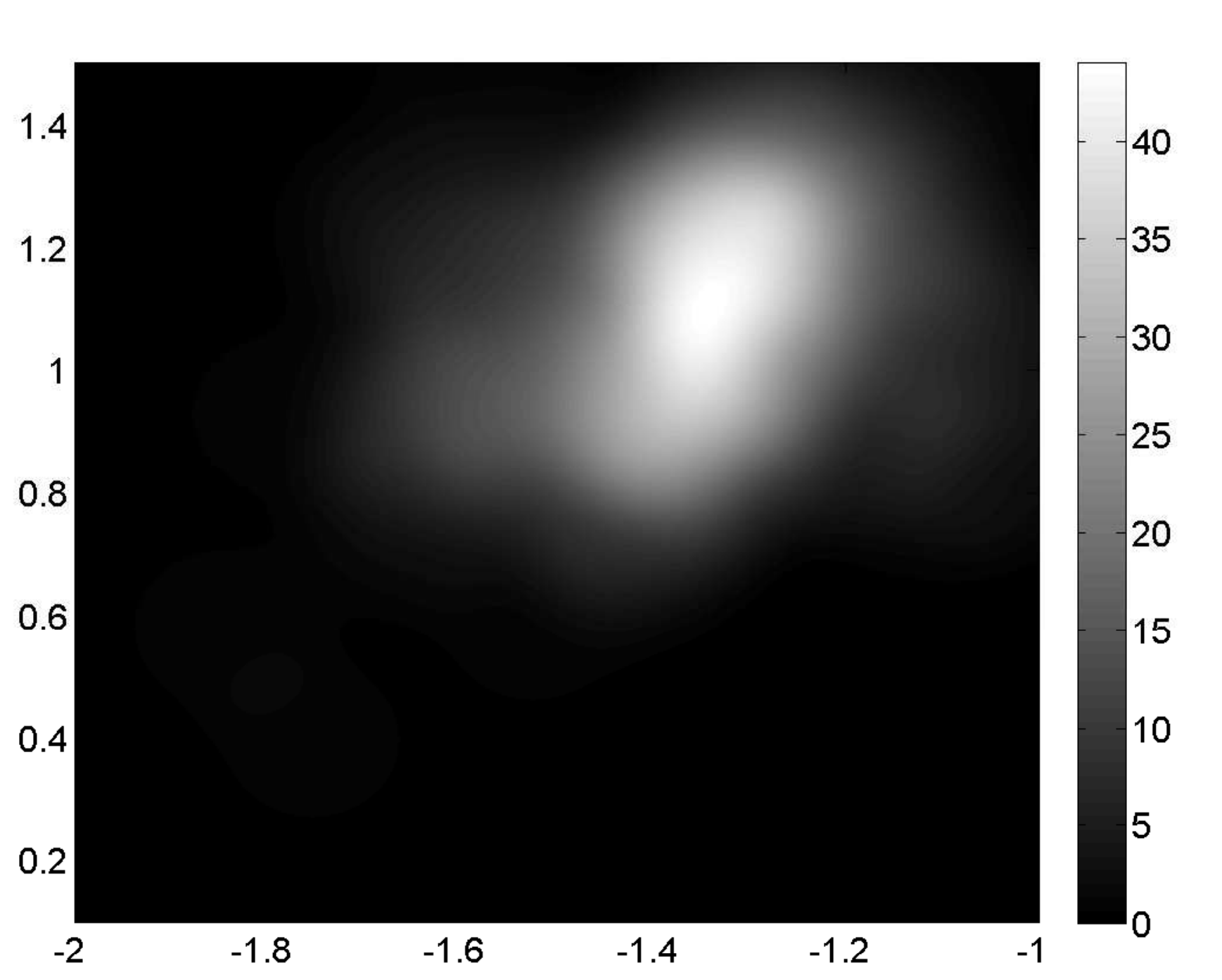} 
\includegraphics[width=0.20\textwidth,clip=true,trim=0cm 0cm 0cm 0cm]{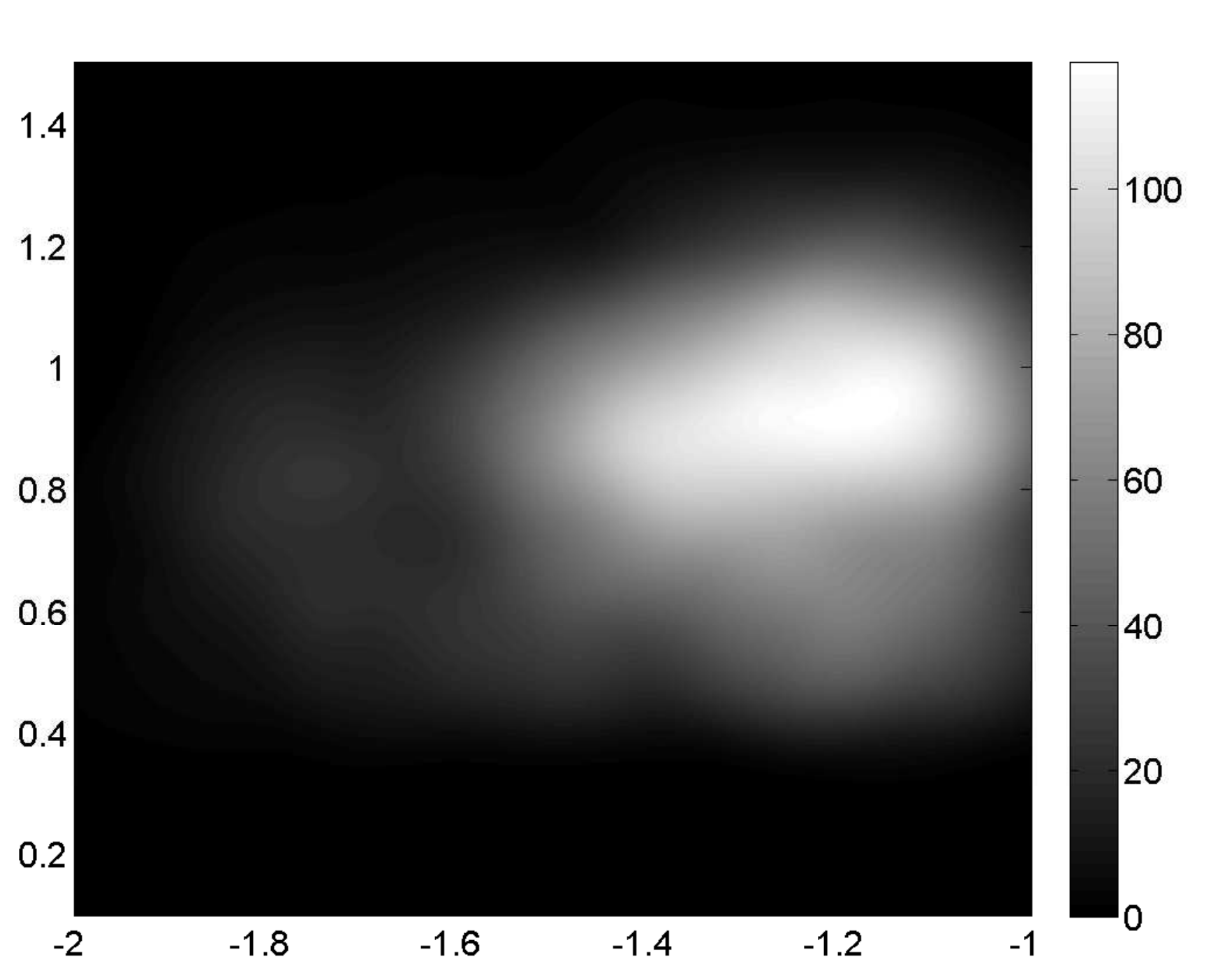} 
\includegraphics[width=0.20\textwidth,clip=true,trim=0cm 0cm 0cm 0cm]{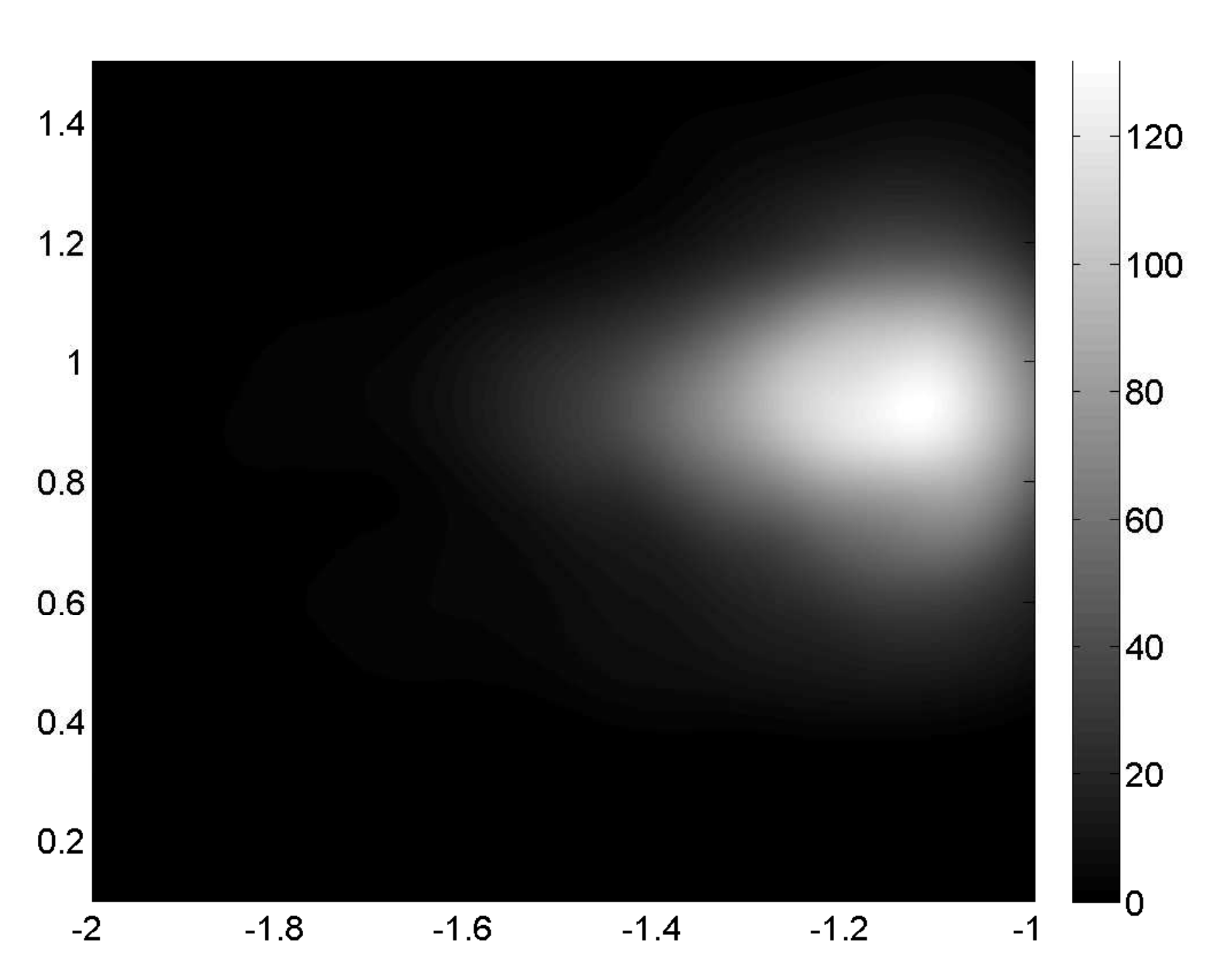} 
\includegraphics[width=0.20\textwidth,clip=true,trim=0cm 0cm 0cm 0cm]{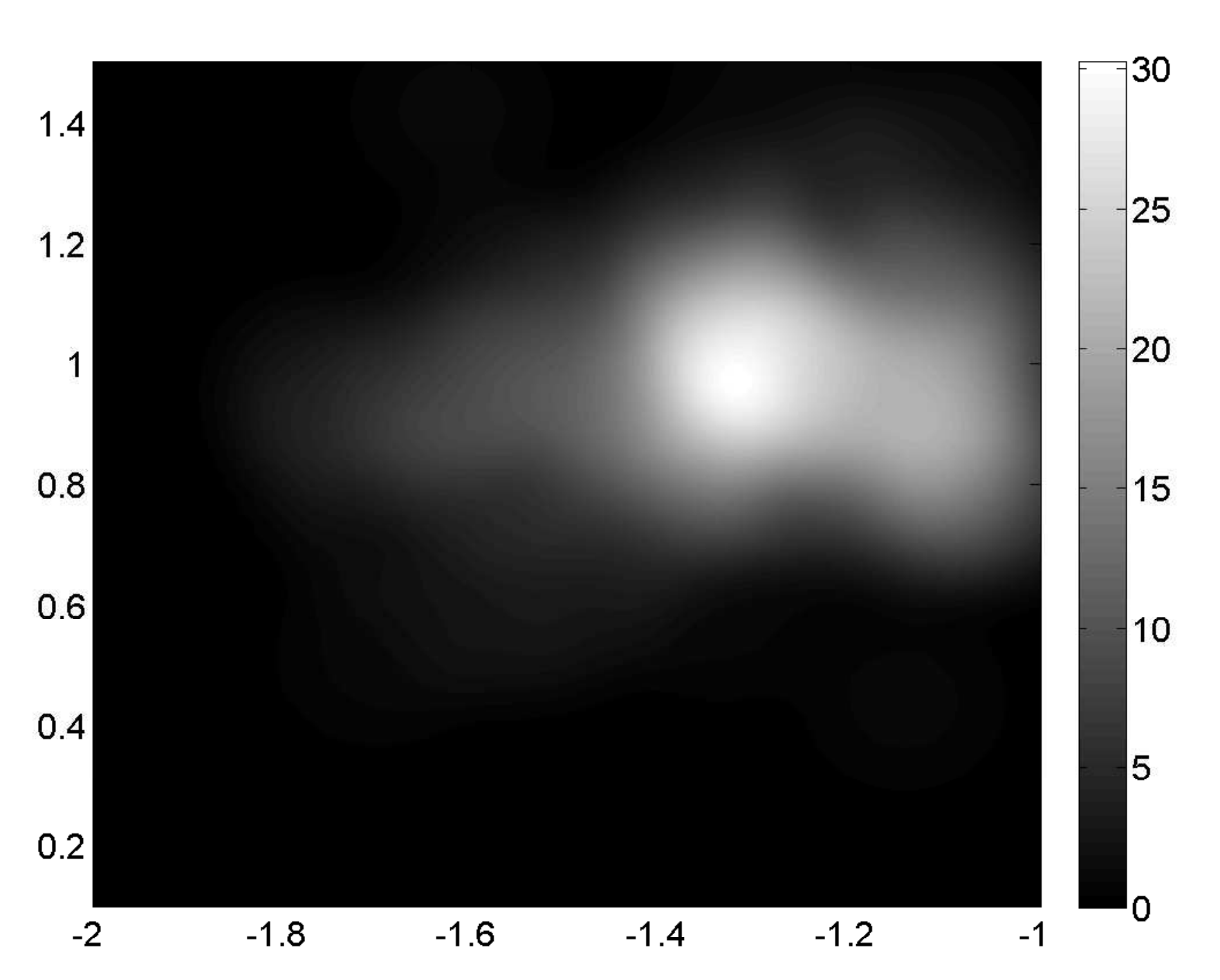} \\
\includegraphics[width=0.20\textwidth,clip=true,trim=0cm 0cm 0cm 0cm]{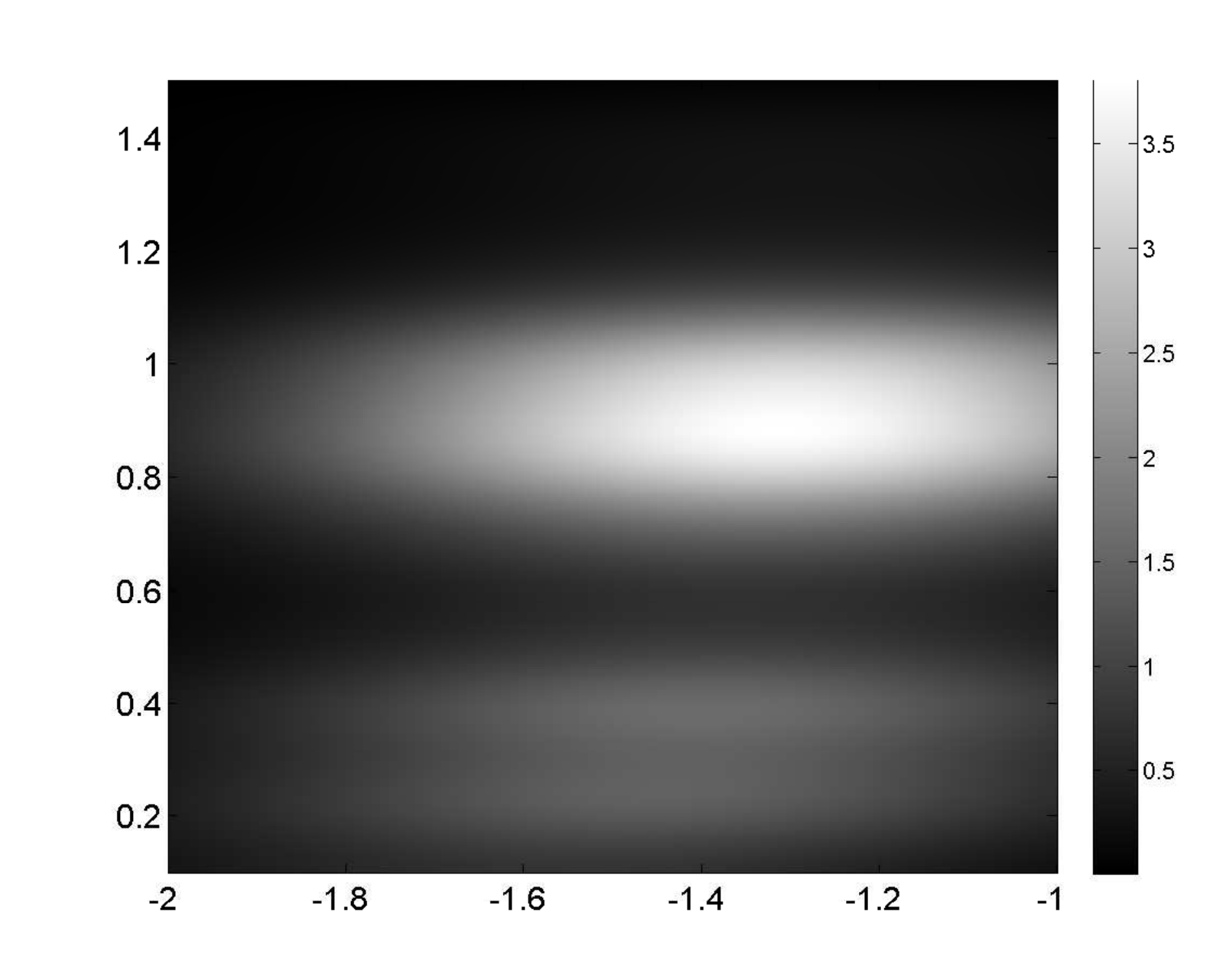} 
\includegraphics[width=0.20\textwidth,clip=true,trim=0cm 0cm 0cm 0cm]{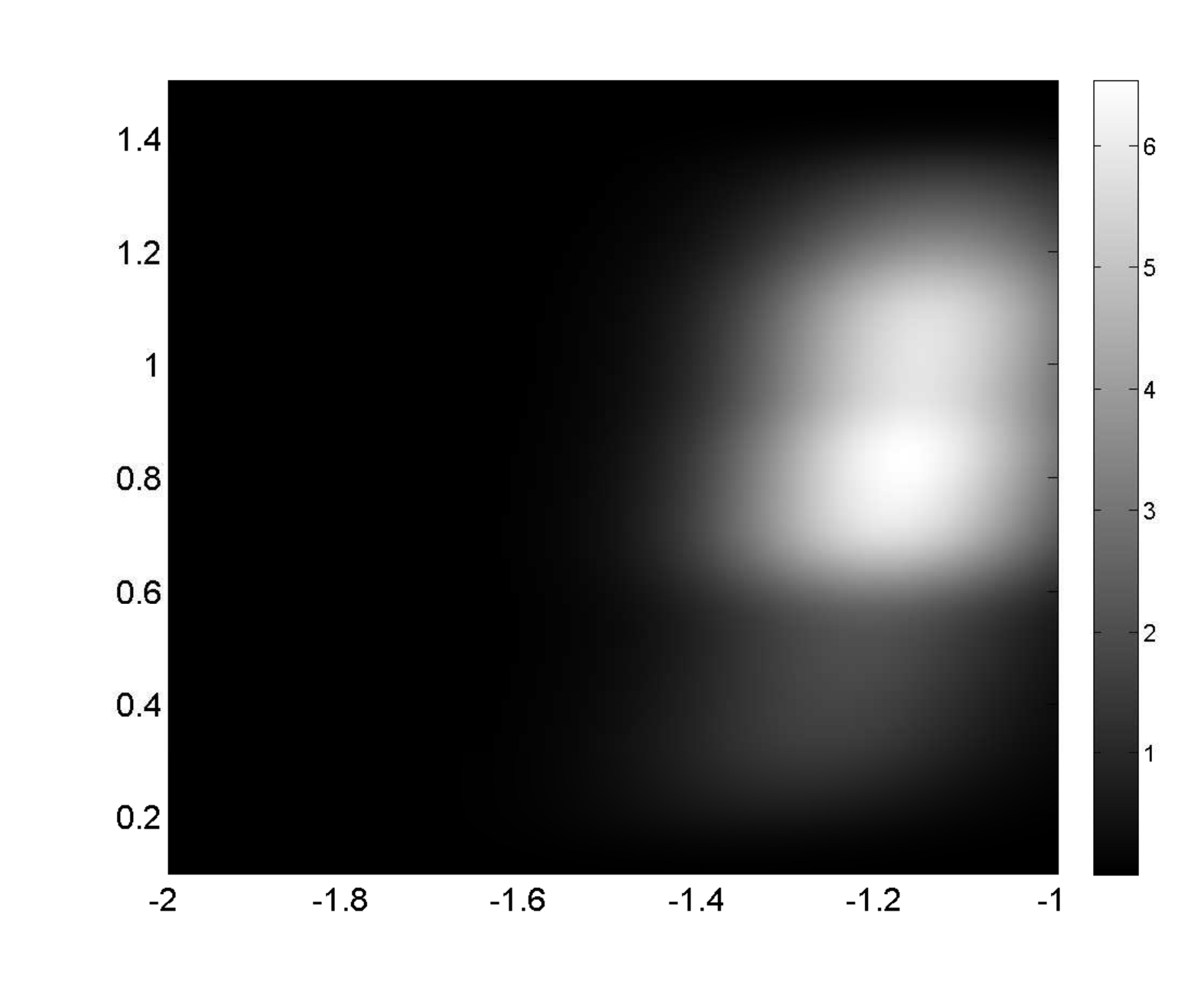} 
\includegraphics[width=0.20\textwidth,clip=true,trim=0cm 0cm 0cm 0cm]{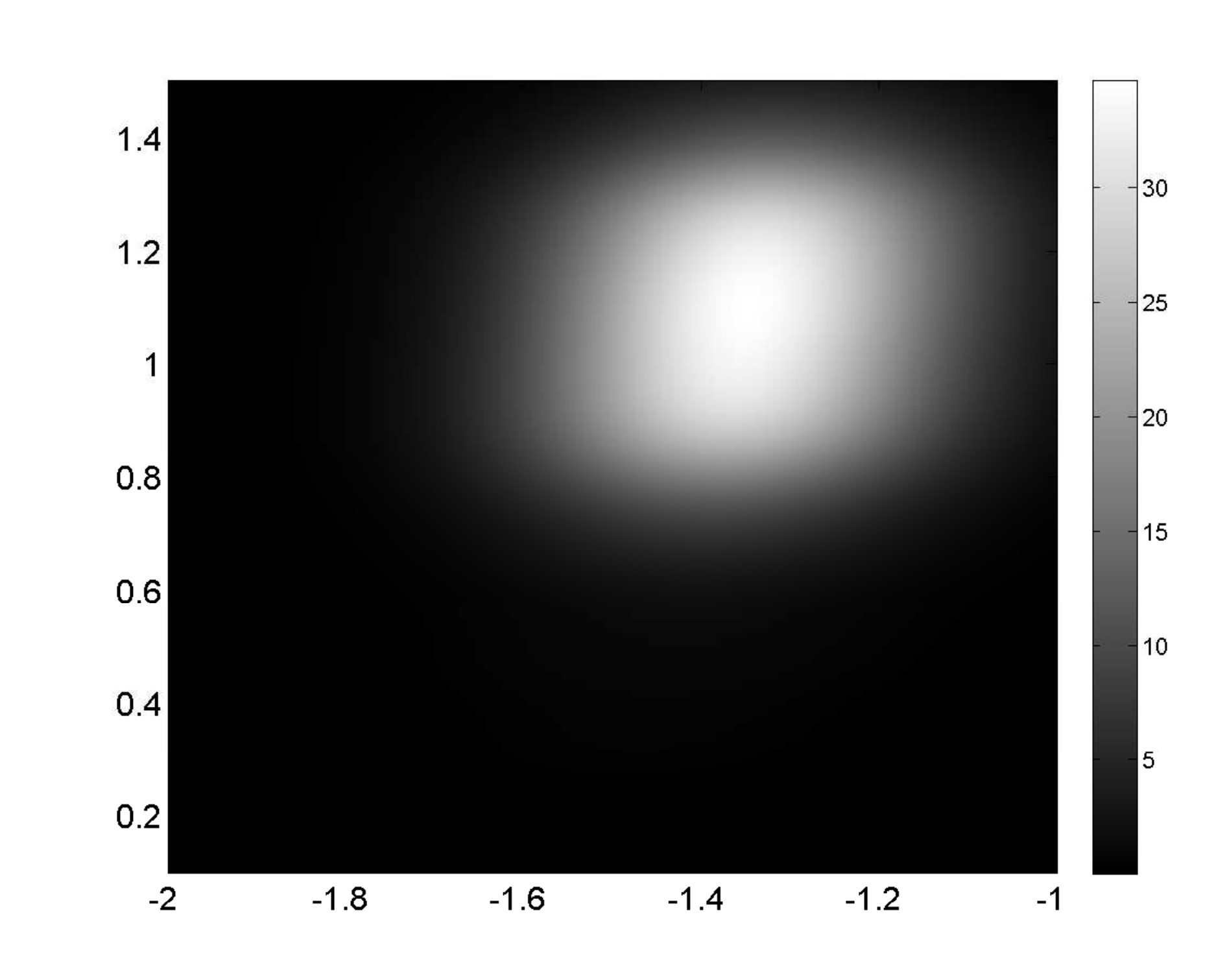}
\includegraphics[width=0.20\textwidth,clip=true,trim=0cm 0cm 0cm 0cm]{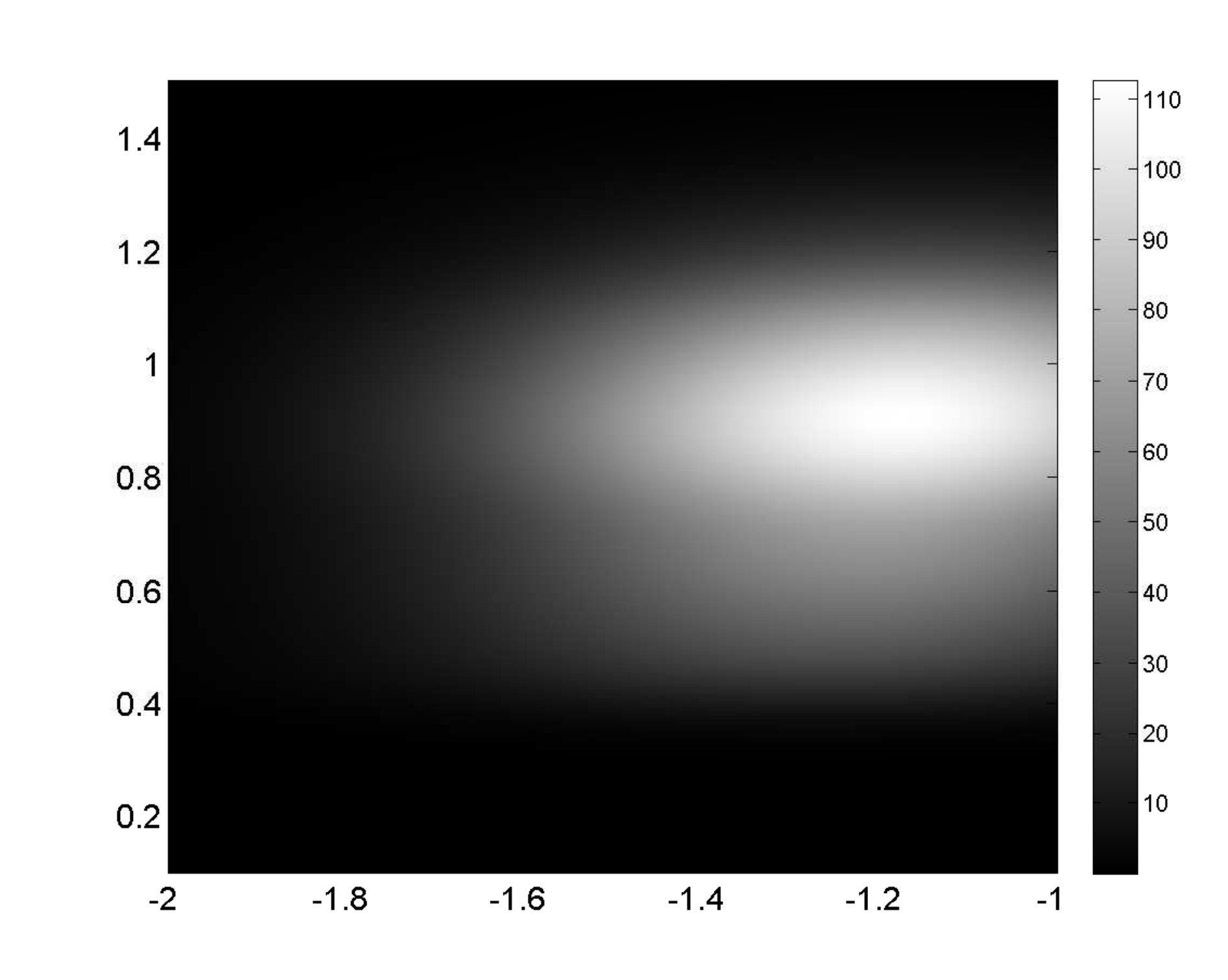} 
\includegraphics[width=0.20\textwidth,clip=true,trim=0cm 0cm 0cm 0cm]{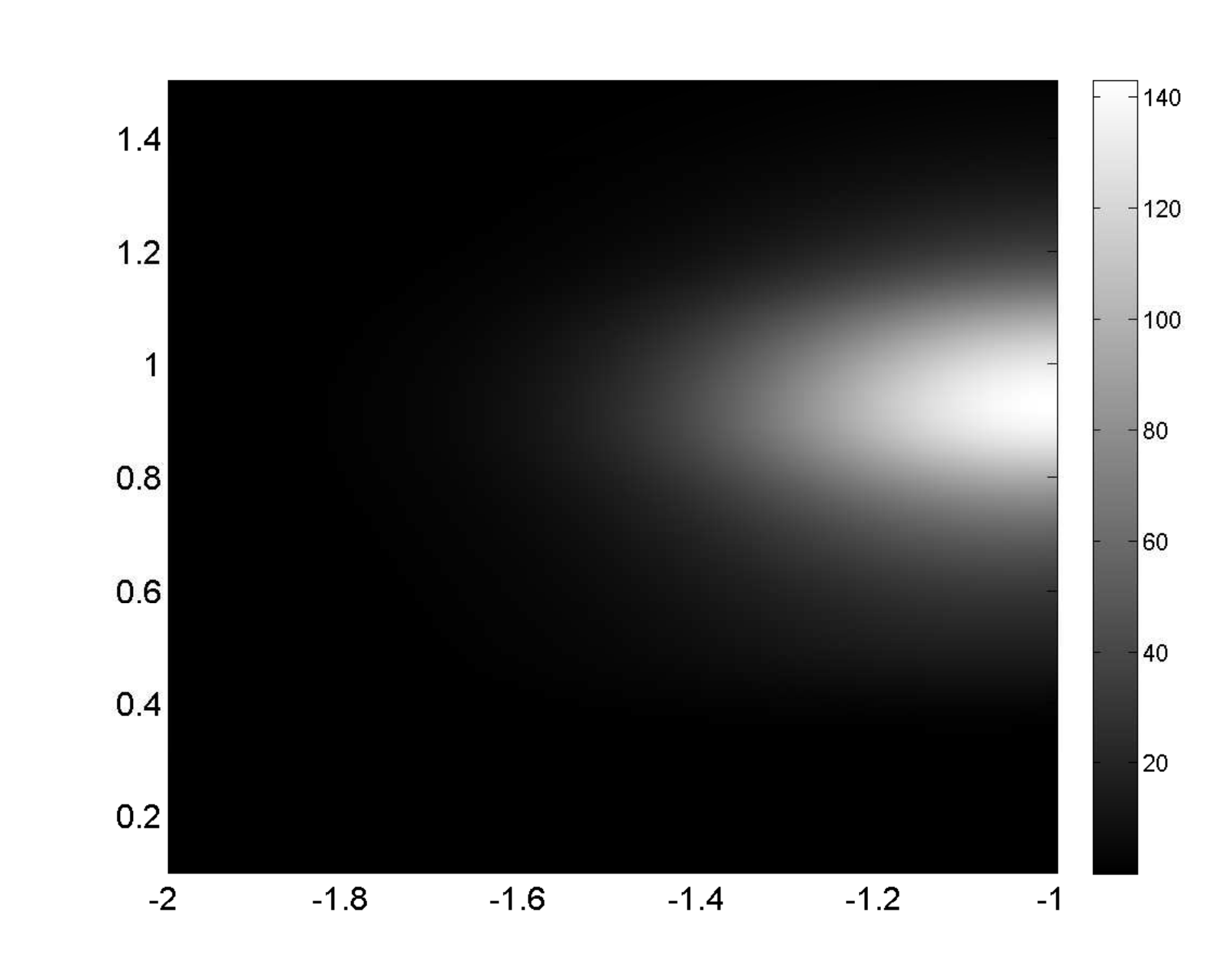} 
\includegraphics[width=0.20\textwidth,clip=true,trim=0cm 0cm 0cm 0cm]{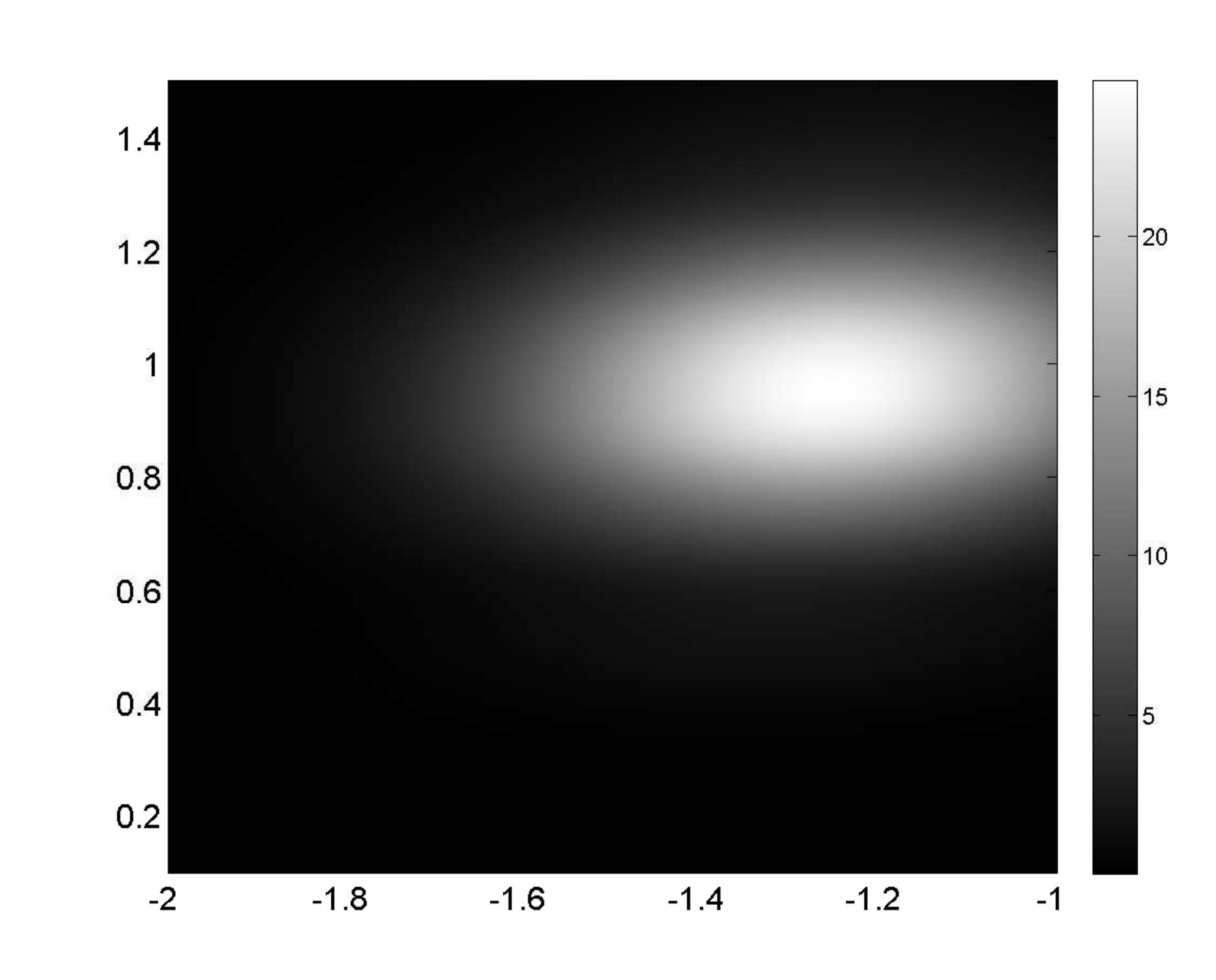} \\
\includegraphics[width=0.20\textwidth,clip=true,trim=0cm 0cm 0cm 0cm]{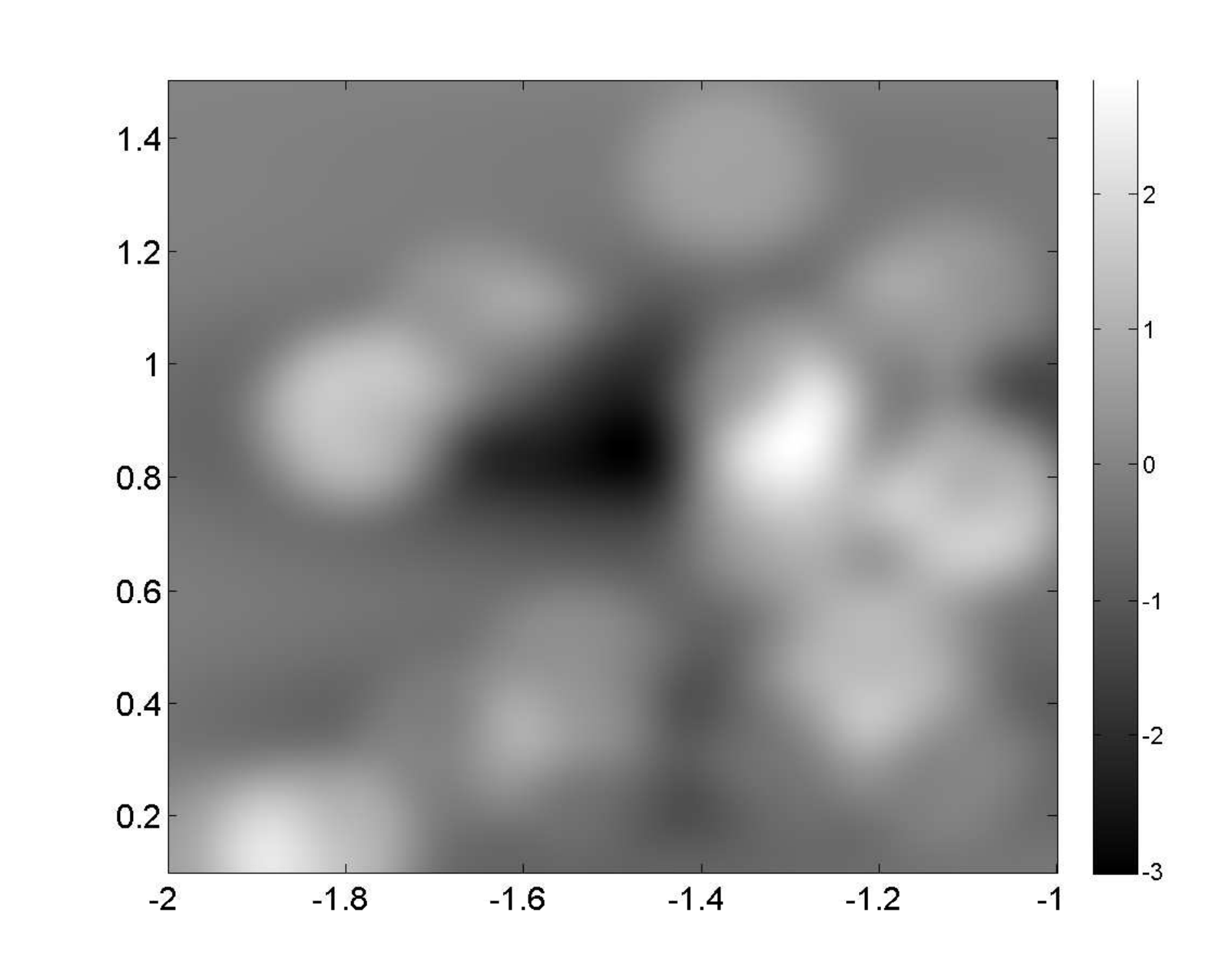} 
\includegraphics[width=0.20\textwidth,clip=true,trim=0cm 0cm 0cm 0cm]{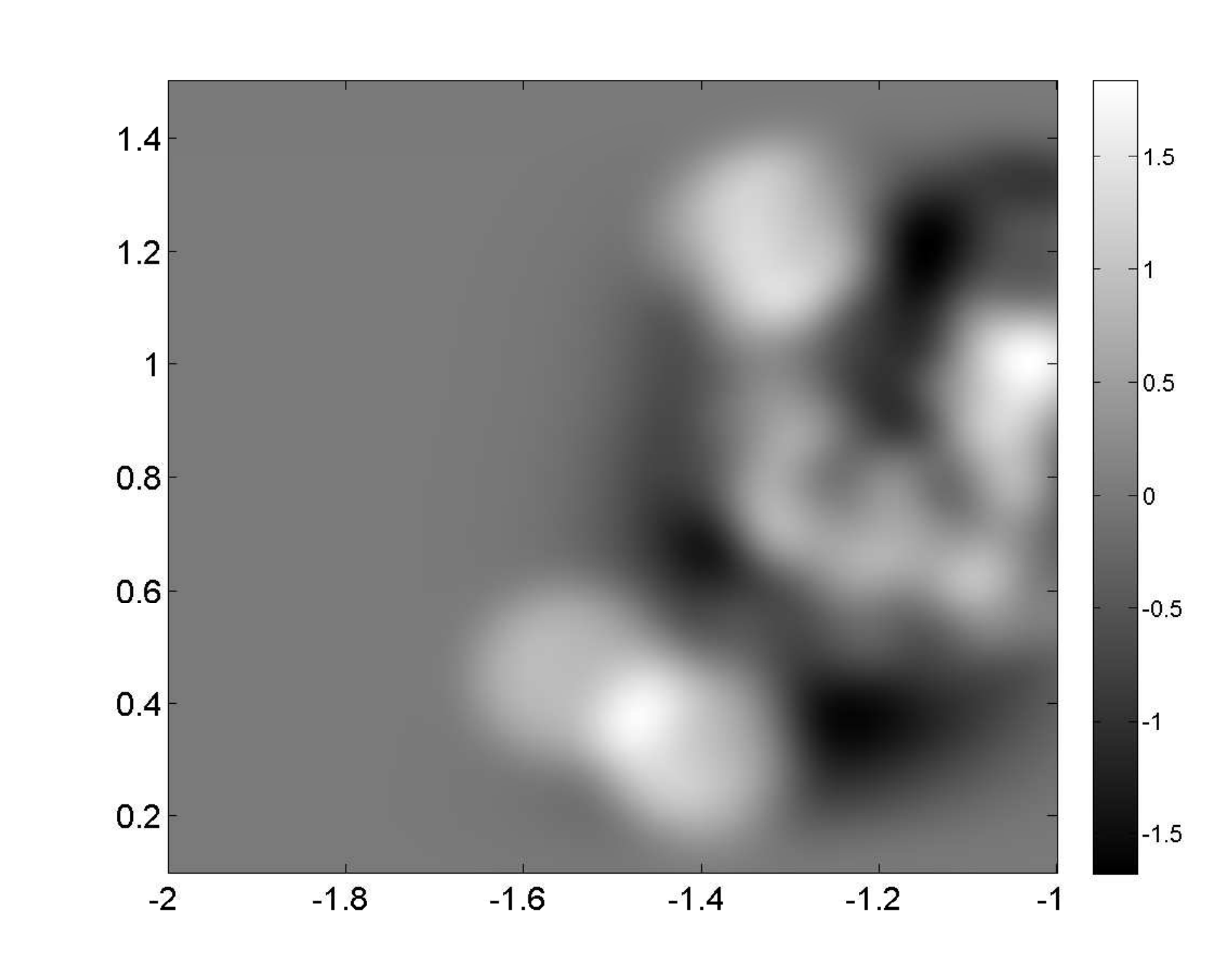} 
\includegraphics[width=0.20\textwidth,clip=true,trim=0cm 0cm 0cm 0cm]{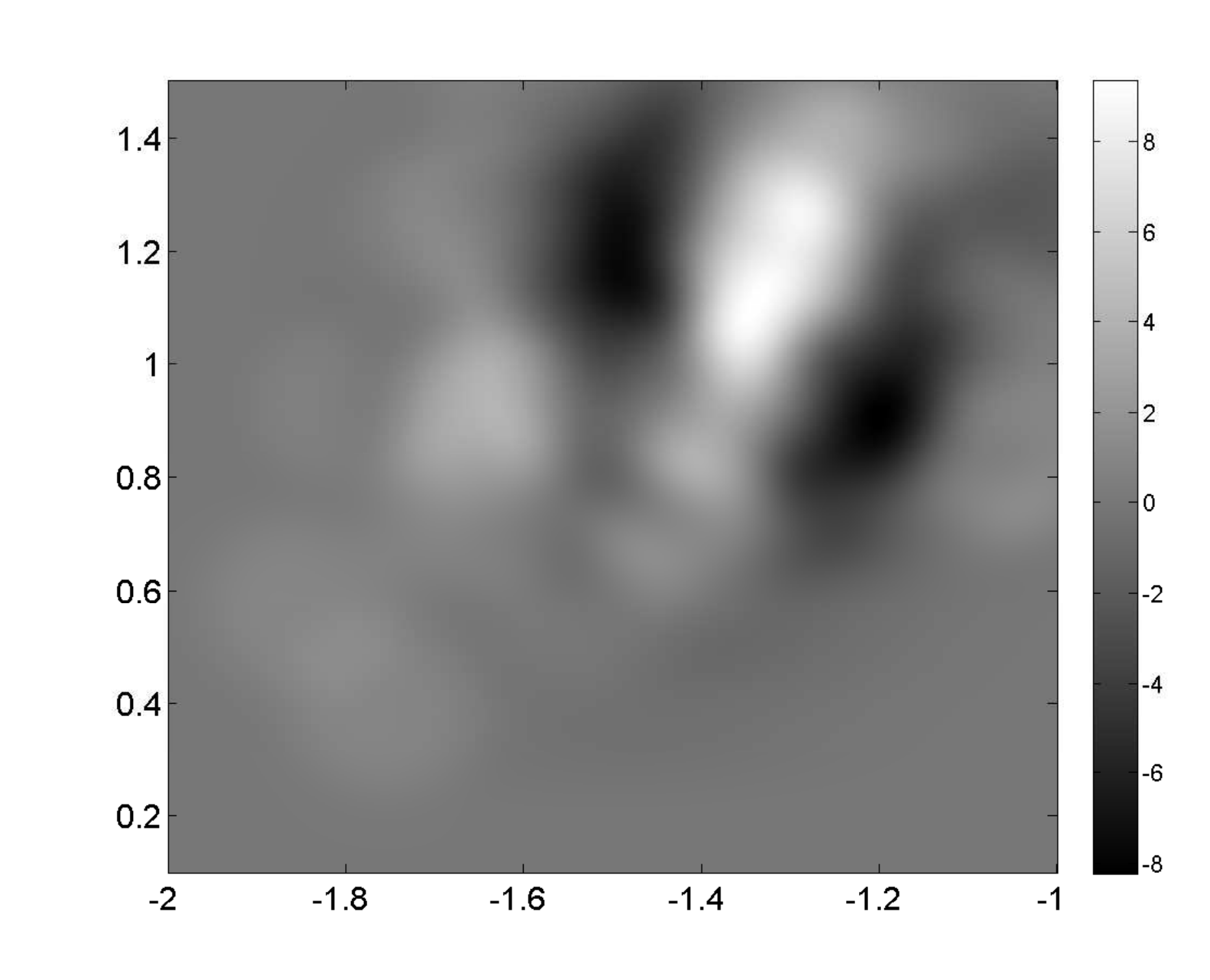} 
\includegraphics[width=0.20\textwidth,clip=true,trim=0cm 0cm 0cm 0cm]{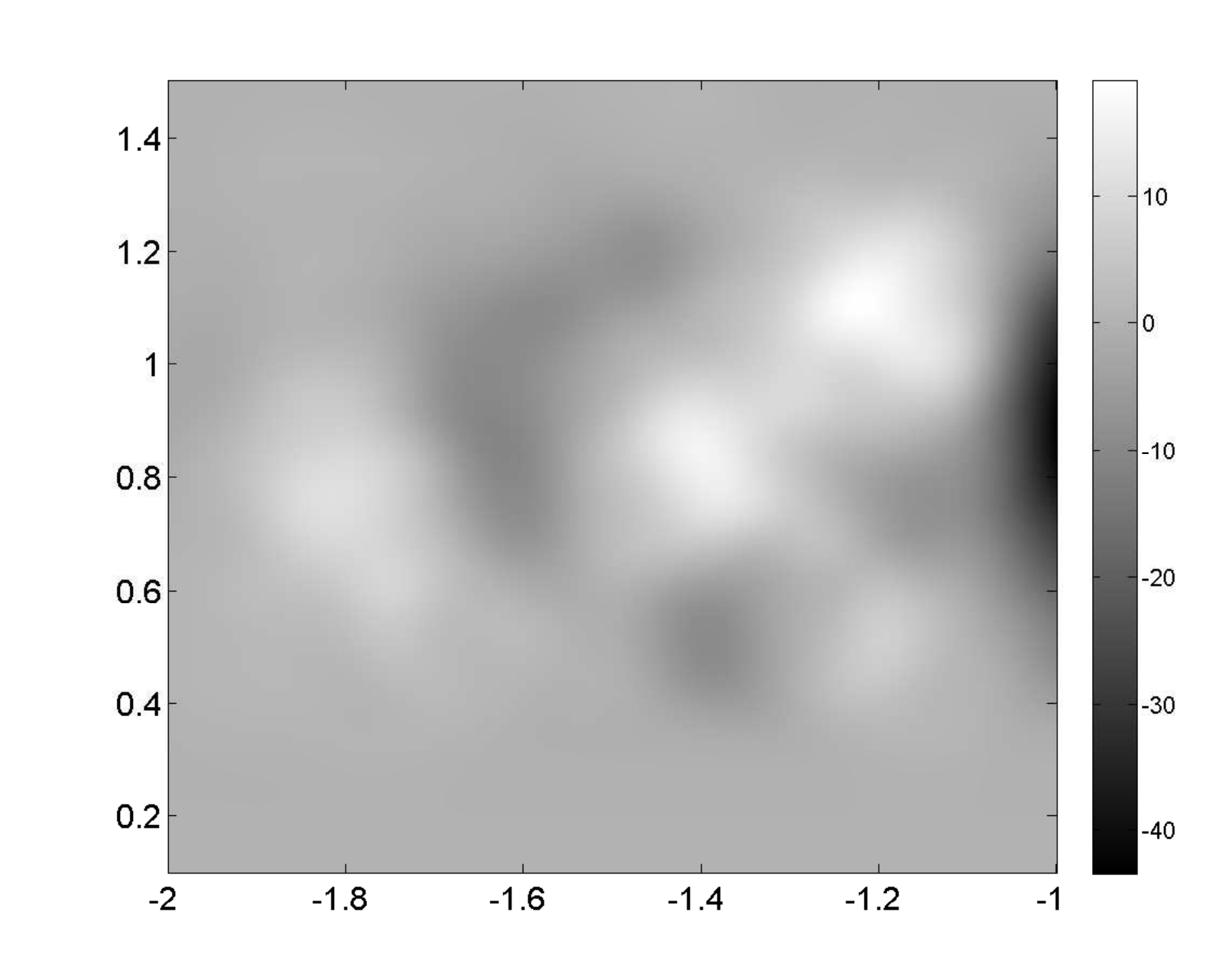} 
\includegraphics[width=0.20\textwidth,clip=true,trim=0cm 0cm 0cm 0cm]{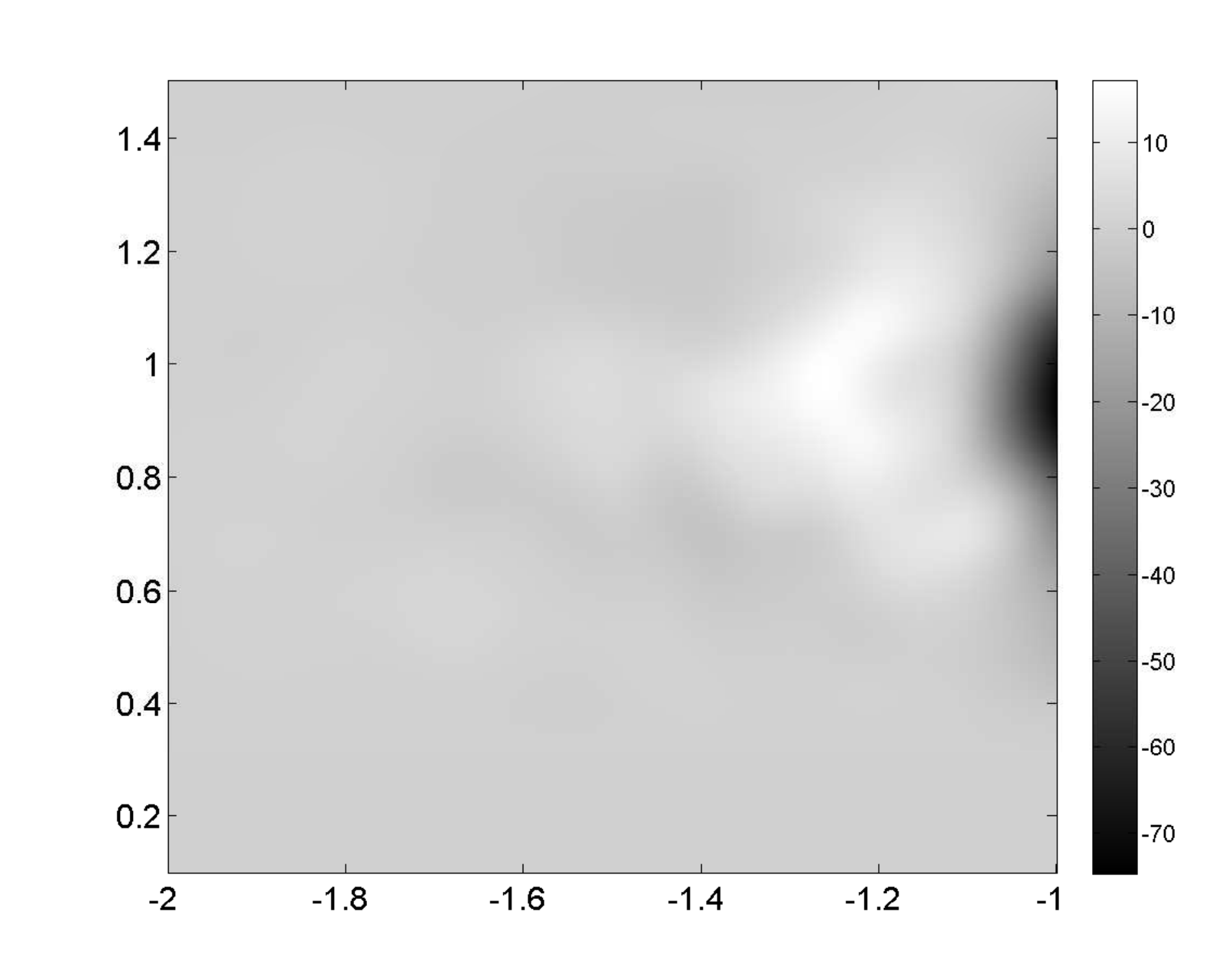} 
\includegraphics[width=0.20\textwidth,clip=true,trim=0cm 0cm 0cm 0cm]{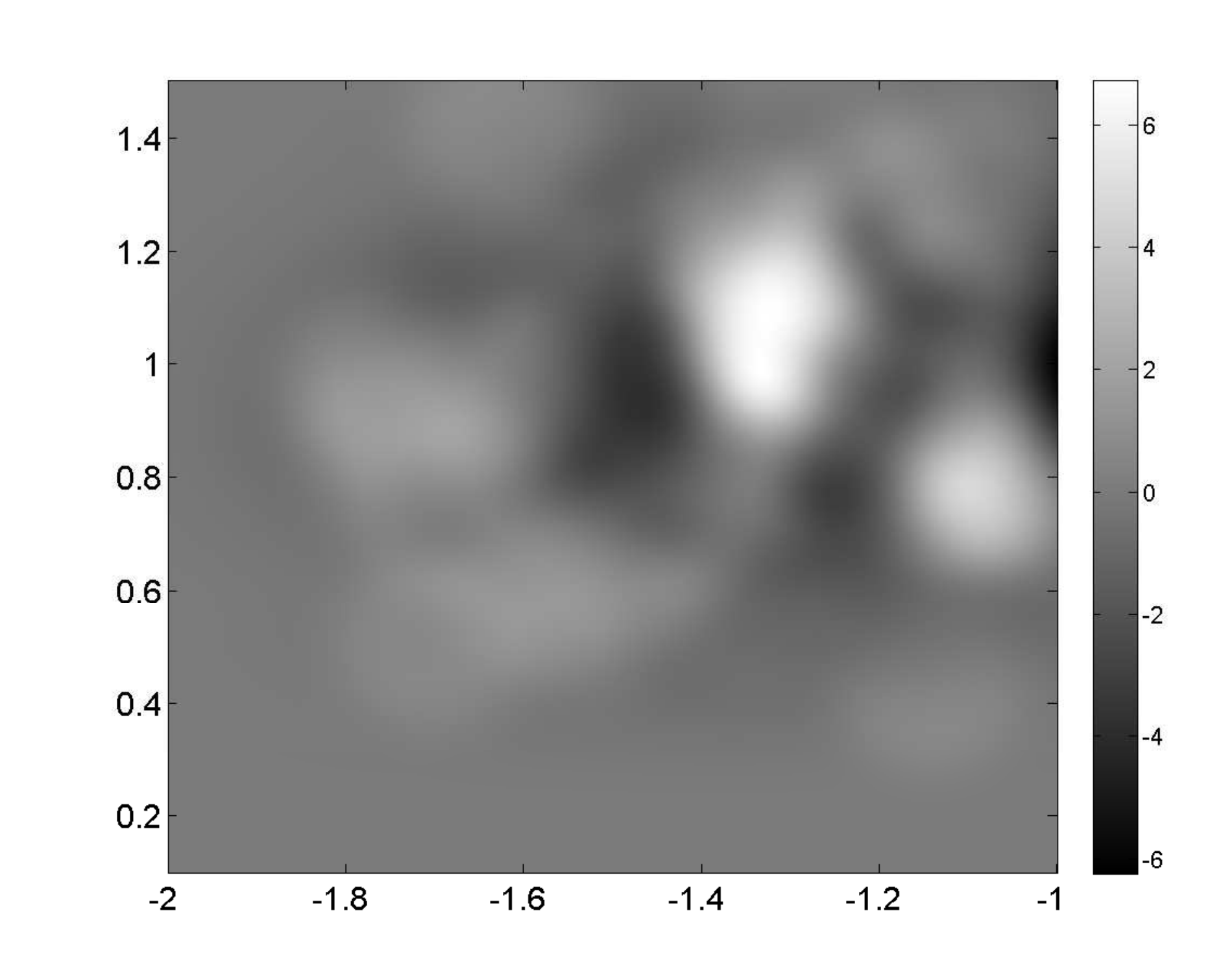}  
\caption{Same as Figure 2 for the a $\propto$ M$^{3/13}$ tidal halting model.}
\end{figure}
\end{landscape}

\clearpage
\begin{landscape}
\begin{figure}
\centering
\includegraphics[width=0.20\textwidth,clip=true,trim=0cm 0cm 0cm 0cm]{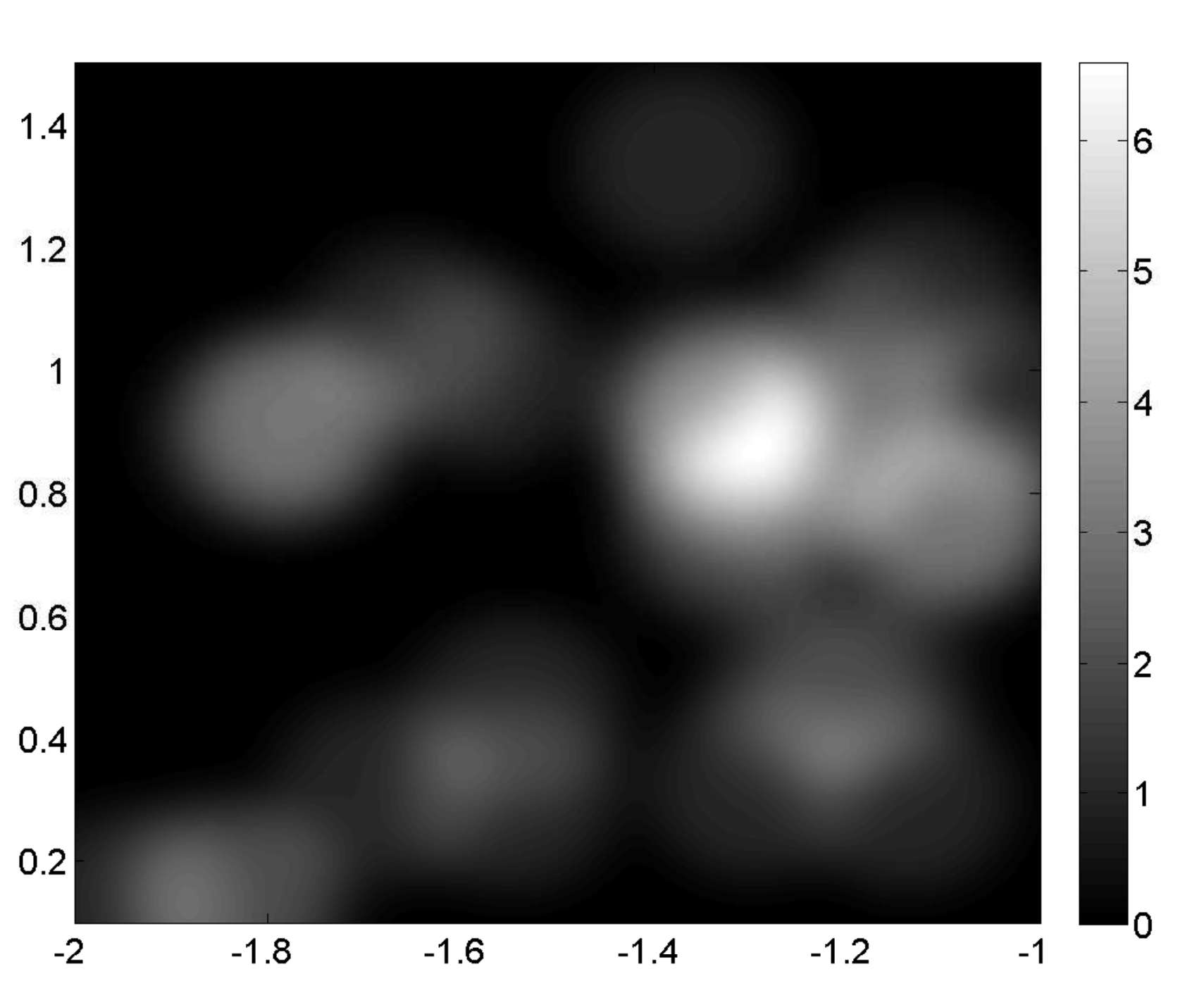} 
\includegraphics[width=0.20\textwidth,clip=true,trim=0cm 0cm 0cm 0cm]{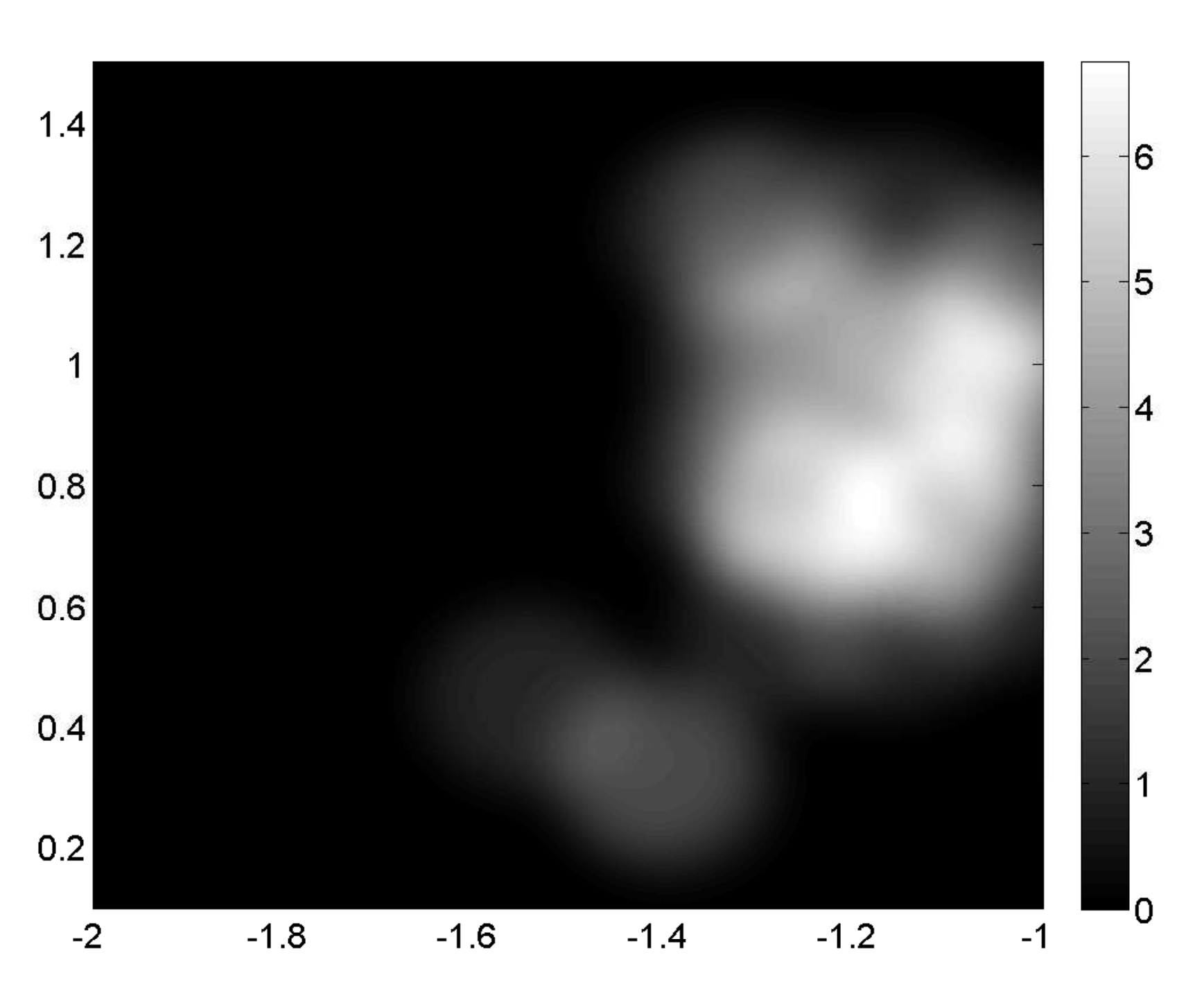} 
\includegraphics[width=0.20\textwidth,clip=true,trim=0cm 0cm 0cm 0cm]{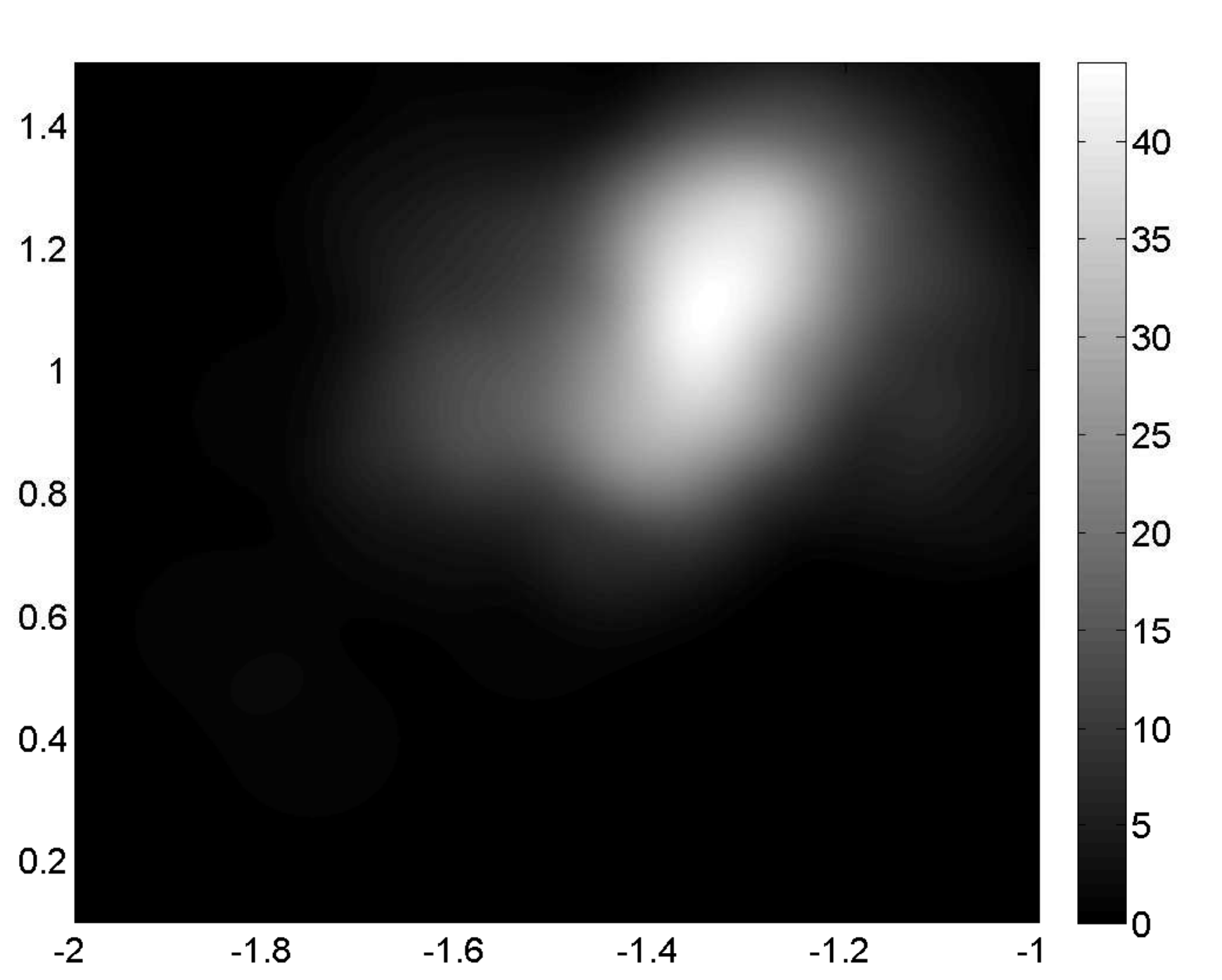} 
\includegraphics[width=0.20\textwidth,clip=true,trim=0cm 0cm 0cm 0cm]{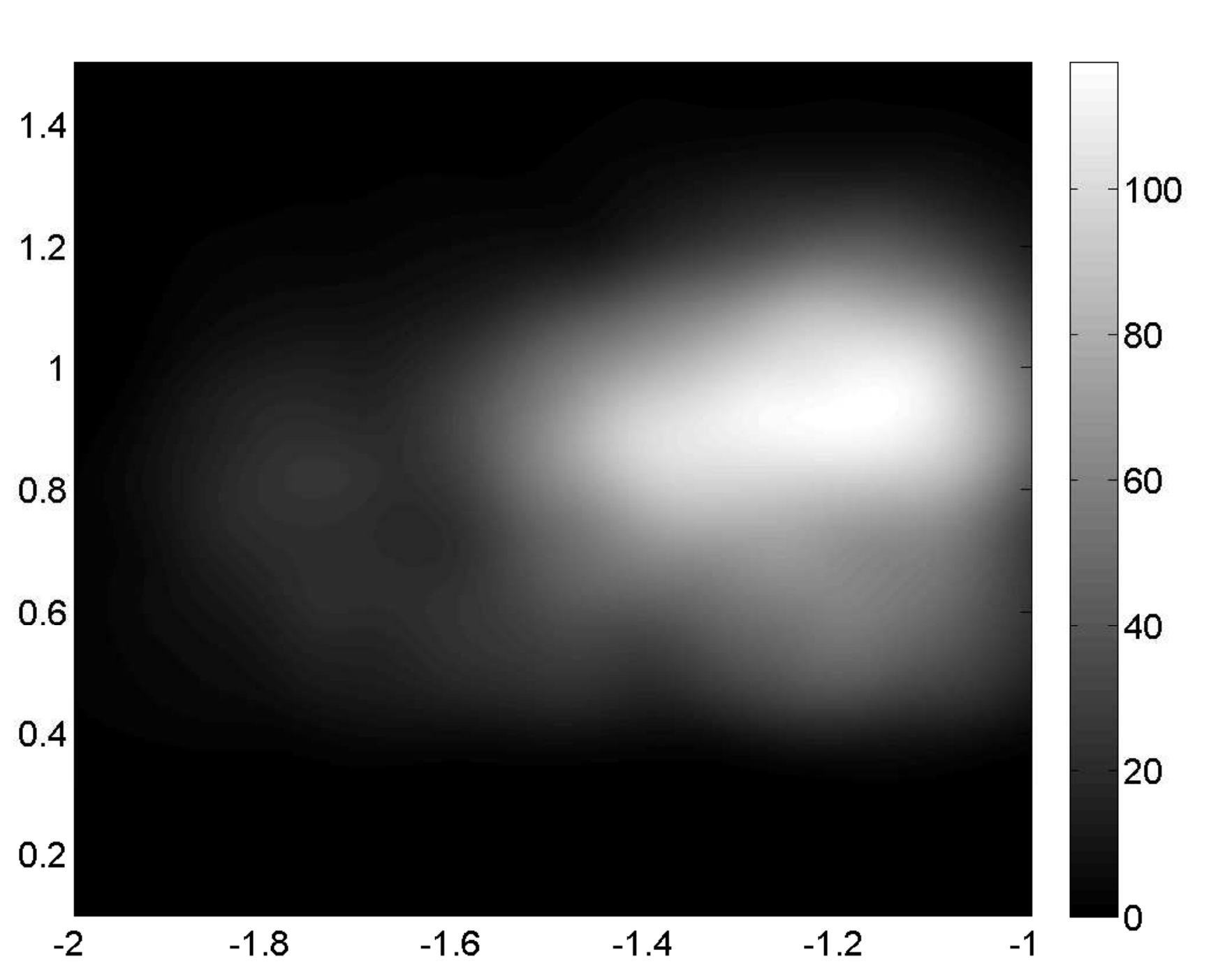} 
\includegraphics[width=0.20\textwidth,clip=true,trim=0cm 0cm 0cm 0cm]{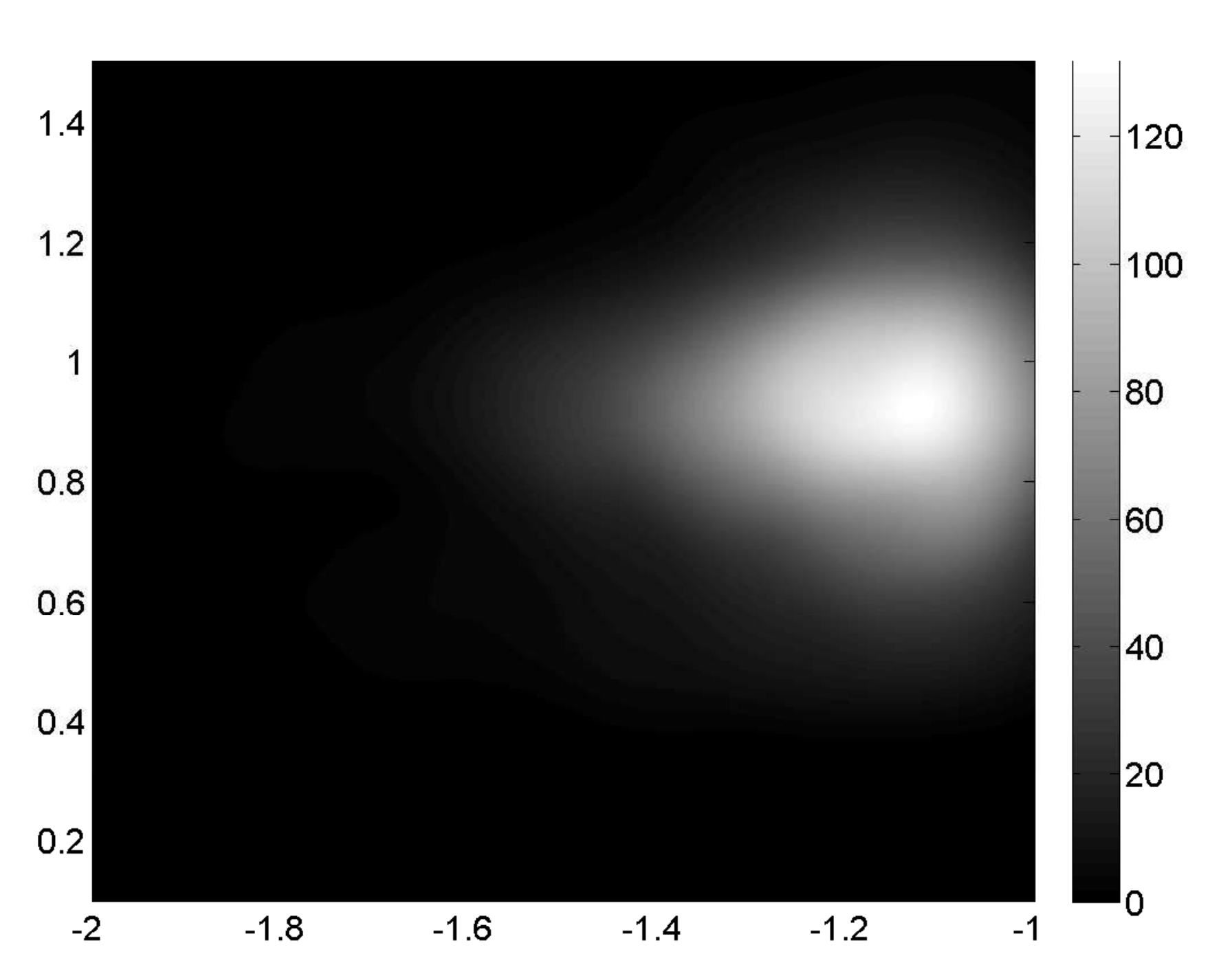} 
\includegraphics[width=0.20\textwidth,clip=true,trim=0cm 0cm 0cm 0cm]{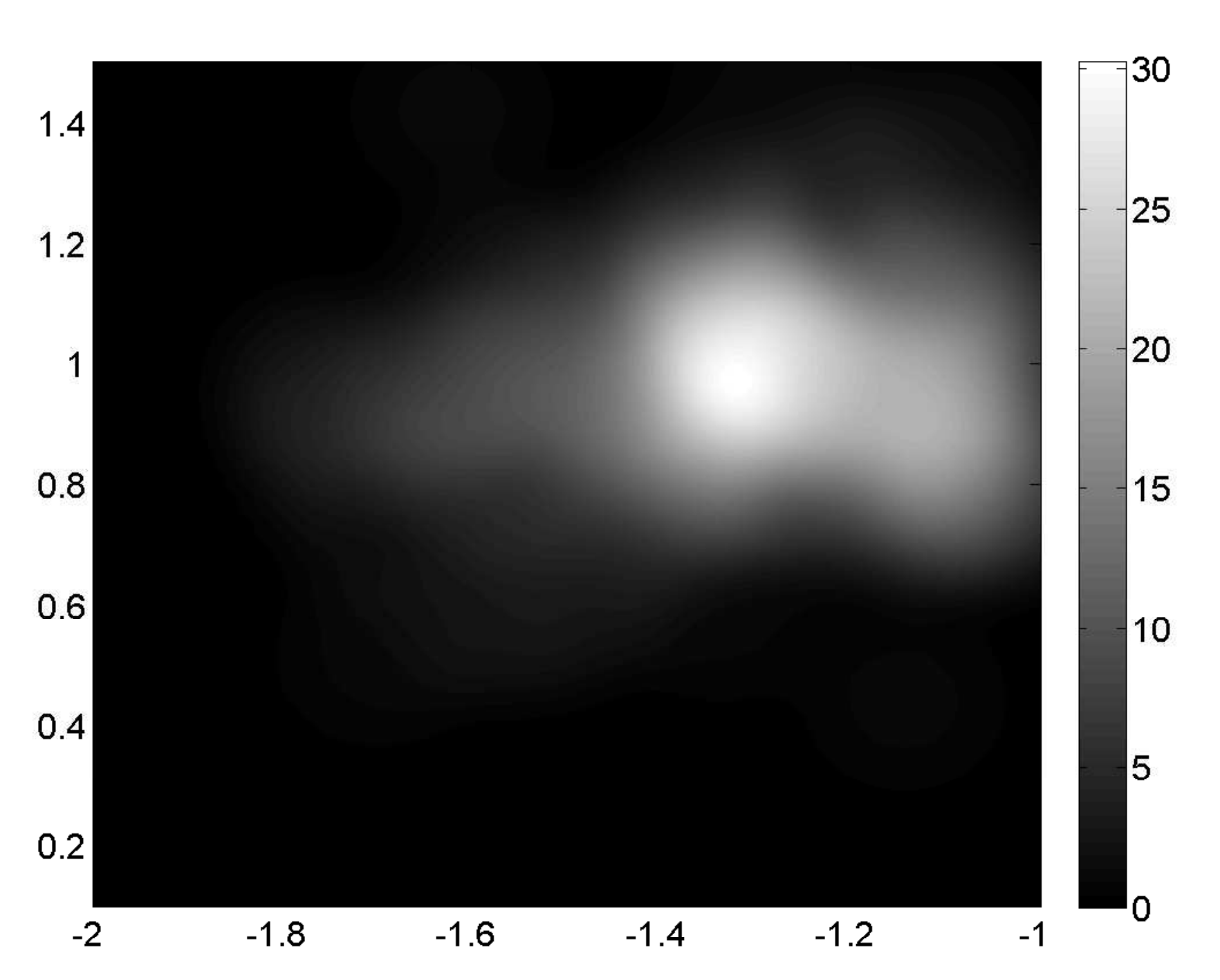} \\
\includegraphics[width=0.20\textwidth,clip=true,trim=0cm 0cm 0cm 0cm]{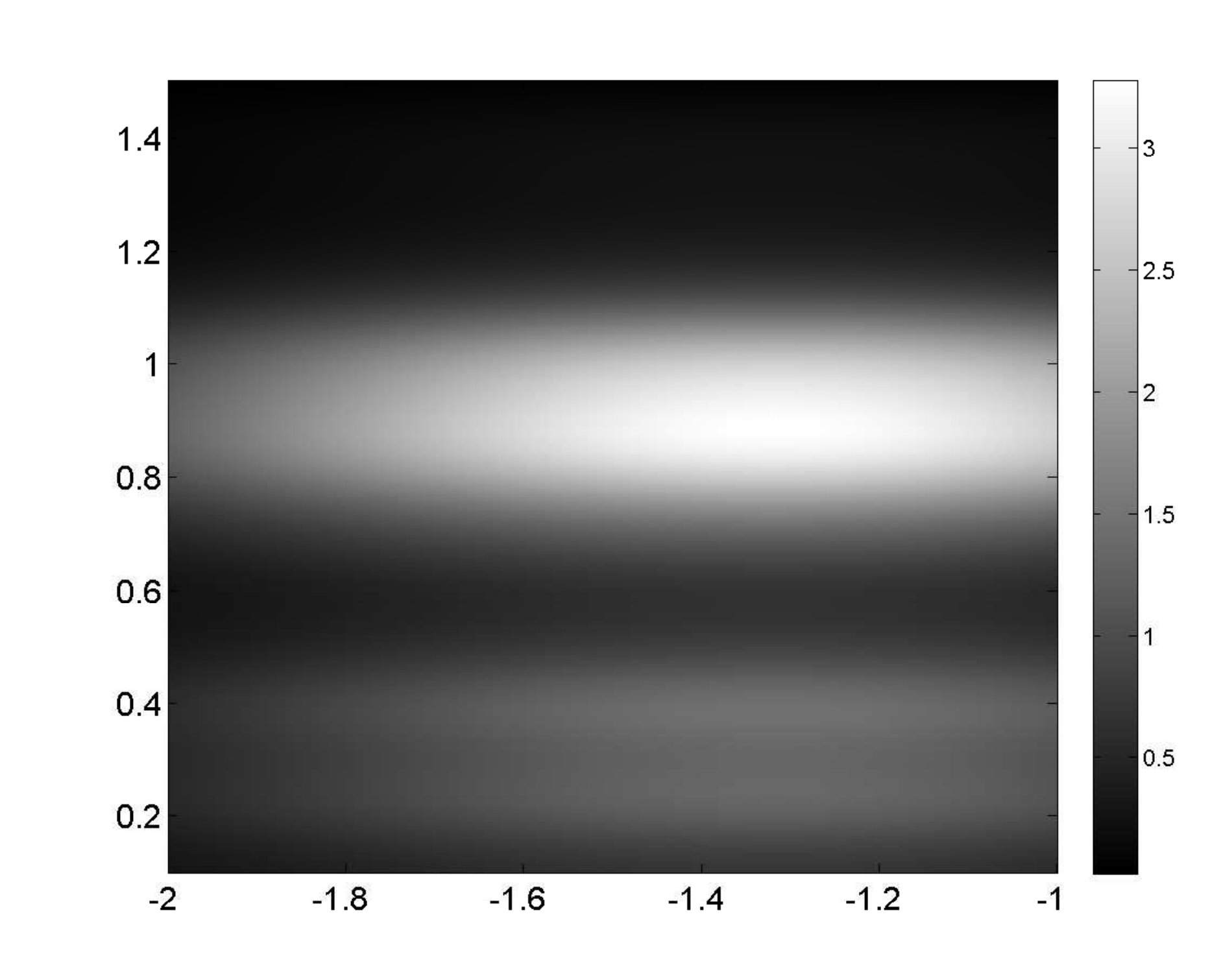} 
\includegraphics[width=0.20\textwidth,clip=true,trim=0cm 0cm 0cm 0cm]{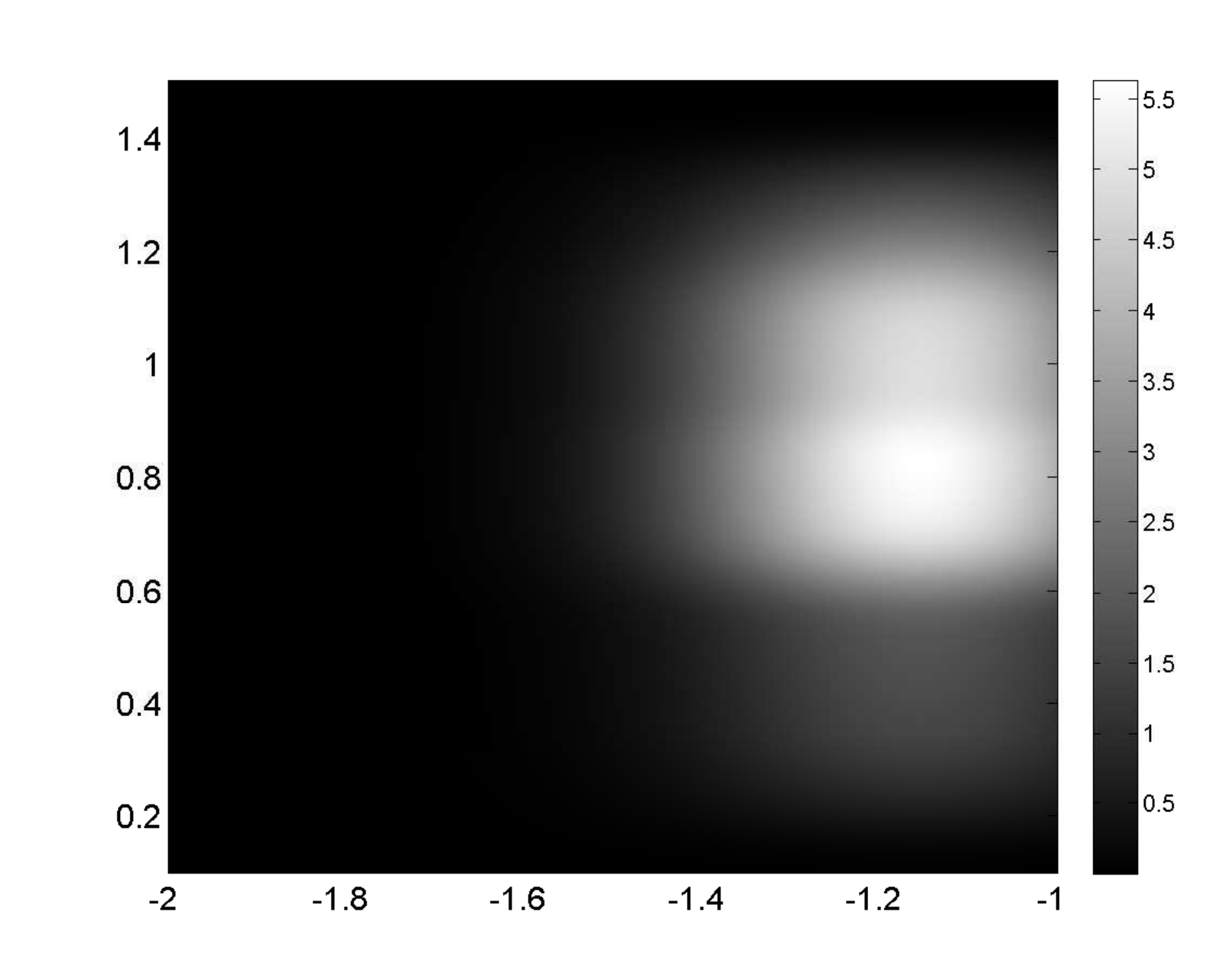} 
\includegraphics[width=0.20\textwidth,clip=true,trim=0cm 0cm 0cm 0cm]{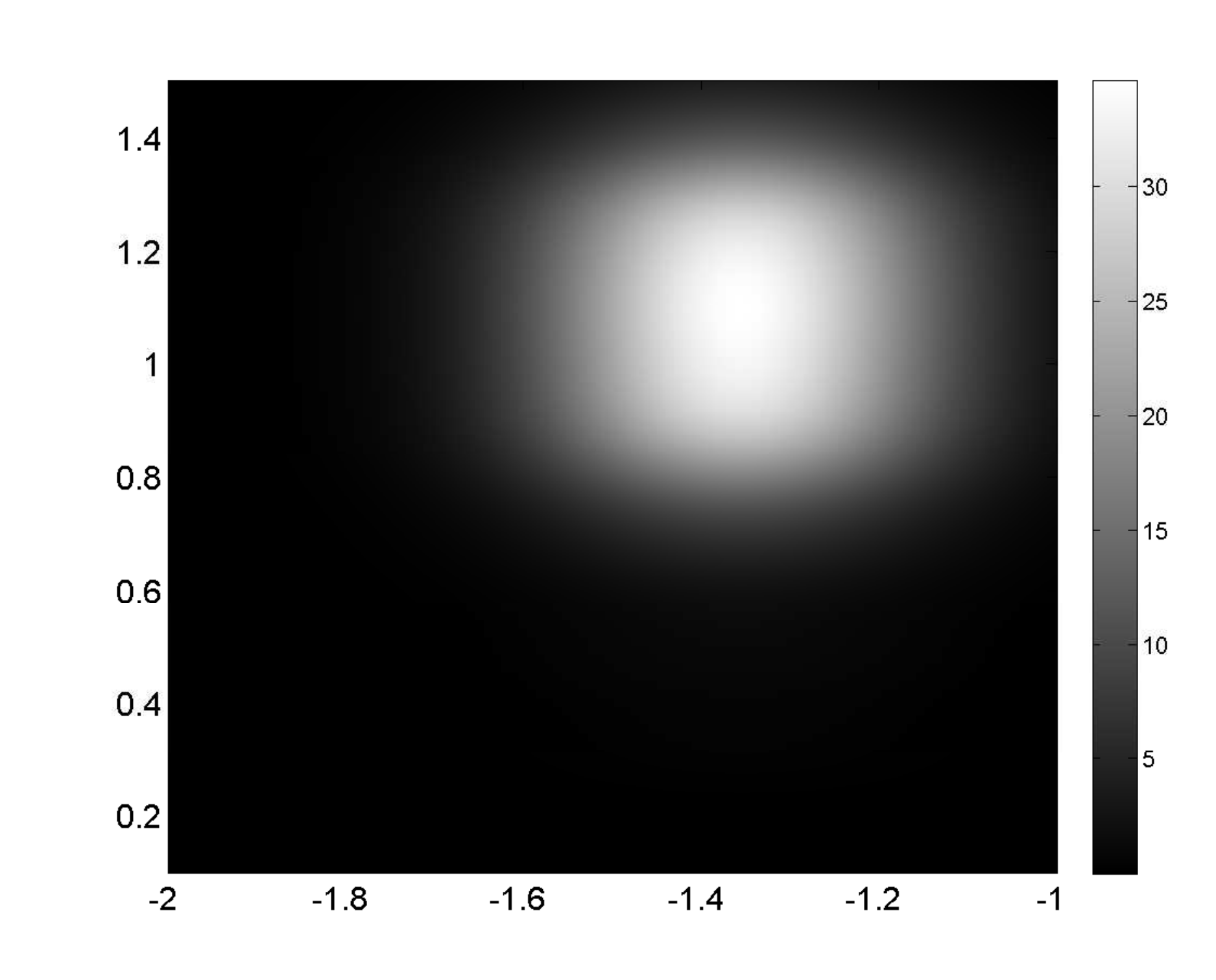}
\includegraphics[width=0.20\textwidth,clip=true,trim=0cm 0cm 0cm 0cm]{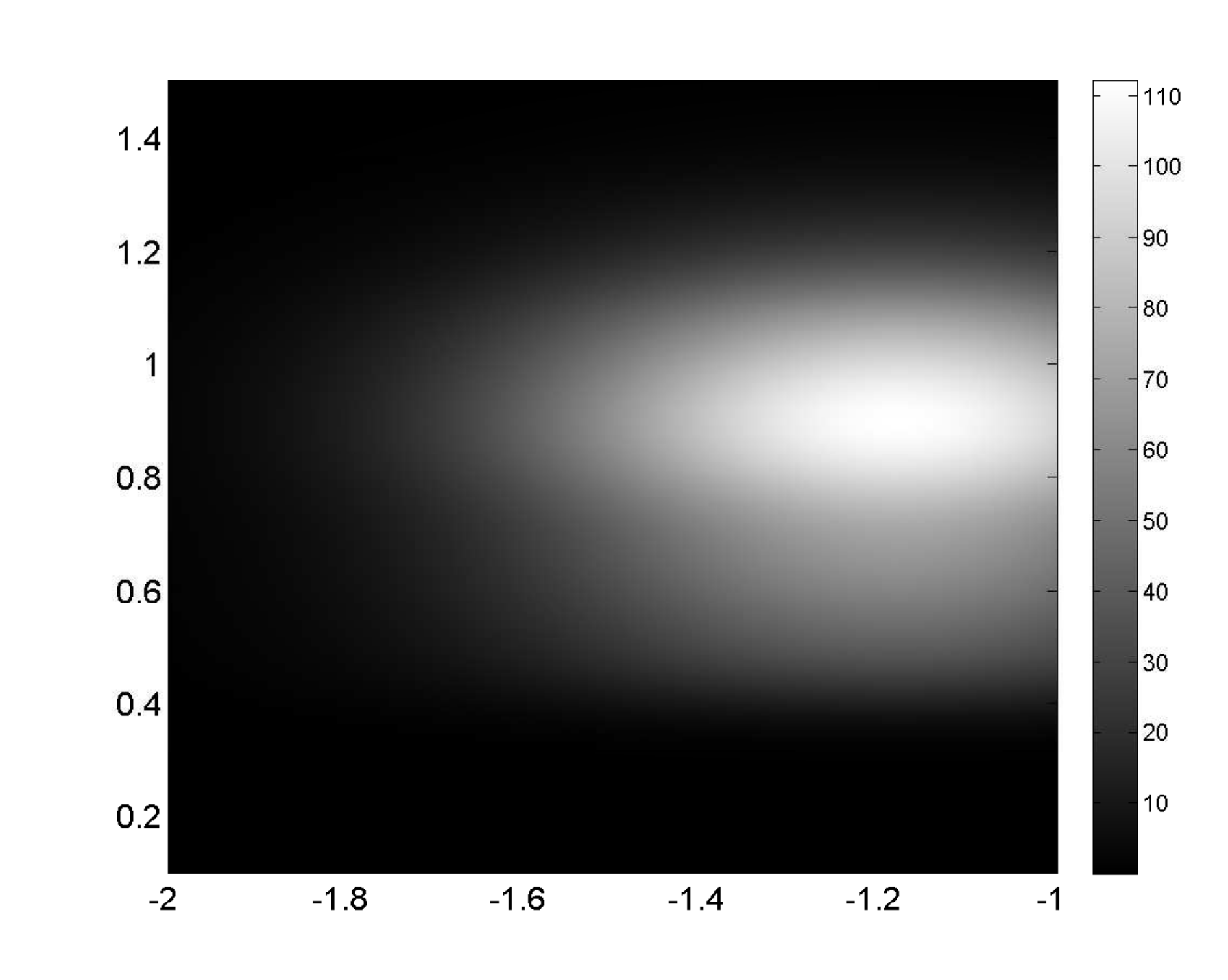} 
\includegraphics[width=0.20\textwidth,clip=true,trim=0cm 0cm 0cm 0cm]{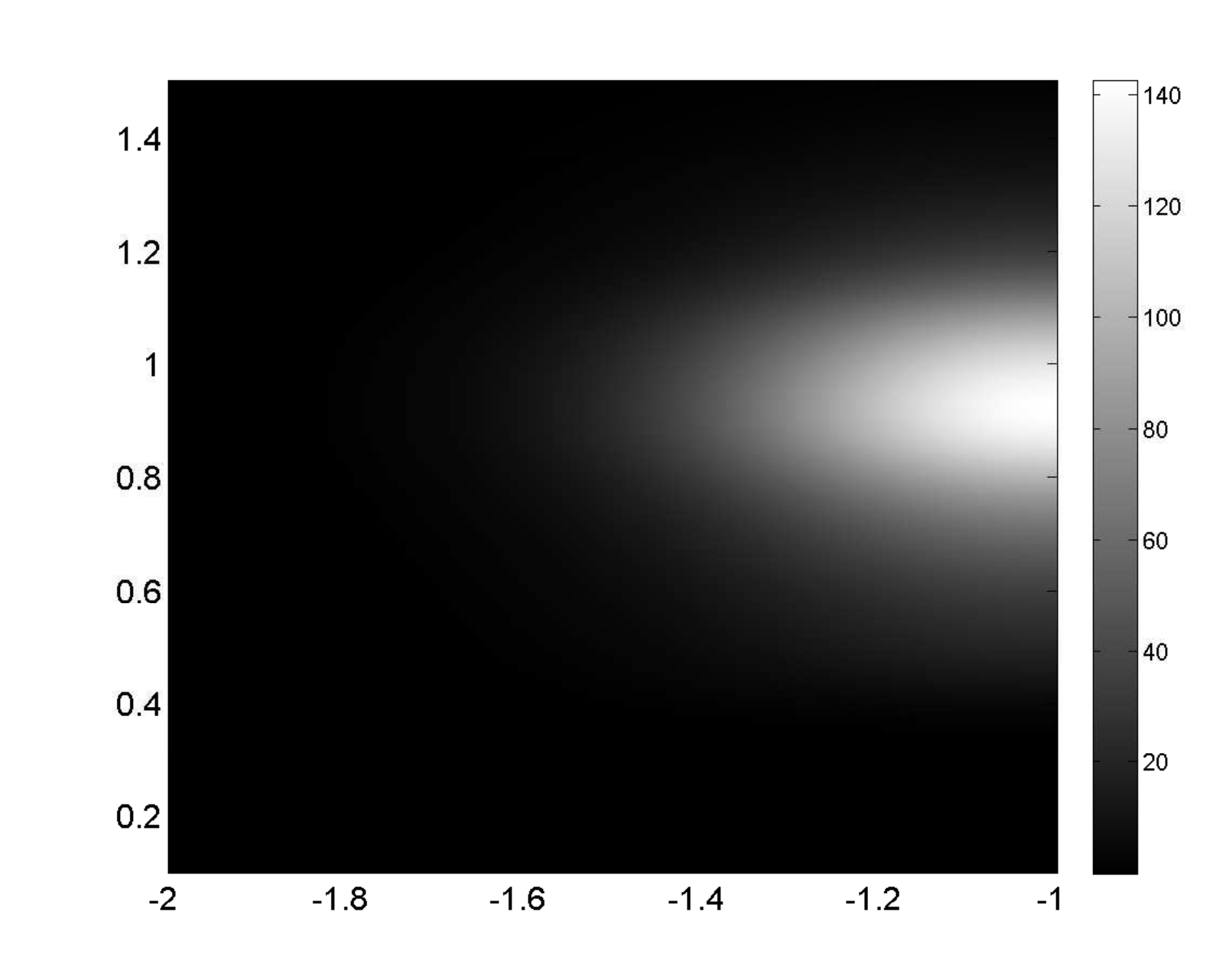} 
\includegraphics[width=0.20\textwidth,clip=true,trim=0cm 0cm 0cm 0cm]{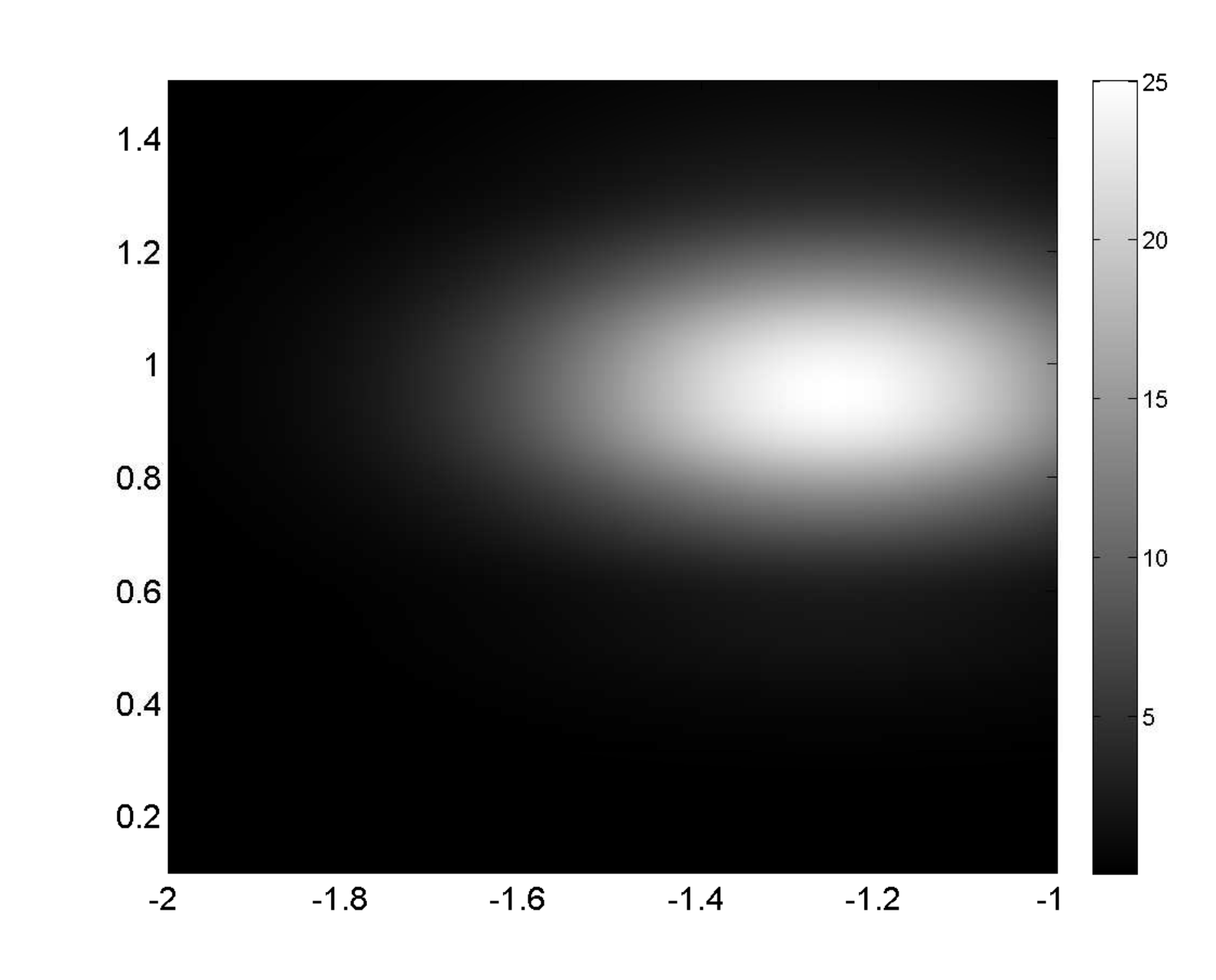} \\
\includegraphics[width=0.20\textwidth,clip=true,trim=0cm 0cm 0cm 0cm]{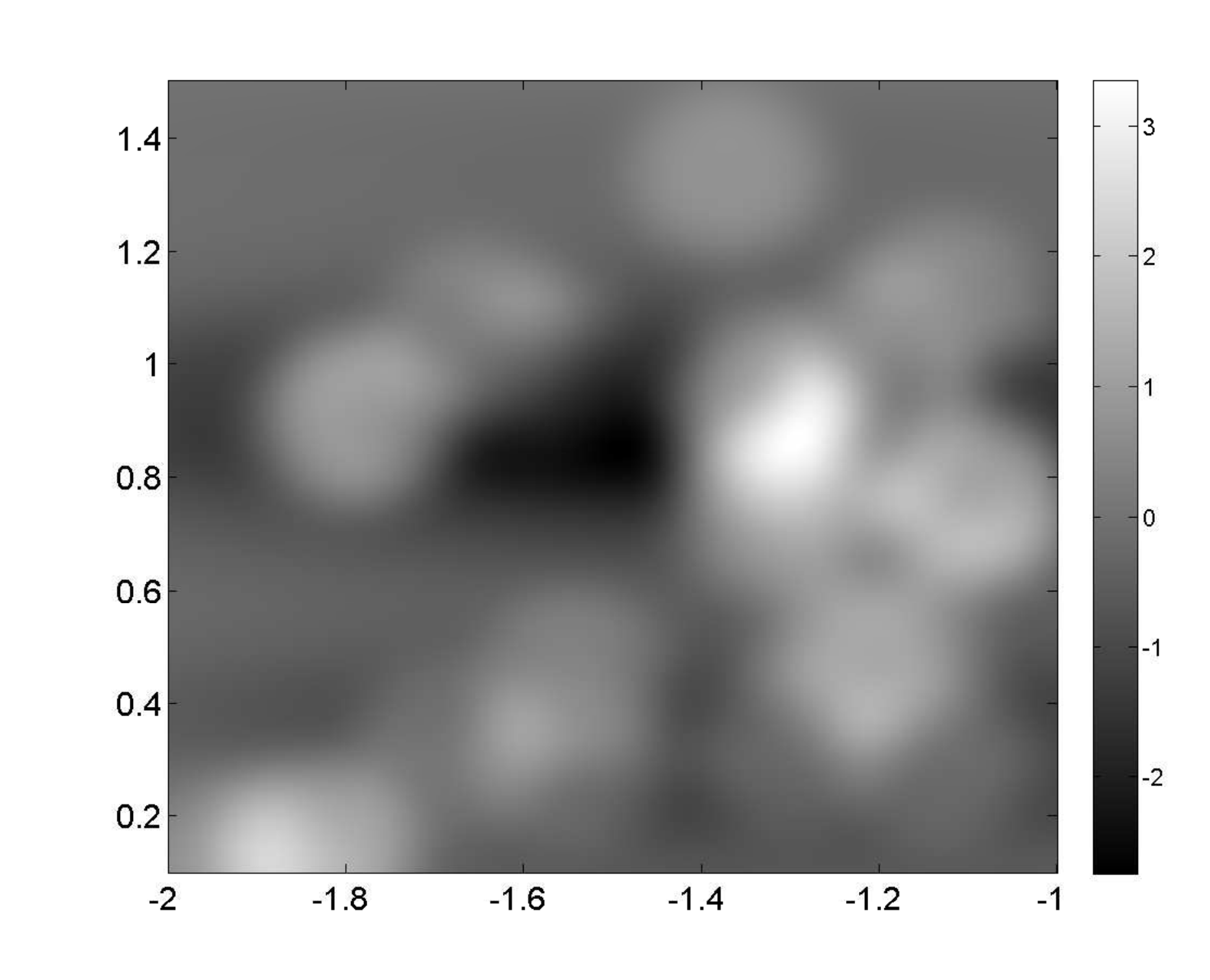} 
\includegraphics[width=0.20\textwidth,clip=true,trim=0cm 0cm 0cm 0cm]{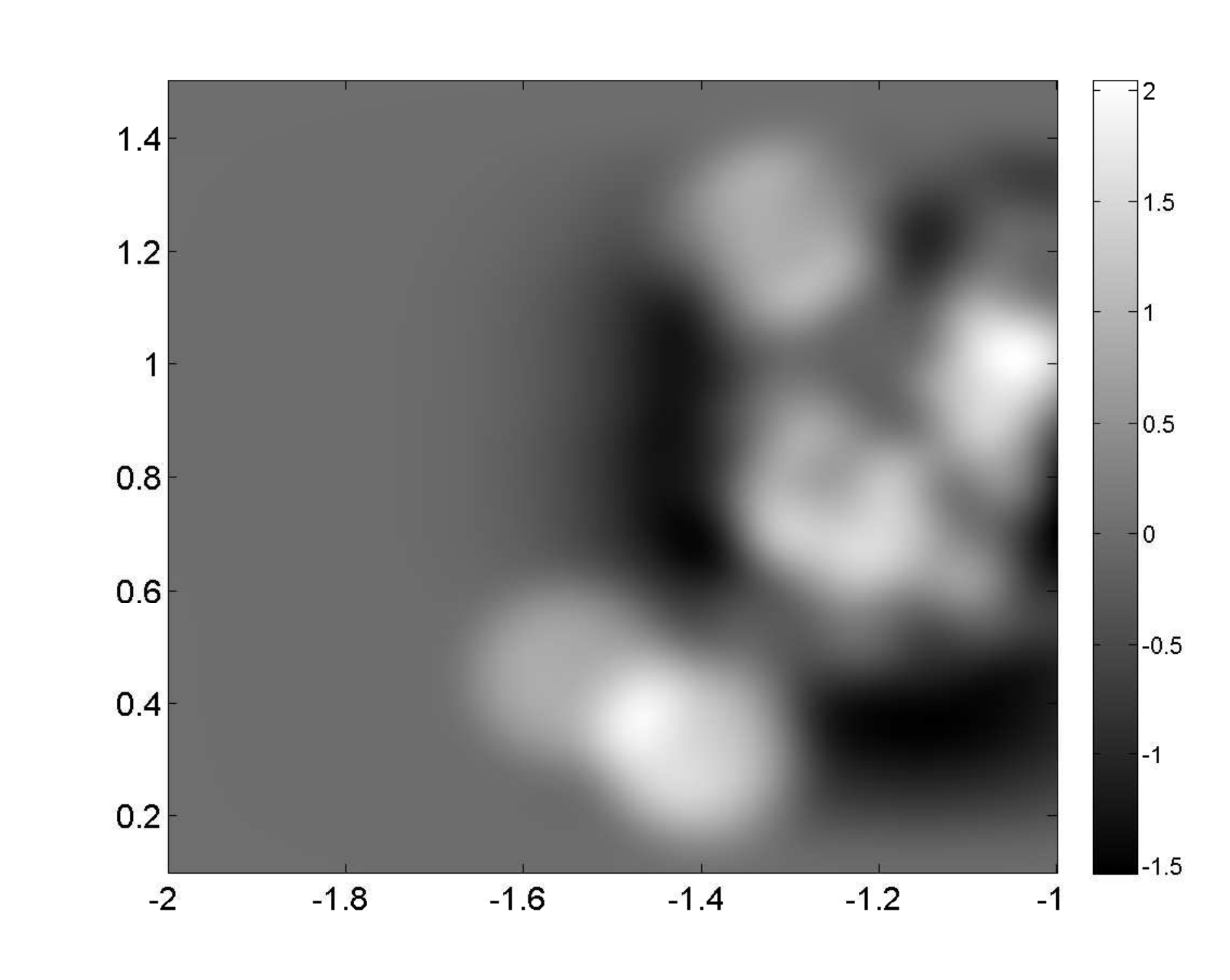} 
\includegraphics[width=0.20\textwidth,clip=true,trim=0cm 0cm 0cm 0cm]{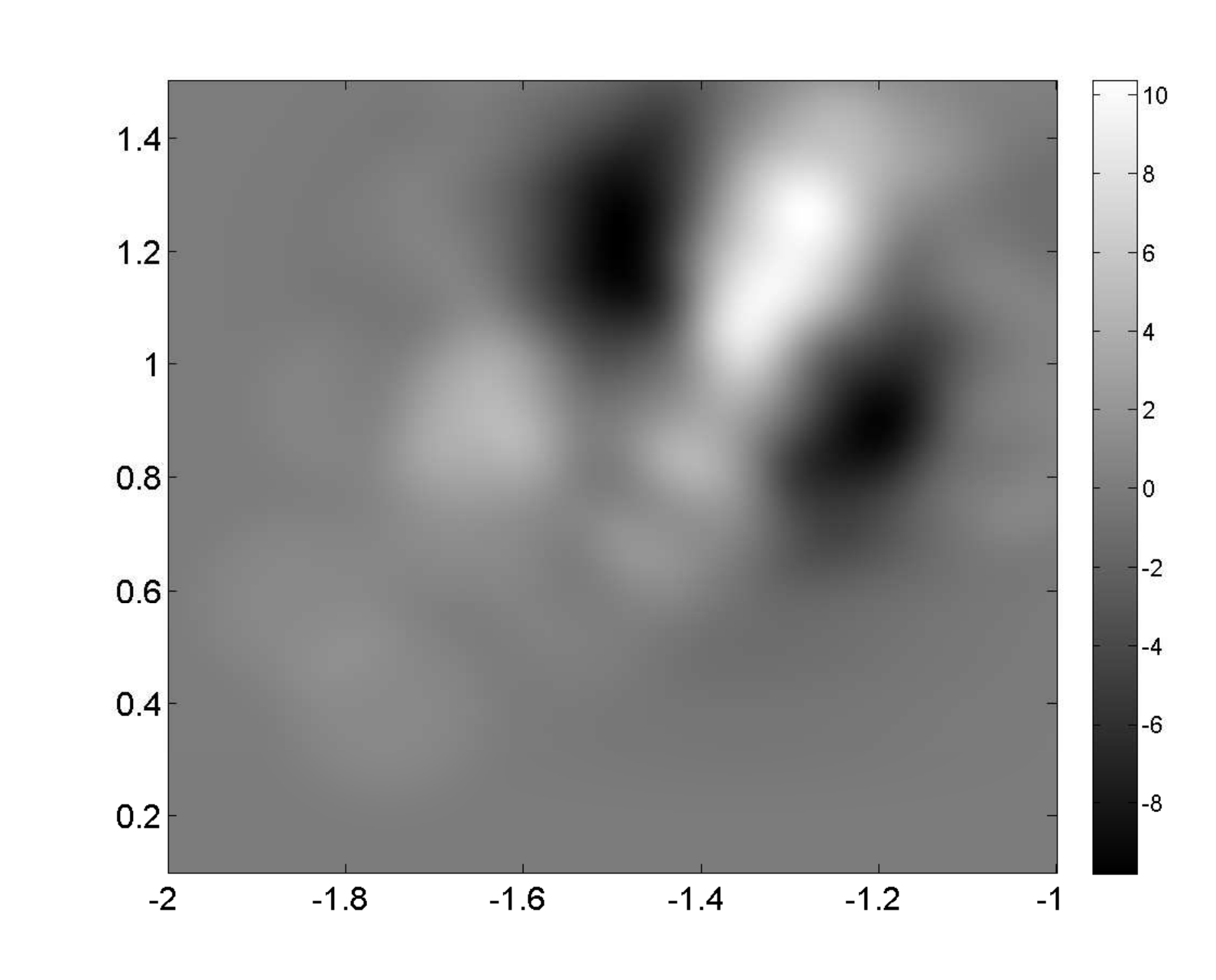} 
\includegraphics[width=0.20\textwidth,clip=true,trim=0cm 0cm 0cm 0cm]{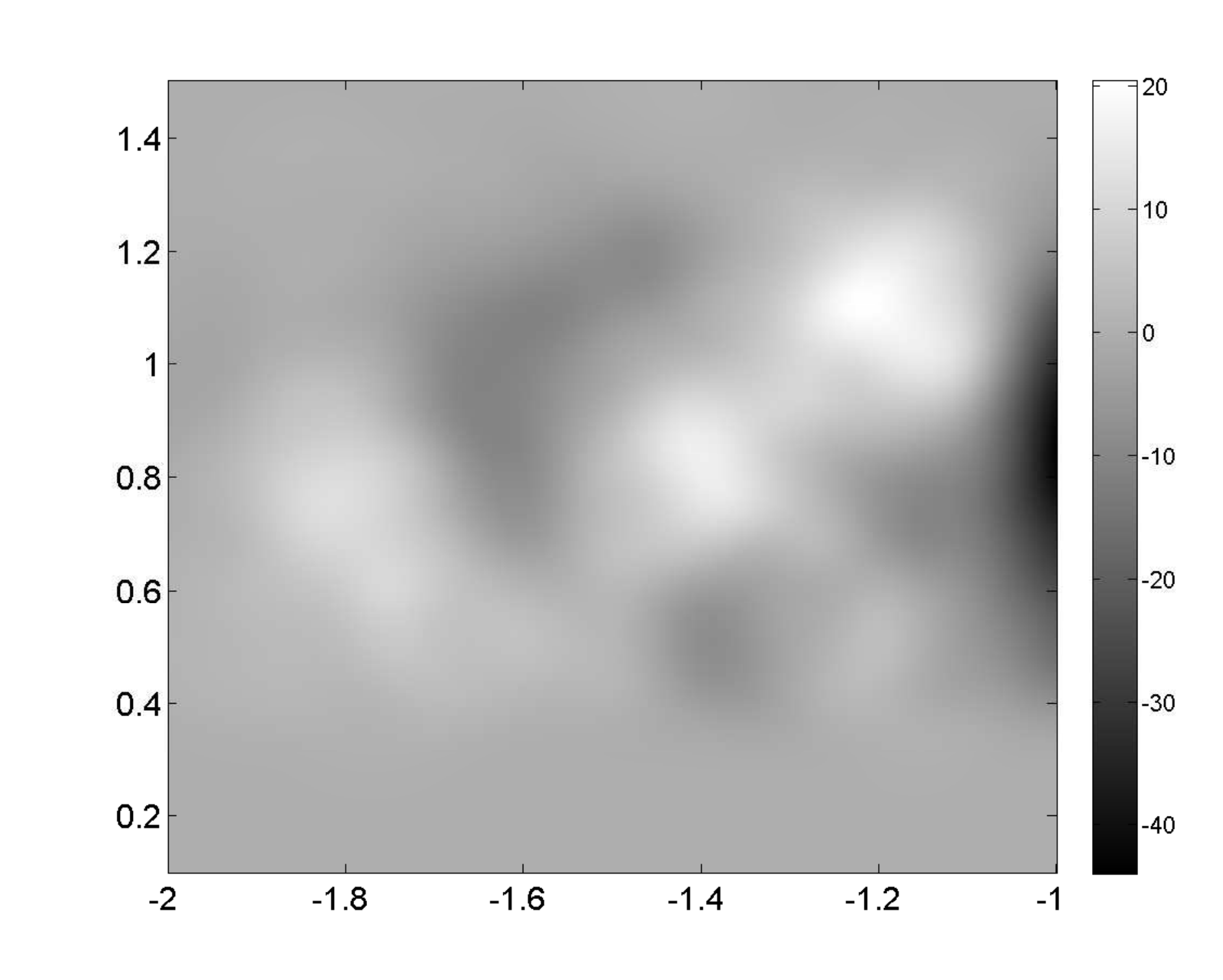} 
\includegraphics[width=0.20\textwidth,clip=true,trim=0cm 0cm 0cm 0cm]{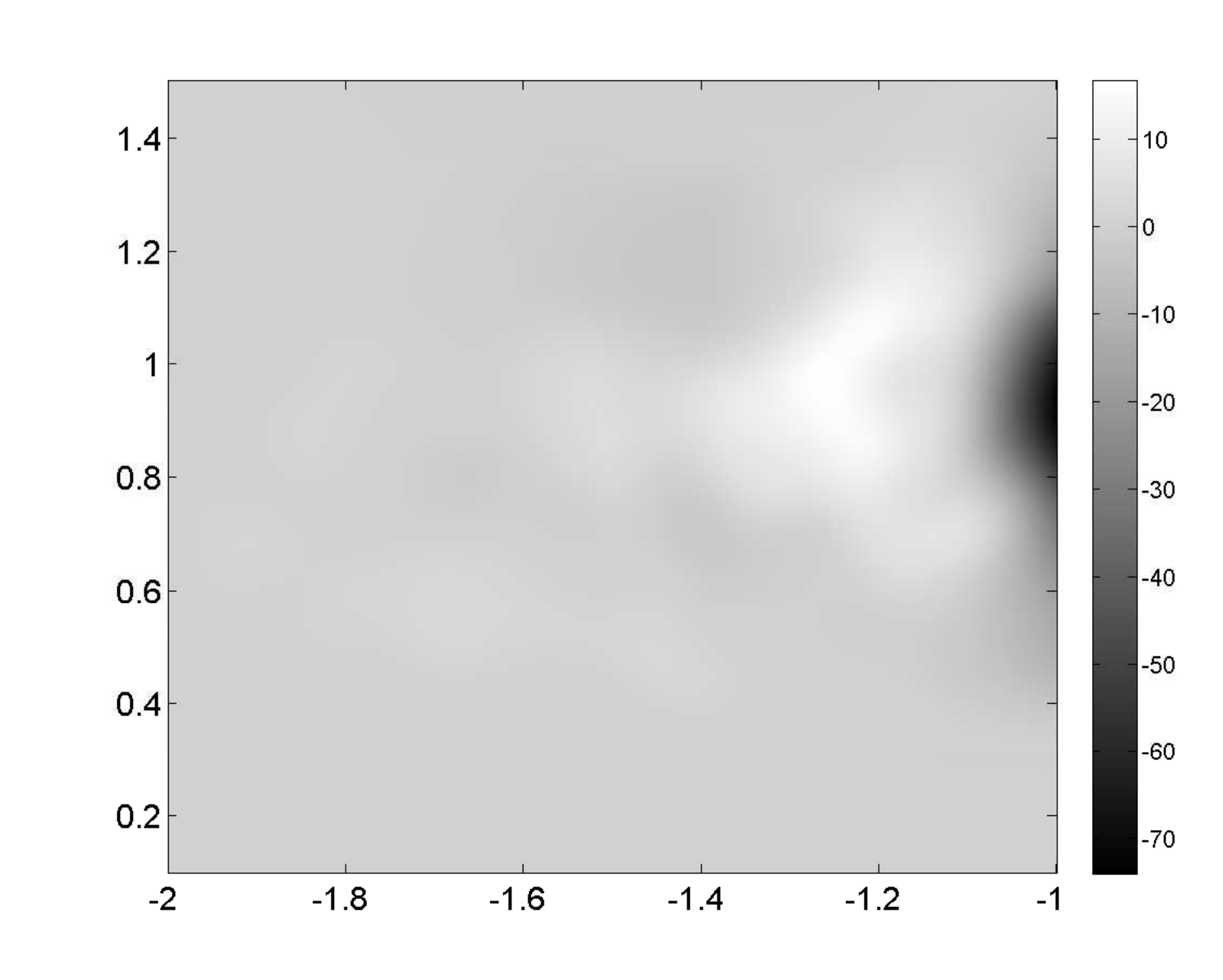} 
\includegraphics[width=0.20\textwidth,clip=true,trim=0cm 0cm 0cm 0cm]{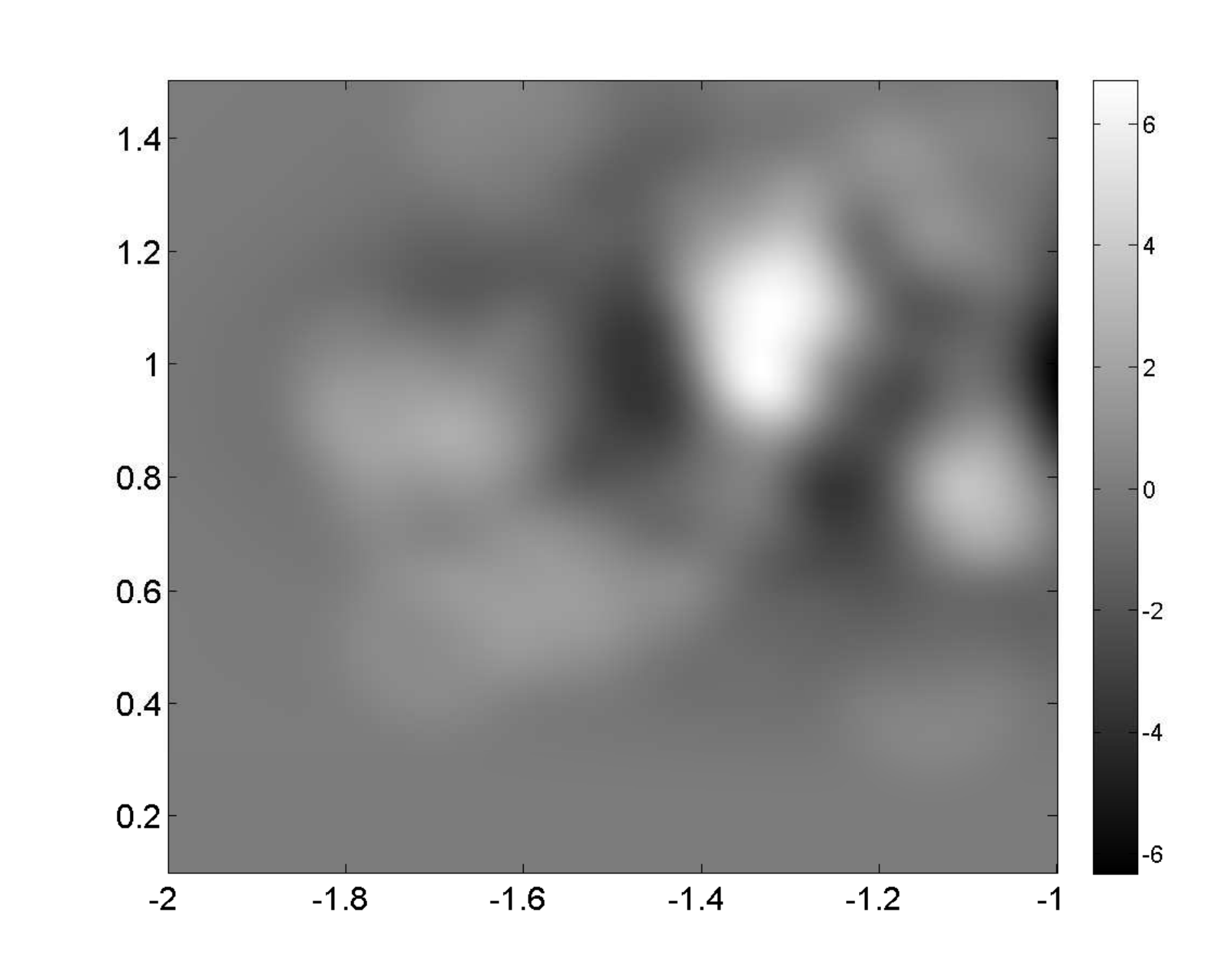} 
\caption{Same as Figure 2 for the stellar-mass independent halting model at a constant semi-major axis.}
\end{figure}
\end{landscape}

\clearpage
\begin{landscape}
\begin{figure}
\centering
\includegraphics[width=0.20\textwidth,clip=true,trim=0cm 0cm 0cm 0cm]{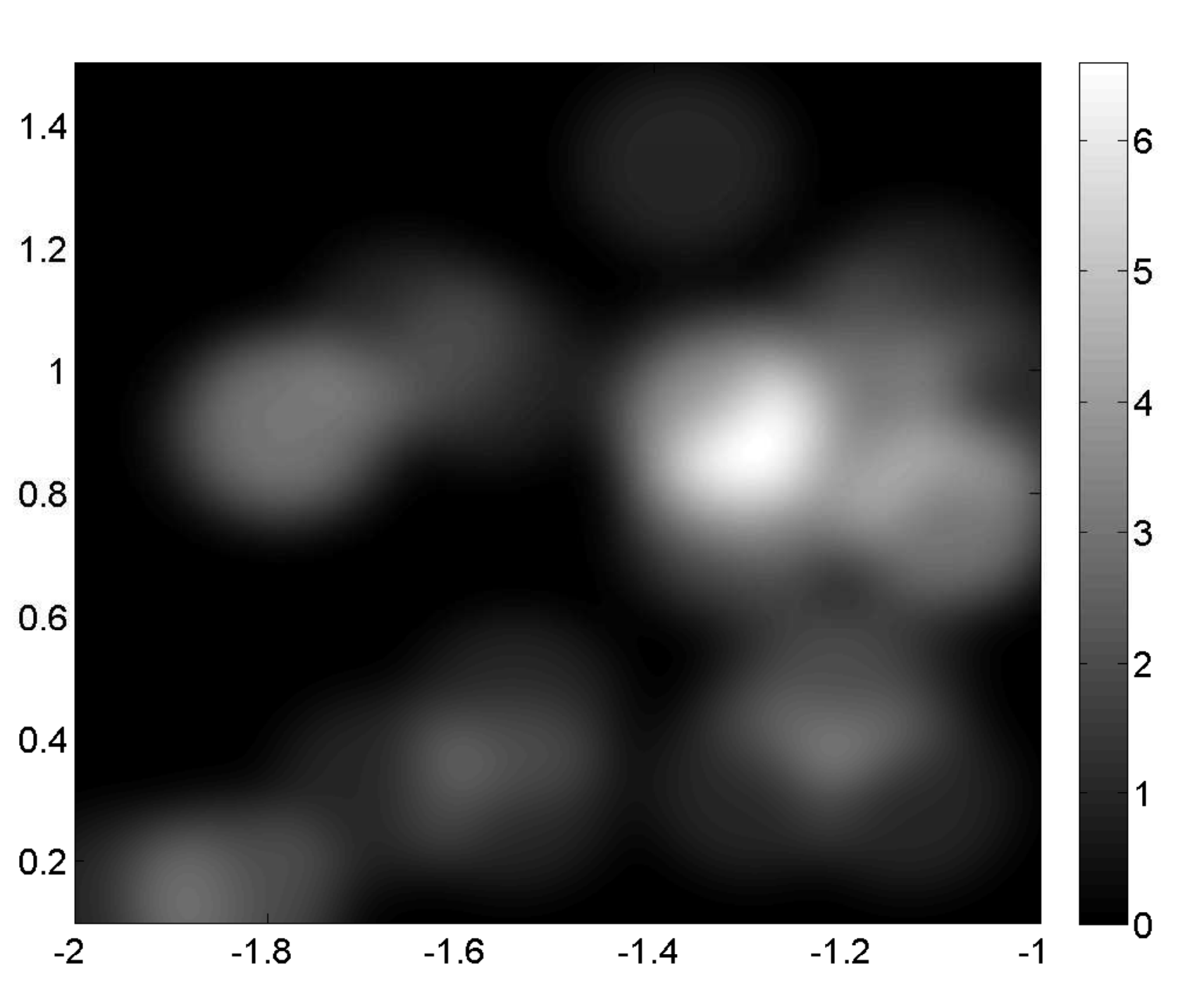} 
\includegraphics[width=0.20\textwidth,clip=true,trim=0cm 0cm 0cm 0cm]{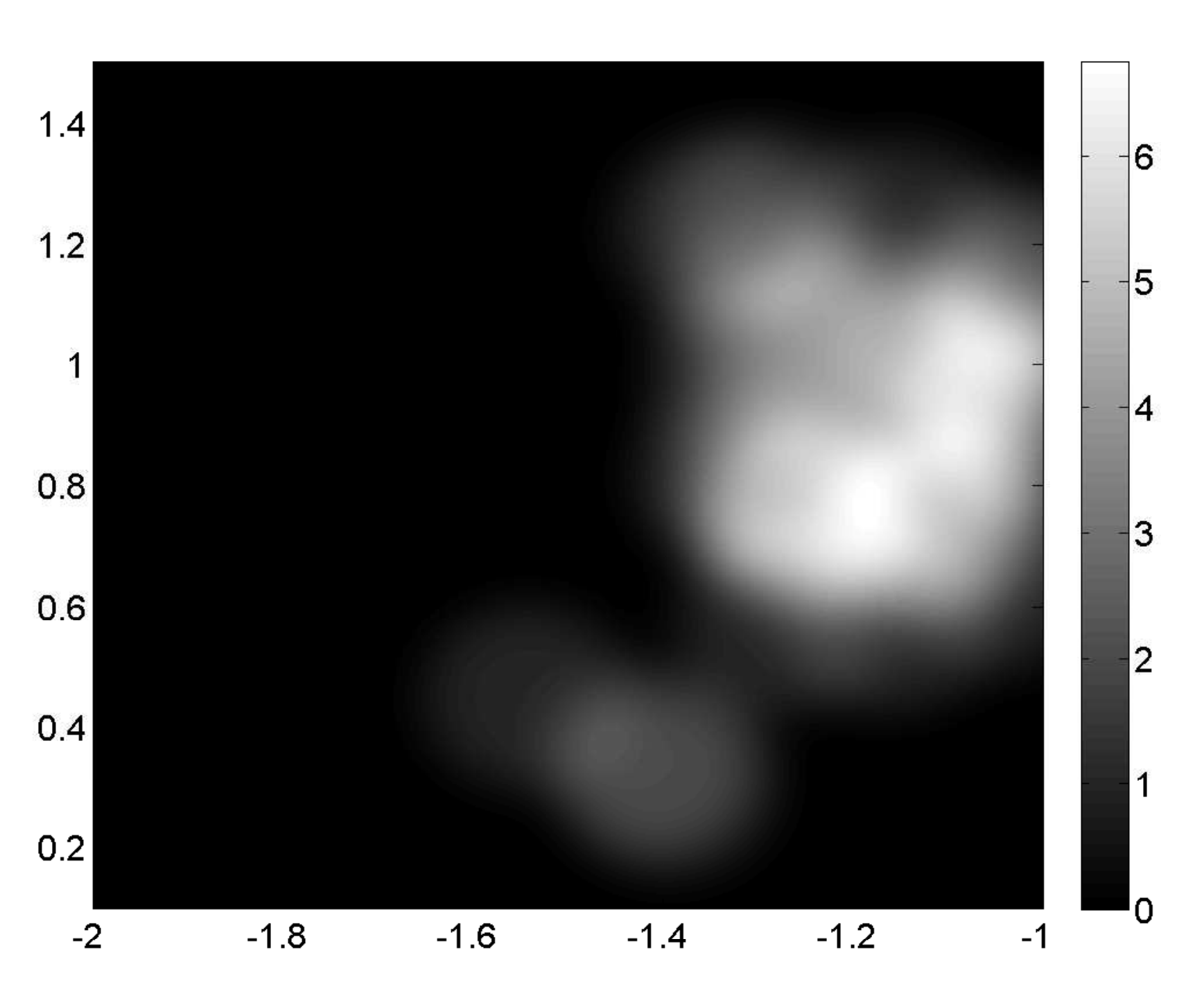} 
\includegraphics[width=0.20\textwidth,clip=true,trim=0cm 0cm 0cm 0cm]{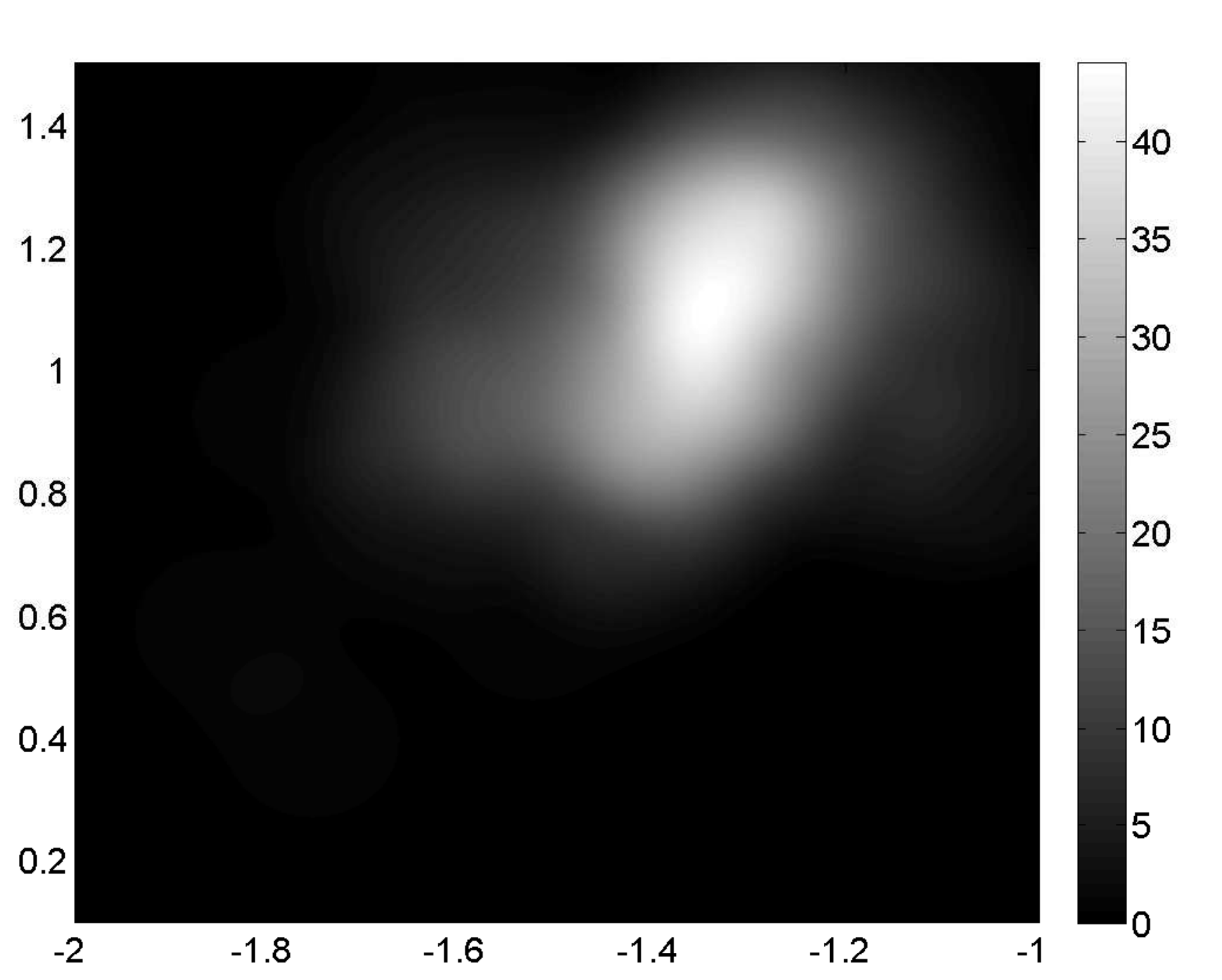} 
\includegraphics[width=0.20\textwidth,clip=true,trim=0cm 0cm 0cm 0cm]{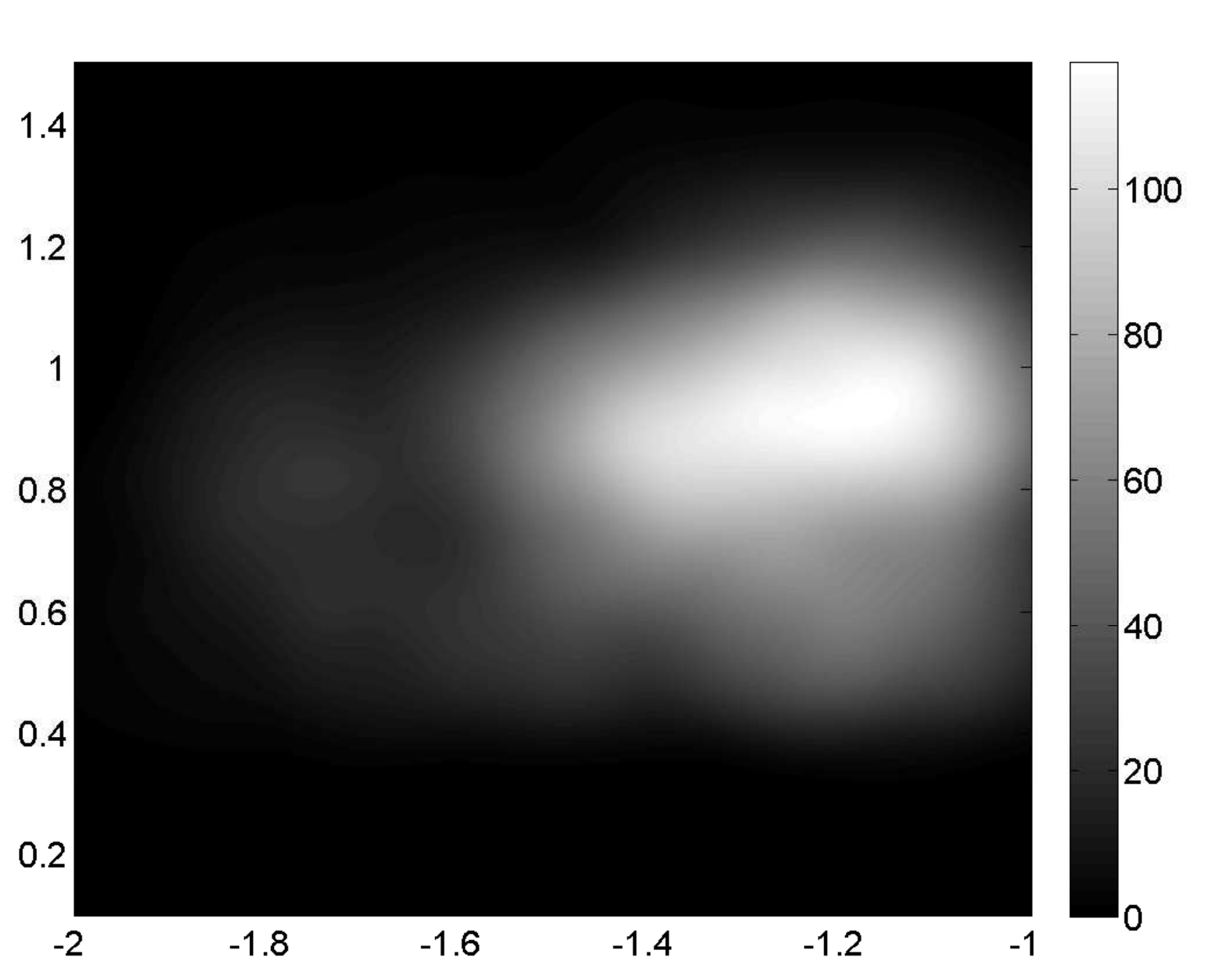} 
\includegraphics[width=0.20\textwidth,clip=true,trim=0cm 0cm 0cm 0cm]{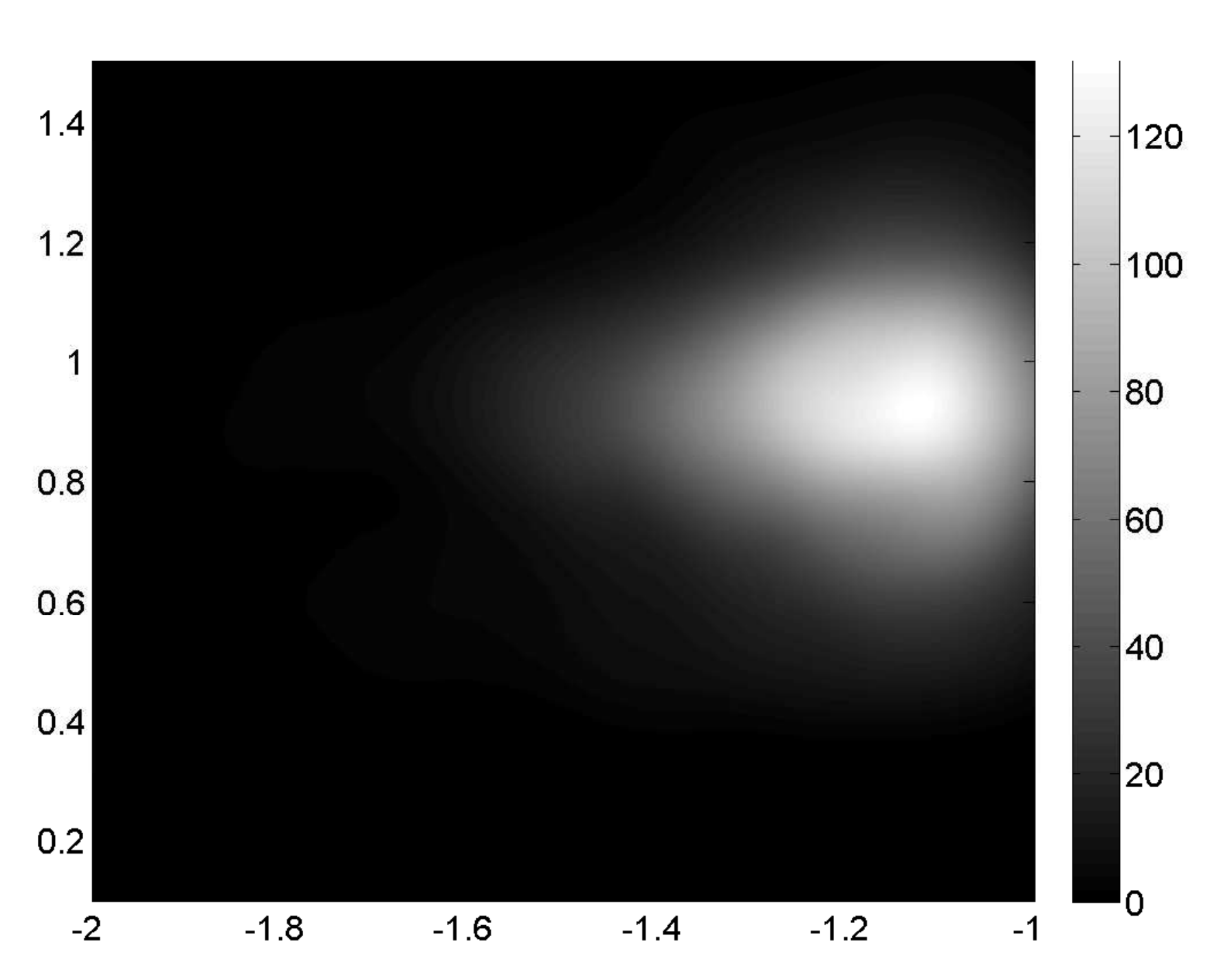} 
\includegraphics[width=0.20\textwidth,clip=true,trim=0cm 0cm 0cm 0cm]{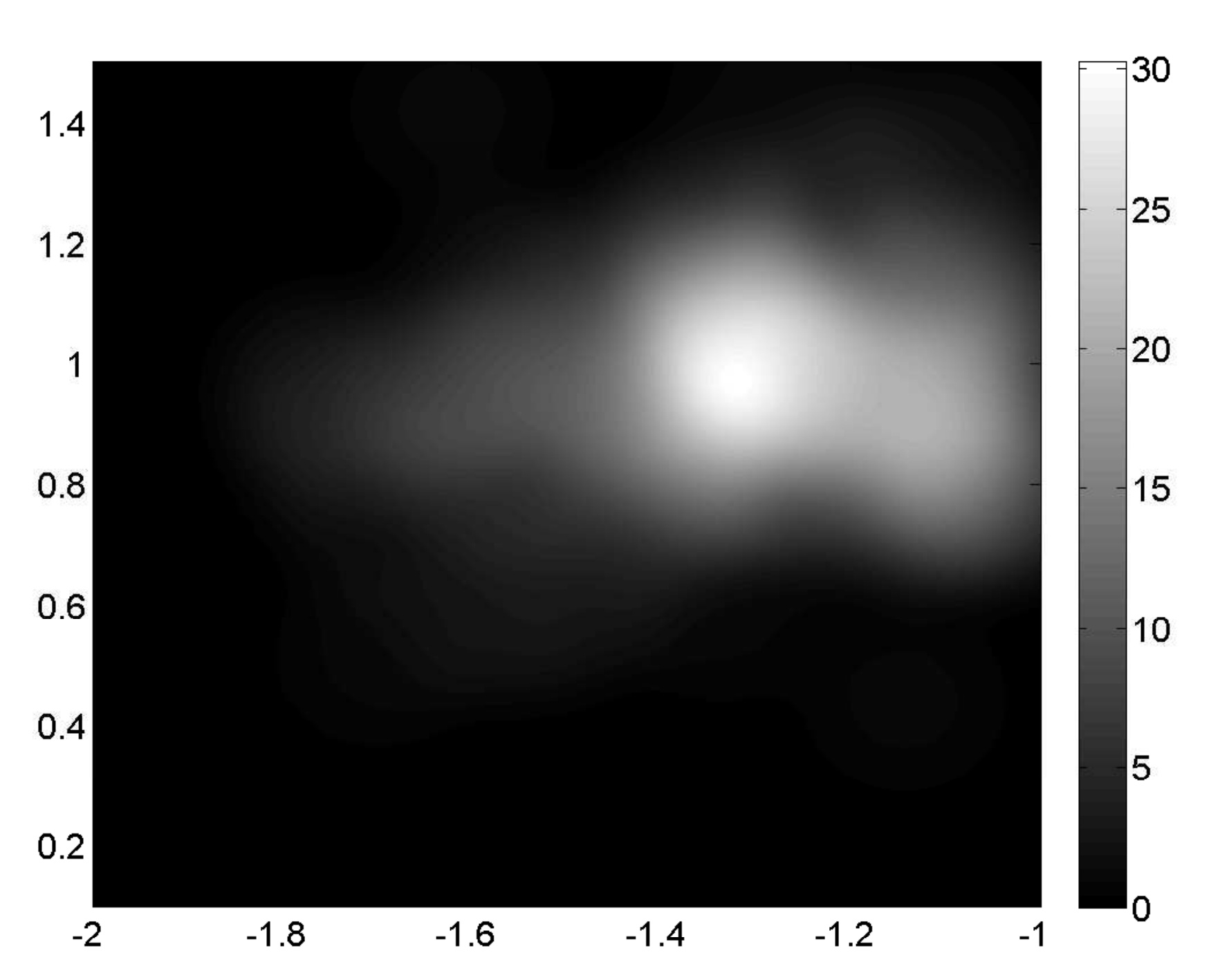} \\
\includegraphics[width=0.20\textwidth,clip=true,trim=0cm 0cm 0cm 0cm]{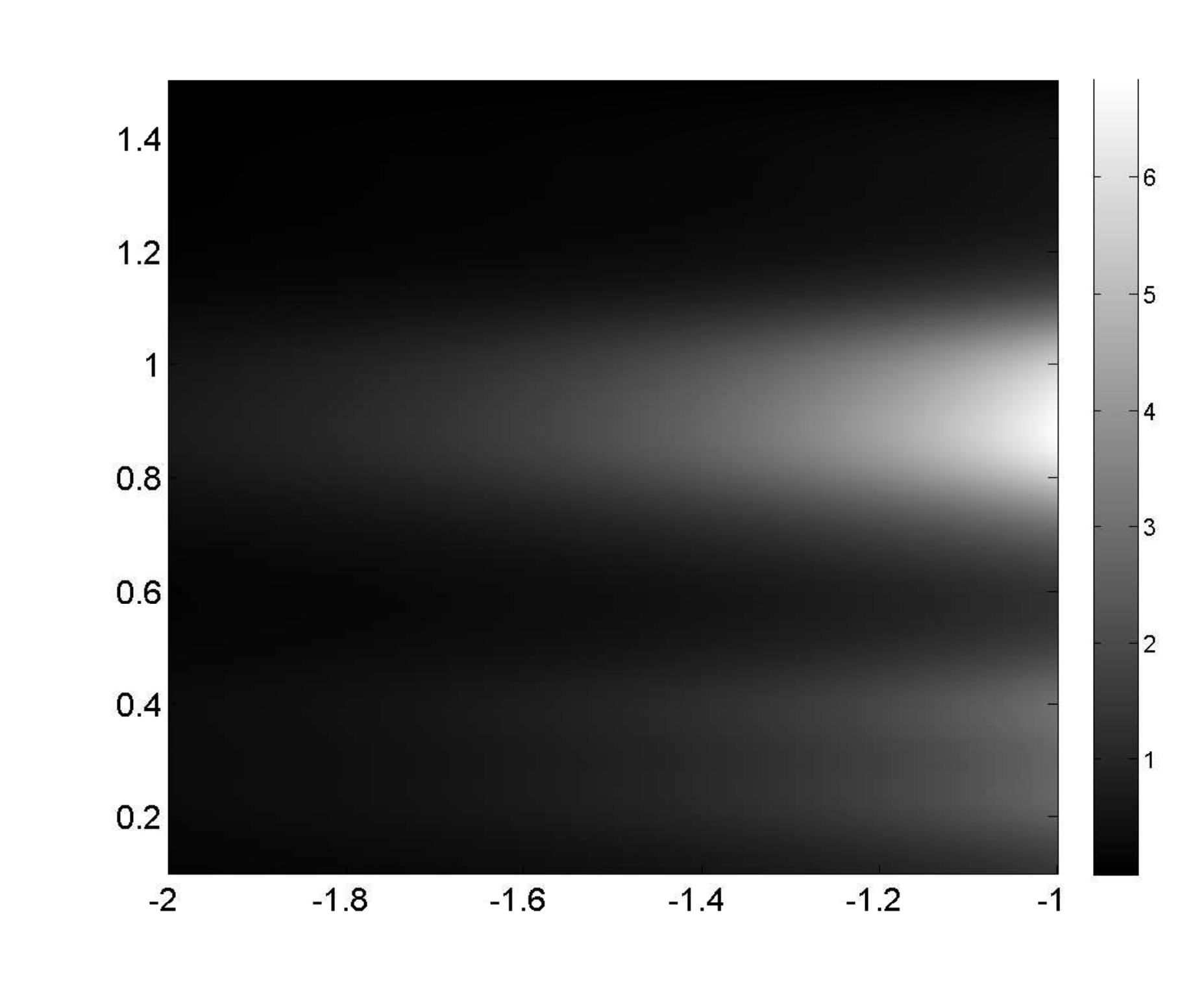} 
\includegraphics[width=0.20\textwidth,clip=true,trim=0cm 0cm 0cm 0cm]{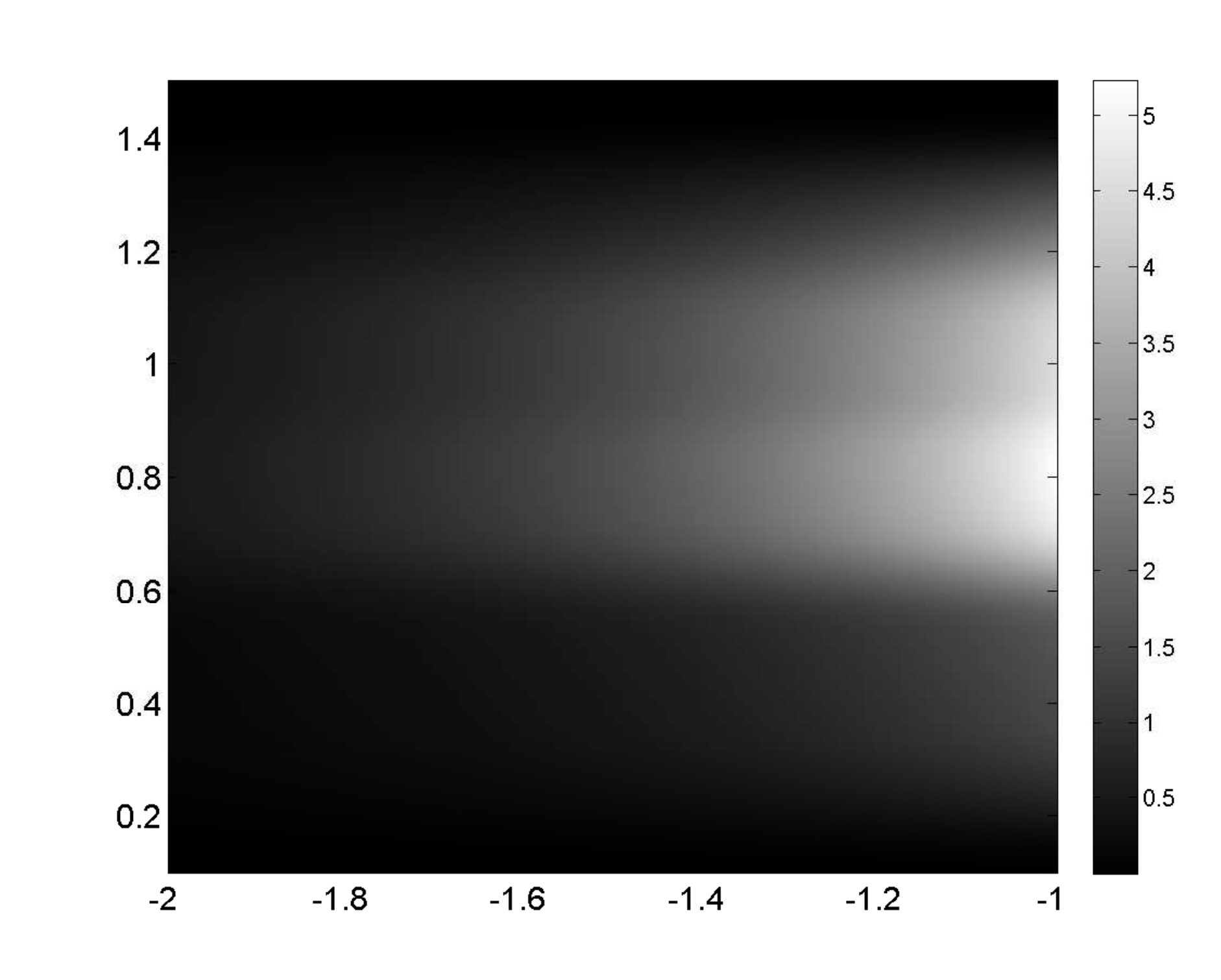} 
\includegraphics[width=0.20\textwidth,clip=true,trim=0cm 0cm 0cm 0cm]{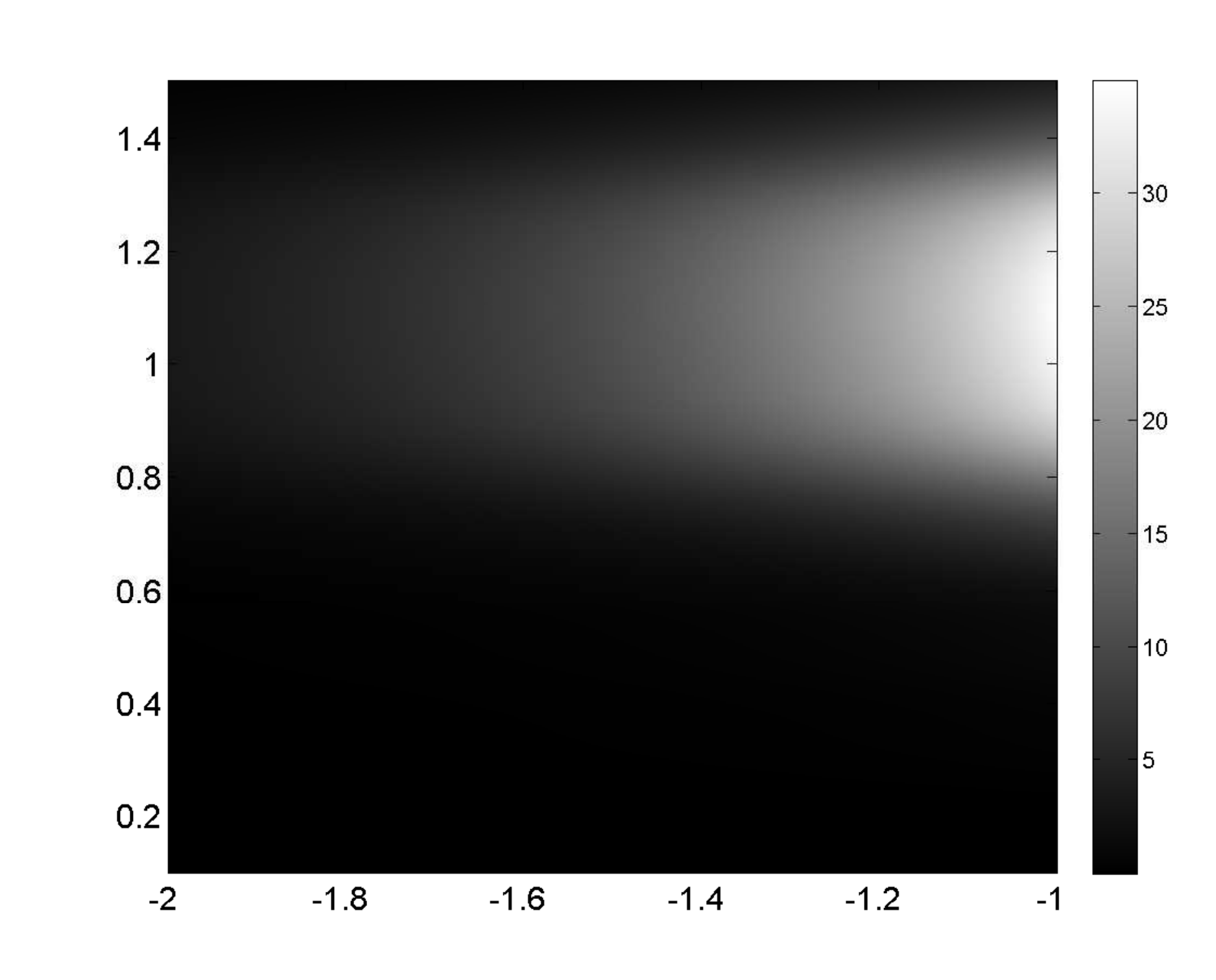}
\includegraphics[width=0.20\textwidth,clip=true,trim=0cm 0cm 0cm 0cm]{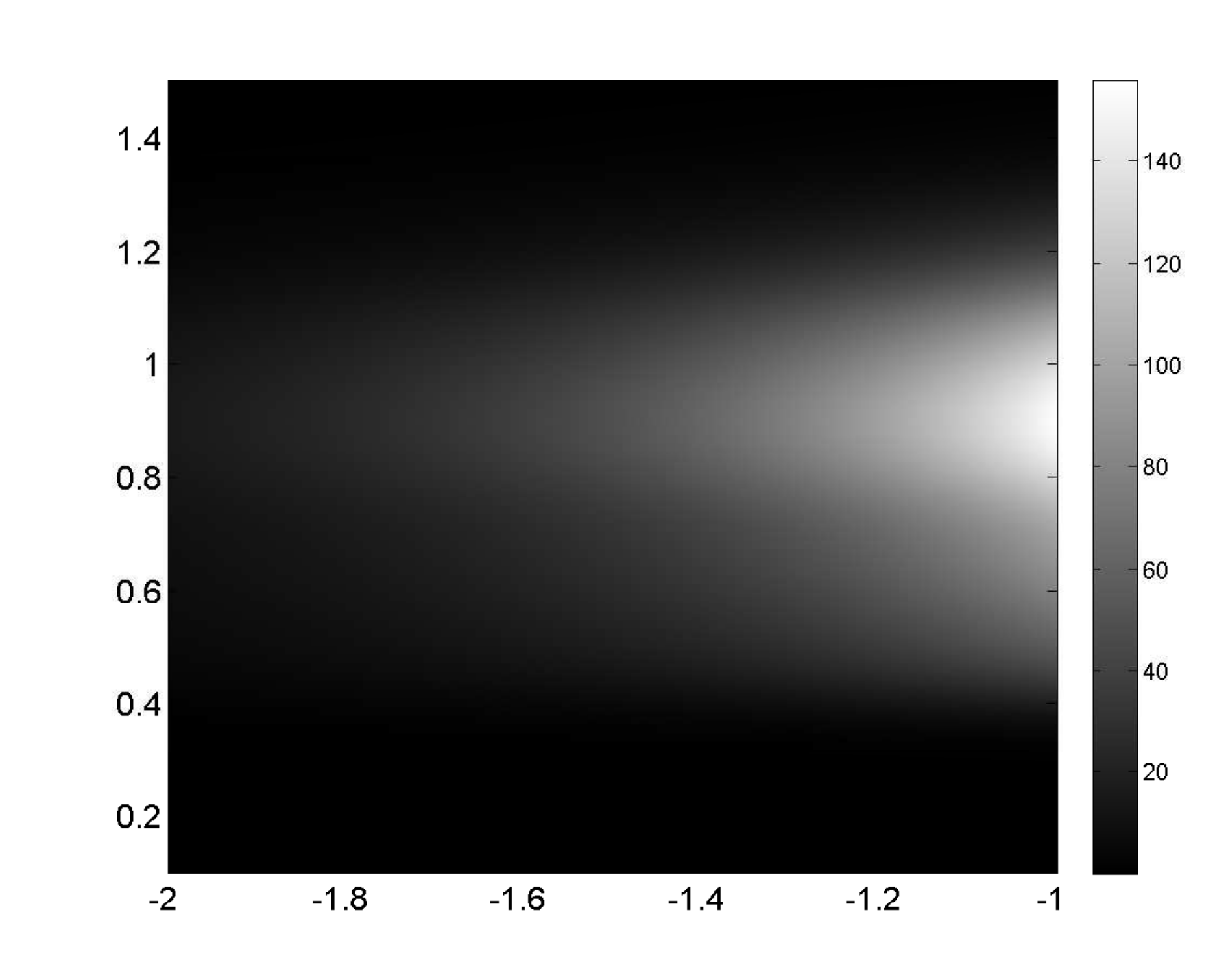} 
\includegraphics[width=0.20\textwidth,clip=true,trim=0cm 0cm 0cm 0cm]{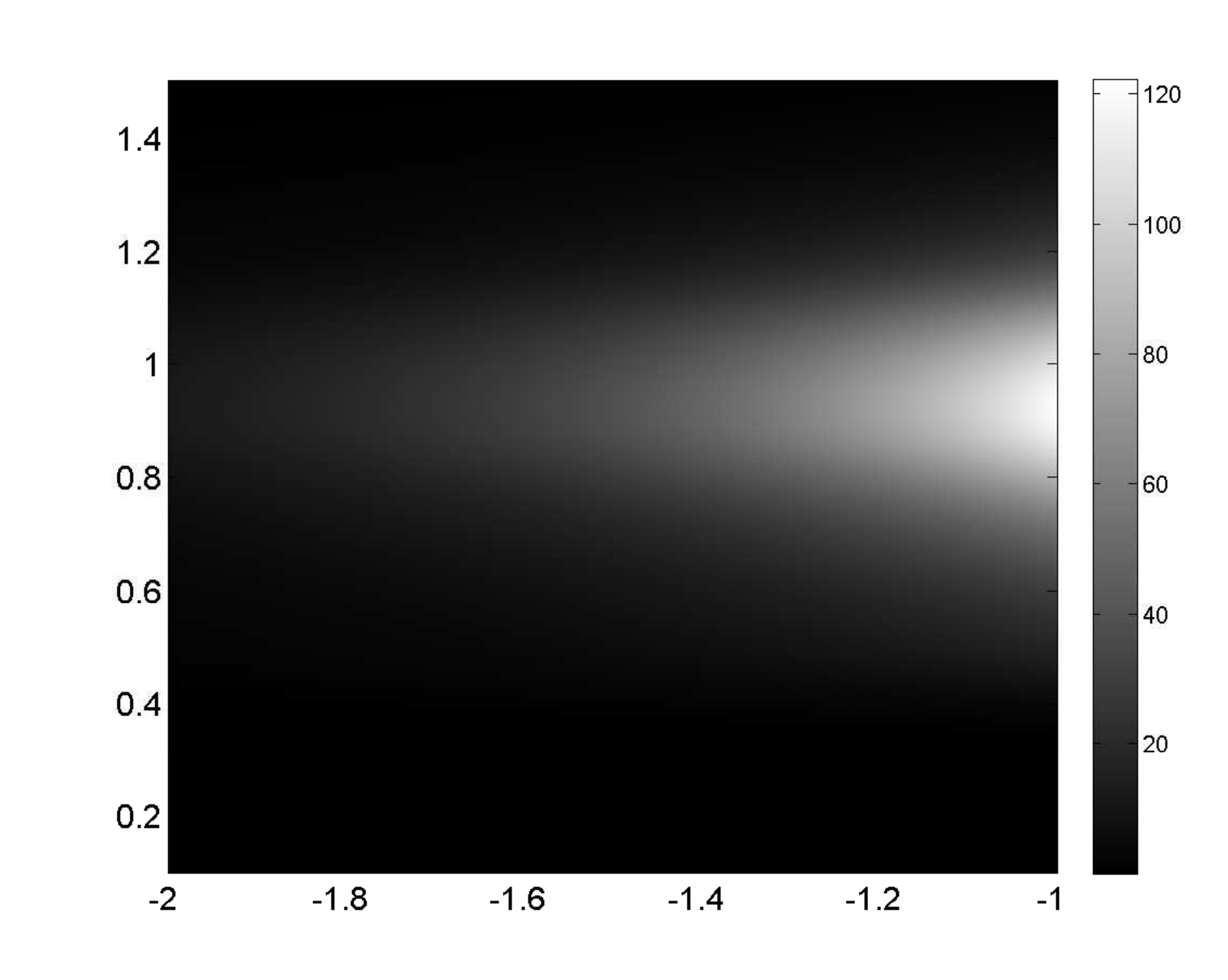} 
\includegraphics[width=0.20\textwidth,clip=true,trim=0cm 0cm 0cm 0cm]{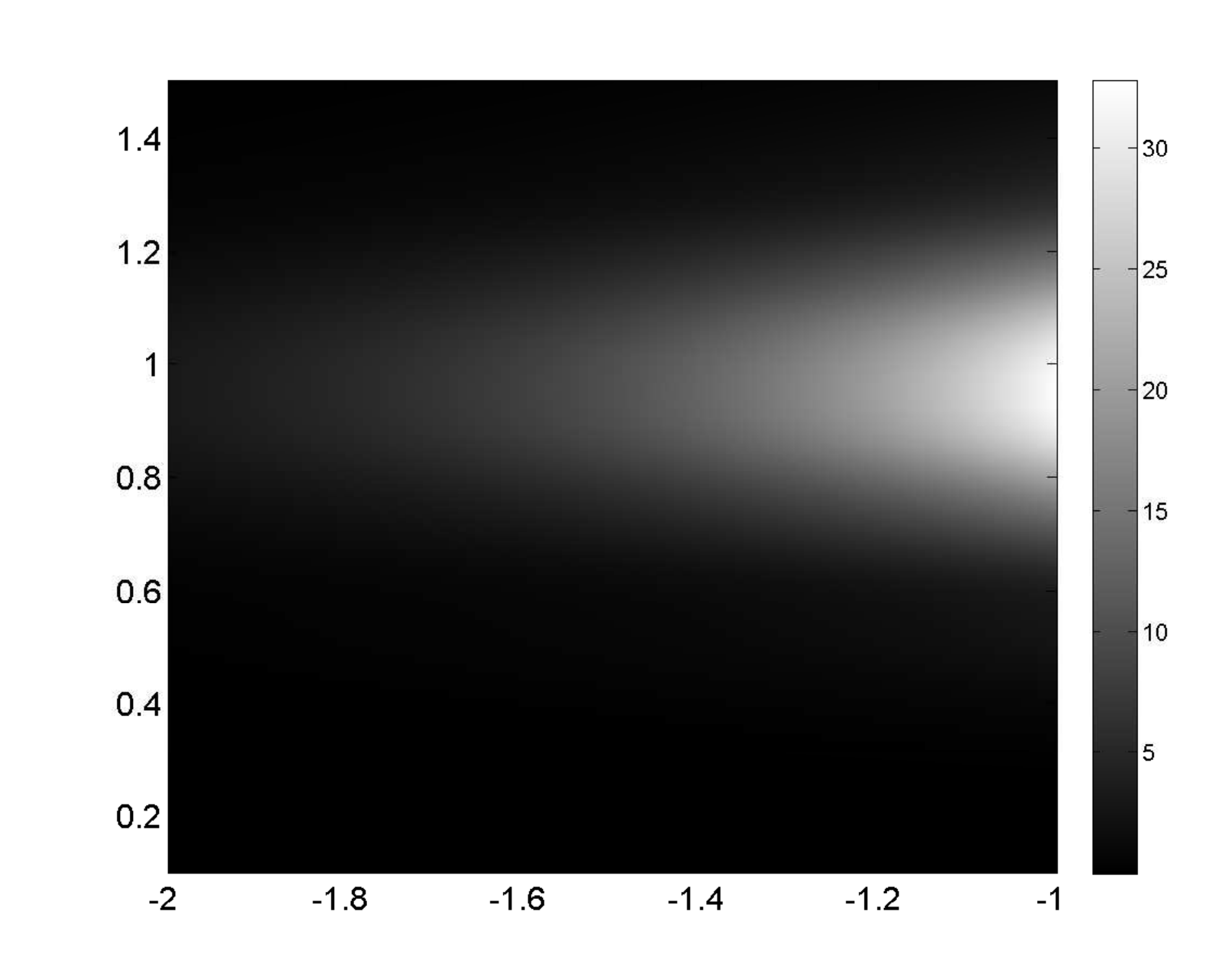} \\
\includegraphics[width=0.20\textwidth,clip=true,trim=0cm 0cm 0cm 0cm]{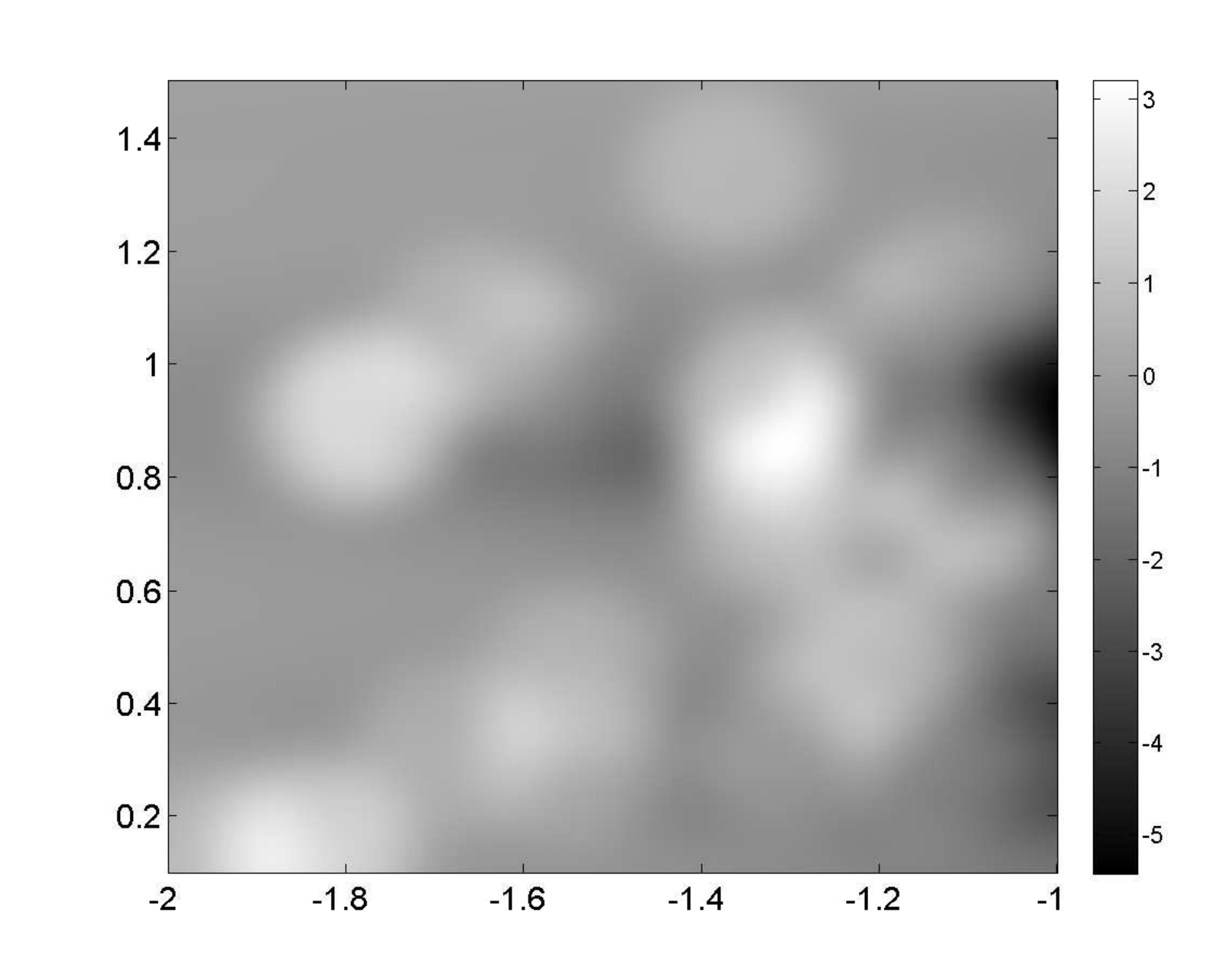} 
\includegraphics[width=0.20\textwidth,clip=true,trim=0cm 0cm 0cm 0cm]{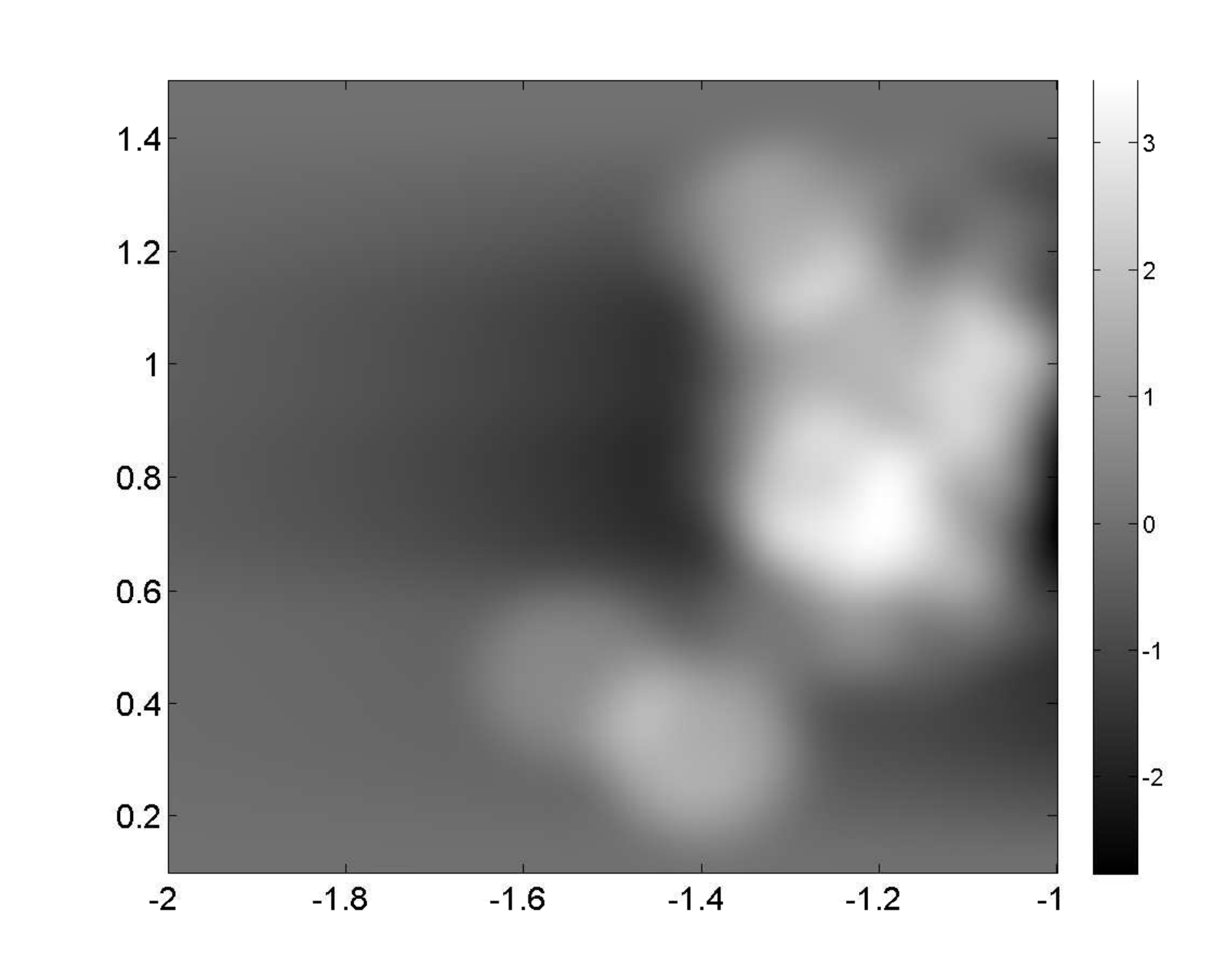} 
\includegraphics[width=0.20\textwidth,clip=true,trim=0cm 0cm 0cm 0cm]{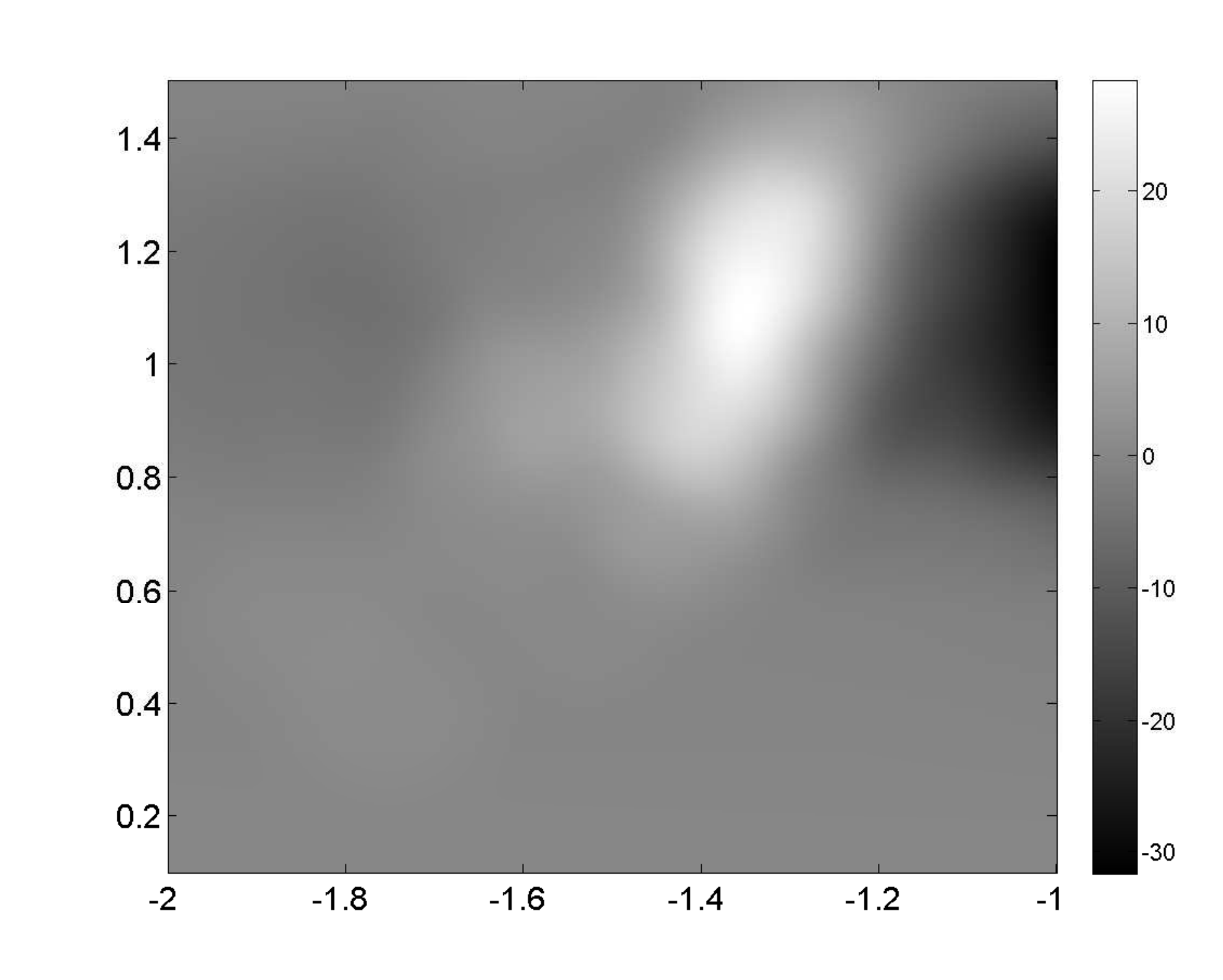} 
\includegraphics[width=0.20\textwidth,clip=true,trim=0cm 0cm 0cm 0cm]{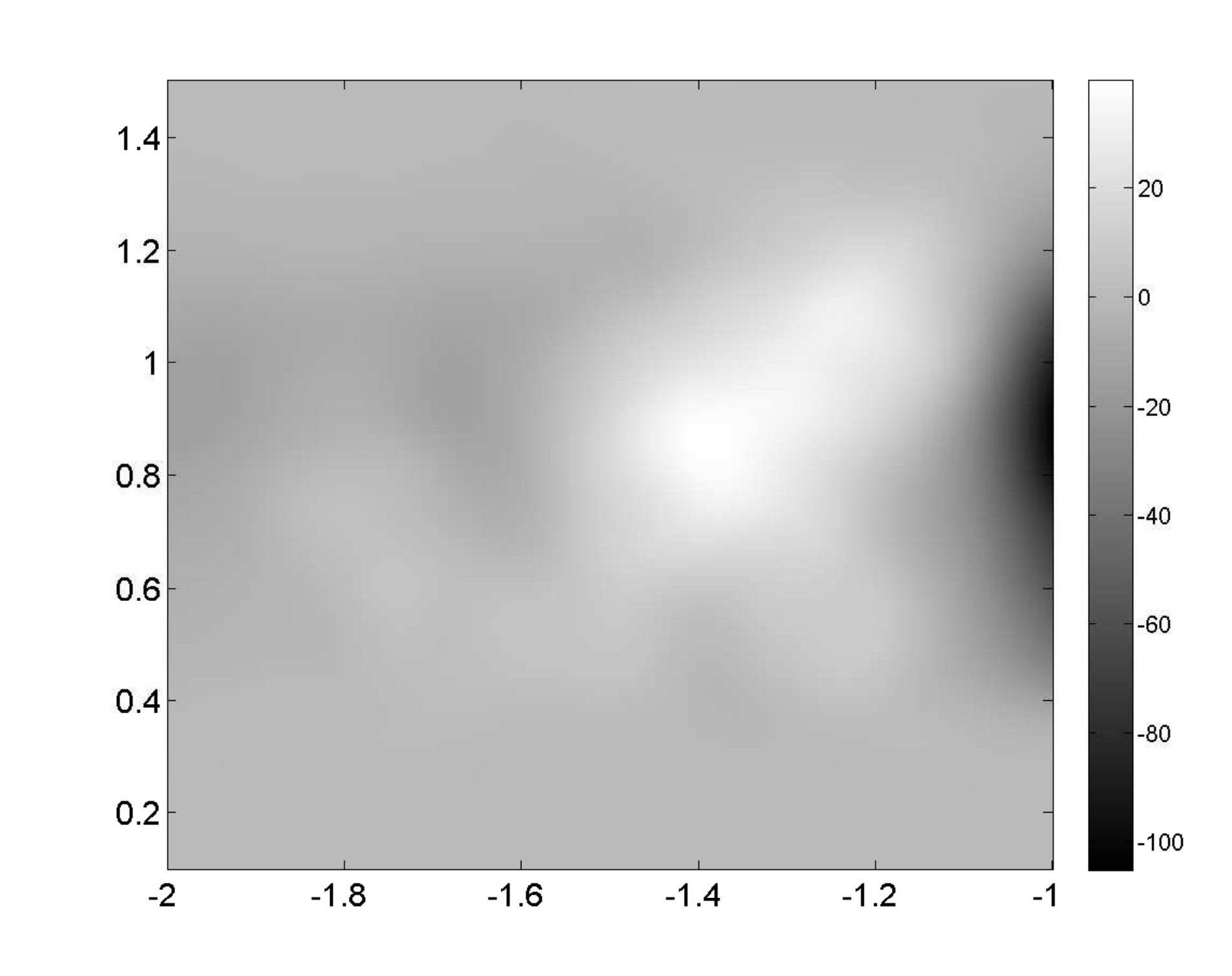} 
\includegraphics[width=0.20\textwidth,clip=true,trim=0cm 0cm 0cm 0cm]{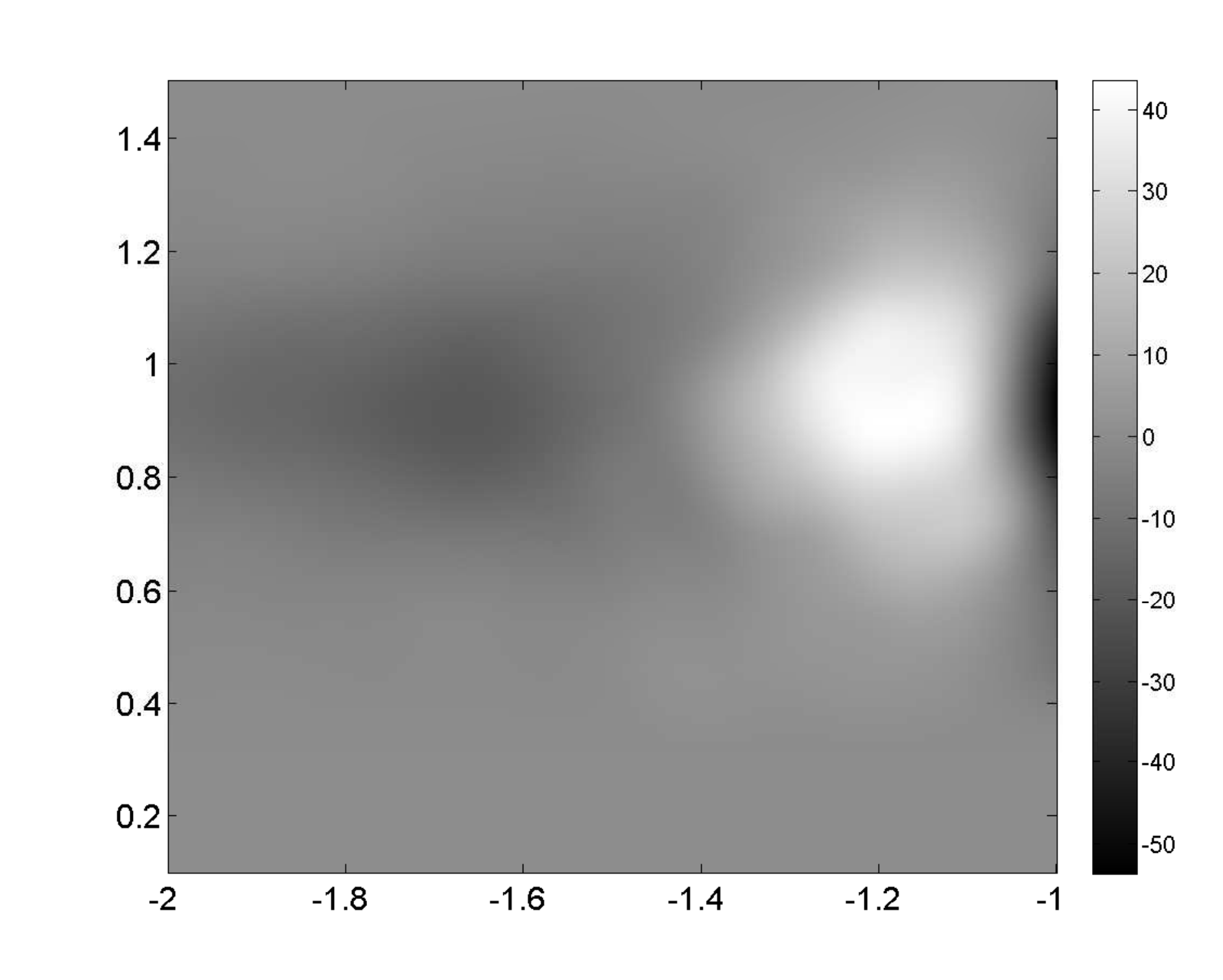} 
\includegraphics[width=0.20\textwidth,clip=true,trim=0cm 0cm 0cm 0cm]{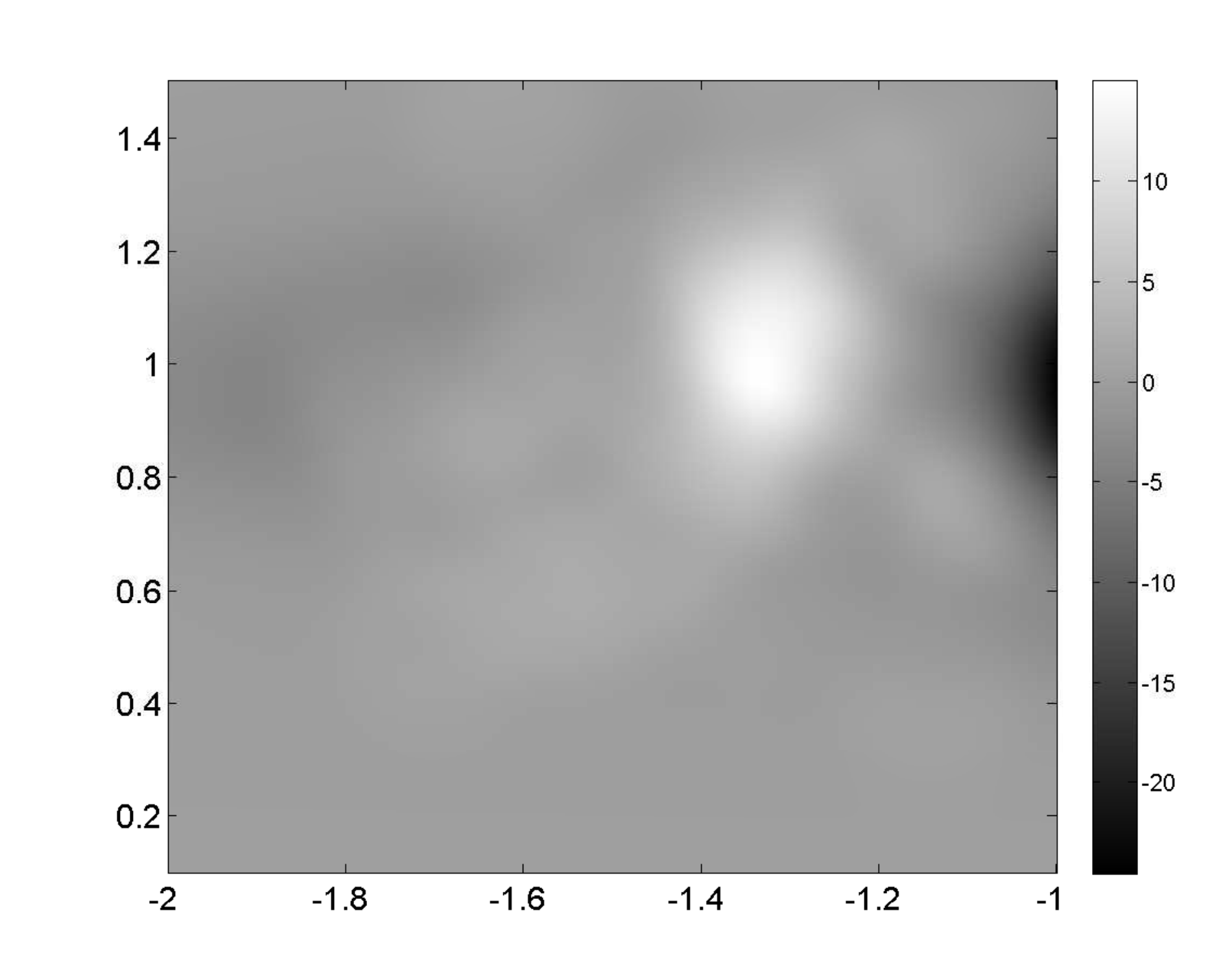} 
\caption{Same as Figure 2 for the uniform random halting model.}
\end{figure}
\end{landscape}

\clearpage
\begin{landscape}
\begin{figure}
\centering
\includegraphics[width=0.20\textwidth,clip=true,trim=0cm 0cm 0cm 0cm]{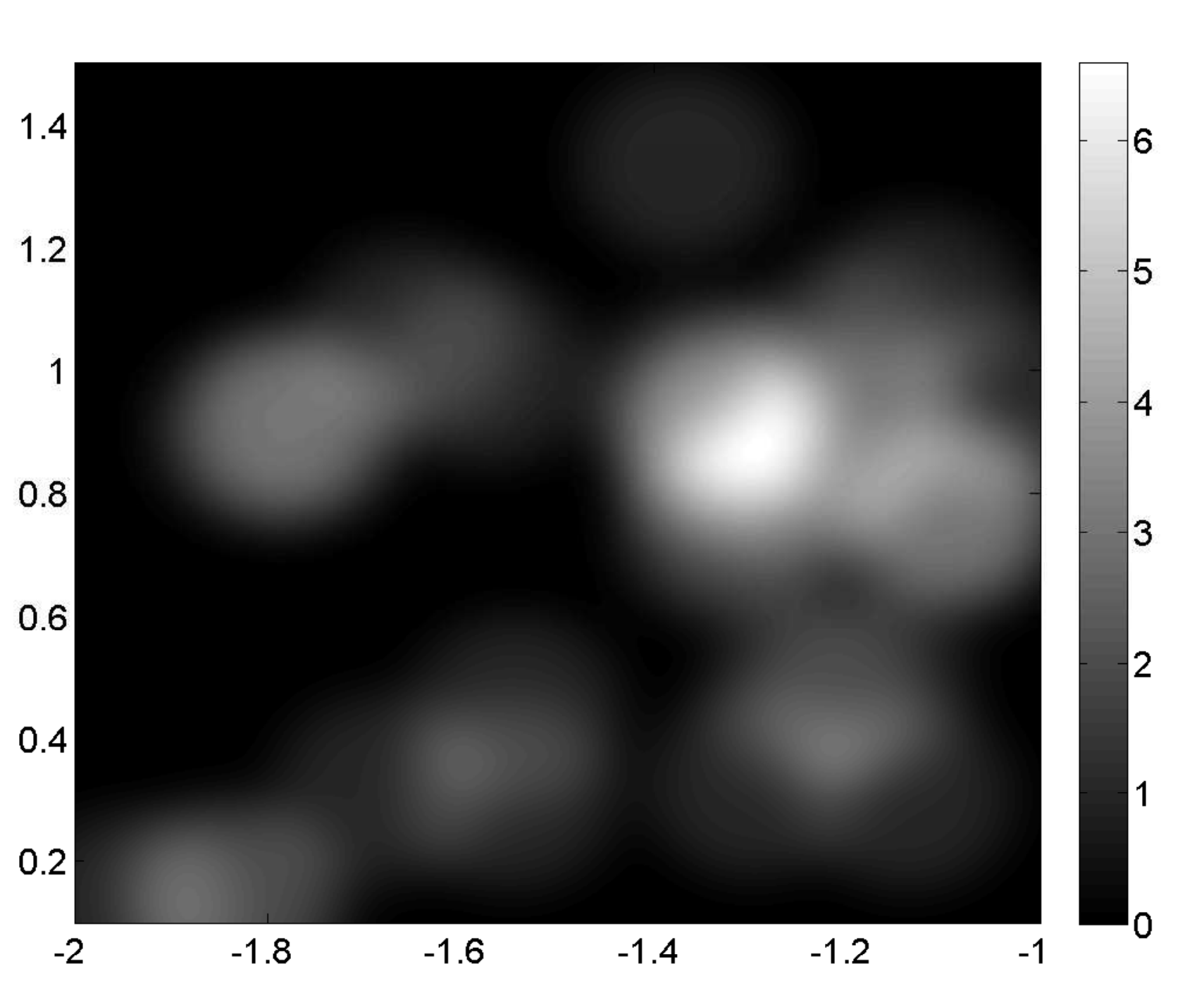} 
\includegraphics[width=0.20\textwidth,clip=true,trim=0cm 0cm 0cm 0cm]{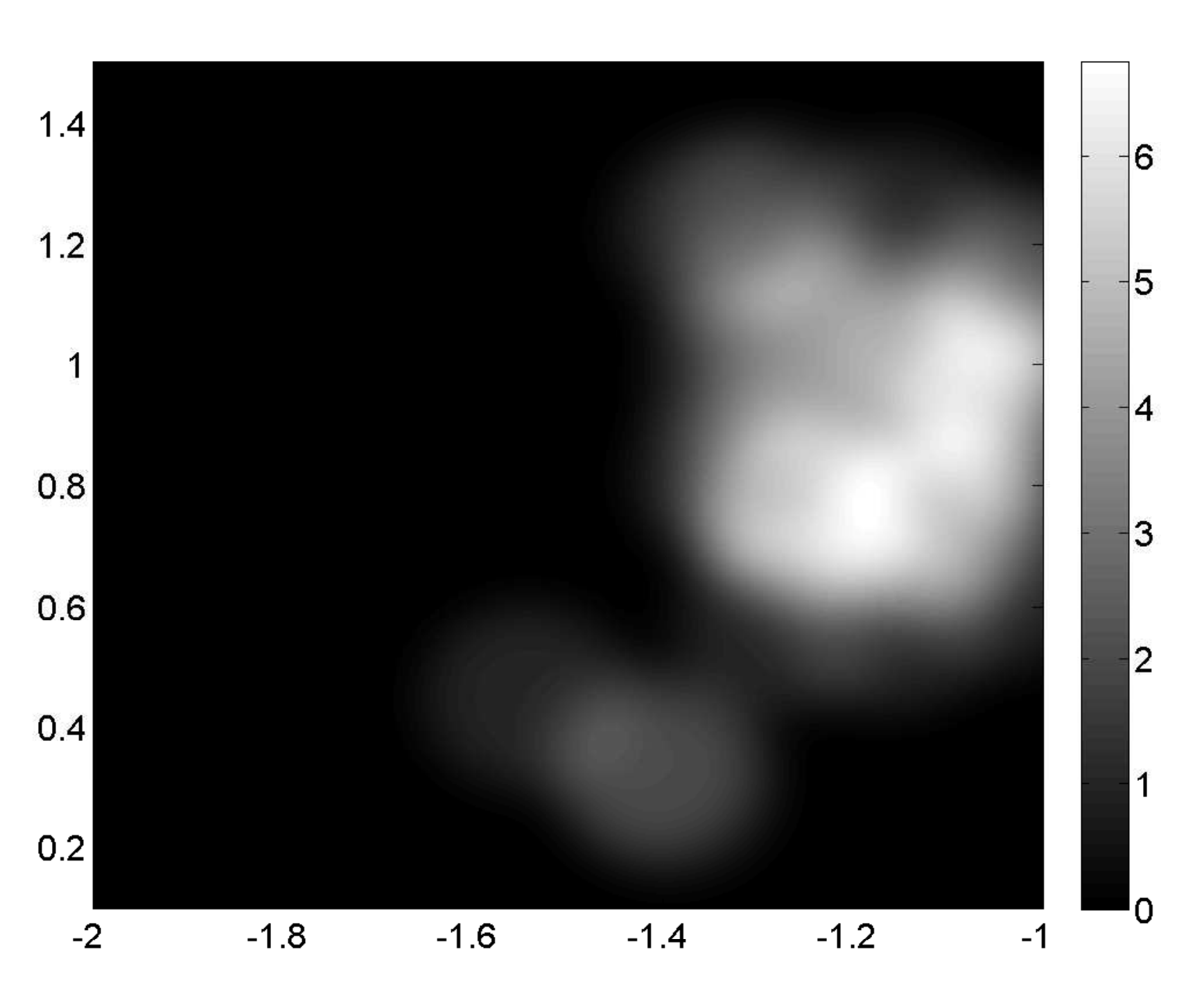} 
\includegraphics[width=0.20\textwidth,clip=true,trim=0cm 0cm 0cm 0cm]{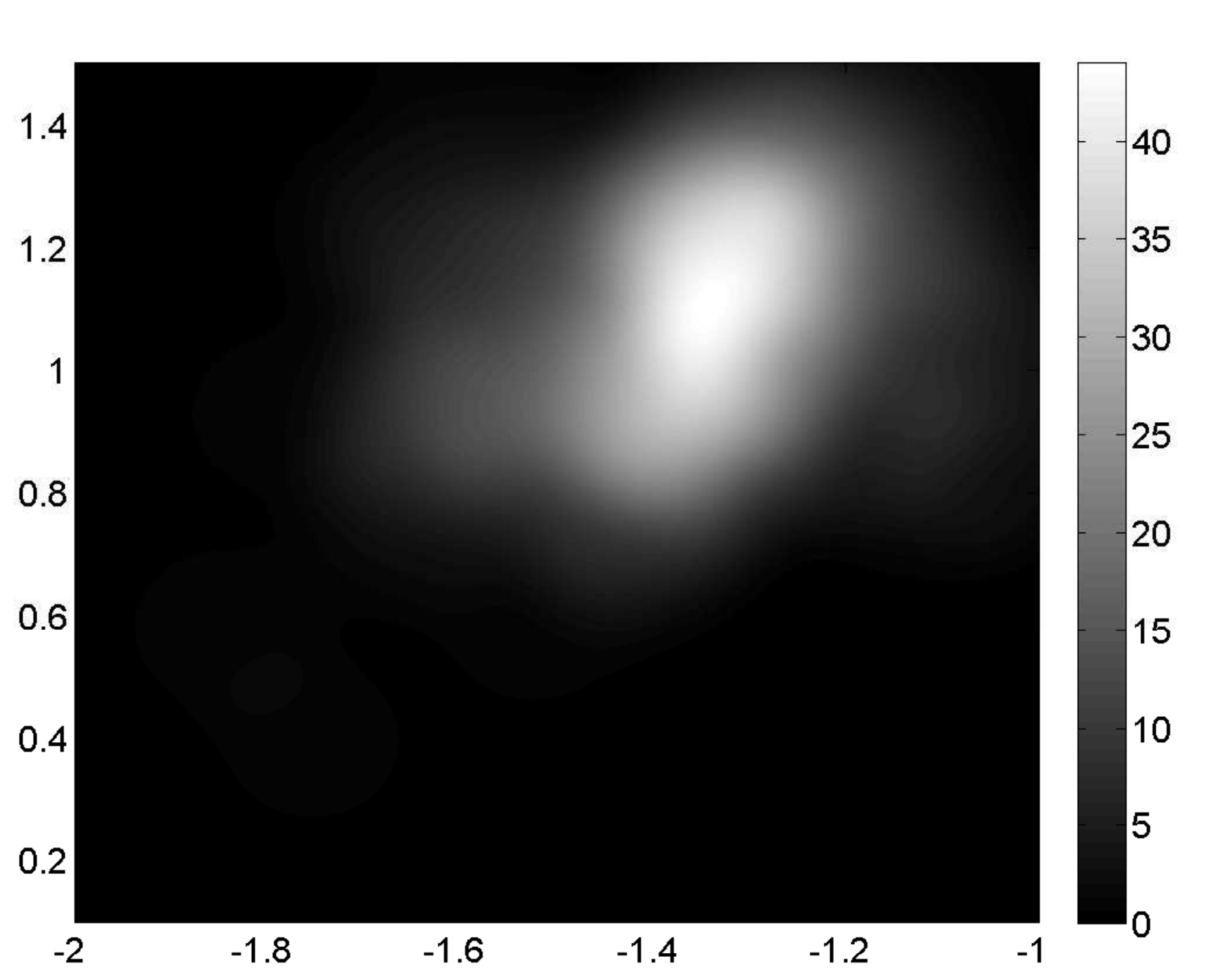} 
\includegraphics[width=0.20\textwidth,clip=true,trim=0cm 0cm 0cm 0cm]{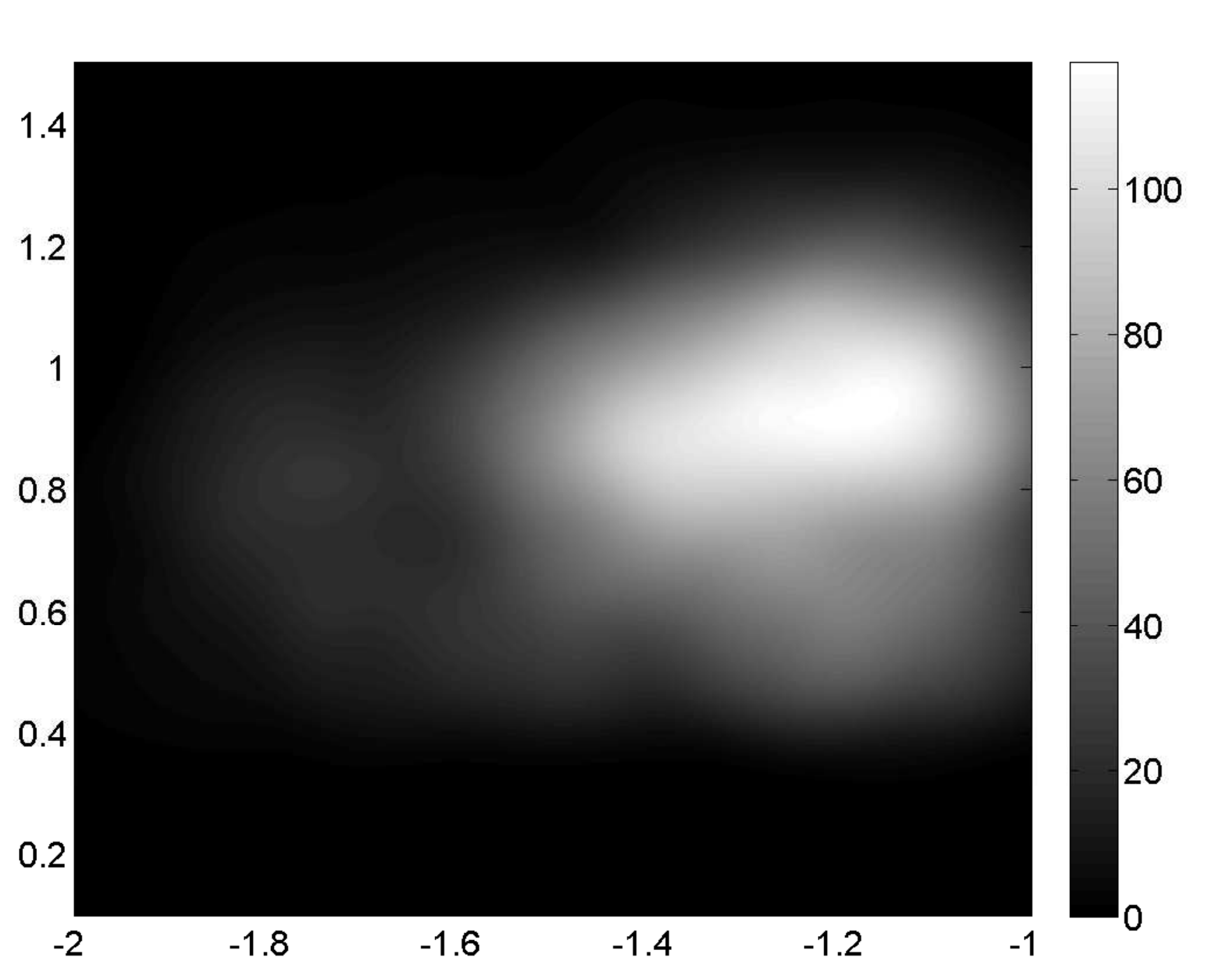} 
\includegraphics[width=0.20\textwidth,clip=true,trim=0cm 0cm 0cm 0cm]{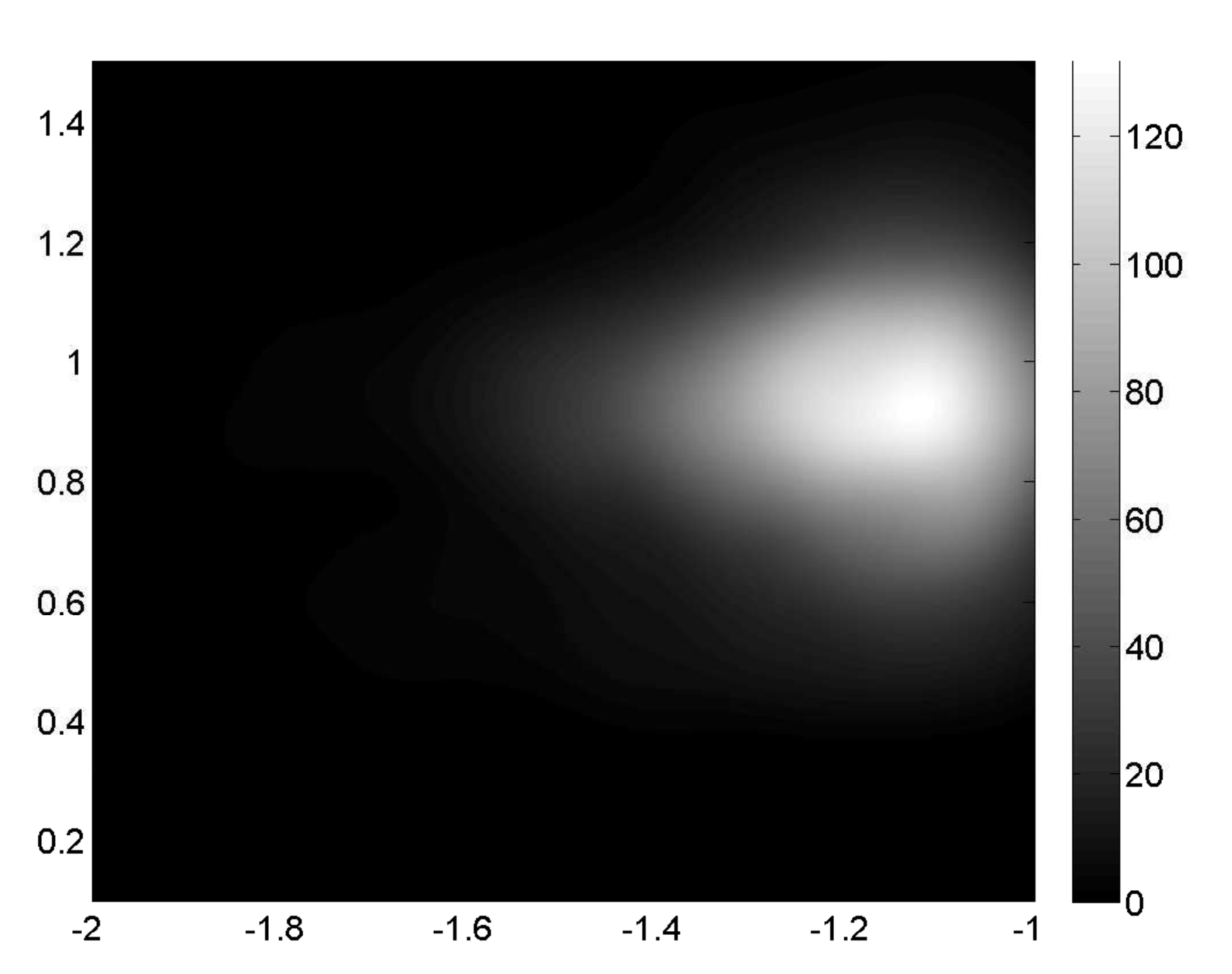} 
\includegraphics[width=0.20\textwidth,clip=true,trim=0cm 0cm 0cm 0cm]{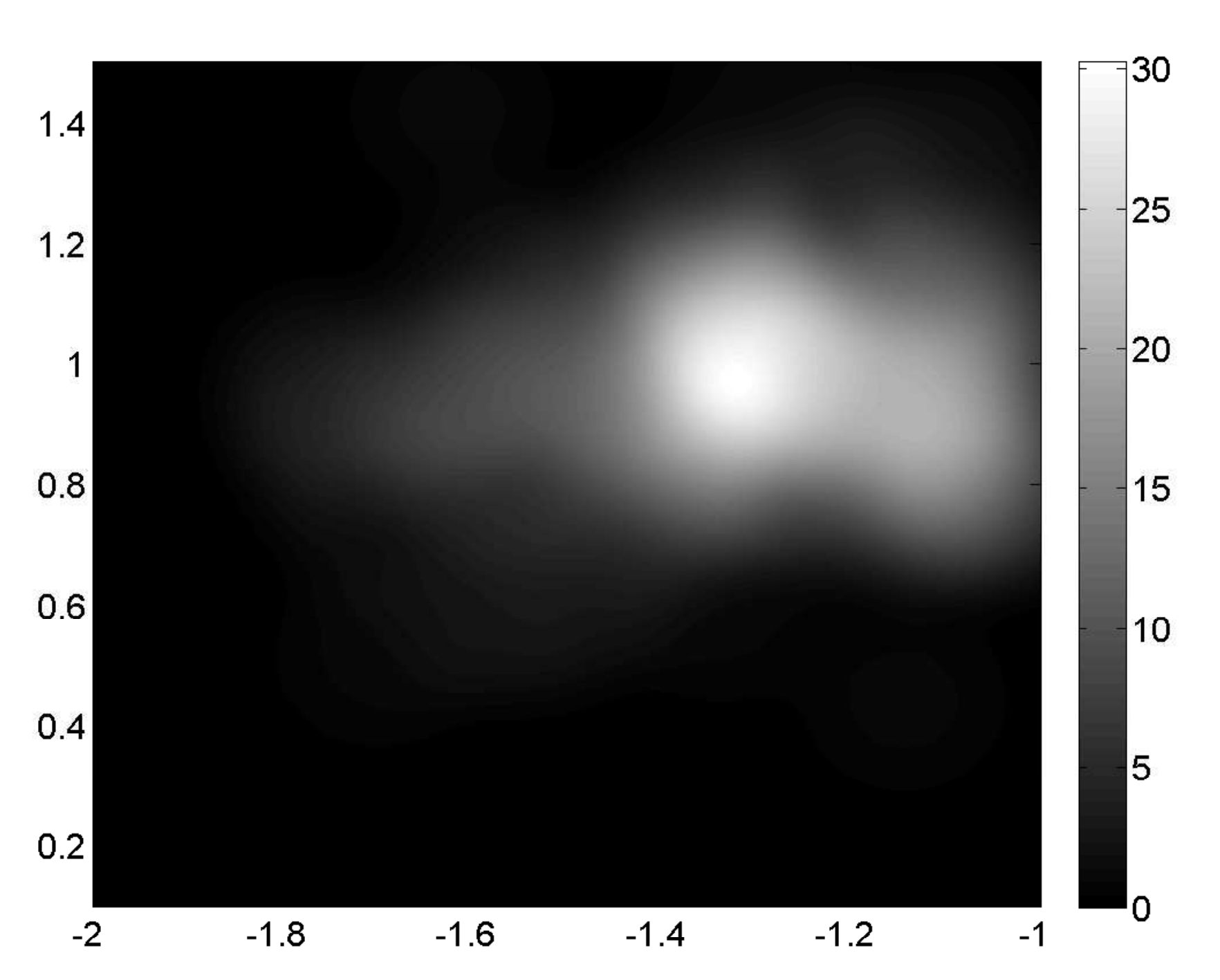} \\
\includegraphics[width=0.20\textwidth,clip=true,trim=0cm 0cm 0cm 0cm]{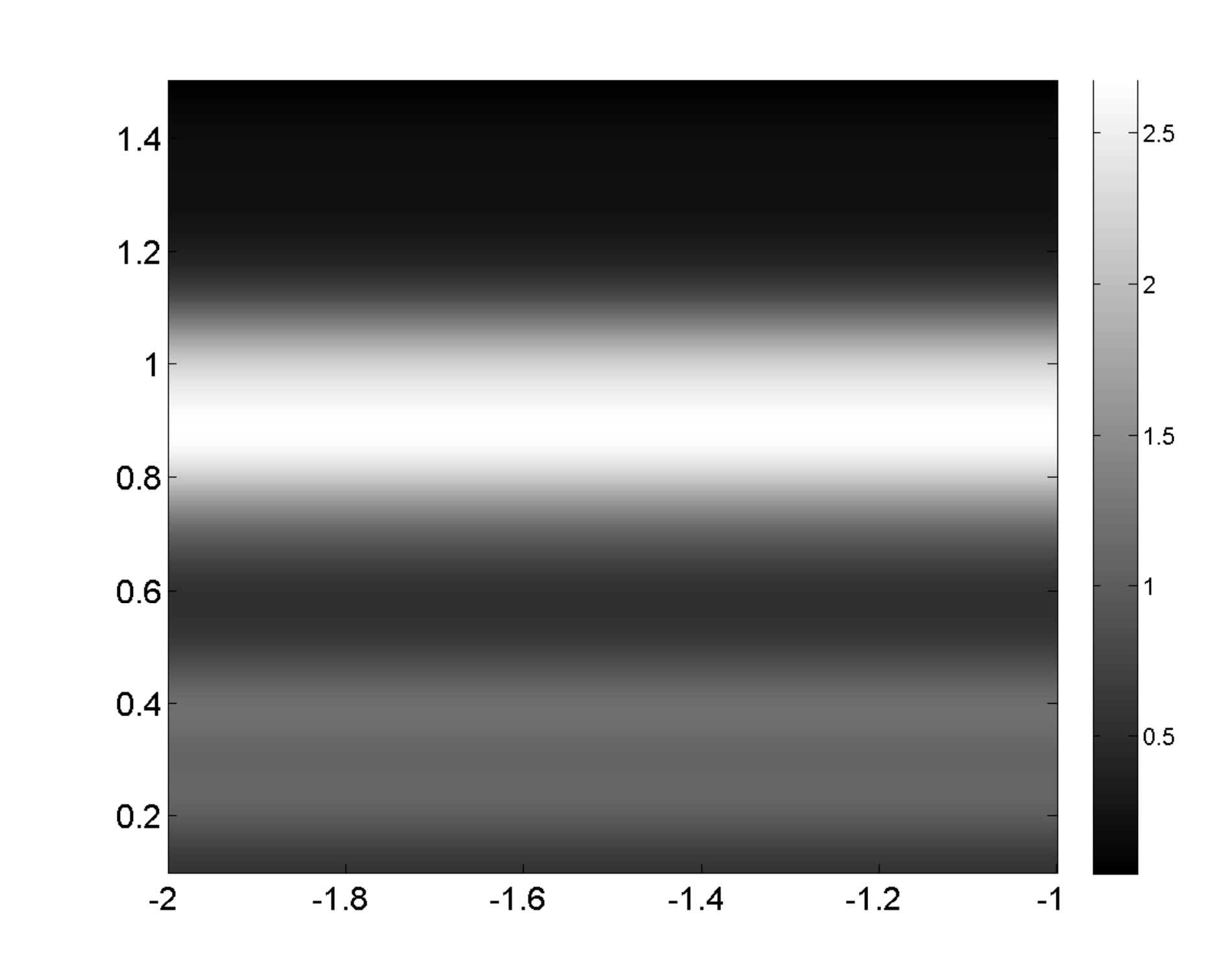} 
\includegraphics[width=0.20\textwidth,clip=true,trim=0cm 0cm 0cm 0cm]{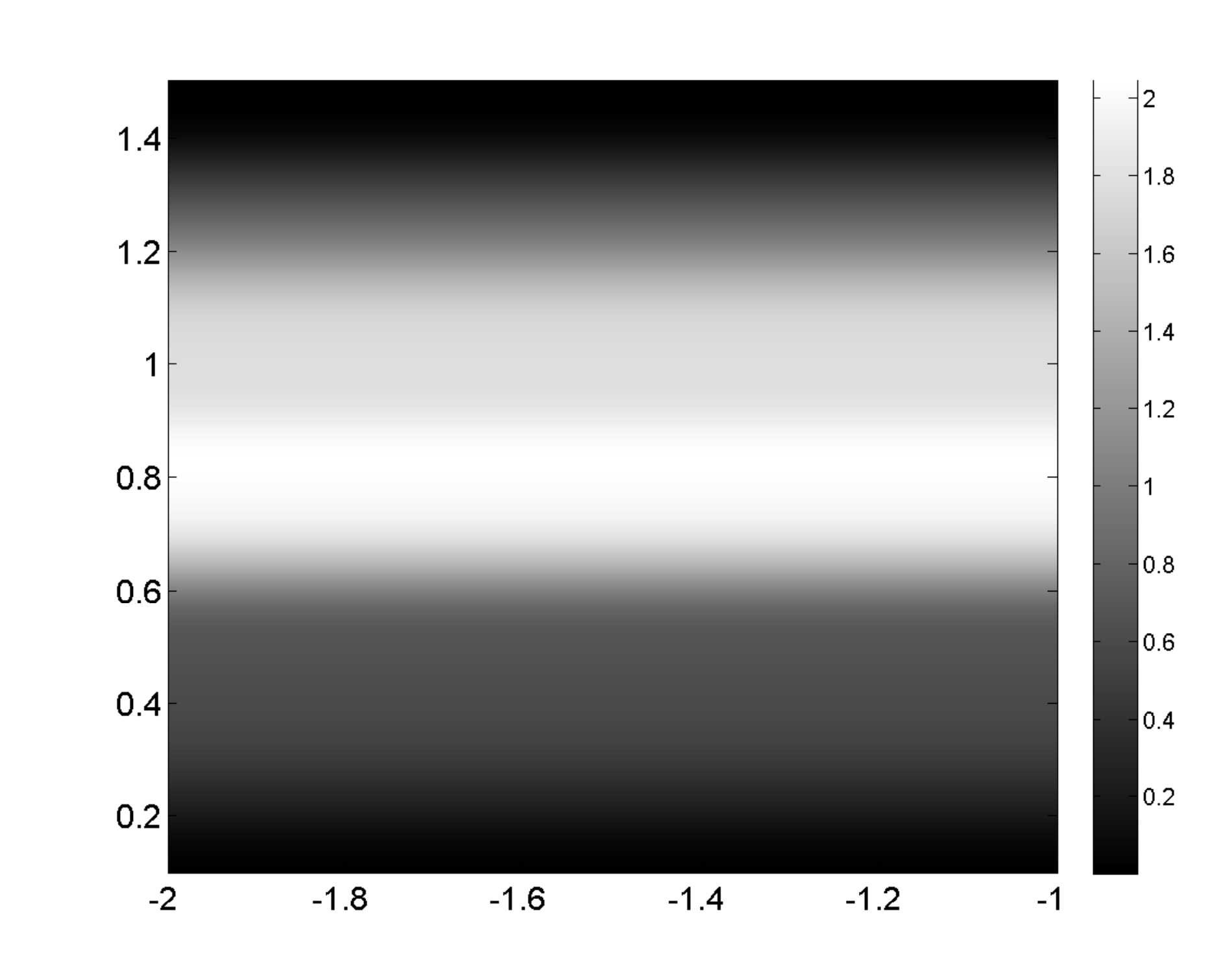} 
\includegraphics[width=0.20\textwidth,clip=true,trim=0cm 0cm 0cm 0cm]{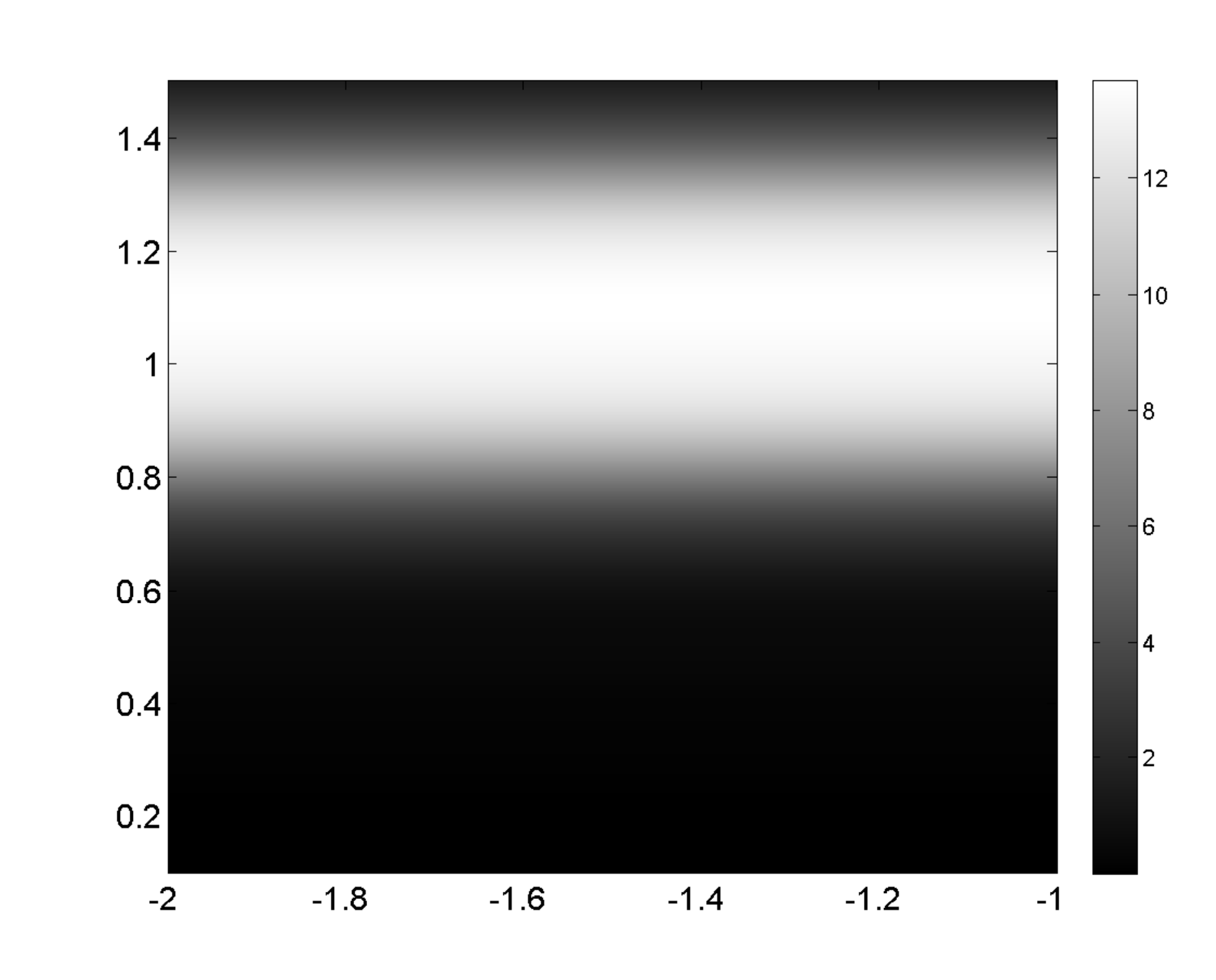}
\includegraphics[width=0.20\textwidth,clip=true,trim=0cm 0cm 0cm 0cm]{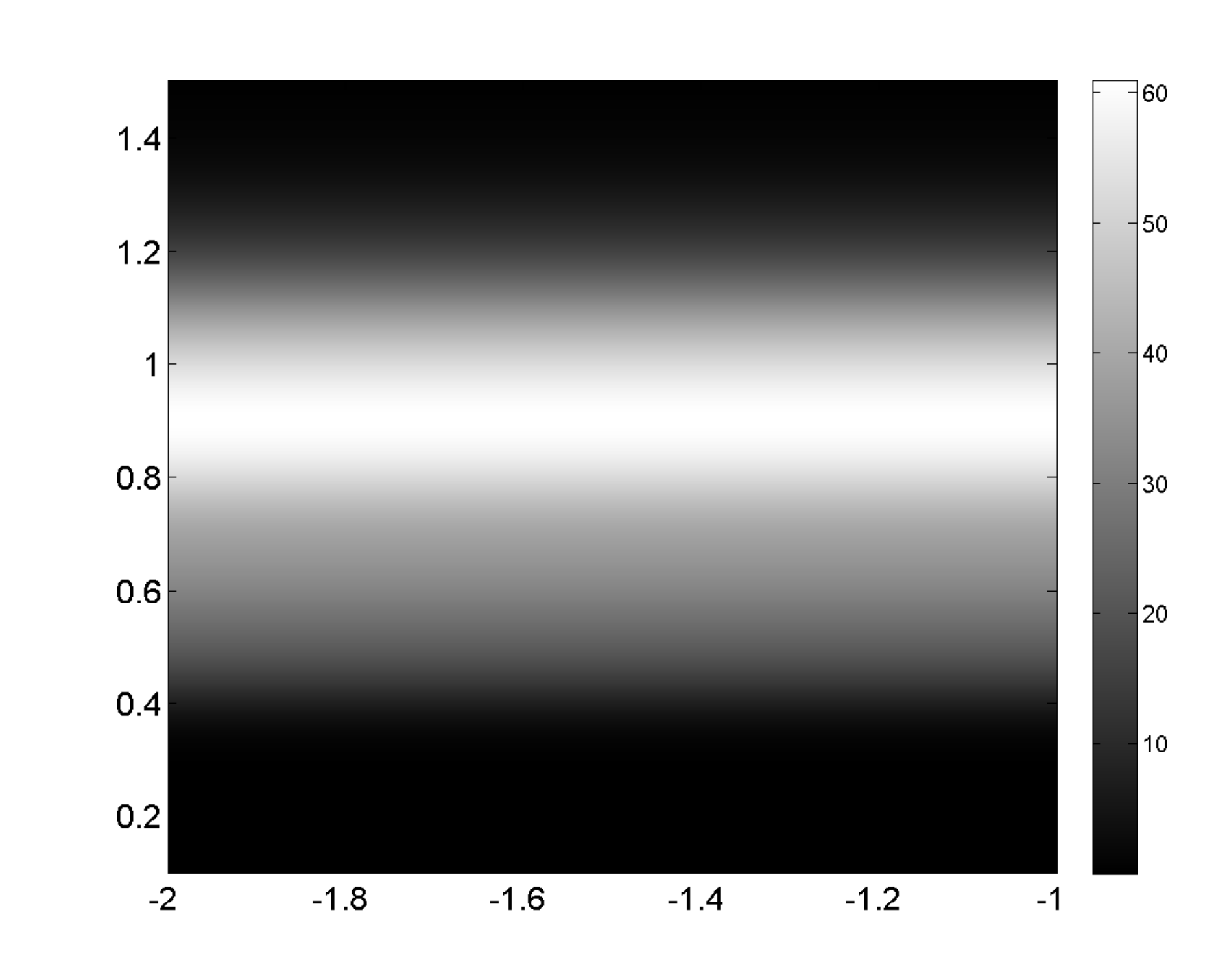} 
\includegraphics[width=0.20\textwidth,clip=true,trim=0cm 0cm 0cm 0cm]{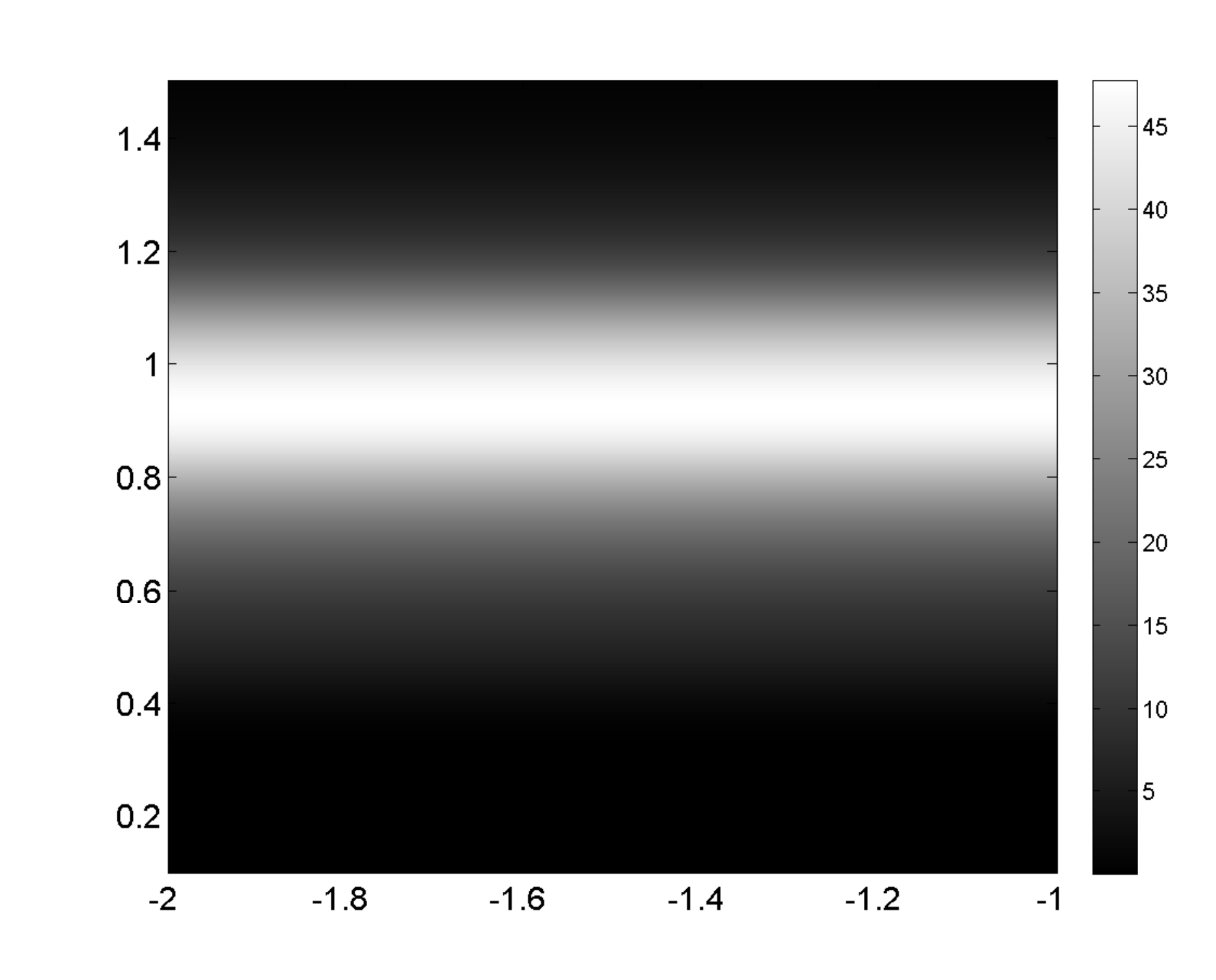} 
\includegraphics[width=0.20\textwidth,clip=true,trim=0cm 0cm 0cm 0cm]{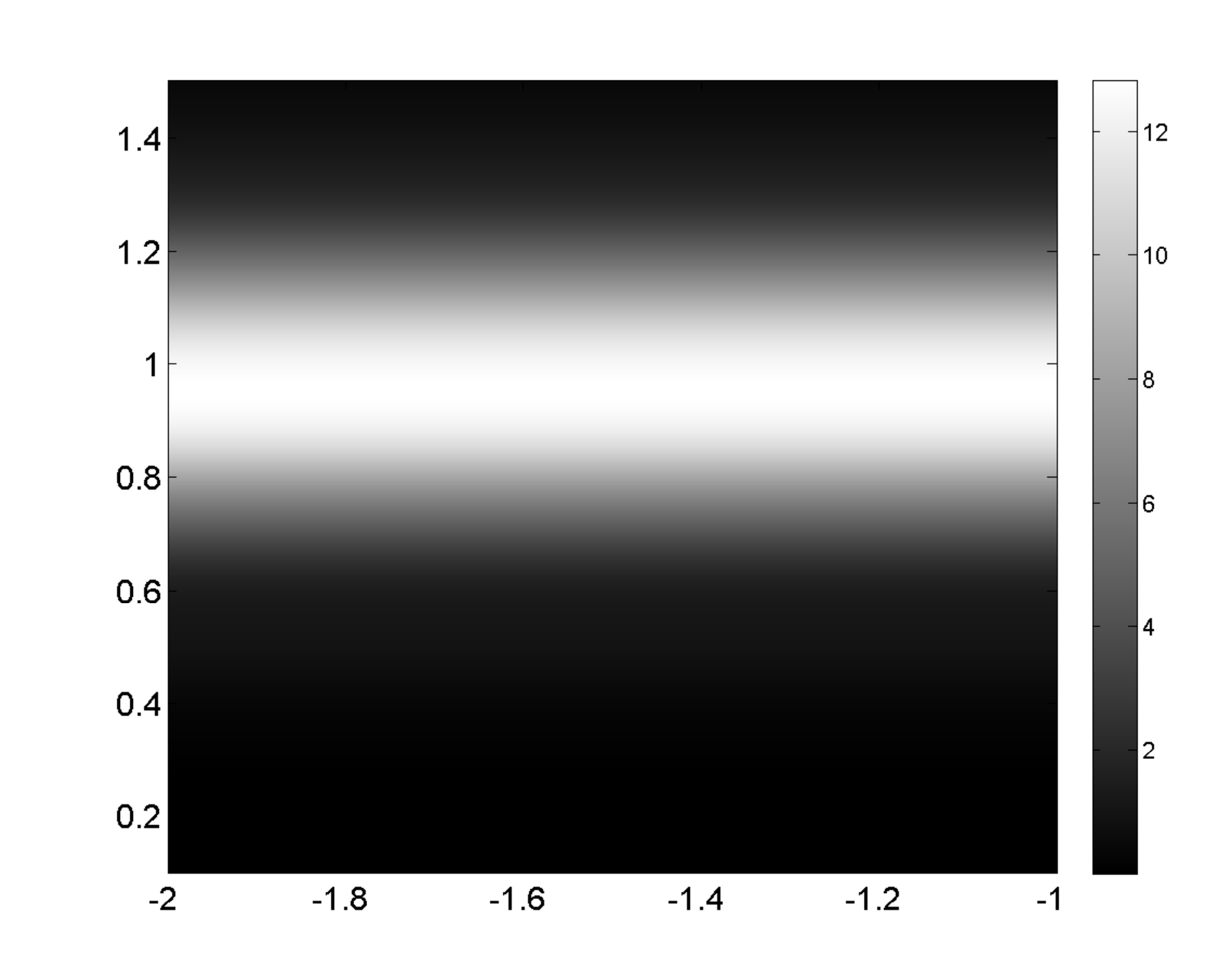} \\
\includegraphics[width=0.20\textwidth,clip=true,trim=0cm 0cm 0cm 0cm]{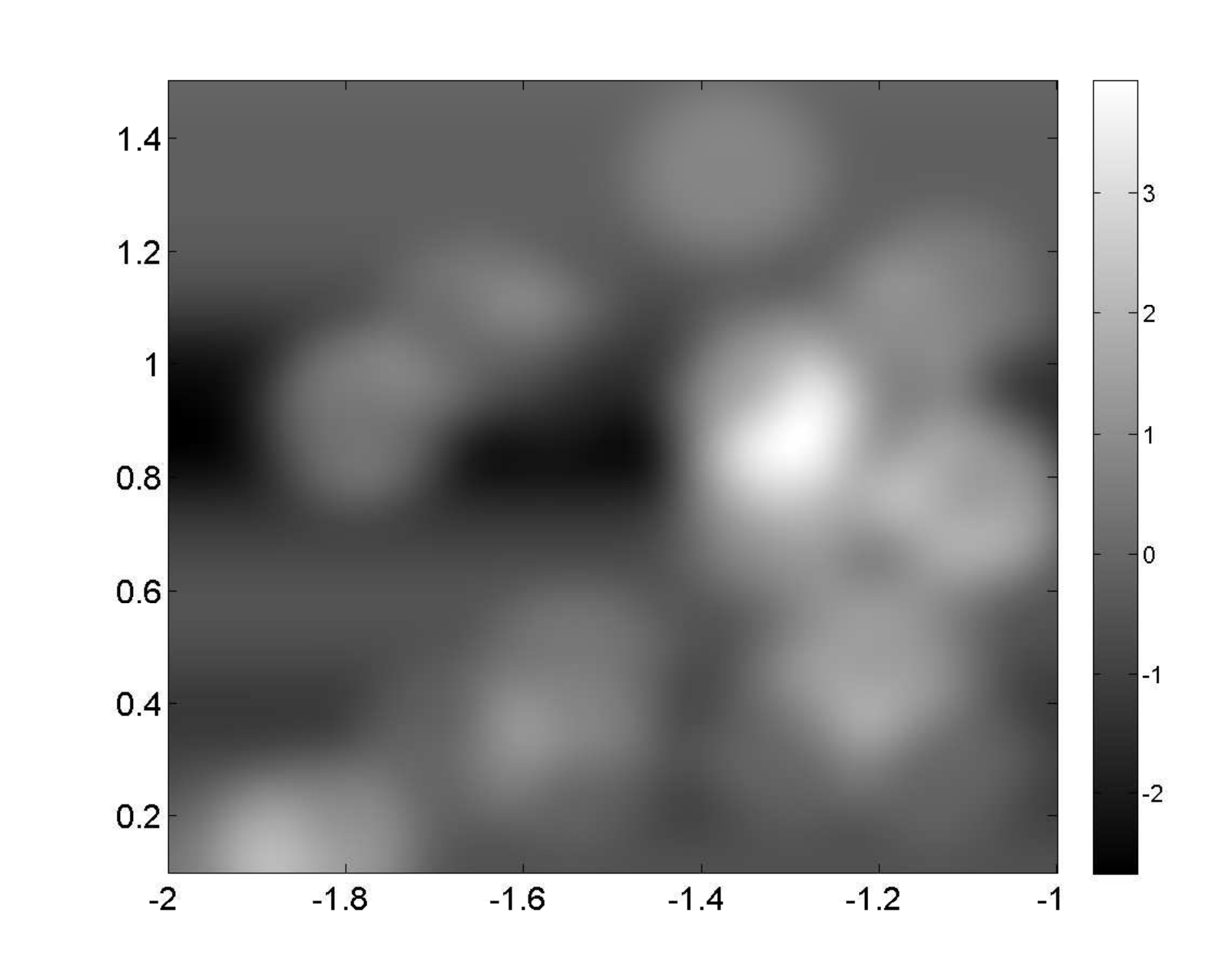} 
\includegraphics[width=0.20\textwidth,clip=true,trim=0cm 0cm 0cm 0cm]{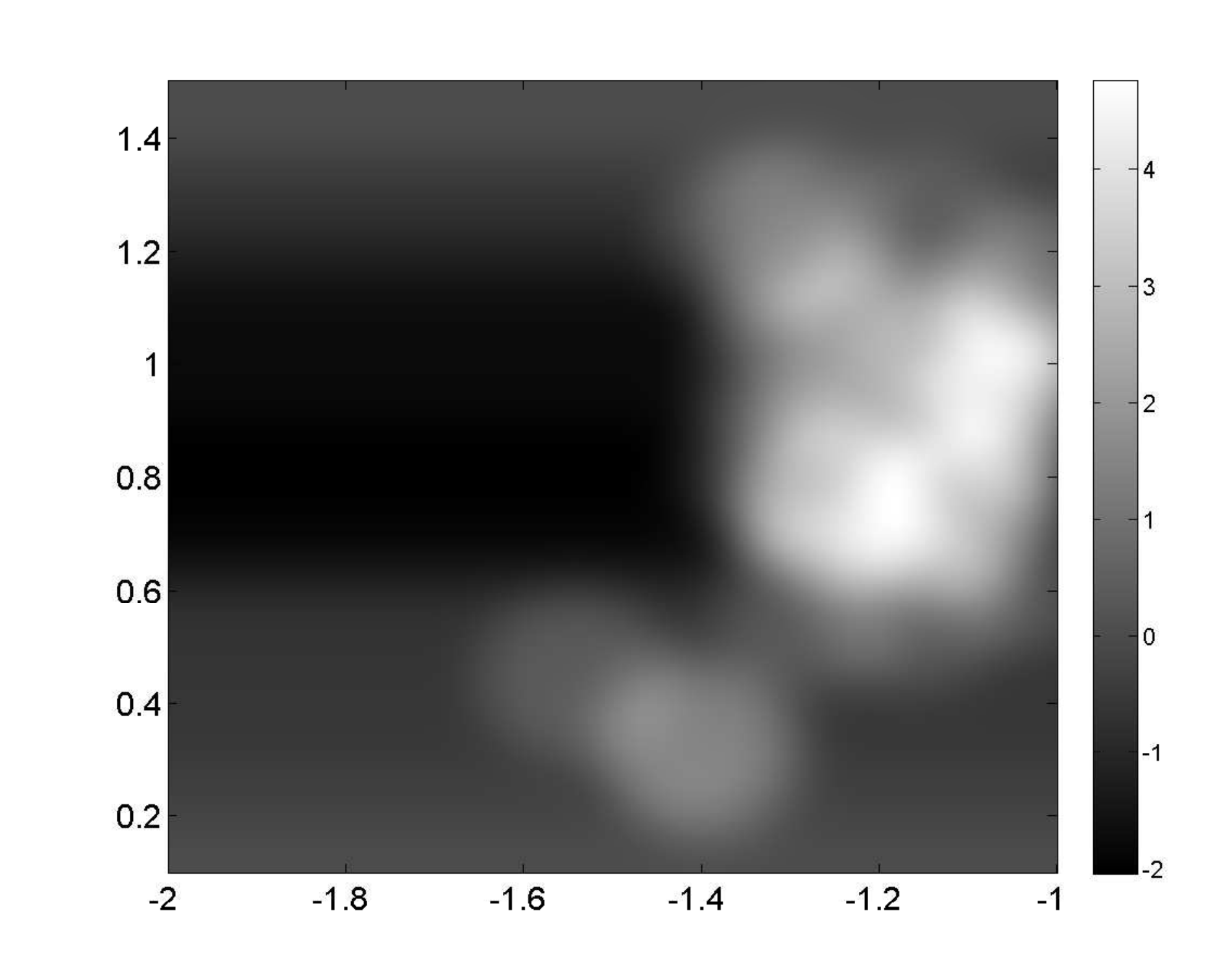} 
\includegraphics[width=0.20\textwidth,clip=true,trim=0cm 0cm 0cm 0cm]{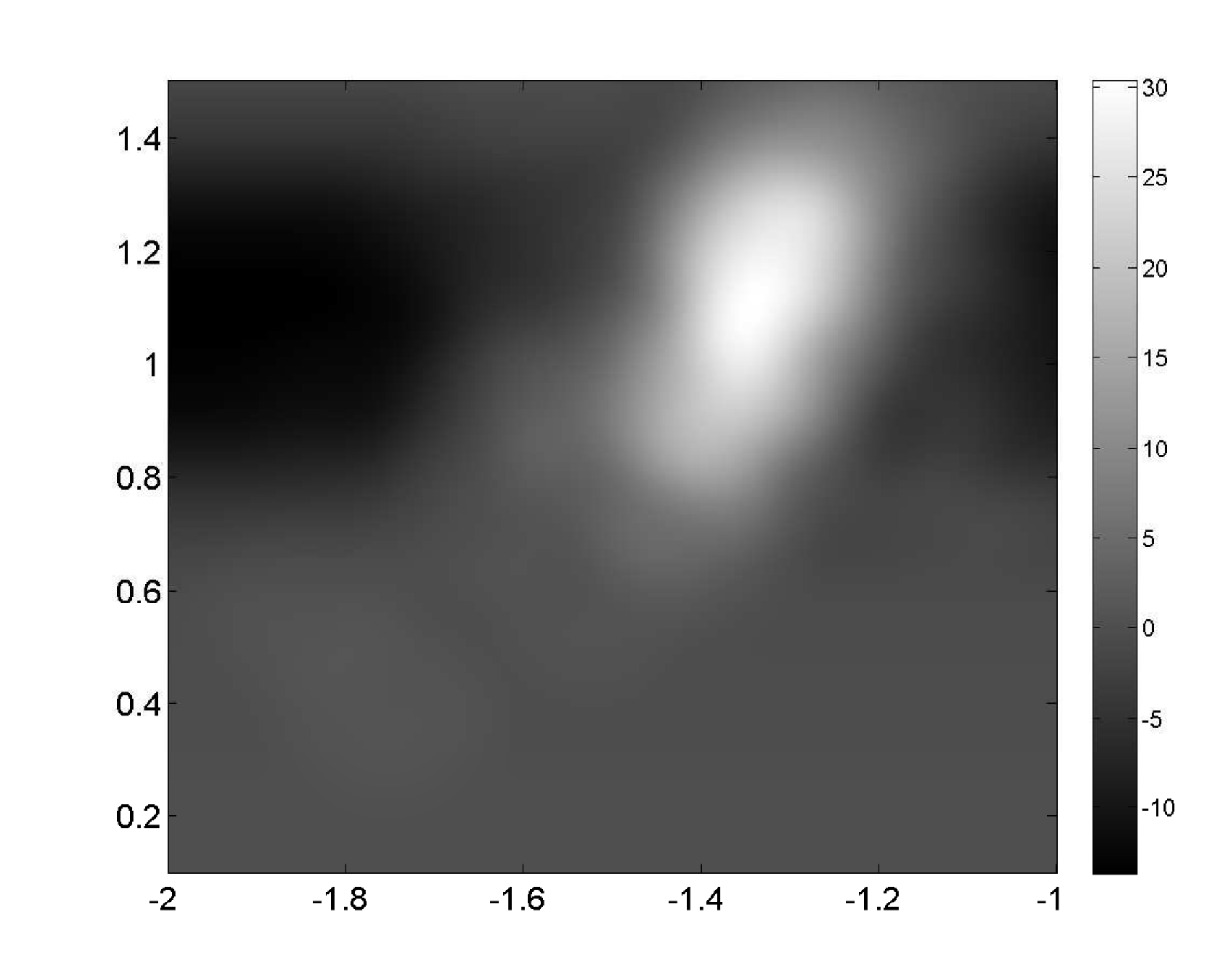} 
\includegraphics[width=0.20\textwidth,clip=true,trim=0cm 0cm 0cm 0cm]{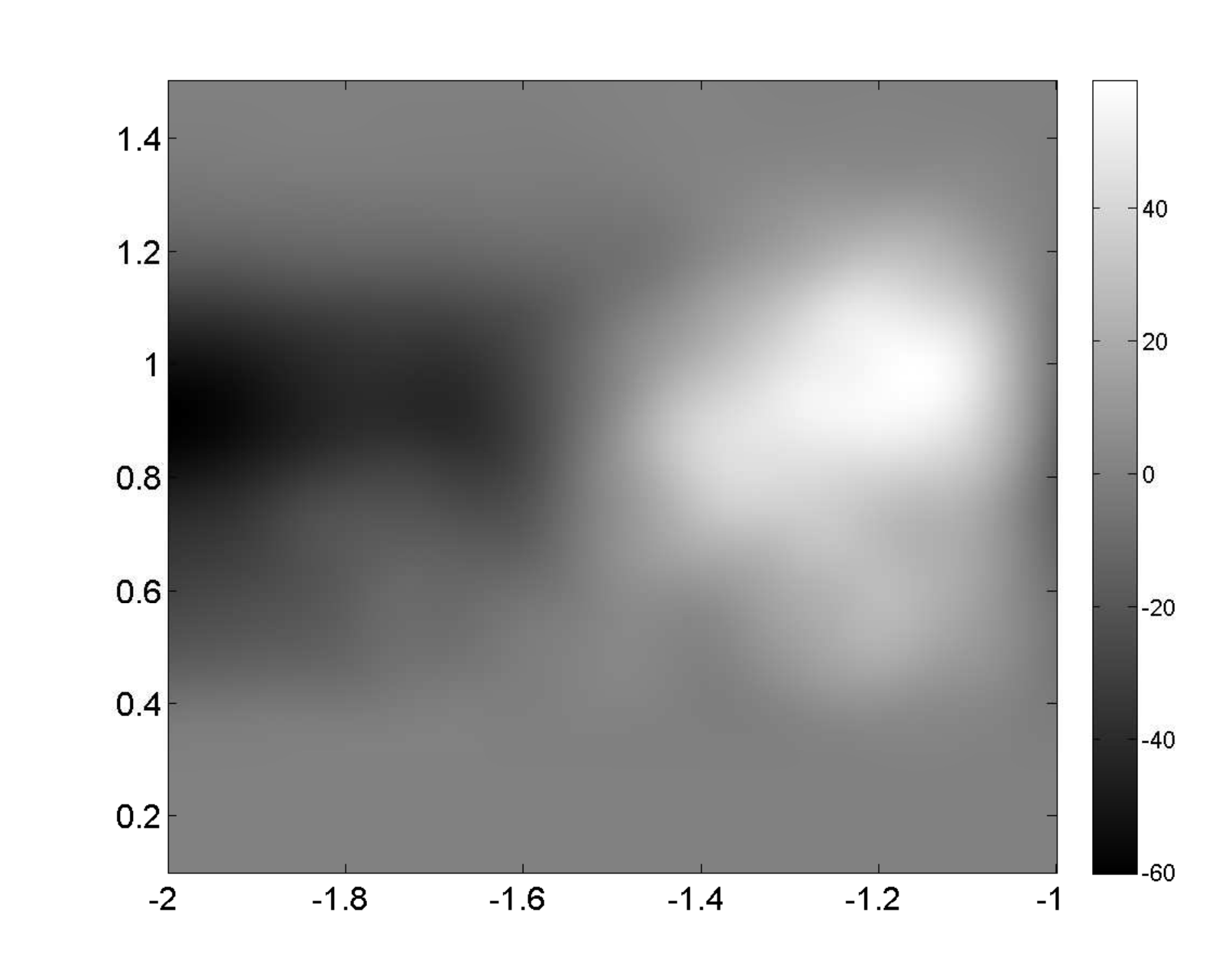} 
\includegraphics[width=0.20\textwidth,clip=true,trim=0cm 0cm 0cm 0cm]{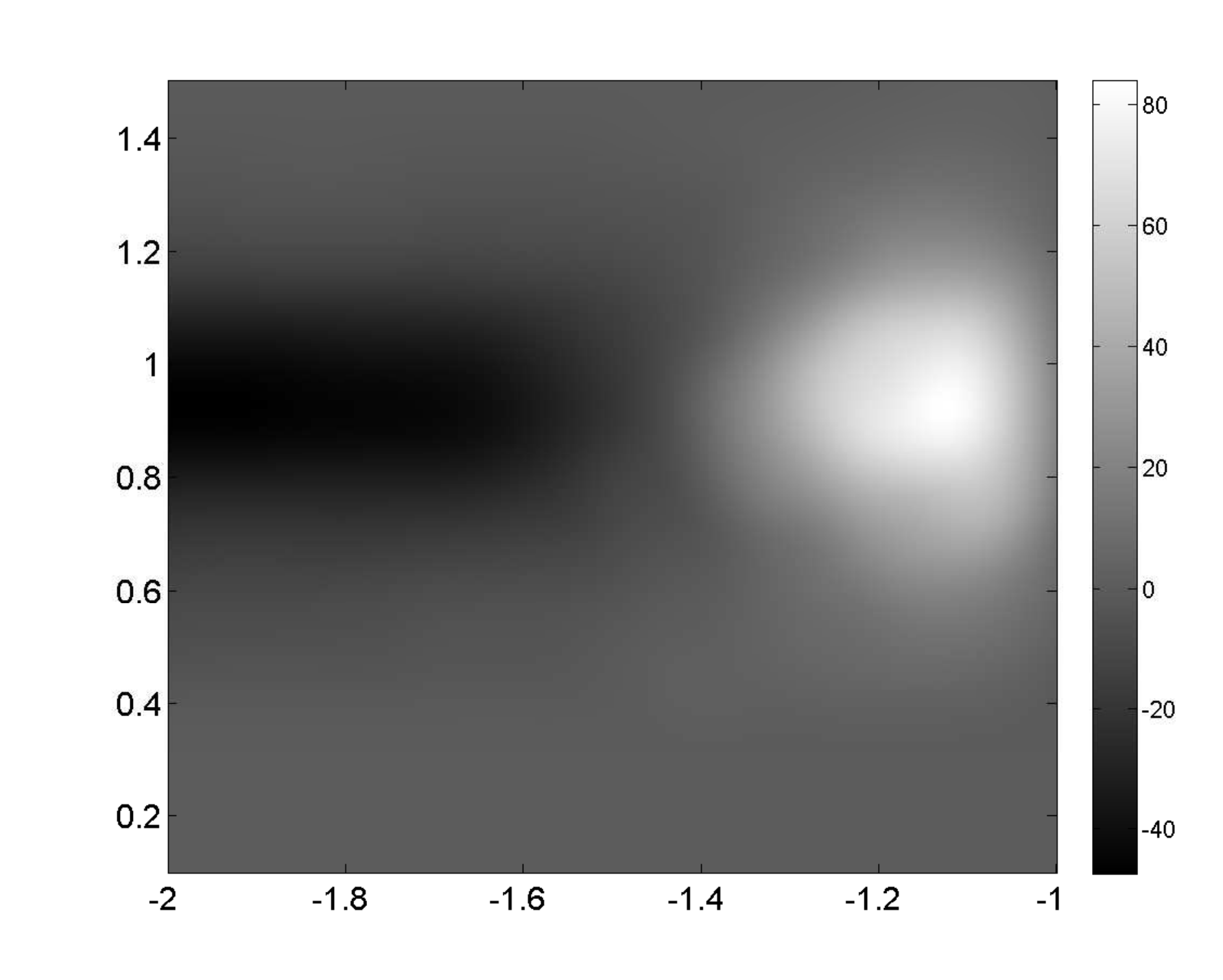} 
\includegraphics[width=0.20\textwidth,clip=true,trim=0cm 0cm 0cm 0cm]{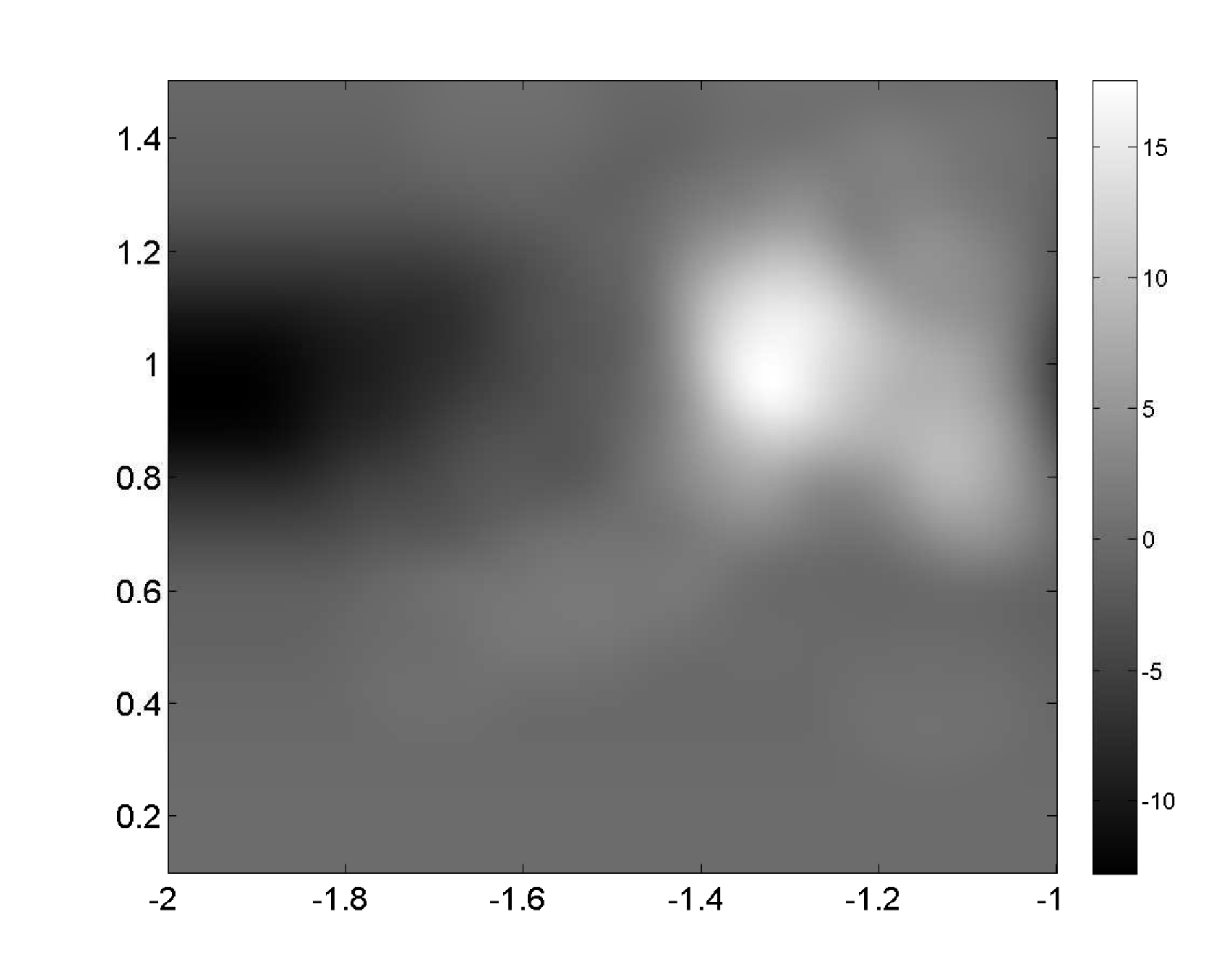} 
\caption{Same as Figure 2 for the exponential decay with semi-major axis tidal halting model.}
\end{figure}
\end{landscape}

\clearpage
\begin{landscape}
\begin{figure}
\centering
\includegraphics[width=0.20\textwidth,clip=true,trim=0cm 0cm 0cm 0cm]{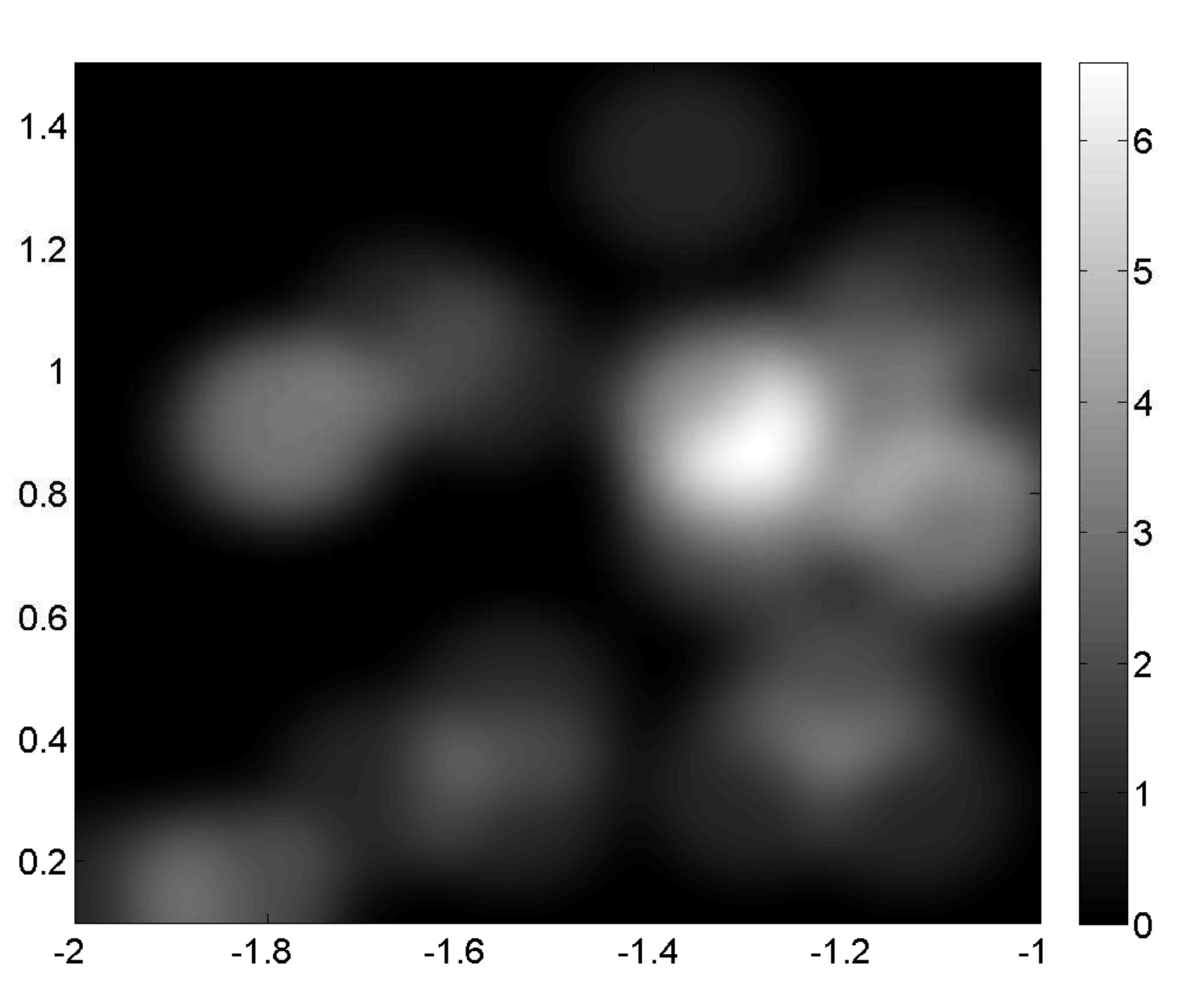} 
\includegraphics[width=0.20\textwidth,clip=true,trim=0cm 0cm 0cm 0cm]{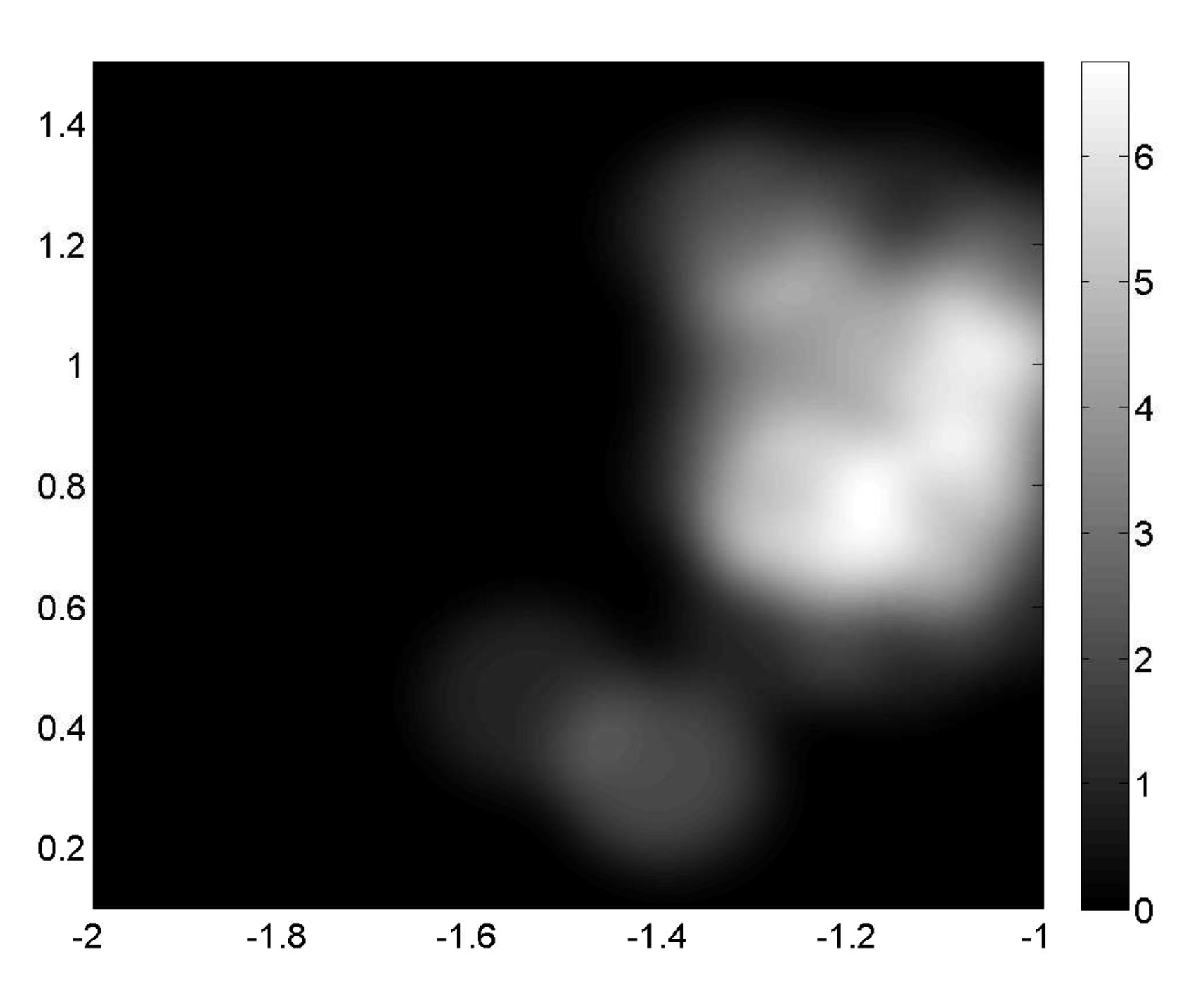} 
\includegraphics[width=0.20\textwidth,clip=true,trim=0cm 0cm 0cm 0cm]{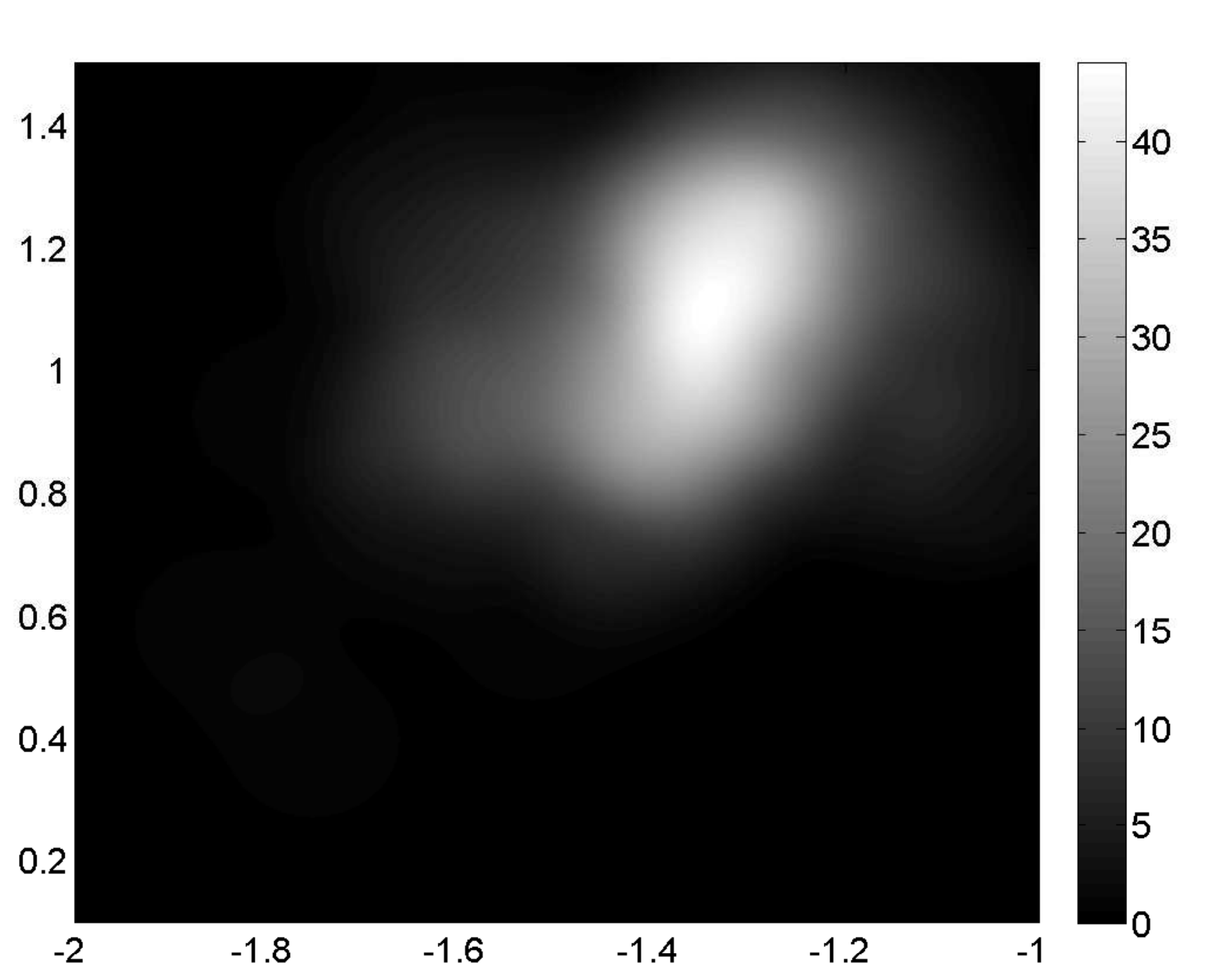} 
\includegraphics[width=0.20\textwidth,clip=true,trim=0cm 0cm 0cm 0cm]{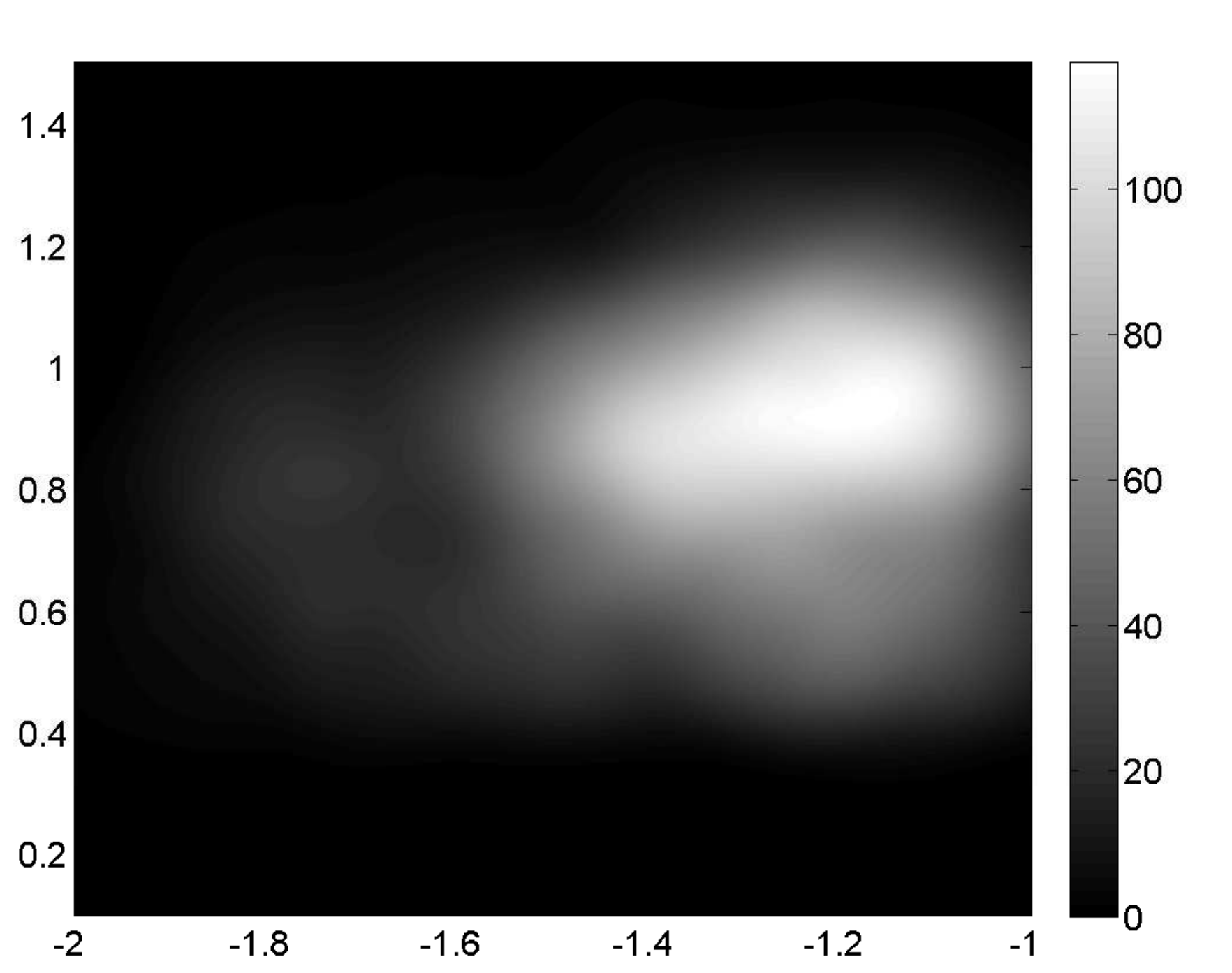} 
\includegraphics[width=0.20\textwidth,clip=true,trim=0cm 0cm 0cm 0cm]{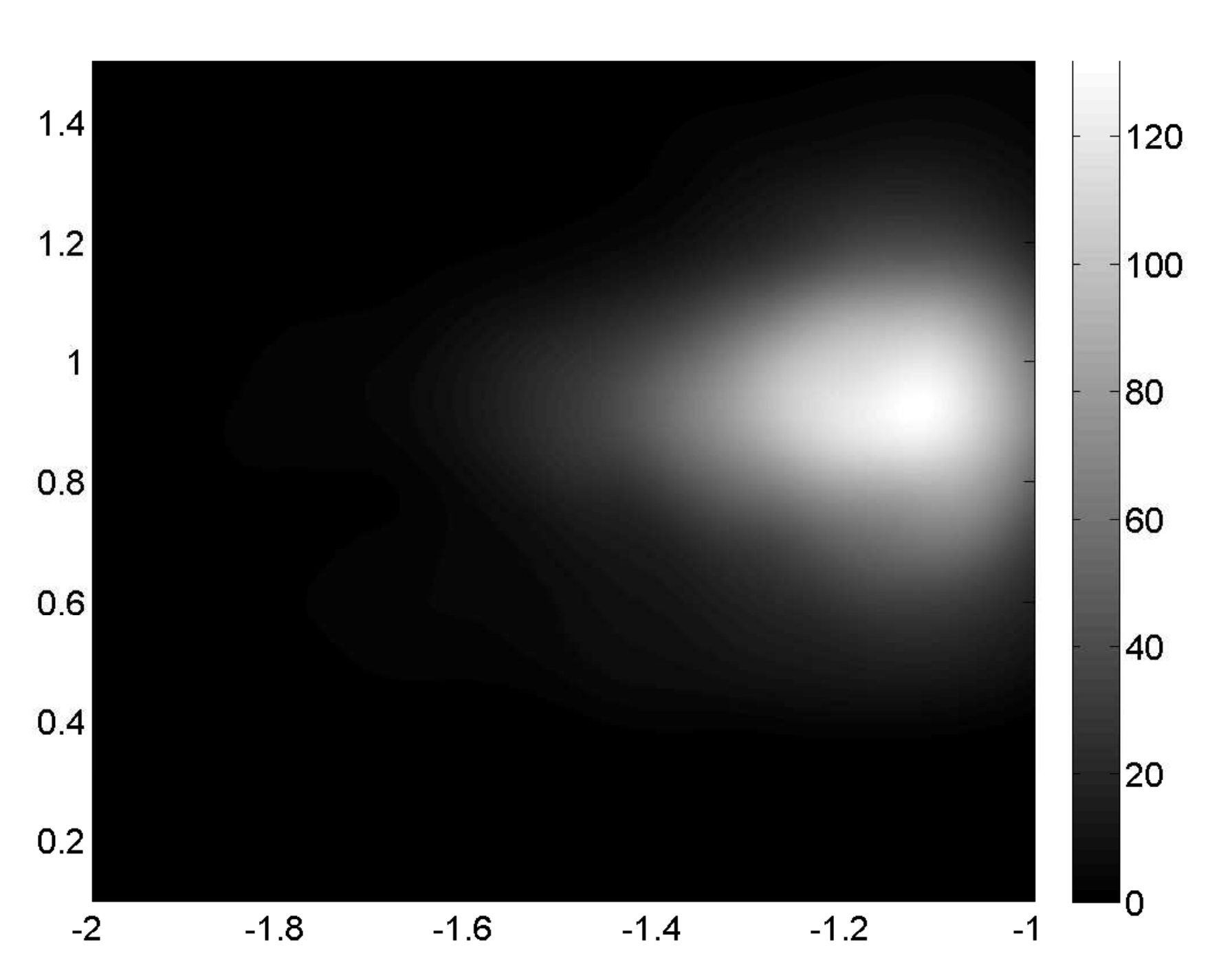} 
\includegraphics[width=0.20\textwidth,clip=true,trim=0cm 0cm 0cm 0cm]{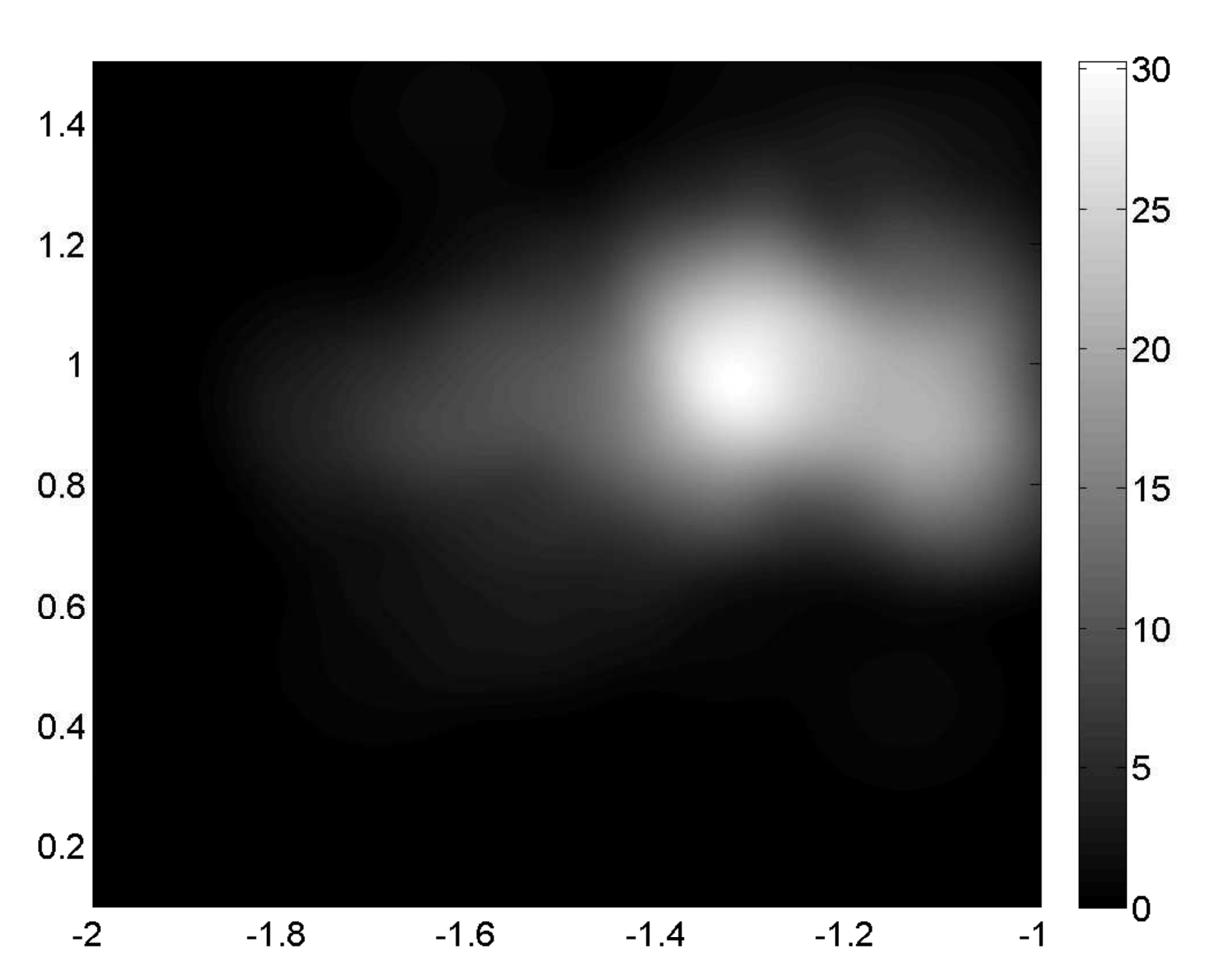} \\
\includegraphics[width=0.20\textwidth,clip=true,trim=0cm 0cm 0cm 0cm]{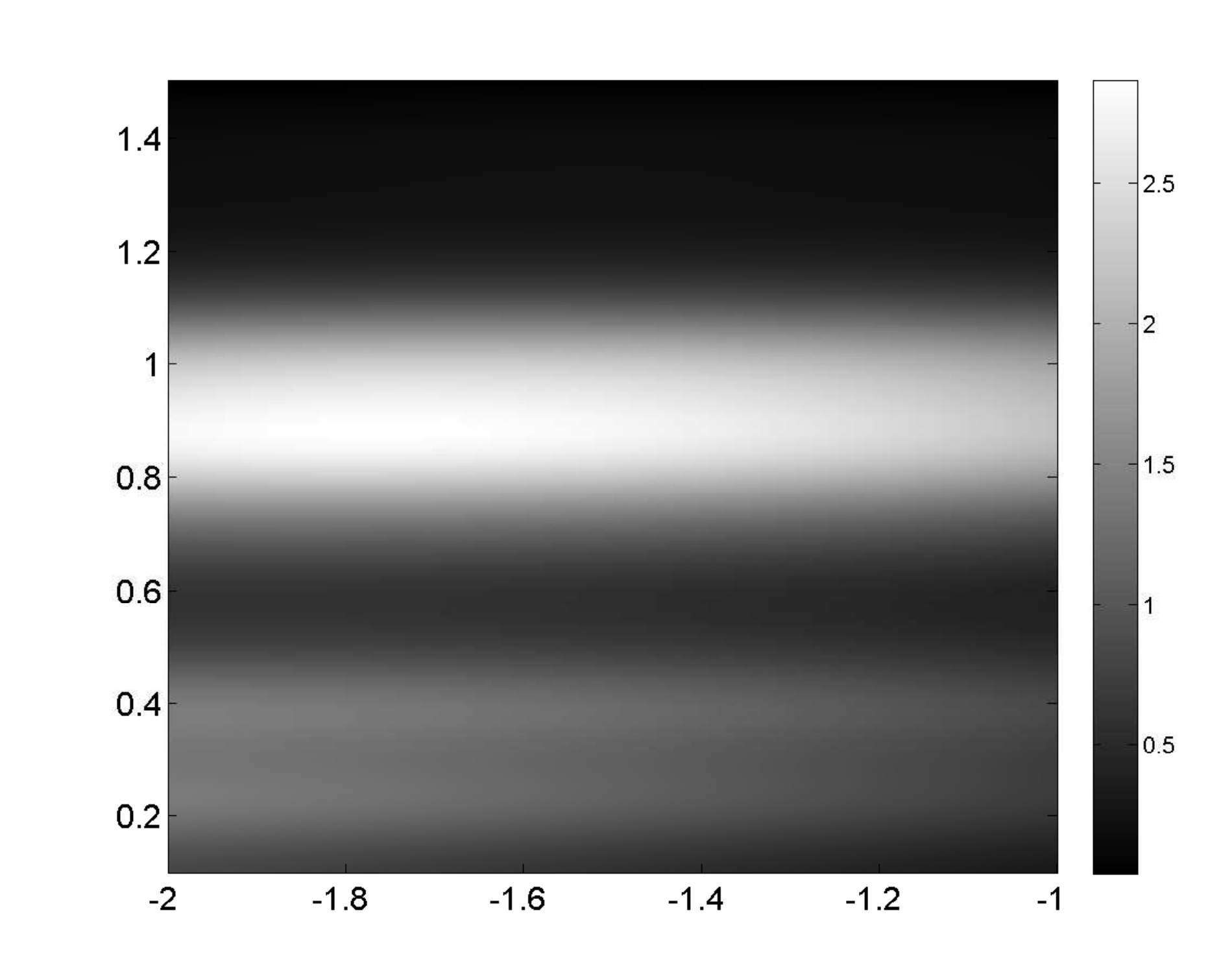} 
\includegraphics[width=0.20\textwidth,clip=true,trim=0cm 0cm 0cm 0cm]{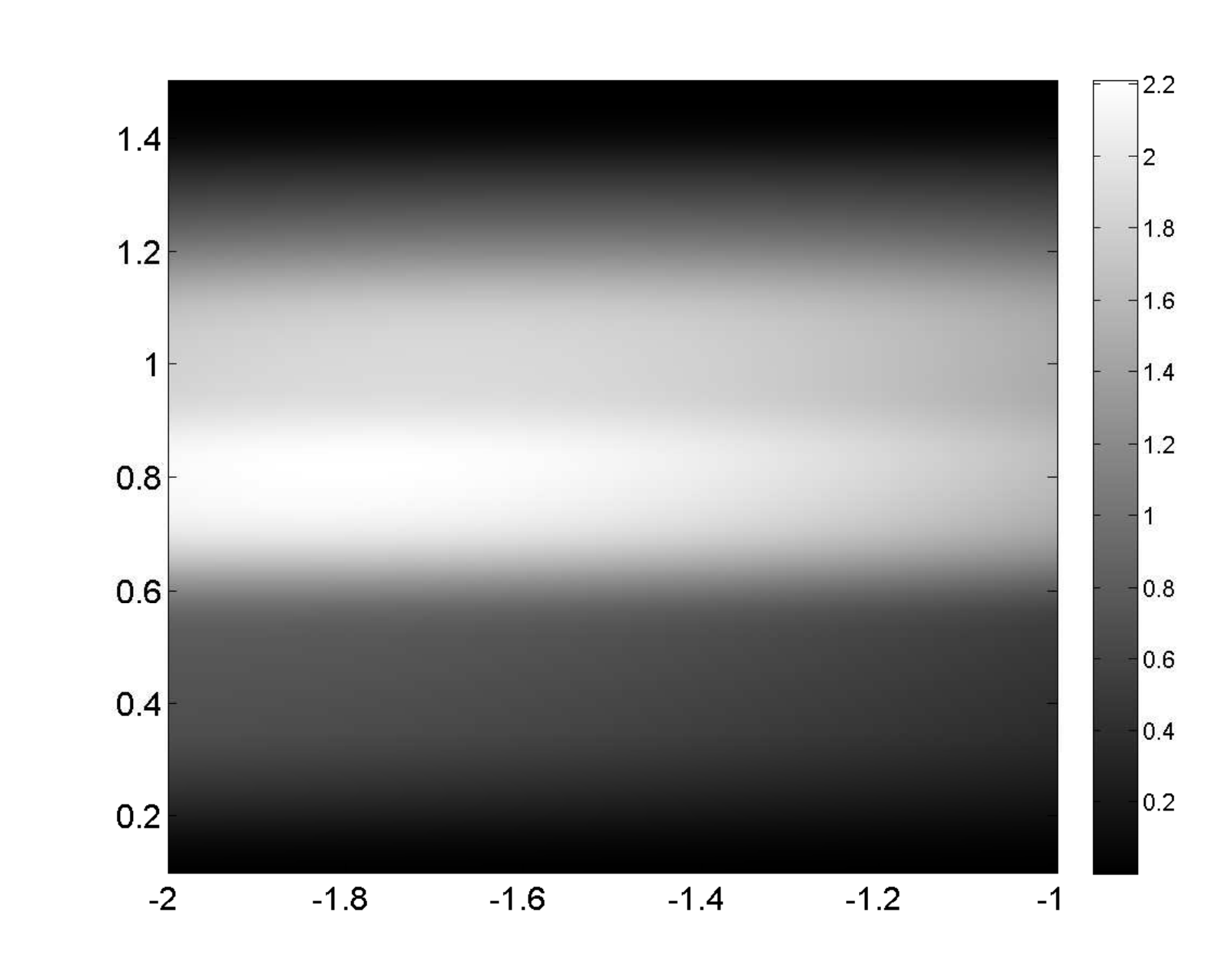} 
\includegraphics[width=0.20\textwidth,clip=true,trim=0cm 0cm 0cm 0cm]{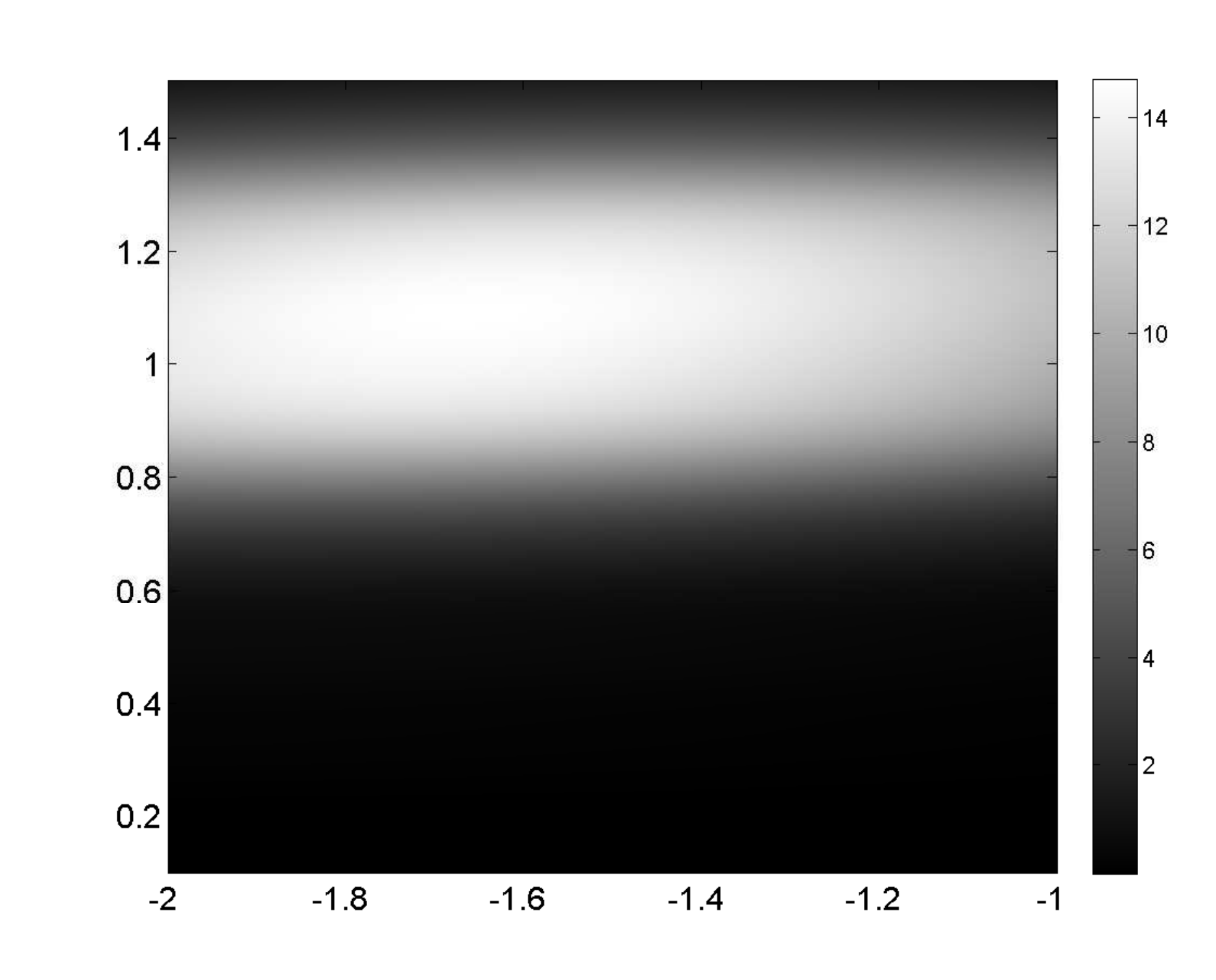}
\includegraphics[width=0.20\textwidth,clip=true,trim=0cm 0cm 0cm 0cm]{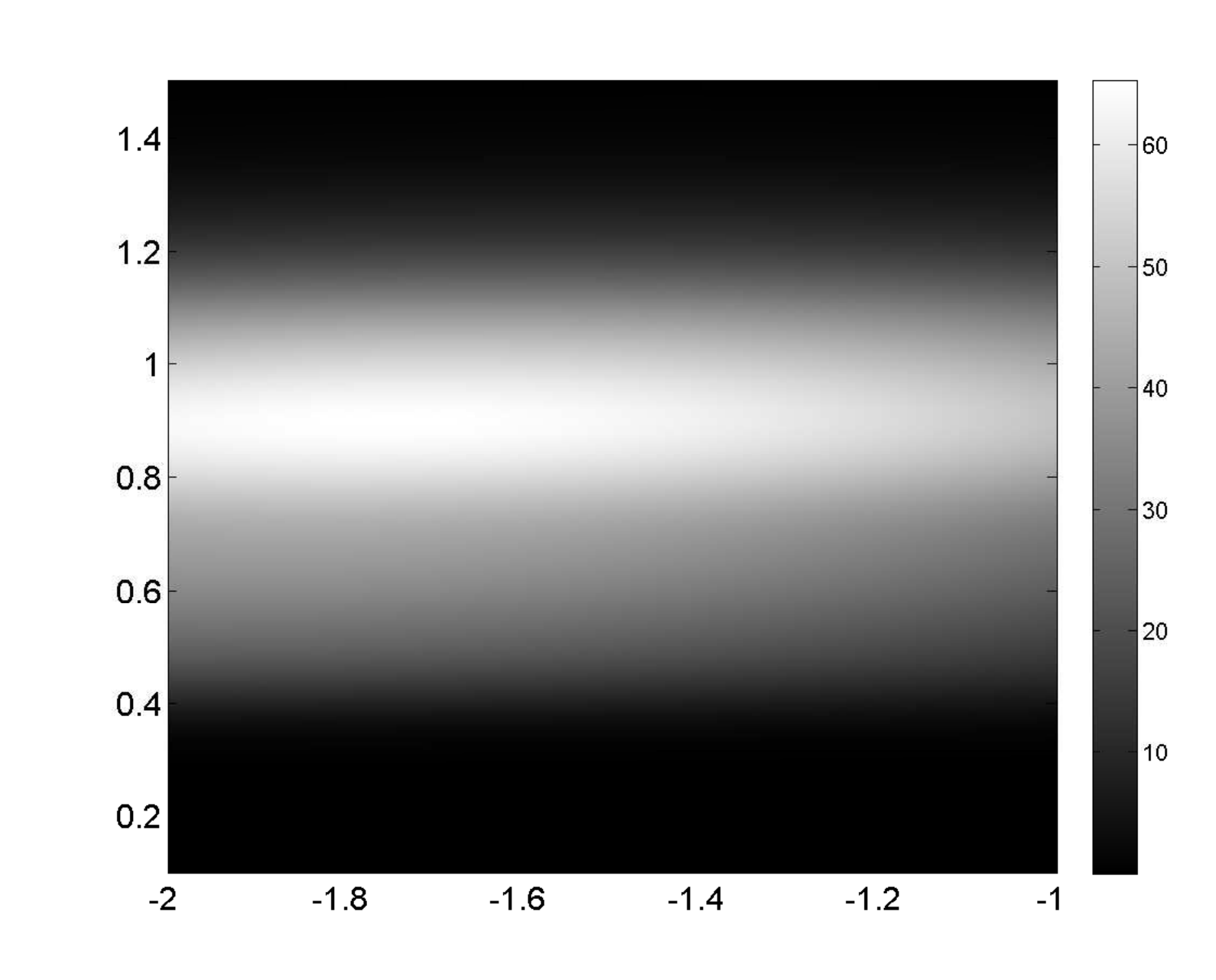} 
\includegraphics[width=0.20\textwidth,clip=true,trim=0cm 0cm 0cm 0cm]{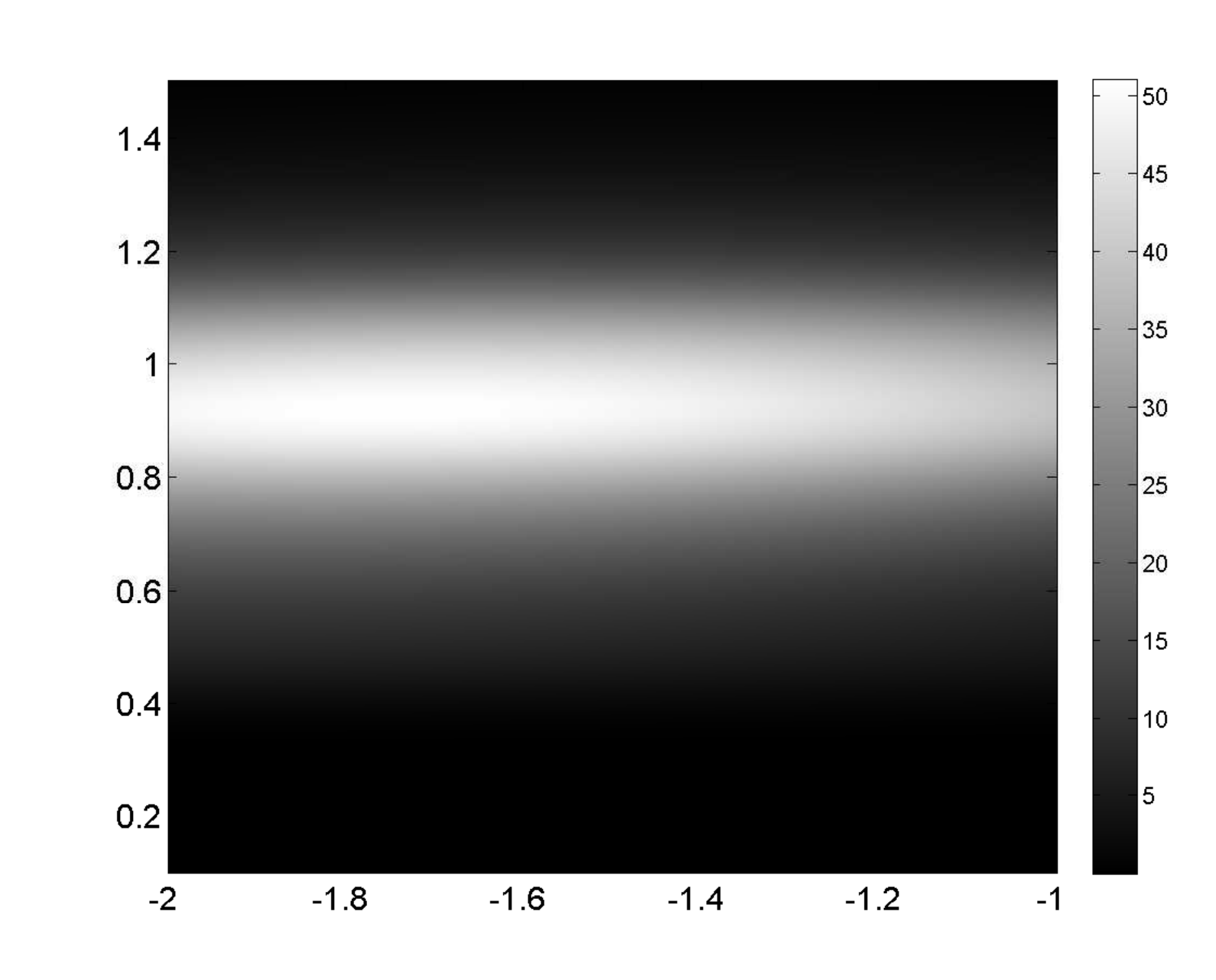} 
\includegraphics[width=0.20\textwidth,clip=true,trim=0cm 0cm 0cm 0cm]{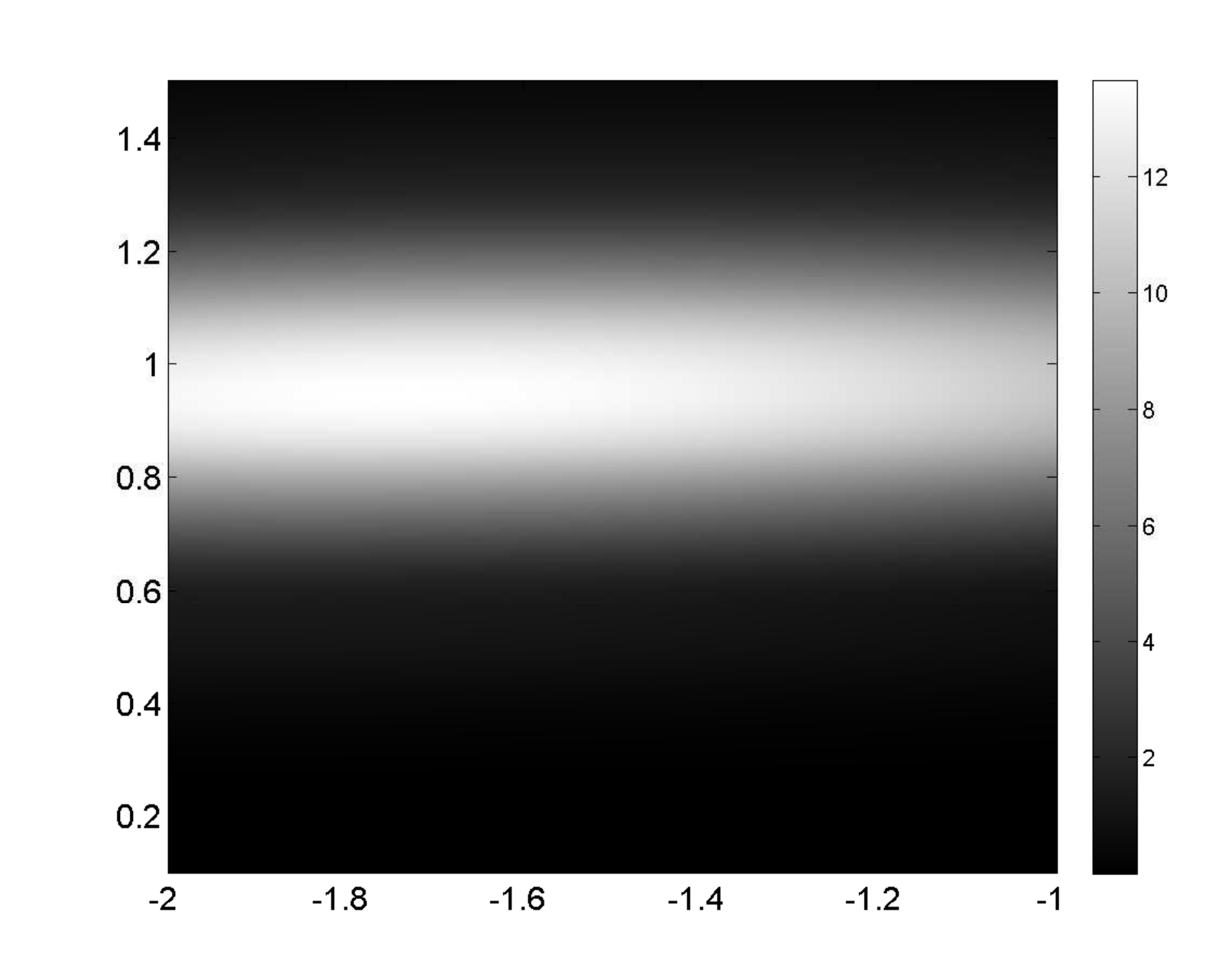} \\
\includegraphics[width=0.20\textwidth,clip=true,trim=0cm 0cm 0cm 0cm]{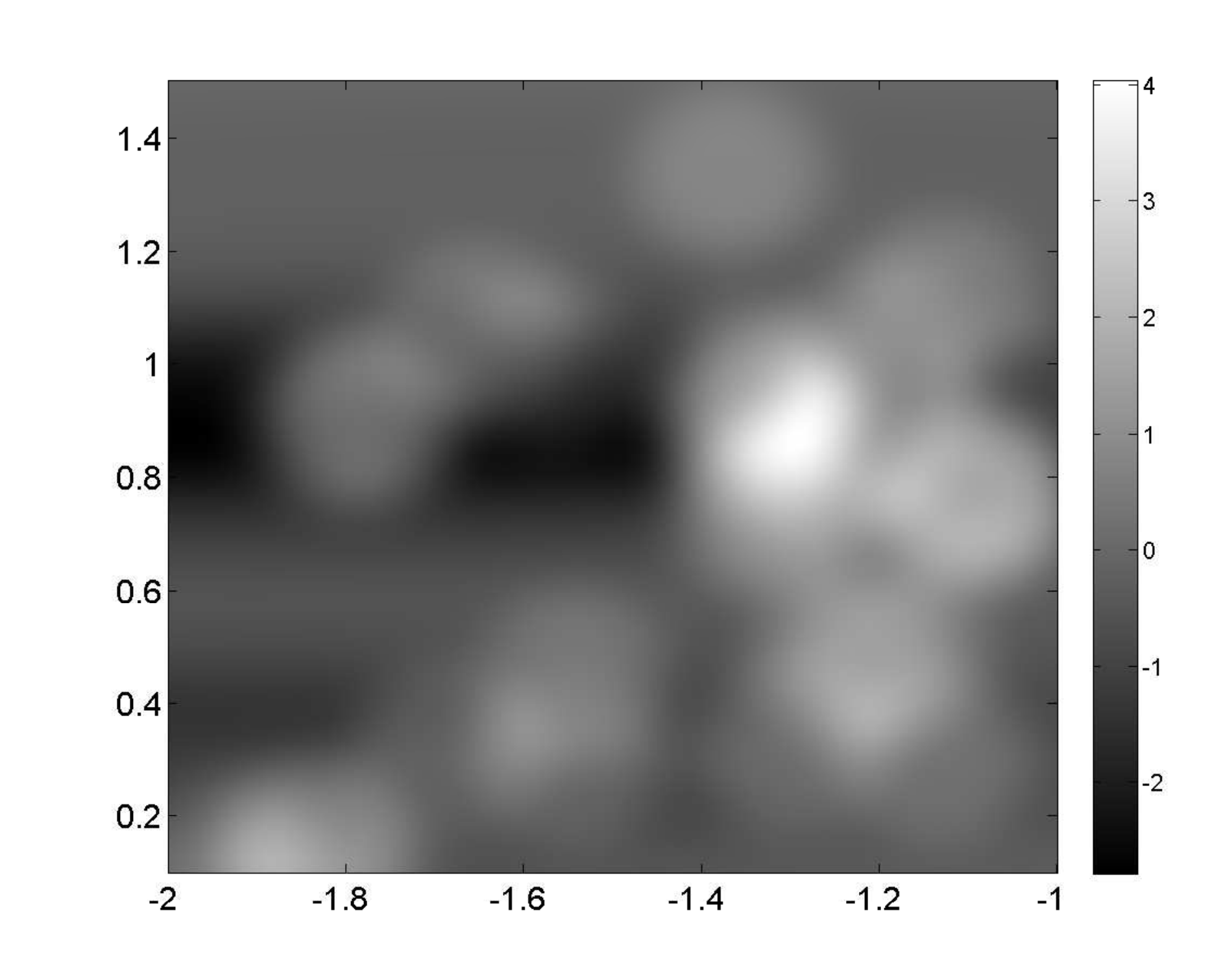} 
\includegraphics[width=0.20\textwidth,clip=true,trim=0cm 0cm 0cm 0cm]{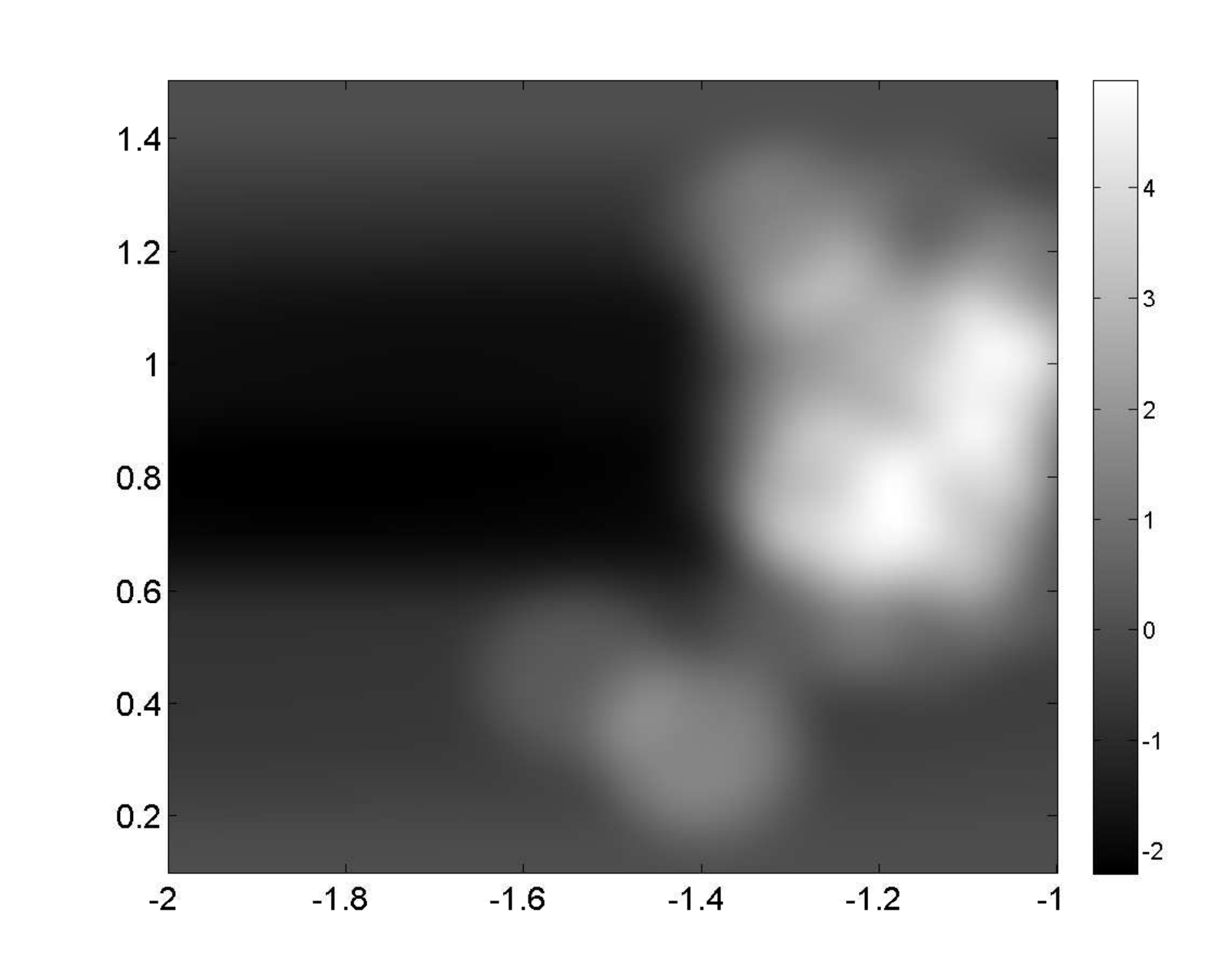} 
\includegraphics[width=0.20\textwidth,clip=true,trim=0cm 0cm 0cm 0cm]{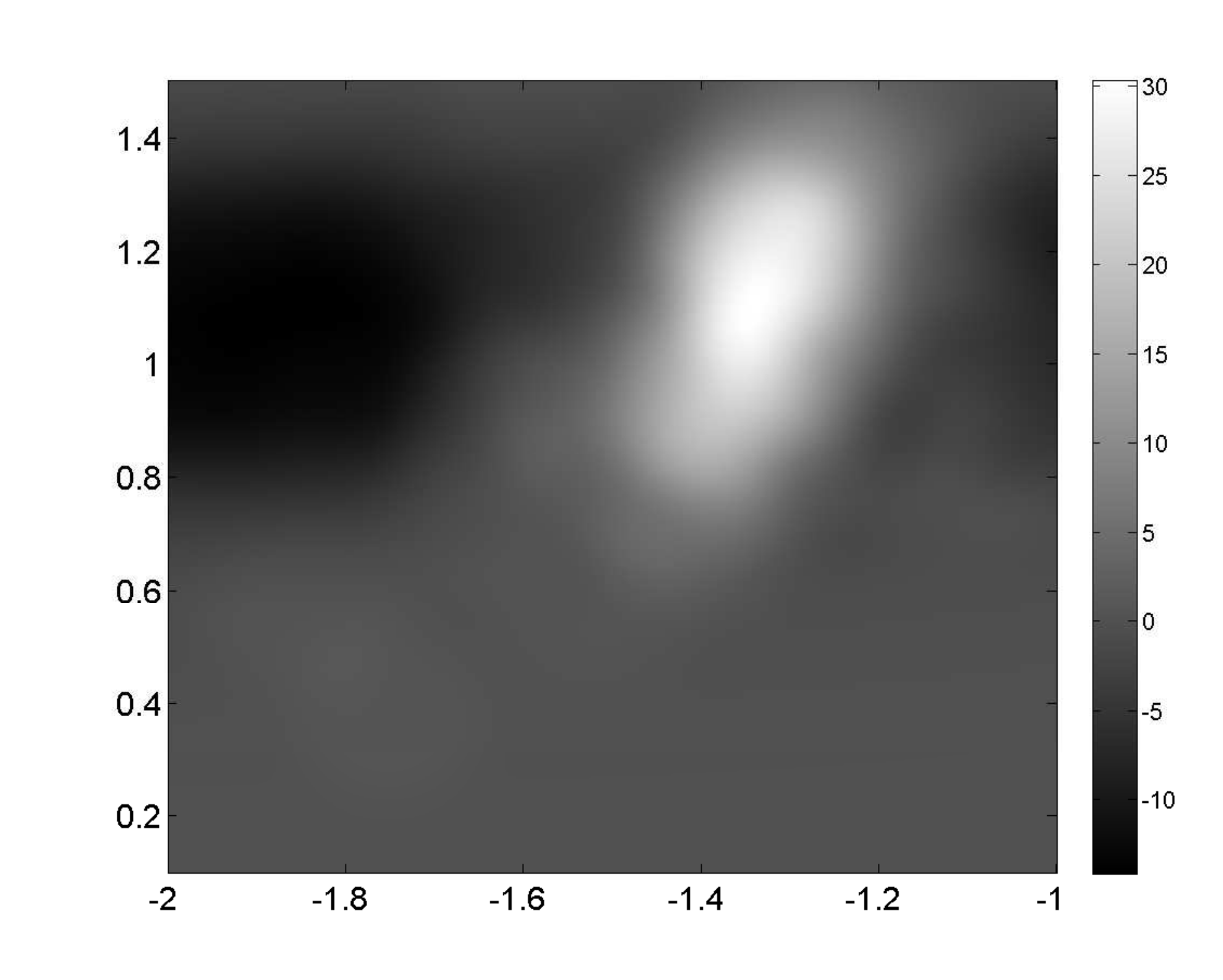} 
\includegraphics[width=0.20\textwidth,clip=true,trim=0cm 0cm 0cm 0cm]{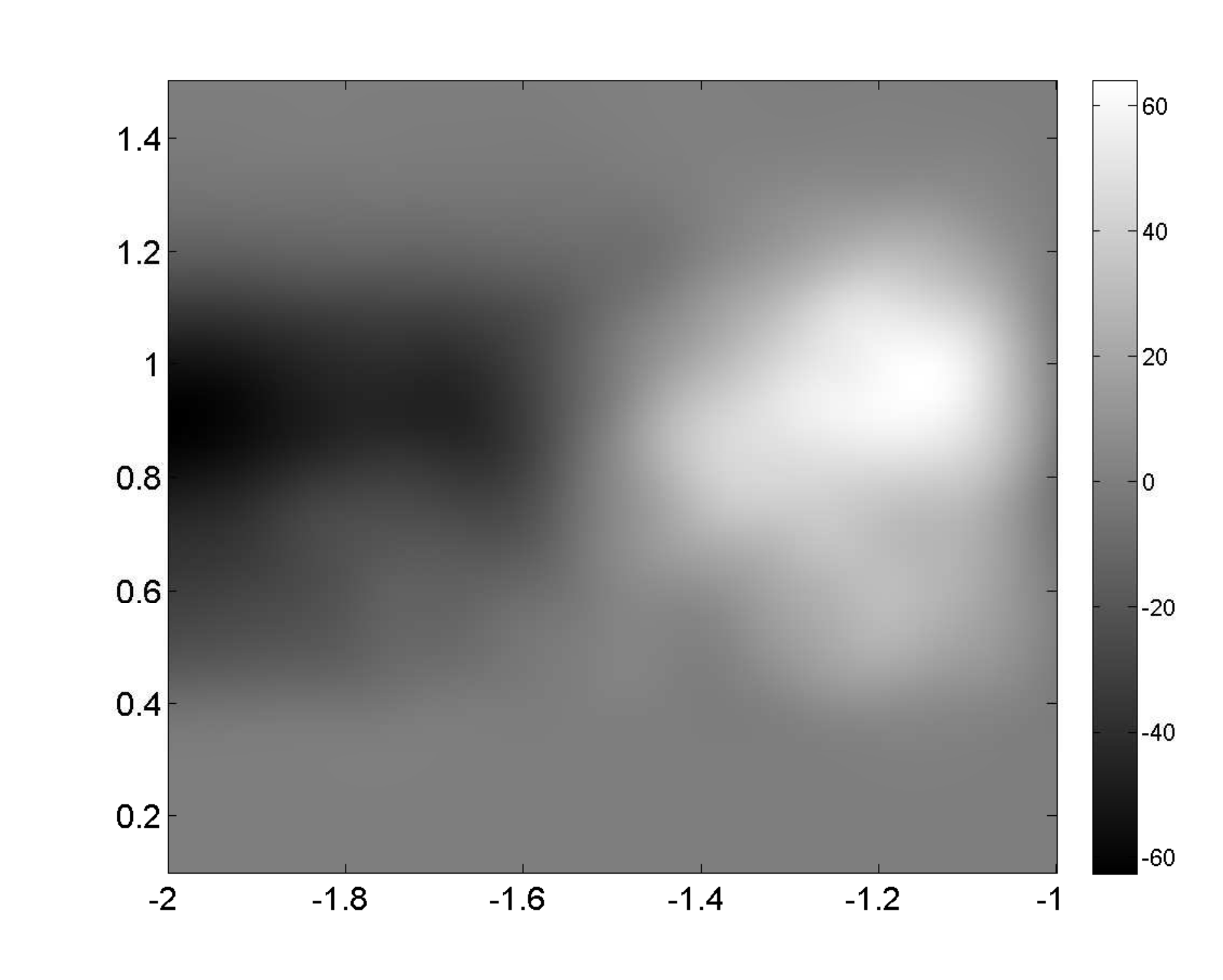} 
\includegraphics[width=0.20\textwidth,clip=true,trim=0cm 0cm 0cm 0cm]{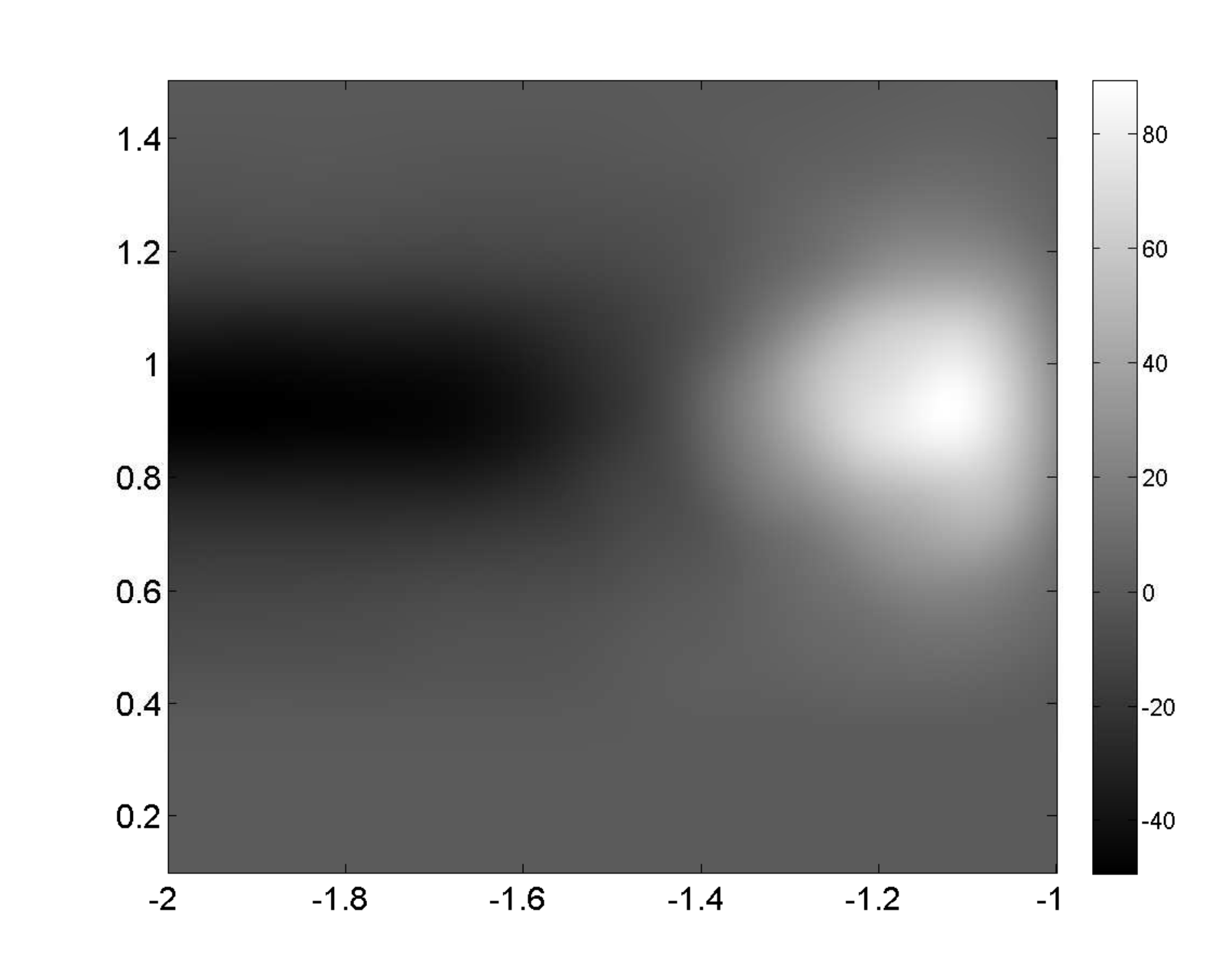} 
\includegraphics[width=0.20\textwidth,clip=true,trim=0cm 0cm 0cm 0cm]{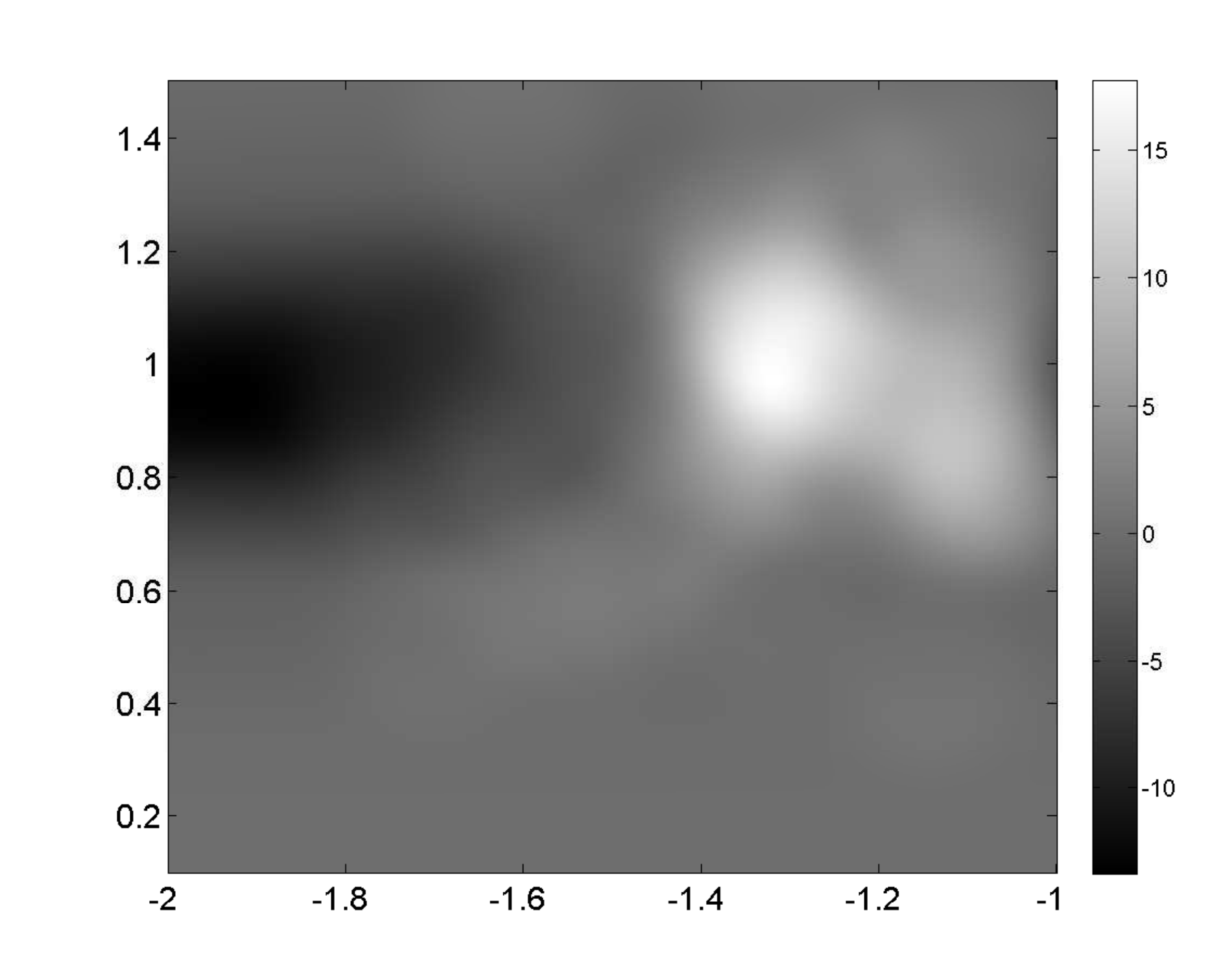} 
\caption{Same as Figure 2 for the halting model at the interior 1:2 orbital resonance with the dust sublimation radius.}
\end{figure}
\end{landscape}

\end{document}